\documentclass[11pt,oneside,a4paper,openright,english]{book}
\usepackage{hyperref}
\usepackage{amsmath, amssymb, braket, amsthm, array}
\providecommand{\tabularnewline}{\\}
\usepackage[square, colon]{natbib}
\usepackage{graphicx, subcaption,color}
\usepackage[acronym,nomain,nonumberlist]{glossaries}
\usepackage[font = small, labelfont = bf]{caption}
\usepackage{epstopdf}
\usepackage{booktabs}
\usepackage{mathptmx}

\usepackage[T1]{fontenc}
\usepackage[latin9]{inputenc}
\usepackage[a4paper]{geometry}
\usepackage{xcolor}
\usepackage{float}
\usepackage{setspace}
\usepackage{esint}
\usepackage{fancyhdr}    

\def\@chapter[#1]#2{\ifnum \c@secnumdepth >\m@ne
                       \if@mainmatter
                         \refstepcounter{chapter}%
                         \typeout{\@chapapp\space\thechapter.}%
                         \addcontentsline{toc}{chapter}%
                                   {\protect\numberline{\thechapter}#1}%
                       \else
                         \addcontentsline{toc}{chapter}{#1}%
                       \fi
                    \else
                      \addcontentsline{toc}{chapter}{#1}%
                    \fi
                    \chaptermark{#1}%
                    \addtocontents{lof}{\protect\addvspace{10\p@}}%
                    \addtocontents{lot}{\protect\addvspace{10\p@}}%
                    \if@twocolumn
                      \@topnewpage[\@makechapterhead{#2}]%
                    \else
                      \@makechapterhead{#2}%
                      \@afterheading
                    \fi}

\begin{document}
	
	\frontmatter
	\begin{titlepage}
		\begin{flushright}
			\Huge{\textbf{An investigation on the nonclassical and quantum phase
properties of a family of engineered quantum states}}\\
			\vspace{.8 in}
			\large{A Thesis submitted by}\\
			\Large{\textbf{Priya Malpani}}\\
			\vspace{.8 in}
			\large{in partial fulfillment of the requirements for the award of the degree of}\\
			\Large{\textbf{Doctor of Philosophy}}\\
			\vspace{3.0 in}
			\includegraphics[width=30mm]{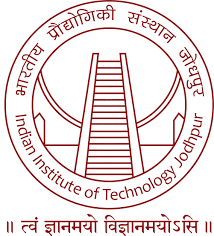}\\
			\Large{\textbf{Indian Institute of Technology Jodhpur}}\\
			\Large{\textbf{Department of Physics}}\\
			\Large{\textit{September, 2020}}\\
		\end{flushright}
	\end{titlepage}
\chapter*{Certificate}
\vspace{-33pt}
\par This is to certify that the thesis titled \textit{An investigation on the nonclassical
and quantum phase properties of a family of engineered quantum states}, submitted by \textit{Priya Malpani (P14EN001)} to the Indian Institute of Technology Jodhpur for the award of the degree of \textit{Doctor of Philosophy}, is a bonafide record of the research work done by her under my supervision. To the best of my knowledge, the contents of this report, in full or in parts, have not been submitted to any other Institute or University for the award of any degree or diploma.
\vspace{22pt}
\begin{center}
\flushright 
\normalfont\sffamily\itshape{Dr. V. Narayanan} \\
Ph.D.Thesis Supervisor
\end{center}
\begin{center}
\chapter*{Acronyms}
\end{center}

\vspace{-33pt}
~
\begin{center}
	~%
	\begin{tabular}{ll}
		BS & Binomial state\tabularnewline
		DFS & Displaced Fock state\tabularnewline
		ECS & Even coherent state\tabularnewline
		HOA & Higher-order antibunching\tabularnewline
		HOSPS & Higher-order sub-Poissionian photon statistics\tabularnewline
		HOS & Higher-order squeezing\tabularnewline
		KS & Kerr state\tabularnewline
		LE & Linear entropy\tabularnewline
		PADFS & Photon added displaced Fock state\tabularnewline
		PSDFS & Photon subtracted displaced Fock state\tabularnewline
		PABS & Photon added binomial state\tabularnewline
		PAECS & Photon added even coherent state\tabularnewline
		PAKS & Photon added Kerr state\tabularnewline
		PASDFS & Photon added then subtracted displaced Fock state\tabularnewline
		VFBS & Vacuum filtered binomial state \tabularnewline
		VFECS & Vacuum filtered even coherent state\tabularnewline
		VFKS & Vacuum filtered Kerr state\tabularnewline
	\end{tabular}
	\par\end{center}

\chapter*{Declaration}
\vspace{-33pt}
\par I hereby declare that the work presented in this thesis entitled \textit{An investigation on the nonclassical
and quantum phase properties of a family of engineered quantum states} submitted to the Indian Institute of Technology Jodhpur in partial fulfillment of the requirements for the award of the degree of Doctor of Philosophy, is a bonafide record of the research work carried out under the supervision of Dr. V. Narayanan. The contents of this thesis in full or in parts, have not been submitted to, and will not be submitted by me to, any other Institute or University in India or abroad for the award of any degree or diploma.
\vspace{22pt}
\begin{center}
\flushright 
\normalfont\sffamily\itshape{Priya Malpani} \\
P14EN001
\end{center}
\chapter*{Acknowledgment}
\addcontentsline{toc}{chapter}{Acknowledgment}
	
\small
It is my great pleasure to express my gratitude to the person (Prof.
Anirban Pathak) without whom my Ph.D. journey was impossible. He has
not only guided me in my thesis but also inspired my life with his
valuable suggestions and magical words \textquotedblleft Tension mat
lo..sab ho jayega..\textquotedblright ., I was in the dark with no
hope then Pathak Sir was candle of hope. Thank you for returning to
me faith in myself. Nothing better than these words can express my
feelings \textquotedblleft koi mujhko yun mila hai..jaise banjaare
ko ghar..\textquotedblright .

Special Thanks to my supervisor Dr. V. Narayanan. Thank you sir for
all the support and being one of the best human being in research
field. I am grateful to Dr. Subhashish Banerjee for his helpful discussion
and motivation. I am thankful to ample number of people involved in
my research activity. I am deeply indebtied to Dr. Kishore Thapliyal
(Google Scholar- apko jitna bhi thanku bolu bahut kam hi hoga!! Thankyou
Sir for ignoring my silly mistakes, your continuous support, motivation
and everlasting inspiration) and Dr. Nasir Alam (Thank you Sir for
all valuable help), who had always supported me and I feel blessed
having seniors like you both. I acknowledge my colleagues, Sanjoy
Chatterjee (Dada, the best lab-mate), Javid Naikoo (Best Advisor \dots without whom the journey of  Ph.D. was next to impossible), Khushboo Dixit,
Vandana Dahiya (Cutest friend-Jatni\dots yaara teri yari ko maine
to khuda mana.. thanku for all the support, without your help survival
in IITJ was not possible), Shilpa Pandey (Best roommate), Mitali Sisodia
(thank you for listening my endless drama), Swarn Rajpoot (naam hi
kafi hai), Satish Sangwan (Tau- the best buddy), Vishwadeepak Kumar
(Rockstar), Ashwin Saxena, and other research scholars, teaching and
non-teaching staff in IITJ department, who had become part of my life
during this period of time. Now, I would like to thank everyone who
had played a prominent role in my life. Special thanks to my parents
for their unconditional love and support, without your support it
was not possible for me to reach this level. I would like to thank
my grandpa (Baba). I will thank my bua (Sheelam Bua) for her continuous
moral support. A big thanks to my lovely Sadi, Vaibhav (no words can
express my thank you to you bro. Thank you for handling all my moods
with extreme patience), Meghai (Friendster), Doll (my soul sister),
Labbu (blessed to have you in my life bro), Yashivangi (strong pillers
of my life) and a big thanks to someone who has completely changed
my way of thinking. I would also like to acknowledge the financial
helps I received in different phases of my research work from IITJ,
JIIT and MHRD without which this work would not have been possible. 
\small
	
	\vspace{22pt}
	\begin{center}
		\flushright 
		\normalfont\sffamily\itshape{Priya Malpani.}
	\end{center}
\chapter*{Abstract}
The main focus of this thesis is to study the nonclassical
and phase properties of a family of engineered quantum states, most
of which show various nonclassical features. The beauty of these states
is that these states can be used to establish quantum supremacy. Earlier,
a considerable amount of works has been reported on various types
of quantum states and their nonclassical properties. Here, complementing
the earlier works, the effect of non-Gaussianity inducing operators
on the nonclassical and phase properties of displaced Fock states
have been studied. This thesis includes 6 chapters. In Chapter \ref{cha:Introduction1},
motivation behind performing the present work is stated explicitly,
also the basic concepts of quantum optics are discussed with a specific
attention on the witnesses and measures of nonclassicality. In Chapter
\ref{cha:PADFS-PSDFS}, nonclassical properties of photon added and
subtracted displaced Fock states have been studied using various witnesses
of lower- and higher-order nonclassicality which are introduced in
Chapter 1. In Chapter \ref{cha:phase}, we have continued our investigation
on photon added and subtracted displaced Fock states (and their limiting
cases). In this chapter, quantum phase properties of these states
are investigated from a number of perspectives, and it is shown that
the quantum phase properties are dependent on the quantum state engineering
operations performed. In Chapter \ref{cha:PASDFS}, we have continued
our investigation on the impact of non-Gaussianity inducing operators
on the nonclassical and phase properties of the displaced Fock states.
In Chapter \ref{cha:QSE-1}, we have performed a comparison between to process
that are used in quantum state engineering to induce nonclassical
features. Finally, this thesis is concluded in Chapter \ref{cha:Conclusions-and-Scope},
where we have summarized the findings of this thesis and have also
described scope of the future works.
\addcontentsline{toc}{chapter}{Abstract}

	\vspace{500cm}
	\begin{center}
	\bfseries{\itshape{}}
	\end{center}
	{\bfseries{\itshape{Dedicated to my parents for their unconditional love and support.}}}
	\tableofcontents
	\listoffigures
	\listoftables

	

%
%
%
%

\newpage

\mainmatter

\pagenumbering{arabic}\setcounter{page}{1}

\chapter{Introduction\textsc{\label{cha:Introduction1}}}


\section{Introduction}

As the title of the thesis suggests, in this thesis, we aim to study
the nonclassical and phase properties of a family of engineered quantum
states. Before, we introduce such states and properties, it would
be apt to lucidly introduce the notion of nonclassical and engineered
quantum states. By nonclassical state we refer to a quantum state
having no classical analogue. Such states are characterized by the
negative values of Glauber-Sudarshan $P$-function or
$P$-function more singular than Dirac delta function \cite{sudarshan1963equivalence,glauber1963coherent}
and witnessed by various operational criteria (to be described in
Section \ref{subsec:Witnesses-of-nonclassicality}). To visualize
the relevance of nonclassical states we may first note that quantum
supremacy refers to the ability of performing a task using quantum
resources in such a manner that either the task itself cannot be performed
using classical resources or the speed/efficiency achieved using quantum
resources cannot be achieved in the classical world \cite{grover1997quantum}.
A recent experiment performed by Google aimed at estabilishing quantum
supremacy has drawn much of public attention \cite{courtland2017google}.
The relevance of the present study lies in the fact that to establish
quantum supremacy or to perform a fundamental test of quantum mechanics,
we would require a state having some features that would not be present
in any classical state. As we have already mentioned, such a state
having no classical analogue is referred to as the nonclassical state.
Frequently used examples of nonclassical states include squeezed,
antibunched, entangled, steered, and Bell nonlocal states. The relevance
of the states having nonclassical features has already been established
in the various domains of physics. For example, we may mention, teleportation
of coherent states \cite{furusawa1998unconditional}, continuous variable
quantum cryptography \cite{hillery2000quantum}, quantum radar \cite{lanzagorta2011quantum},
and many more. Further, we may note that the art of generating and
manipulating quantum states as per need is referred to as the ``quantum
state engineering'' \cite{dakna1998quantum,sperling2014quantum,vogel1993quantum,miranowicz2004dissipation,dell2006multiphoton}.
Particularly interesting examples of such engineered quantum states
are Fock state, photon added/subtracted coherent state \cite{agarwal1991nonclassical},
displaced Fock state (DFS) which is also referred to as generalized
coherent state and displaced number state \cite{satyanarayana1985generalized,wunsche1991displaced,ziesel2013experimental,zavatta2004quantum,de1990properties,malpani2019lower},
photon added DFS (PADFS) \cite{malpani2019lower}, and photon subtracted
DFS (PSDFS) \cite{malpani2019lower}. In what follows, we will state
the relevance of such engineered quantum states in the implementation
of different tasks exploiting their nonclassical and phase properties.  The relatively new area of research on quantum state engineering
has drawn much attention of the scientific community because of its
success in experimentally producing various quantum states \cite{zavatta2004quantum,torres2003preparation,rauschenbeutel2000step,gao2010experimental,lu2007experimental}
having nonclassical properties and applications in realizing quantum
information processing tasks, like quantum key distribution \cite{bennett1984quantum}
and quantum teleportation \cite{brassard1998teleportation,chen2015bidirectional}.
Engineered quantum states, such as cat states, Fock state and superposition
of Fock states, are known to play a crucial role in performing fundamental
tests of quantum mechanics and in establishing quantum supremacy in
the context of quantum computation and communication (\cite{kues2017chip}
and references therein).

As mentioned in the previous paragraph, with the advent of quantum
state engineering \cite{vogel1993quantum,sperling2014quantum,miranowicz2004dissipation,marchiolli2004engineering}
and quantum information processing (\cite{pathak2013elements} and
references therein), the study of nonclassical properties of engineered
quantum states have become a very important field. This is so because
only the presence of nonclassical features in a quantum state can
provide quantum supremacy \cite{grover1997quantum}. In the recent past, various techniques for quantum state engineering
have been developed \cite{vogel1993quantum,sperling2014quantum,miranowicz2004dissipation,agarwal1991nonclassical,lee2010quantum,marchiolli2004engineering}.
If we restrict ourselves to optics, these techniques are primarily
based on the clever use of beam splitters, detectors, and measurements
with post selection, etc. Such techniques are useful in creating holes
in the photon number distribution \cite{escher2004controlled} and
in generating finite dimensional quantum states \cite{miranowicz2004dissipation},
both of which are nonclassical \cite{pathak2018classical}. The above
said techniques are also useful in realizing non-Gaussianity inducing
operations, like photon addition and subtraction \cite{zavatta2004quantum,podoshvedov2014extraction}.
Motivated by the above, in this thesis, we aim to study the nonclassical
properties of a set of engineered quantum states such as photon added,
photon subtracted, and photon added then subtracted displaced Fock
states which can be produced by using the above mentioned techniques.
In the present thesis, we also wish to investigate the phase properties
of the above mentioned engineered quantum states for the reasons explained
below.

The impossibility of writing a Hermitian operator for quantum phase
is a longstanding problem (see \cite{perinova1998phase,carruthers1968phase,lynch1987phase}
for review). Early efforts of Dirac \cite{dirac1927quantum} to introduce
a Hermitian quantum phase operator were not successful, but led to
many interesting proposals \cite{susskind1964quantum,pegg1989phase,barnett1986phase}.
Specifically, Susskind-Glogower \cite{susskind1964quantum}, Pegg-Barnett
\cite{pegg1988unitary,pegg1989phase,barnett1990quantum}, and Barnett-Pegg
\cite{barnett1986phase} formalisms played very important role in
the studies of phase properties and the phase fluctuation \cite{imry1971relevance}.
Thereafter, phase properties of various quantum states have been reported
using these formalisms \cite{sanders1986bc,gerry1987phase,yao1987phase,carruthers1968phase,vaccaro1989phase,pathak2000phase,alam2016quantum,alam2017quantum,verma2009reduction}.
Other approaches have also been used for the study of the phase properties.
For example, quantum phase distribution is defined using phase states
\cite{agarwal1992classical}, while Wigner \cite{garraway1992quantum}
and $Q$ \cite{leonhardt1993phase,leonhardt1995canonical} phase distributions
are obtained by integrating over radial parameter of the corresponding
quasidistribution function. In experiments, the phase measurement
is performed by averaging the field amplitudes of the $Q$ function
\cite{noh1991measurement,noh1992operational}; Pegg-Barnett and Wigner
phase distributions are also reported with the help of reconstructed
density matrix \cite{smithey1993complete}. Further, quantum phase
distribution under the effect of the environment was also studied
in the past leading to phase diffusion \cite{banerjee2007phase,banerjee2007phaseQND,abdel2010anabiosis,banerjee2018open}.
A measure of phase fluctuation named phase dispersion using quantum
phase distribution has also been proposed in the past \cite{perinova1998phase,banerjee2007phase}.
Recently, quantum phase fluctuation \cite{zheng1992fluctuation} and
Pancharatnam phase \cite{mendas1993pancharatnam} have been studied
for DFS. The quantum phase fluctuation in parametric down-conversion
\cite{gantsog1991quantum} and its revival \cite{gantsog1992collapses}
are also reported. Experiments on phase super-resolution without using
entanglement \cite{resch2007time} and role of photon subtraction
in concentration of phase information \cite{usuga2010noise} are also
performed. Optimal phase estimation \cite{sanders1995optimal} using
different quantum states \cite{higgins2007entanglement} (including
NOON and other entangled states and unentangled single-photon states)
has long been the focus of quantum metrology \cite{giovannetti2006quantum,giovannetti2011advances}. Nonclassicality measure based on the shortening of the regular distribution
defined on phase difference interval broadbands due to nonclassicality
is also proposed in the recent past \cite{perina2019quasidistribution,thapliyal2019quasidistribution}.
In brief, quantum phase properties are of intense interest of the
community since long (see \cite{pathak2002quantum,perinova1998phase}
and references therein), and the interest in it has been further enlightened
in the recent past as many new applications of quantum phase distribution
and quantum phase fluctuation have been realized.

To be specific, this work is also motivated by the fact that recently
several applications of nonclassical states and quantum phase properties
have been reported. Specifically, squeezed states have played an important
role in {{} the studies related to phase diffusion} \cite{banerjee2007phase,banerjee2007phaseQND},
the detection of gravitational waves in LIGO experiments \cite{abbasi2013thermal,abbott2016gw151226,abbott2016observation}.
The rising demand for a single photon source can be fulfilled by an
antibunched light source \cite{yuan2002electrically}. {The study
of quantum correlations is important both from the perspective of
pure and mixed states} \cite{chakrabarty2010study,dhar2013controllable,banerjee2010dynamics,banerjee2010entanglement}.
Entangled states are found to be useful in both secure \cite{ekert1991quantum}
and insecure \cite{bennett1992communication,bennett1993teleporting}
quantum communication schemes. Stronger quantum correlation present
in the steerable states are used to ensure the security against all
the side-channel attacks on devices used in one-side (i.e., either
preparation or detector side) for quantum cryptography \cite{branciard2012one}.
Quantum supremacy in computation is established due to quantum algorithms
for unsorted database search \cite{grover1997quantum}, factorization
and discrete logarithm problems \cite{shor1999polynomial}, and machine
learning \cite{biamonte2017quantum} using essentially nonclassical
states. We may further stress on the recently reported applications
of quantum phase distribution and quantum phase fluctuation by noting
that these have applications in quantum random number generation \cite{xu2012ultrafast,raffaelli2018soi},
cryptanalysis of squeezed state based continuous variable quantum
cryptography \cite{horak2004role}, generation of solitons in a Bose-Einstein
condensate \cite{denschlag2000generating}, storage and retrieval
of information from Rydberg atom \cite{ahn2000information}, in phase
encoding quantum cryptography \cite{gisin2002quantum}, phase imaging
of cells and tissues for biomedical application \cite{park2018quantitative};
as well as have importance in determining the value of transition
temperature for superconductors \cite{emery1995importance}.

Now to achieve the above advantages of the nonclassical states, we
need to produce these states via the schemes of quantum state engineering.
For the same, there are some distinct theoretical tools, like quantum
scissoring \cite{miranowicz2001quantum}, hole-burning \cite{escher2004controlled,gerry2002hole,malpani2019filter}
or filtering out a particular Fock state from the photon number distribution
\cite{Meher2018}, applying non-Gaussianity inducing operations \cite{agarwal2013quantum}.
However, these distinct mechanisms are experimentally realized primarily
by appropriately using beam splitters, mirrors, and single photon
detectors or single photon counting module. Without going into finer
details of the optical realization of quantum state engineering tools,
we may note that these tools can be used to generate various nonclassical
states, e.g., DFS \cite{de1990properties}, PADFS \cite{malpani2019lower},
PSDFS \cite{malpani2019lower}, photon added squeezed coherent state
\cite{thapliyal2017comparison}, photon subtracted squeezed coherent
state \cite{thapliyal2017comparison}, number state filtered coherent
state \cite{Meher2018}. Some of these states, like photon added coherent
state, have already been realized experimentally \cite{zavatta2004quantum}.

Many of the above mentioned engineered quantum states have already
been studied in detail. Primarily, three types of investigations have
been performed on the engineered quantum states- (i) study of various
nonclassical features of these states (and their variation with the
state parameters) as reflected through different witnesses of nonclassicality.
Initially, such studies were restricted to the lower-order nonclassical
features. In the recent past, various higher-order nonclassical features
have been predicted theoretically \cite{alam2018higher,alam2018higher1,pathak2006control,pathak2010recent,verma2008higher,thapliyal2017comparison}
and confirmed experimentally (\cite{hamar2014non,perina2017higher}
and references therein) in quantum states generated in nonlinear optical
processes. (ii) Phase properties of the nonclassical states have been
studied \cite{el2000phase} by computing quantum
phase fluctuations, phase dispersion, phase distribution functions,
etc., under various formalisms, like Susskind and Glogower \cite{susskind1964quantum},
Pegg-Barnett \cite{pegg1989phase} and Barnett-Pegg \cite{barnett1986phase}
formalisms. (iii) Various applications of the engineered quantum states
have been designed. Some of them have already been mentioned.

Motivated by the above observations, in this thesis, we would like
to perform an investigation on nonclassical and phase properties of
a particularly interesting set of engineered quantum states which
would have the flavor of the first two facets of the studies mentioned
above. Applications of the engineered quantum states will also be
discussed briefly, but will not be investigated in detail. To begin with we would like to briefly describe the physical and
mathematical concepts used in this thesis, and that will be the focus
of the rest of this chapter.

The rest of this chapter is organized as follows. In Section \ref{sec:Quantum-theory-of-radiation},
we will briefly discuss quantum theory of radiation and introduce
annihilation, creation, and number operators as well as Fock and coherent
states. In Section \ref{sec:Quantum-states-of-4} this will be followed
by an introduction to a set of quantum states which will be studied
in this thesis. After this, the notion of nonclassicality will be
introduced mathematically in Section \ref{sec:The-notion-of nonclassicality},
and a set of operational criteria for observing nonclassical properties
will be introduced in Section \ref{sec:Nonclassical-states:-witnesses}.
Subsequently, the parameters used for the study of quantum phase properties
will be introduced in Section \ref{sec:Analytic-tools-forphase}.
These witnesses of nonclassicality and the parameters for the study
of phase properties will be used in the subsequent chapters, to investigate
the nonclassical and phase properties of the quantum states discussed
in Section \ref{sec:Quantum-states-of-4}. Finally, the structure
of the rest of the thesis will be provided in Section \ref{sec:Structure-of-the thesis}. 

\section{Quantum theory of radiation field\label{sec:Quantum-theory-of-radiation}}

Historically, quantum physics started with the ideas related to quanta
of radiation. To be precise, Planck's work on black body radiation
\cite{planck1901law} and Einstein's explanation of photoelectric
effect \cite{einstein1905tragheit} involved a notion of quantized
radiation field. In Planck's work, light was considered to be emitted
from  and absorbed by a black body in quanta; and
in Einstein's work, it was also considered that the radiation field
propagates from one point to another as quanta. These initial works
contributed a lot in the development of quantum mechanics, but after
the introduction of quantum mechanics, in 1920s, in the initial days,
most of the attention was given to the quantization of matter. A quantum
theory of radiation was introduced by Dirac in 1927 \cite{dirac1927quantum}.
In what follows, we will describe it briefly as this would form the
backbone of the present thesis.

\subsection{Creation and annihilation operator}

Maxwell gave a classical description of electromagnetic field. But
here the objective is to study light and its properties apart from Maxwell's equations. So, to begin with, it would be reasonable to
write Maxwell's equations in free space:

\begin{equation}
\nabla\,.\,E=0,\label{eq:Maxwell1}
\end{equation}

\begin{equation}
\nabla\,.\,B=0,\label{Maxwell2}
\end{equation}

\begin{equation}
\nabla\times E=-\frac{\partial B}{\partial t},\label{Maxwell3}
\end{equation}
and

\begin{equation}
\nabla\times B=\frac{1}{c^{2}}\frac{\partial E}{\partial t},\label{Maxwell4}
\end{equation}
where $c=\frac{1}{\sqrt{\mu_{0}\epsilon_{0}}}$ is the speed of light
without any medium (i.e. vacuum). Using the set of above equations
one can express magnetic and electric fields in the form of the solution
of wave equations, like

\begin{equation}
\nabla^{2}E-\frac{1}{c^{2}}\frac{\partial^{2}E}{\partial t^{2}}=0.\label{wave-equantion}
\end{equation}
The quantization of radiation field can be done by assuming a cavity
of length $L$ having a linearly polarized electric field whose direction
of propagation is in $z$ direction. Because
of the linearity of the wave equation (\ref{wave-equantion}), we
are allowed to write the electric field in the form of linear combination
of all the normal modes as

\begin{equation}
E_{x}\left(z,t\right)=\sum_{n}A_{n}q_{n}\left(t\right){\rm sin}\left(k_{n}z\right),\label{eq:electric}
\end{equation}
where $q_{n}$ being the amplitude of the $n$th normal mode with
$k_{n}=\frac{n\pi}{L},$$V$ is the volume of the resonator, $A_{n}=\frac{2m_{n}\nu_{n}^{2}}{\epsilon_{0}V}$
with $\nu_{n}=ck_{n}$, and $m_{n}$ is a constant (in the units of
mass). With the help of this, we try to form
an analogy of radiation with mechanical oscillator. In analogy to
Eq. (\ref{eq:electric}) we are able to write the corresponding magnetic
field equation in cavity as

\begin{equation}
B_{y}\left(z,t\right)=\sum_{n}A_{n}\left(\frac{\dot{q_{n}}}{c^{2}k_{n}}\right){\rm cos}\left(k_{n}z\right).\label{eq:magnetic}
\end{equation}
So, the total energy of the field can be written as a classical Hamiltonian

\begin{equation}
H=\frac{1}{2}\int_{V}d\tau\left(\epsilon_{0}E_{x}^{2}+\frac{1}{\mu_{0}}B_{y}^{2}\right),\label{eq:hamiltonian}
\end{equation}

\begin{equation}
H=\frac{1}{2}\sum_{n}\left(m_{n}\nu_{n}^{2}q_{n}^{2}+m_{n}\dot{q}_{n}^{2}\right)=\frac{1}{2}\sum_{n}\left(m_{n}\nu_{n}^{2}q_{n}^{2}+\frac{p_{n}^{2}}{m_{n}}\right),\label{eq:hamiltonian2}
\end{equation}
 Substituting position and momentum variables by
corresponding operators to obtain the quantum mechanical Hamiltonian,
where $p_{n}=m_{n}\dot{q_{n}}.$ The position ($q_{n}$) and momentum
($p_{n}$) operators follow the commutation relations

\[
\left[q_{n},p_{m}\right]=\iota\hbar\delta_{nm},\,\,\,\,\,\,\,\,\,\,\,\,\,\,\,\,\,\,\,\,\left[q_{n},q_{m}\right]=\left[p_{n},p_{m}\right]=0,
\]
where $\hbar$ is the reduced Planck's constant. Using these one
may define a new set of operators which can be analytically written
as

\begin{equation}
\hat{a_{n}}{\rm exp\left[-\iota\nu_{n}t\right]=\frac{1}{\sqrt{2\hbar\text{\ensuremath{m_{n}}}\nu_{n}}}}\left(\text{\ensuremath{m_{n}}}\nu_{n}q_{n}+\iota p_{n}\right)\label{annihilation}
\end{equation}
and 
\begin{equation}
\hat{a}_{n}^{\dagger}{\rm exp\left[\iota\nu_{n}t\right]=\frac{1}{\sqrt{2\hbar\text{\ensuremath{m_{n}}}\nu_{n}}}}\left(\text{\ensuremath{m_{n}}}\nu_{n}q_{n}-\iota p_{n}\right).\label{eq:creation}
\end{equation}
Thus, the Hamiltonian can be written as

\begin{equation}
H=\sum_{n}\hbar\nu_{n}\left(\hat{a_{n}}^{\dagger}\hat{a_{n}}+\frac{1}{2}\right),\label{eq:hamiltonian2-1}
\end{equation}
and the commutation relations

\[
\left[\hat{a}_{n},\hat{a}_{m}^{\dagger}\right]=\delta_{nm},\,\,\,\,\,\,\,\,\,\,\,\,\,\,\,\,\,\,\,\,\left[\hat{a}_{n},\hat{a}_{m}\right]=\left[\hat{a}_{n}^{\dagger},\hat{a}_{m}^{\dagger}\right]=0,
\]
with corresponding electric and magnetic fields,   as given by Eq. 1.1.27 in~ \cite{scully1997quantum} 

\[
E\left(\overrightarrow{r},t\right)=\sum_{k}\hat{\epsilon_{k}}\xi_{k}\hat{a}_{k}{\rm exp}\left[-\iota\nu_{k}t+\iota k.\overrightarrow{r}\right]+{\rm H.c}.
\]
and

\[
B\left(\overrightarrow{r},t\right)=\sum_{k}\frac{k\times\hat{\epsilon_{k}}}{\nu_{k}}\xi_{k}\hat{a}_{k}{\rm exp}\left[-\iota\nu_{k}t+\iota k.\overrightarrow{r}\right]+{\rm H.c}.,
\]
where $\xi_{k}=\left(\frac{\hbar\nu_{k}}{\epsilon_{0}V}\right)^{1/2}$
is a constant, and $\hat{\epsilon_{k}}$ is a unit polarization vector
with the wave vector $k$.

The above analysis shows that a single-mode field is identical to
harmonic oscillator. So, in the domain of quantum optics, harmonic
oscillator system plays an important role.

Notice that the quantum treatment of electromagnetic radiation hinges
on annihilation $\hat{a}$ (depletes photon) and creation $\hat{a}^{\dagger}$
(creates photon) operators. The annihilation operator $\hat{a}$ depletes
one quantum of energy and thus lowers down the system from harmonic
oscillator level $\left|n\right\rangle $ to $\left|n-1\right\rangle $,
given by

\begin{equation}
\hat{a}\left|n\right\rangle =\sqrt{n}\left|n-1\right\rangle .\label{eq:annihilationop}
\end{equation}
Here, $\left|n\right\rangle $ is called Fock or number state. Further,
an application of annihilation operator on vacuum leads to $0$, i.e.,
$\hat{a}\left|0\right\rangle =0$. The creation operator $\hat{a}^{\dagger}$
creates one quantum of energy by raising the state from $\left|n\right\rangle $
to $\left|n+1\right\rangle $. Therefore, the creation operator in
the number state can be represented as 
\begin{equation}
\hat{a}^{\dagger}\left|n\right\rangle =\sqrt{n+1}\left|n+1\right\rangle .\label{eq:creationop}
\end{equation}
If creation operator is applied to vacuum it creates a photon so these
operators enables one to write a Fock state ($\left|n\right\rangle $)
in terms of the vacuum state as 
\[
\left|n\right\rangle =\frac{\left(\hat{a}^{\dagger}\right)^{n}}{\sqrt{n!}}\left|0\right\rangle .
\]
In the above, we have seen that annihilation and creation operators
are the important field operators and are required for the quantum
description of radiation. These operators can induce nonclassicality
and non-Gaussianity when applied on classical states \cite{zavatta2004quantum,agarwal2013quantum}.
In the present thesis, we study the role of these non-Gaussianity
inducing operations in controlling the nonclassicality of the quantum
states which are often already nonclassical. For
instance, enhacement in squeezing in a nonclassical state does not
ensure advantage with respect to use as a single photon source and
vice-versa. In the following subsection, we will introduce a set
of other operators which can be expressed in terms of annihilation
and creation operators and which play a crucial role in our understanding
of the quantum states of radiation field.

\subsection{Some more quantum operators of relevance }

\label{SMQO}

So far we have introduced some non-unitary operations (operations
$\hat{O}$ which are not norm preserving and do not satisfy $\hat{O}^{\dagger}=\hat{O}^{-1}$,
where $\hat{O}^{\dagger}$ and $\hat{O}^{-1}$ are the Hermitian conjugate
and inverse of $\hat{O}$, respectively), namely photon addition and
subtraction. We now aim to introduce some more unitary operations
important in the domain of quantum state engineering in general, and
in this thesis in particular. To begin with let us describe displacement
operator.

\subsubsection{Displacement operator}

Displacement operator is a unitary operator. The mathematical from
of displacement operator is given as

\begin{equation}
\hat{D}(\alpha)=\exp\left(\alpha\hat{a}^{\dagger}-\alpha^{\star}\hat{a}\right).\label{eq:Displacement}
\end{equation}
This operator can be used as a tool to generate coherent state from
vacuum. Specifically, a coherent state $\left|\alpha\right\rangle $
is defined as $\left|\alpha\right\rangle =\hat{D}(\alpha)\left|0\right\rangle .$

\subsubsection{Squeezing operator}

\label{SqOpt}

\begin{figure}
\begin{centering}
\begin{tabular}{c}
\includegraphics[width=110mm]{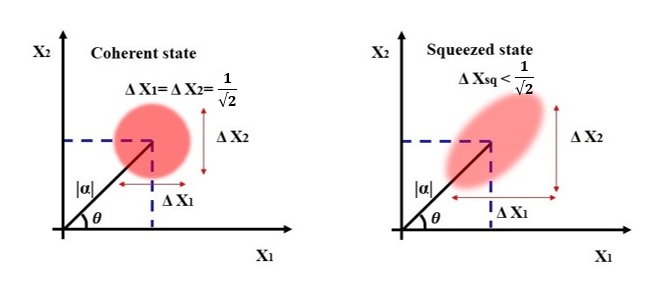}\tabularnewline
\tabularnewline
\end{tabular}
\par\end{centering}
\caption{\label{fig:coh-sq} Phase picture for coherent state and squeezed
state. }
\end{figure}

The squeezing operator for a single mode of electromagnetic field
is 
\begin{equation}
\hat{S}(z)=\exp\left(\frac{1}{2}\left(z^{\star}\hat{a}^{2}-z\hat{a}^{\dagger2}\right)\right).\label{eq:squeezed}
\end{equation}
The description of light is given by two quadratures namely phase
$(X_{1})$ and amplitude $(X_{2})$ in the domain of quantum optics,
mathematically defined as  
\[
\hat{X_{\theta}}=\frac{1}{\sqrt{2}}\left( i \hat{a}^\dagger \exp\left[\iota\theta \right] - \iota \hat{a}\exp\left[-\iota\theta\right]\right),
\]
The corresponding uncertainty of these two quadratures is observed
by relation $\Delta X_{1}\Delta X_{2} \geq \hbar\,/2$,
where $\Delta X_{1}$ $\left(\Delta X_{2}\right)$ is variance in
the measured values of quadrature  $\hat{X_{1}}=\hat{X_{1}}\left(\theta=0\right)$
$\left(\hat{X_{2}}=\hat{X_{2}}\left(\theta=\frac{\pi}{2}\right)\right)$.
Specifically, $\Delta X_{i}=\sqrt{\left\langle X_{i}^{2}\right\rangle -\left\langle X_{i}\right\rangle ^{2}},$where
$\left\langle \hat{A}\right\rangle =\left\langle \psi\left|\hat{A}\right|\psi\right\rangle $
is the expectation value of the operator $A$ with respect to the
quantum state $\left|\psi\right\rangle $. Coherent state has an equal
uncertainty in both quadratures so they form a circle in the phase
picture (shown in Fig. \ref{fig:coh-sq}).  Least value
of the variance for suitable $\theta$ is studied as principle squeezing.
With the advent of nonlinear optics, a very special branch of optics,
the uncertainty  in one of the quadratures can be
reduced at the cost of increment in other quadrature's uncertainty,
which means that the circle can be squeezed.

\subsection{Eigen states of the field operators}

Here, we will discuss eigen states of some of the operators we have
introduced. The eigenvalue equation can be defined as $\hat{A}\lambda=a\lambda$
with eigen operator $\hat{A}$, eigenvalue $a$, and eigen function
$\lambda$. For example, Schrodinger equation $H\psi_{i}=E_{i}\psi_{i}$
has Hamiltonian $H$ as eigen operator with eigen functions $\psi_{i}$
and eigenvalues as allowed energy levels.

\subsubsection{Fock state: Eigen state of the number operator}

In case of quantum optics or quantized light picture, photon number
state is known as number state. The single-mode photon number states
are known Fock states, and its ground state is defined as vacuum state. As the set of number states are a full set of orthonormal basis so
any quantum state can be written in terms of these basis. The method
of representing a quantum state as superposition of number states
is known as number state representation. Now using Eq. (\ref{eq:annihilationop})
and (\ref{eq:creationop}), we can introduce an operator 
\[
\hat{N}=\hat{a}^{\dagger}\hat{a}.
\]
which would satisfy the following eigen value equation 
\[
\hat{N}\left|n\right\rangle =n\left|n\right\rangle .
\]
Clearly Fock states are the eigen states of the number operators and
in consistency with what has already been told, a Fock state $\left|n\right\rangle $
represent a $n$ photon state.

\subsubsection{Coherent state: Eigen state of the annihilation operator \label{subsec:Coherent-state}}

Coherent state \cite{fox2006quantum} is considered as a state of
the quantized electromagnetic field which shows classical behaviour
(specifically, beavior closest to classical states). According to
Erwin Schrodinger it is a minimum uncertainty state, having same uncertainity
in position and momentum \cite{schrodinger1926stetige}. According
to Glauber, any of three mathematical definitions described below
can define coherent state:

(i) Eigen vectors of annihilation operator $\hat{a}\left|\alpha\right\rangle =\alpha\left|\alpha\right\rangle $,
$\alpha$ being a complex number.

(ii) Quantum states having minimum uncertainty $\Delta X_{1}=\Delta X_{2}=1/\sqrt{2}$,
with $X_{2}$ and $X_{1}$ as momentum and position operators.

(iii) States realized by the application of the displacement operator
$D(\alpha)$ on the vacuum state. Thus, is also known as displaced
vacuum state and can be expressed as 
\[
\left|\alpha\right\rangle =D\left(\alpha\right)\left|0\right\rangle .
\]
In Fock basis, it is expressed as infinite superposition of Fock state
as 
\begin{equation}
\left|\alpha\right\rangle =\exp\left[-\frac{\mid\alpha\mid^{2}}{2}\right]\sum\limits _{n=0}^{\infty}\frac{\alpha^{n}}{\sqrt{n!}}|n\rangle,\label{eq:CS}
\end{equation}
where $\alpha$ is a complex number. Experimentally established state
very close to this coherent state was possible only after the successful
development of laser. Finally, one can easily see that $\hat{a}\left|\alpha\right\rangle =\alpha\left|\alpha\right\rangle $
implies $\left\langle \alpha\right|\hat{a}^{\dagger}=\left\langle \alpha\right|\alpha^{\star}$
and consequently $\left\langle \alpha\right|\hat{a}^{\dagger}\hat{a}\left|\alpha\right\rangle =\left\langle \alpha\right|\hat{N}\left|\alpha\right\rangle =N=\left|\alpha\right|^{2}$
or average photon number in a coherent state is $\left|\alpha\right|^{2}$.

\section{Quantum states of our interest\label{sec:Quantum-states-of-4}}

In this section, we provide basic mathematical details of the set
of engineered quantum states studied in the present thesis.

\subsection{Displaced Fock state}

Displaced Fock state \cite{satyanarayana1985generalized} are formed
by applying displacement operator on Fock state and thus a DFS is
defined as 
\[
\left|\phi\right\rangle =D\left(\alpha\right)\left|n\right\rangle .
\]
Analytically it is given as 
\begin{equation}
|\phi(n,\alpha)\rangle=\frac{1}{\sqrt{n!}}\sum_{p=0}^{n}{n \choose p}(-\alpha^{\star})^{(n-p)}\exp\left(-\frac{\mid\alpha\mid^{2}}{2}\right)\sum_{m=0}^{\infty}\frac{\alpha^{m}}{m!}\sqrt{(m+p)!}|m+p\rangle.\label{eq:GCS}
\end{equation}
Various nonclassical properties of DFS are reported in literature
\cite{de1990properties,el2000phase,lvovsky2002synthesis,mendas1993pancharatnam}.

\subsection{Photon added and photon subtracted displaced Fock state\label{subsec:Photon-added-and sub}}

Using DFS, we can define a $u$ photon added DFS (i.e., a PADFS) as
\begin{eqnarray}
|\psi_{+}(u,n,\alpha)\rangle & = & N_{+}\hat{a}^{\dagger u}|\phi(n,\alpha)\rangle=\frac{N_{+}}{\sqrt{n!}}\sum_{p=0}^{n}{n \choose p}(-\alpha^{\star})^{(n-p)}\exp\left(-\frac{\mid\alpha\mid^{2}}{2}\right)\sum_{m=0}^{\infty}\frac{\alpha^{m}}{m!}\nonumber \\
 & \times & \sqrt{(m+p+u)!}|m+p+u\rangle.\label{eq:PADFS}
\end{eqnarray}
Similarly, a $v$ photon subtracted DFS (i.e., a PSDFS) can be expressed
as 
\begin{eqnarray}
|\psi_{-}(v,n,\alpha)\rangle & = & N_{-}\hat{a}^{v}|\phi(n,\alpha)\rangle=\frac{N_{-}}{\sqrt{n!}}\sum_{p=0}^{n}{n \choose p}(-\alpha^{\star})^{(n-p)}\exp\left(-\frac{\mid\alpha\mid^{2}}{2}\right)\sum_{m=0}^{\infty}\frac{\alpha^{m}}{m!}\nonumber \\
 & \times & \dfrac{(m+p)!}{\sqrt{(m+p-v)!}}|m+p-v\rangle,\label{eq:PSDFS}
\end{eqnarray}
where $m$ and $p$ are the real integers. Here, {\small{}
\begin{eqnarray}
N_{+}=\left[\frac{1}{n!}\sum_{p,p'=0}^{n}{n \choose p}{n \choose p'}(-\alpha^{\star})^{(n-p)}(-\alpha)^{(n-p')}\exp\left[-\mid\alpha\mid^{2}\right]\sum_{m=0}^{\infty}\frac{\alpha^{m}(\alpha^{\star})^{m+p-p'}(m+p+u)!}{m!(m+p-p')!}\right]^{-\frac{1}{2}}\label{eq:norP}
\end{eqnarray}
and 
\begin{equation}
N_{-}=\left[\frac{1}{n!}\sum_{p,p'=0}^{n}{n \choose p}{n \choose p'}\left(\alpha^{\star}\right)^{(n-p)}\left(-\alpha\right){}^{(n-p')}\exp\left[-\mid\alpha\mid^{2}\right]\sum_{m=0}^{\infty}\frac{\alpha^{m}\left(\alpha^{\star}\right){}^{m+p-p'}(m+p)!}{m!(m+p-p')!(m+p-v)!}\right]^{-\frac{1}{2}}.\label{eq:norS}
\end{equation}
}are the normalization constants, and subscripts $+$ and $-$ correspond
to photon addition and subtraction. Thus, $|\psi_{+}(u,n,\alpha)\rangle$
and $|\psi_{-}(v,n,\alpha)\rangle$ represent $u$ photon added DFS
and $v$ photon subtracted DFS, respectively, for the DFS which has
been produced by displacing the Fock state $|n\rangle$ by a displacement
operator $D(\alpha)$ characterized by the complex parameter $\alpha.$
Clearly, the addition and the subtraction of photons on the DFS can
be mathematically viewed as application of the creation and annihilation
operators from the left on the Eq. (\ref{eq:GCS}). Here, it may be
noted that different well-known states can be obtained as special
cases of these two states. For example, using the notation introduced
above to define PADFS and PSDFS, we can describe a coherent state
$|\alpha\rangle$ as $|\alpha\rangle=|\psi_{+}(0,\alpha,0)\rangle=|\psi_{-}(0,\alpha,0)\rangle$,
naturally, coherent state can be viewed as a special case of both
PADFS and PSDFS. Similarly, we can describe a single photon added
coherent state as $|\psi\rangle_{+1}=|\psi_{+}(1,\alpha,0)\rangle$,
a Fock state as $|n\rangle=|\psi_{+}(0,0,n)\rangle=|\psi_{-}(0,0,n)\rangle$
and a DFS as $|\psi\rangle_{{\rm DFS}}=|\psi_{+}(0,\alpha,n)\rangle=|\psi_{-}(0,\alpha,n)\rangle.$

\subsection{Photon added then subtracted displaced Fock state \label{subsec:PASDFS}}

A PASDFS can be obtained by sequentially applying appropriate number
of annihilation (photon subtraction) and creation (photon addition)
operators on a DFS. Analytical expression for PASDFS (specifically,
a $k$ photon added and then $q$ photon subtracted DFS) in Fock basis
can be shown to be

\begin{eqnarray}
|\psi(k,q,n,\alpha)\rangle & = & N\hat{a}^{q}\hat{a}^{\dagger k}|\psi(n,\alpha)\rangle=\frac{N}{\sqrt{n!}}\sum_{p=0}^{n}{n \choose p}(-\alpha^{\star})^{(n-p)}\exp\left(-\frac{\mid\alpha\mid^{2}}{2}\right)\nonumber \\
 & \times & \sum_{m=0}^{\infty}\frac{\alpha^{m}\left(m+p+k\right)!}{m!\sqrt{(m+p+k-q)!}}|m+p+k-q\rangle,\label{eq:PADFS-1}
\end{eqnarray}
where {\small{}
\[
N=\left[\frac{1}{n!}\sum_{p,p'=0}^{n}{n \choose p}{n \choose p'}(-\alpha^{\star})^{(n-p)}(-\alpha)^{(n-p')}\exp\left[-\mid\alpha\mid^{2}\right]\right]^{-\frac{1}{2}}
\]
}is the normalization factor.

\subsection{Even coherent state and states generated by holeburning on it}

Even coherent state can be defined as the superposition of two coherent
states having opposite phase ($|\phi(\alpha)\rangle\propto|\alpha\rangle+|-\alpha\rangle$).
The analytical expression for ECS in number basis can be written as
\begin{equation}
\begin{array}{lcl}
|\phi(\alpha)\rangle & = & \frac{\,\exp\left[-\frac{\mid\alpha\mid^{2}}{2}\right]}{\sqrt{2\left(1+\exp\left[-2\mid\alpha\mid^{2}\right]\right)}}\sum\limits _{n=0}^{\infty}\frac{\alpha^{n}}{\sqrt{n!}}\left(1+\left(-1\right)^{n}\right)|n\rangle.\end{array}\label{eq:ECS}
\end{equation}
The parameter $\alpha=|\alpha|\exp(i\theta)$, in Eq. (\ref{eq:ECS}),
is complex in general and $\theta$ is phase angle in the complex
plane. Various schemes to generate ECS are reported in \cite{brune1992manipulation,ourjoumtsev2007generation,gerry1993non}.
The nonclassical properties (witnessed through the antibunching and
squeezing criteria, $Q$ function, Wigner function, and photon number
distribution, etc.) of ECS have been studied in the recent past \cite{gerry1993non}.

\subsubsection{Vacuum filtered even coherent state}

As mentioned above, experimentally, an ECS or
a cat state can be generated in various ways, and the same can be
further engineered to produce a hole at vacuum in its photon number
distribution. Specifically, filtration of vacuum will burn a hole
at $n=0$ and produce VFECS, which can be described in Fock basis
as 
\begin{equation}
\begin{array}{lcl}
|\phi_{1}(\alpha)\rangle & = & N_{{\rm VFECS}}\sum\limits _{n=0,\,n\neq0}^{\infty}\frac{\alpha^{n}}{\sqrt{n!}}\left(1+\left(-1\right)^{n}\right)|n\rangle,\end{array}\label{eq:VFECS}
\end{equation}
where 
\begin{equation}
\begin{array}{lcl}
N_{{\rm VFECS}} & = & \{4{\rm cosh}\left(\mid\alpha\mid^{2}\right)-1\}^{-1/2}\end{array}
\end{equation}
is the normalization constant. For simplicity, we may expand Eq. (\ref{eq:VFECS})
as a superposition of a standard ECS and a vacuum state as follows
\begin{equation}
\begin{array}{lcl}
|\phi_{1}(\alpha)\rangle & = & N_{{\rm VFECS}}\left(\sum\limits _{n=0}^{\infty}\frac{\alpha^{n}}{\sqrt{n!}}\left(1+\left(-1\right)^{n}\right)|n\rangle-2|0\rangle\right).\end{array}\label{eq:VFECS-EXPANDED}
\end{equation}
In what follows, Eq. (\ref{eq:VFECS-EXPANDED}) will be used to explore
various nonclassical features that may exist in VFECS.

\subsubsection{Photon added even coherent state}

One can define a single photon added ECS as 
\begin{equation}
|\phi_{2}(\alpha)\rangle=N_{{\rm PAECS}}\hat{a}^{\dagger}|\phi(\alpha)\rangle=N_{{\rm PAECS}}\sum\limits _{n=0}^{\infty}\frac{\alpha^{n}}{\sqrt{n!}}\left(1+\left(-1\right)^{n}\right)\sqrt{n+1}|n+1\rangle,\label{eq:PAECS}
\end{equation}
where 
\begin{equation}
\begin{array}{lcl}
N_{{\rm PAECS}} & = & \{{\rm cosh}\left(\mid\alpha\mid^{2}\right)+\mid\alpha\mid^{2}{\rm sinh}\left(\mid\alpha\mid^{2}\right)\}^{-1/2}/2\end{array}
\end{equation}
is the normalization constant for PAECS.

\subsection{Binomial state and the states generated by holeburning on it}

Binomial state is a finite superposition of Fock states having binomial
photon number distribution. It is quite similar to the coherent state
which is the linear combination of Fock states having the Poissonian
photon number distribution \cite{stoler1985binomial}. BS can be defined
as 
\begin{equation}
\begin{array}{lcl}
|p,M\rangle & = & \sum\limits _{n=0}^{M}\left[\frac{M!}{n!(M-n)!}p^{n}\left(1-p\right)^{M-n}\right]^{1/2}|n\rangle.\end{array}\label{eq:BS}
\end{equation}
The binomial coefficient describes the presence of $n$ photons with
probability $p$ in $M$ number of ways. Recently, the study of nonclassical
properties of BS, specifically, antibunching, squeezing, HOSPS \cite{verma2008higher,verma2010generalized,bazrafkan2004tomography},
etc., have been studied very extensively. However, no effort has yet
been made to study the nonclassical properties of VFBS and PABS.

\subsubsection{Vacuum filtered binomial state}

The vacuum filtration of BS can be obtained by simply eliminating
vacuum state from the BS as 
\begin{equation}
\begin{array}{lcl}
|p,M\rangle_{1} & = & N_{{\rm VFBS}}\sum\limits _{n=0}^{M}\left[\frac{M!}{n!(M-n)!}p^{n}\left(1-p\right)^{M-n}\right]^{1/2}|n\rangle-N_{VFBS}\left[\left(1-p\right)^{M}\right]^{1/2}|0\rangle,\end{array}\label{eq:VFBS}
\end{equation}
where 
\begin{equation}
\begin{array}{lcl}
N_{{\rm VFBS}} & = & \{1-\left(1-p\right)^{M}\}^{-1/2}\end{array}
\end{equation}
is the normalization constant for the VFBS.

\subsubsection{Photon added binomial state}

A hole at $n=0$ at a BS can also be introduced by the addition of
a single photon on the BS. A few steps of computation yield the desired
expression for PABS as 
\begin{equation}
\begin{array}{lcl}
|p,M\rangle_{2} & =N_{{\rm PABS}} & \sum\limits _{n=0}^{M}\left[\frac{M!(n+1)!}{\left(n!\right)^{2}(M-n)!}p^{n}\left(1-p\right)^{M-n}\right]^{1/2}|n+1\rangle,\end{array}\label{eq:PABS}
\end{equation}
where 
\begin{equation}
\begin{array}{lcl}
N_{{\rm PABS}} & = & \left(1+Mp\right)^{-1/2}\end{array}
\end{equation}
is the normalization constant for single photon added BS.

\subsection{Kerr state and the states generated by holeburning on it}

A KS can be obtained when electromagnetic field
in a coherent state interacts with nonlinear medium with Kerr type
nonlinearity \cite{gerry1994statistical}. This interaction generates
phase shifts which depend on the intensity. The Hamiltonian involved
in this process is given as 
\begin{equation}
H=\hbar\omega\hat{a}^{\dagger}\hat{a}+\hbar\chi\left(\hat{a}^{\dagger}\right)^{2}\left(\hat{a}\right)^{2},
\end{equation}
where $\chi$ depends on the third-order susceptibility of Kerr medium.   Explicit contribution of $H$ is ${\rm exp}\left[-i\chi n(n-1)\right]$. Thus, the compact analytic form
of the KS in the Fock basis can be given as 
\begin{equation}
\begin{array}{lcl}
|\psi_{K}\left(n\right)\rangle & = & \sum\limits _{n=0}^{\infty}\frac{\alpha^{n}}{\sqrt{n!}}\exp\left(-\frac{\mid\alpha\mid^{2}}{2}\right)\exp\left(-\iota\chi n\left(n-1\right)\right)|n\rangle.\end{array}\label{eq:KS}
\end{equation}

\subsubsection{Vacuum filtered Kerr state}

Similarly, a VFKS, which can be obtained using the same quantum state
engineering process that leads to VFECS and VFBS, is given by 
\begin{equation}
\begin{array}{lcl}
|\psi_{K}\left(n\right)\rangle_{1} & = & N_{{\rm VFKS}}\left[\sum\limits _{n=0}^{\infty}\frac{\alpha^{n}}{\sqrt{n!}}\exp\left(-\iota\chi n\left(n-1\right)\right)|n\rangle-|0\rangle\right]\end{array},\label{eq:VFKS-expanded}
\end{equation}
where 
\begin{equation}
\begin{array}{lcl}
N_{{\rm VFKS}} & = & {{\left(\exp\left[\mid\alpha\mid^{2}\right]-1\right)}^{-1/2}}\end{array}\label{eq:NVFKS-expanded}
\end{equation}
is the normalization constant for the VFKS.

\subsubsection{Photon added Kerr state}

An addition of a photon to Kerr state would yield PAKS which can be
expanded in Fock basis as 
\begin{equation}
\begin{array}{lcl}
|\psi_{K}\left(n\right)\rangle_{2} & =N_{{\rm PAKS}} & \sum\limits _{n=0}^{\infty}\frac{\alpha^{n}}{\sqrt{n!}}\exp\left(-\iota\chi n\left(n-1\right)\right)\sqrt{\left(n+1\right)}|n+1\rangle,\end{array}\label{eq:PAKS}
\end{equation}
where 
\begin{equation}
\begin{array}{lcl}
N_{{\rm PAKS}} & = & {\left(\exp\left[\mid\alpha\mid^{2}\right]\left(1+\mid\alpha\mid^{2}\right)\right)^{-1/2}}\end{array}\label{eq:NPAKS}
\end{equation}
is the normalization constant for the PAKS.

\section{The notion of nonclassical states \label{sec:The-notion-of nonclassicality}}

Quantum states which do not have any classical analogue have been
referred to as nonclassical states \cite{agarwal2013quantum}. In
other words, states having their $P$-distribution more singular than
delta function or having negative values are referred to as nonclassical
states \cite{dodonov2003classicality}. This idea was possible only
when Glauber and Sudarshan published papers in 1963 \cite{sudarshan1963equivalence,glauber1963photon,glauber1963coherent}.
Sudarshan found a mathematical form to represent any state in the
coherent basis, mathematically given as 
\[
\rho=\int P\left(\alpha\right)\left|\alpha\right\rangle \left\langle \alpha\right|d^{2}\alpha,
\]
where $P\,(\alpha)$ is known as Glauber-Sudarshan $P$-function,
which follows normalization condition as $\int dP(\alpha)=1$, but it may have negative values. Thus, it is defined as quasidistribution
function or quasiprobability distribution. When
$P\,(\alpha)$ attains a positive probability density function, immediately
it indicates that the state is classical. This leads to the definition
of nonclassicality. If an arbitrary quantum state is failed to represent
as mixture of coherent states, that is known as nonclassical state.
To establish quantum supremacy, these nonclassical states play very
essential role, for instance in theses states are useful in establishing
quantum supremacy of quantum information processing, quantum communication,
etc. Although $P$-function is not reconstructable for any arbitrary
state yet it has been of major interest as it provides an important
signature of nonclassicality. The negativity (positivity or non-negativity)
of the $P$-function essentially provides the nonclassical (classical)
behavior of the state under consideration. The experimental difficulty
associated with the easurement of $P$-function in its reconstruction
led to various feasible substitutes as nonclassicality witnesses.
Here, we list some of those nonclassicalty witnesses. These witnesses
can be viewed as operational criterion of nonclassicality.

\section{Nonclassical states: witnesses and measures \label{sec:Nonclassical-states:-witnesses}}

Using nonclassical states, the essence of quantum theory of light
can be understood. There are various tools for characterization of
nonclassical states. In this section, some tools are described which
are used to characterize such states. If historically seen, the first
such approach was aimed to check the deviation from Poissonian photon
statistics, the second is to evaluate the volume of the negative part
of the quasiprobability distribution in the phase space, etc. An infinite
set of moments based criteria is available in literature which is
used as witness of nonclassicality equivalent to $P$-function \cite{shchukin2005nonclassical}.
Any subset of this infinite set may detect nonclassicality or fail
to do so. Example of these witnesses are lower- and higher-order antibunching,
sub-Poissonian photon statistics, squeezing as well as Mandel $Q_{M}$
parameter, Klyshko's, Vogel's, and Agarwal-Tara's criteria, $Q$ function,
etc. To quantify the amount of nonclassicality, a number of measures
have been proposed, like linear entropy, Wigner volume, concurrence
and many more. A small description of these criteria is given here.

\subsection{Witnesses of nonclassicality\label{subsec:Witnesses-of-nonclassicality}}

\subsubsection{Lower- and higher-order antibunching}

\begin{figure}
\centering{}%
\begin{tabular}{c}
\includegraphics[width=100mm]{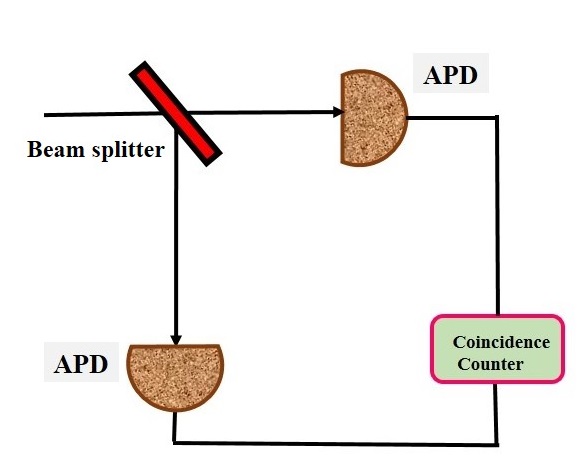}\tabularnewline
\tabularnewline
\end{tabular}\caption{Hanbury Brown and Twiss setup. Here, APD is avalanche
photo diode.}
\label{fig:HBT} 
\end{figure}

In this section, we study lower- and higher-order antibunching. To
do so, we use the following criterion of $(l-1)$th order antibunching
(\cite{pathak2006control} and references therein)  in
terms of nonclassicality witness ($d(l-1)$) as
\begin{equation}
d(l-1)=\langle\hat{a}^{\dagger l}\hat{a}^{l}\rangle-\langle\hat{a}^{\dagger}\hat{a}\rangle^{l}<0.\label{eq:HOA-1}
\end{equation}
This nonclassical feature characterizes suitability of the quantum
state to be used as single photon source as the negative values of
$d(l-1)$ parameter show that the probability of photons coming bunched
is less compared to that of coming independently. The signature of
lower-order antibunching can be obtained as a special case of Eq.
(\ref{eq:HOA-1}) for $l=2$, and that for $l\geq3$, the negative
values of $d(l-1)$ correspond to higher-order antibunching of $(l-1)$th
order. Figure \ref{fig:HBT} illustrates the scheme for studying antibunching
experimentally    (corresponds to $l=2$). For higher
values of $l$, we require more beamsplitters and APDs. On these cascaded
beamplitters signal is mixed with vacuum and measured higher-order
correlation ~\cite{avenhaus2010accessing}.

\subsubsection{Lower- and higher-order sub-Poissionian photon statistics}

The lower-order counterparts of antibunching and sub-Poissonian photon
statistics are closely associated as the presence of latter ensures
the possibility of observing former (see \cite{thapliyal2014higher,thapliyal2017comparison}
for a detailed discussion). However, these two nonclassical features
were shown to be independent phenomena in the past (\cite{thapliyal2014higher,thapliyal2017comparison}
and references therein). Higher-order counterpart of sub-Poissonian
photon statistics can be introduced as

\begin{equation}
\begin{array}{lcccc}
\mathcal{D}_{h}(l-1) & = & \sum\limits _{e=0}^{l}\sum\limits _{f=1}^{e}S_{2}(e,\,f)\,^{l}C_{e}\,\left(-1\right)^{e}d(f-1)\langle N\rangle^{l-e} & < & 0,\end{array}\label{eq:hosps22-1}
\end{equation}
where $S_{2}(e,\,f)$ is the Stirling number of second kind, {and
$\,^{l}C_{e}$ is the usual binomial coefficient}.

\subsubsection{Higher-order squeezing}

As mentioned beforehand, the squeezing of quadrature is defined in
terms of variance in the measured values of the quadrature (say, position
or momentum) below the corresponding value for the coherent state,
i.e., minimum uncertainty state. The higher-order counterpart of squeezing
is studied in two ways, namely Hong-Mandel and Hillery-type squeezing
\cite{hong1985higher,hong1985generation,hillery1987amplitude}. Specifically,
the idea of the higher-order squeezing originated from the pioneering
work of Hong and Mandel \cite{hong1985higher,hong1985generation},
who generalized the lower-order squeezing using the higher-order moments
of field quadrature. According to the Hong-Mandel criterion, the $l$th
order squeezing can be observed if the $l$th moment (for even values
of $l>2$) of a field quadrature operator is less than the corresponding
coherent state value. The condition of Hong-Mandel type higher-order
squeezing is given as follows \cite{hong1985higher,hong1985generation}

\begin{equation}
S(l)=\frac{\langle(\Delta X)^{l}\rangle-\left(\frac{1}{2}\right)_{\frac{l}{2}}}{\left(\frac{1}{2}\right)_{\frac{l}{2}}}<0,\label{eq:Hong-Def}
\end{equation}
Here, $S_{(l)}$ is higher-order squeezing, $\Delta X$ is the quadrature
( $\Delta X_{1}$) as defined in Section \ref{SMQO}. Further, $(x)_{l}$
is conventional Pochhammer symbol. The inequality in Eq. (\ref{eq:Hong-Def})
can also be rewritten as 
\begin{equation}
\begin{array}{lcl}
\langle(\Delta X)^{l}\rangle & < & \left(\frac{1}{2}\right)_{\frac{l}{2}}=\frac{1}{2^{\frac{l}{2}}}(l-1)!!\end{array}\label{eq:Hong-def2-2}
\end{equation}
with 
\begin{equation}
\langle\left(\text{\ensuremath{\Delta}}\text{X}\right)^{l}\rangle=\sum\limits _{r=0}^{l}\sum\limits _{i=0}^{\frac{r}{2}}\sum\limits _{k=0}^{r-2i}\left(-1\right)^{r}\frac{1}{2^{\frac{l}{2}}}\left(2i-1\right)!^{2i}C_{k}{}^{l}C_{r}{}^{r}C_{2i}\langle\hat{a}^{\dagger}+\hat{a}\rangle^{l-r}\langle\hat{a}^{\dagger k}\hat{a}^{r-2i-k}\rangle.\label{eq:cond2.1-1}
\end{equation}

\subsubsection{Klyshko's criterion}

This criterion is relatively simpler as to calculate this witness
of nonclssicality, only three consecutive probability terms are required
rather than all the terms. Negative values of $B(m)$ are symbol of
nonclassicality present in the state. Klyshko introduced this criterion
\cite{klyshko1996observable} to investigate the nonclassical property
using only three successive photon-number {{} probabilities}. In
terms of the photon-number probability $p_{m}=\langle m|\rho|m\rangle$
of the state with density matrix $\rho$, the Klyshko's criterion
in the form of an inequality can be written as 
\begin{equation}
B(m)=(m+2)p_{m}p_{m+2}-(m+1)\left(p_{m+1}\right)^{2}<0.\label{eq:Klyshko-1}
\end{equation}

\subsubsection{Vogel's criterion}

 The moments-based nonclassicality criterion of the
previous subsection was later extended to Vogel's nonclassicality
criterion \cite{shchukin2005nonclassical} in terms of matrix of moments
as

\begin{equation}
v =\left[\begin{array}{ccc}
1 & \langle\hat{a}\rangle & \langle\hat{a}^{\dagger}\rangle\\
\langle\hat{a}^{\dagger}\rangle & \langle\hat{a}^{\dagger}\hat{a}\rangle & \langle\hat{a}^{\dagger2}\rangle\\
\langle\hat{a}\rangle & \langle\hat{a}^{2}\rangle & \langle\hat{a}^{\dagger}\hat{a}\rangle
\end{array}\right].\label{eq:vogel}
\end{equation}
The negative value of the determinant $dv $ of matrix $v$
in Eq. (\ref{eq:vogel}) is signature of nonclassicality.

\subsubsection{Agarwal-Tara's criterion}

There were certain quantum states having negative $P$-function yet
showing no squeezing and sub-Poissonian behavior, to witness the nonclassicality
residing in those particular types of states Agarwal and Tara \cite{agarwal1992nonclassical}
introduced this criterion which is again a moments based criterion.
This can be written in a matrix form and expressed as 
\begin{equation}
A_{3}=\dfrac{\det m^{(3)}}{\det\mu^{(3)}-\det m^{(3)}}<0,\label{eq:Agarwal-1}
\end{equation}
where 
\[
m^{(3)}=\begin{bmatrix}1 & m_{1} & m_{2}\\
m_{1} & m_{2} & m_{3}\\
m_{2} & m_{3} & m_{4}
\end{bmatrix}
\]
and 
\[
\mu^{(3)}=\begin{bmatrix}1 & \mu_{1} & \mu_{2}\\
\mu_{1} & \mu_{2} & \mu_{3}\\
\mu_{2} & \mu_{3} & \mu_{4}
\end{bmatrix}.
\]
The matrix elements are defined as $m_{i}=\langle\hat{a}^{\dagger i}\hat{a}^{i}\rangle$
and $\mu_{j}=(\langle\hat{a}^{\dagger}\hat{a}\rangle)^{j}=(m_{1})^{j}$.

\subsubsection{Mandel $Q_{M}$ parameter}

The Mandel $Q_{M}$ parameter \cite{mandel1979sub} illustrates the
nonclassicality through photon number distribution in a quantum state.
The Mandel $Q_{M}$ parameter is defined as 
\begin{equation}
Q_{M}=\frac{\langle(\hat{a}^{\dagger}\hat{a})^{2}\rangle-\langle\hat{a}^{\dagger}\hat{a}\rangle^{2}-\langle\hat{a}^{\dagger}\hat{a}\rangle}{\langle\hat{a}^{\dagger}\hat{a}\rangle}.\label{eq:MandelQ}
\end{equation}
The negative values of $Q_{M}$ parameter essentially indicate the
negativity for $P$-function and so it gives a witness for nonclassicality.
For the Poissonian statistics it becomes 0, while for the sub-Poissonian
(super-Poissonian) photon statistics it has negative (positive) values.

\subsection{Other quasiprobability distributions}

Inability to give a phase space description of quantum mechanics is
exploited in terms of quasiprobability distributions (\cite{thapliyal2015quasiprobability}
and references therein). Later, it was found that they are useful
as witnesses of nonclassicality. These real and normalized quasiprobability
distributions allow to calculate the expectation value of an operator
as any classical probability distribution. One such quasiprobability
distributions is $Q$ function \cite{husimi1940some}, and zeros of
this function are signature of nonclassicality. Another example is
Wigner \cite{wigner1932quantum} function whose negative values corresponds
to the nonclassicality.

\subsubsection{$Q$ function }

$Q$ function \cite{husimi1940some} is defined as 
\begin{equation}
Q=\dfrac{1}{\pi}\langle\beta|\rho|\beta\rangle,\label{eq:Q-function-1}
\end{equation}
where $|\beta\rangle$ is the coherent state (\ref{eq:CS}).

\subsubsection{Wigner function }

Another quasiprobability distribution is Wigner function formulated
by Wigner in 1932 \cite{wigner1932quantum} in the early stage of
quantum mechanics, the motive was to connect the wavefunction approach
to a probability distribution in phase space. Negativity of Wigner
function represents the nonclassicality present in an arbitrary quantum
state. Also the ability to reconstruct the Wigner function experimentally
makes this approach more impactful than any other approach.  Specifically,
Wigner function obtained through optical tomography can be used to
obtain other quasidistributions, however, Wigner function is stronger
witness of nonclassicality than $Q$ function while is not singular
like $P$-function.  Mathematically, it is expressed as 
\begin{equation}
W\left(\gamma,\gamma^{\star}\right)=A\exp\left[-2\left|\gamma\right|^{2}\right]\int d^{2}\lambda\langle-\lambda|\rho|\lambda\rangle\exp\left[-2\left(\gamma^{\star}\lambda-\gamma\lambda^{\star}\right)\right].\label{eq:wigner-def}
\end{equation}
The zeros of $Q$ function while the negativity of $P$-function and
Wigner function correspond to the nonclassical behavior of any arbitrary
quantum state. It is worth stressing here that only
$P$-function is both necessary and sufficient criterion of nonclassicality,
while rest of the quasidistribution functions are only sufficient.

\subsection{Measures of nonclassicality \label{subsec:Measures-of-nonclassicality}}

In the above section, we have seen that there exist numerous criteria
of nonclassicality. However, most of these criteria only witness the
nonclassicality. They do not provide any quantification of the nonclassicality. Except $P$-function and infinite set of vogel's criteria, all other
criteria are sufficient but not necessary. However,
many efforts have been made for the quantification of nonclassicality, e.g., in 1987, a distance-based measure of nonclassicality was introduced
by Hillery \cite{hillery1987nonclassical}. A trace norm based measure \cite{mari2011directly} was introduced
by Mari et al., for the set of all states having the positive Wigner
function. In 1991, Lee gave a measure of nonclassicality known as
nonclassical depth \cite{lee1991measure}. However, in this work,
we will not study these measures. There are
certain measures those can be exploited in terms of entanglement,
like linear entropy \cite{wei2003maximal}, which
we will use for our calculations and the same is described below.

\subsubsection{Linear entropy }

In 2005, a measure of nonclassicality was proposed as entanglement
potential, which is the amount of entanglement in two output ports
of a beam splitter with the quantum state $\rho_{in}$ and vacuum
$|0\rangle\langle0|$ sent through two input ports \cite{asboth2005computable}.
The amount of entanglement quantifies the amount of nonclassicality
in the input quantum state as classical state can not generate entanglement
in the output. The post beam splitter state can be obtained as $\rho_{out}=U\left(\rho_{in}\otimes|0\rangle\langle0|\right)U^{\dagger}$
with $U=\exp[-iH\theta]$, where $H=(\hat{a}^{\dagger}\hat{b}+\hat{a}\hat{b}^{\dagger})/2$,
and $\hat{a}^{\dagger}\,(\hat{a})$, $\hat{b}^{\dagger}\,(\hat{b})$
are the creation (annihilation) operators of the input modes. For
example, considering quantum state ($|\psi\rangle=\sum\limits _{n=0}^{\infty}c_{n}|n\rangle$)
and a vacuum state $|0\rangle$ as input states, we can write the
analytic expression of the two-mode output state as 
\begin{equation}
|\phi\rangle=U\left(|\psi\rangle\otimes|0\rangle\right)\equiv U|\psi,0\rangle=\sum_{n=0}^{\infty}\,\frac{c_{n}}{2^{n/2}}\sum_{j=0}^{n}\sqrt{^{n}C_{j}}\,\,|j,\,n-j\rangle.\label{eq:inout_psi}
\end{equation}
We can measure the amount of entanglement in the output state to quantify
the amount of input nonclassicality in $|\psi\rangle$. Here, we use
linear entropy of single mode subsystem (obtained after tracing over
the other subsystem) as entanglement potential. The linear entropy
for an arbitrary bipartite state $\rho_{AB}$ is defined as \cite{wei2003maximal}
\begin{equation}
\mathcal{L}=1-{\rm Tr}\left(\rho_{B}^{2}\right),\label{eq:le}
\end{equation}
where $\rho_{B}$ is obtained by tracing over subsystem $A$.

\section{Analytic tools for the study of phase properties of nonclassical
states \label{sec:Analytic-tools-forphase}}

In this section, we aim to introduce the parameters that are used
to study phase properties of a given quantum state under consideration
in this section.

\subsection{Phase distribution function}

A distribution function allows us to calculate expectation values
of an operator analogous to that from the corresponding density matrix.
Phase distribution function for a given density operator \cite{banerjee2007phase,agarwal1992classical}
can be defined as 
\begin{equation}
P_{\theta}=\frac{1}{2\pi}\langle\theta|\varrho|\theta\rangle,\label{eq:Phase-Distridution-1}
\end{equation}
where the phase state $|\theta\rangle$, complementary to the number
state $|n\rangle$, is defined \cite{agarwal1992classical} as 
\begin{equation}
|\theta\rangle=\sum_{n=0}^{\infty}e^{\iota n\theta}|n\rangle.\label{eq:phase-1}
\end{equation}

\subsection{Phase dispersion}

A known application of phase distribution function (\ref{eq:Phase-Distridution-1})
is that it can be used to quantify the quantum phase fluctuation.
Although the variance is also used occasionally as a measure of phase
fluctuation, it has a drawback that it depends on the origin of phase
integration \cite{banerjee2007phase}. A measure of phase fluctuation,
free from this problem, is phase dispersion \cite{perinova1998phase}
defined as 
\begin{equation}
D=1-\left|\intop_{-\pi}^{\pi}d\theta\exp\left[-\iota\theta\right]P_{\theta}\right|^{2}.\label{eq:Dispersion-1}
\end{equation}

\subsection{Angular Q function}

Analogous to the phase distribution $P_{\theta}$, phase distributions
are also defined as radius integrated quasidistribution functions
which are used as the witnesses for quantumness \cite{thapliyal2015quasiprobability}.
One such phase distribution function based on the angular part of
the $Q$ function is studied in \cite{leonhardt1993phase,leonhardt1995canonical}.
Specifically, the angular $Q$ function is defined as 
\begin{equation}
Q_{\theta_{1}}=\intop_{0}^{\infty}Q\left(\beta,\beta^{\star}\right)\left|\beta\right|d\left|\beta\right|,\label{eq:ang-Qf-1}
\end{equation}
where the $Q$ function \cite{husimi1940some} is defined in Eq. (\ref{eq:Q-function-1}).

\subsection{Phase fluctuation}

In attempts to get rid of the limitations of the Hermitian phase operator
of Dirac \cite{dirac1927quantum}, Louisell \cite{louisell1963amplitude}
first mentioned that bare phase operator should be replaced by periodic
functions. As a consequence, sine $(\mathcal{\hat{S}})$ and cosine
$(\hat{\mathcal{C}})$ operators appeared, explicit forms of these
operators were given by Susskind and Glogower \cite{susskind1964quantum},
and further modified by Barnett and Pegg \cite{barnett1986phase}
as 
\begin{equation}
\mathcal{\hat{S}}=\frac{\hat{a}-\hat{a}^{\dagger}}{2\iota\left(\bar{N}+\frac{1}{2}\right)^{\frac{1}{2}}}\label{eq:fluctuation1-1}
\end{equation}
and 
\begin{equation}
\hat{\mathcal{C}}=\frac{\hat{a}+\hat{a}^{\dagger}}{2\left(\bar{N}+\frac{1}{2}\right)^{\frac{1}{2}}}.\label{eq:fluctuation2-1}
\end{equation}
Here, $\bar{N}$ is the average number of photons in the measured
field, and here we refrain our discussion to Barnett and Pegg sine
and cosine operators \cite{barnett1986phase}. Carruthers and Nieto
\cite{carruthers1968phase} have introduced three quantum phase fluctuation
parameters in terms of sine and cosine operators 
\begin{equation}
U=\left(\Delta N\right)^{2}\left[\left(\Delta\mathcal{S}\right)^{2}+\left(\Delta\mathcal{C}\right)^{2}\right]/\left[\langle\mathcal{\hat{S}}\rangle^{2}+\langle\hat{\mathcal{C}}\rangle^{2}\right],\label{eq:fluctuation3-1}
\end{equation}
\begin{equation}
S=\left(\Delta N\right)^{2}\left(\Delta\mathcal{S}\right)^{2},\label{eq:fluctuation4-1}
\end{equation}
and 
\begin{equation}
Q=S/\langle\hat{\mathcal{C}}\rangle^{2}.\label{eq:fluctuation5-1}
\end{equation}
 These three phase fluctuation parameters $U$, $S$
and $Q$ show phase properties of PADFS and PSDFS, while $U$ parameter
is shown relevant as a witness of nonclassicality (antibunching).

\subsection{Quantum phase estimation parameter}

Quantum phase estimation is performed by sending the input state through
a Mach-Zehnder interferometer and applying the phase to be determined
($\phi$) on one of the arms of the interferometer. To study the phase
estimation using Mach-Zehnder interferometer, angular momentum operators
\cite{sanders1995optimal,demkowicz2015quantum}, defined as 
\begin{equation}
\hat{J_{x}}=\frac{1}{2}\left(\hat{a}^{\dagger}\hat{b}+\hat{b}^{\dagger}\hat{a}\right),\label{eq:ang-mom1-1}
\end{equation}
\begin{equation}
\hat{J_{y}}=\frac{\iota}{2}\left(\hat{b}^{\dagger}\hat{a}-\hat{a}^{\dagger}\hat{b}\right),\label{eq:ang-mom2-1}
\end{equation}
and 
\begin{equation}
\hat{J_{z}}=\frac{1}{2}\left(\hat{a}^{\dagger}\hat{a}-\hat{b}^{\dagger}\hat{b}\right),\label{eq:ang-mom3-1}
\end{equation}
are used. Here, $\hat{a}$ and $\hat{b}$ are the annihilation operators
for the modes corresponding to two input ports of the Mach-Zehnder
interferometer. The average value of $\hat{J_{z}}$ operator in the
output of the Mach-Zehnder interferometer, which is one-half of the
difference of photon numbers in the two output ports (\ref{eq:ang-mom3-1}),
can be written as 
\begin{equation}
\langle\hat{J_{z}}\rangle=\cos\text{\ensuremath{\phi}}\langle\hat{J_{z}}\rangle_{in}-\sin\text{\ensuremath{\phi}}\langle\hat{J_{x}}\rangle_{in}.
\end{equation}
Therefore, variance in the measured value of operator $\hat{J_{z}}$
can be computed as 
\begin{equation}
\left(\Delta J_{z}\right)^{2}=\cos^{2}\phi\left(\Delta{J_{z}}\right)_{in}^{2}+\sin^{2}\phi\left(\Delta{J_{x}}\right)_{in}^{2}-2\sin\text{\ensuremath{\phi\,\cos\phi\,}cov}\left(\hat{J_{x}},\hat{J_{z}}\right)_{in},
\end{equation}
where covariance of the two observables is defined as 
\begin{equation}
\text{{\rm cov}}\left(\hat{J_{x}},\hat{J_{z}}\right)=\frac{1}{2}\langle\hat{J_{x}}\hat{J_{z}}+\hat{J_{z}}\hat{J_{x}}\rangle-\langle\hat{J_{x}}\rangle\langle\hat{J_{z}}\rangle.
\end{equation}
This allows us to quantify precision in phase estimation \cite{demkowicz2015quantum}
as 
\begin{equation}
\Delta\phi=\frac{\Delta{J_{z}}}{\left|\frac{d\langle\hat{J_{z}}\rangle}{d\phi}\right|}.\label{eq:PE-1}
\end{equation}
Before we proceed further and conclude this chapter by noting the
structure of the rest of the thesis, it would be apt to to note that
there exist various methods of quantum state engineering (some of
which have already been mentioned) and photon addition, subtraction,
filteration, punching etc., which can be viewed as examples of quantum
state engineering processes. In rest of the thesis these processes
will be studied with detail.

\section{Structure of the rest of the thesis\label{sec:Structure-of-the thesis}}

This thesis has 6 chapters. The next 4 chapters are focused on the
study of nonclassical and phase properties of the engineered quantum
states and the last chapter is dedicated to conclusion. These chapters
and thus the rest of this thesis is organized as follows.

In Chapter \ref{cha:Introduction1}, in Section \ref{sec:Quantum-states-of-4},
quantum states of our interest (i.e., PADFS and PSDFS) have been introduced
in detail. In Chapter \ref{cha:PADFS-PSDFS}, in Section \ref{sec:Nonclassicality-witnesses},
the analytical expressions of various witnesses of nonclassicality
are reported. Further, the existence of various lower- and higher-order
nonclassical features in PADFS and PSDFS are shown through a set of
plots. Finally, we conclude in Section \ref{sec:Conclusions}.

In Chapter \ref{cha:phase}, in Section \ref{sec:phase-witnesses},
we investigate the phase properties of PADFS and PSDFS from a number
of perspectives. Finally, the chapter is concluded in Section \ref{sec:Conclusions-1}.

In Chapter \ref{cha:PASDFS}, we describe the quantum state of interest
(i.e., PASDFS) in Fock basis and calculate the analytic expressions
for the higher-order moments of the relevant field operators for this
state. In Section \ref{sec:Nonclassicality-witnesses-2}, we investigate
the possibilities of witnessing various nonclassical features in PASDFS
and its limiting cases by using a set of moments-based criteria for
nonclassicality. Variations of nonclassical features (witnessed through
different criteria) with various physical parameters are also discussed
here. In Section \ref{sec:Phase-properties-of}, phase properties
of PASDFS are studied. $Q$ function for PASDFS is obtained in Section
\ref{sec:Qfn}. Finally, we conclude in Section \ref{sec:Conclusions-5}.

In Chapter \ref{cha:QSE-1}, in Section \ref{sec:Quantum-states-of-1},
we have introduced the quantum states of our interest which include
ECS, BS, KS, VFECS, VFBS, VFKS, PAECS, PABS, and PAKS. In Section
\ref{sec:Nonclassicality-witnesses-1}, we have investigated the nonclassical
properties of these states using various witnesses of lower- and higher-order
noncassicality as well as a measure of nonclassicality. Specifically,
in this section, we have compared nonclassicality features found in
vacuum filtered and single photon added versions of the states of
our interest using the witnesses of Higher-order antibunching (HOA),
Higher-order squeezing (HOS) and Higher-order sub-Poissonian photon
statistics (HOSPS). Finally, in Section \ref{sec:Conclusion}, the
results are analyzed, and the chapter is concluded.

Finally, the thesis is concluded in Chapter \ref{cha:Conclusions-and-Scope},
where we have summarized the findings reported in Chapter \ref{cha:PADFS-PSDFS}-\ref{cha:QSE-1}
and have emphasized on the main conclusion of the present thesis.
We have also discussed the scopes of future work.

%
%
%
%
\chapter{Lower-and higher-order nonclassical properties of photon added and
subtracted displaced Fock state\textsc{\label{cha:PADFS-PSDFS}}}

In this chapter,   which is based on \cite{malpani2019lower}, we aim to study the nonclassical properties of the
PADFS and PSDFS (which are already introduced in Section \ref{subsec:Photon-added-and sub})
using the witnesses of nonclassicality introduced in Section \ref{subsec:Witnesses-of-nonclassicality}.

\section{Introduction\label{sec:Introduction-chap2} }

As we have mentioned in Chapter \ref{cha:Introduction1}, with the
advent of quantum state engineering \cite{vogel1993quantum,sperling2014quantum,miranowicz2004dissipation,marchiolli2004engineering}
and quantum information processing (\cite{pathak2013elements} and
references therein), the study of nonclassical properties of engineered
quantum states have become a very important field. Quantum state engineering
is helpful in realizing non-Gaussianity inducing operations, like
photon addition and subtraction \cite{zavatta2004quantum,podoshvedov2014extraction}.
Keeping this in mind, in what follows, in this chapter, we aim to
study the nonclassical properties of a set of engineered quantum states
(both photon added and subtracted) which can be produced by using
the above mentioned techniques.

\begin{figure}[h]
\begin{centering}
\begin{tabular}{c}
\includegraphics[width=100mm]{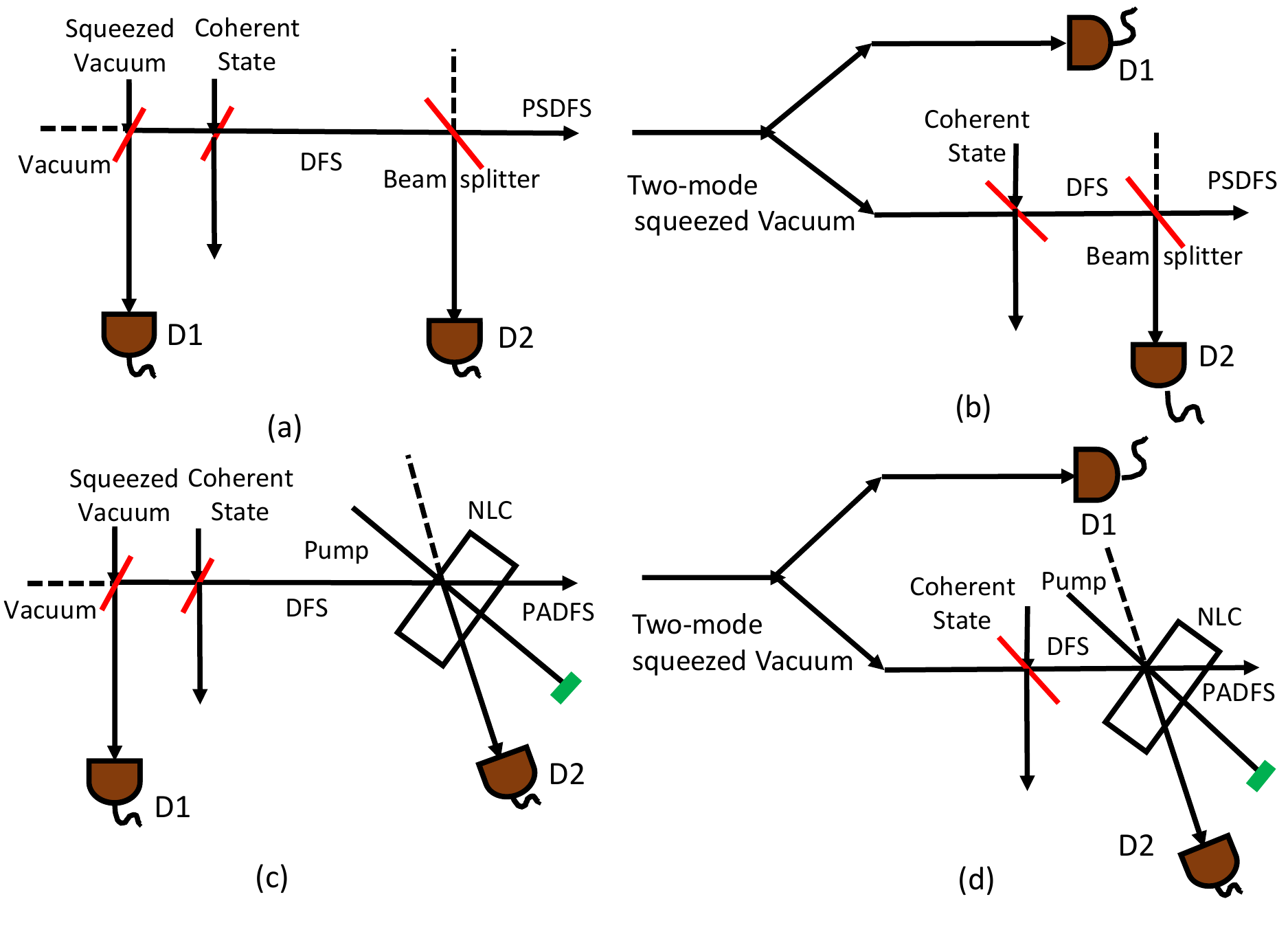}\tabularnewline
\end{tabular}
\par\end{centering}
\caption{\label{DFS} A schematic diagram for the generation of PSDFS (in (a)
and (b)) and PADFS (in (c) and (d)). In (a) and (c) ((b) and (d)),
single-mode (two-mode) squeezed vacuum is used for generation of the
desired state. Here, NLC corresponds to nonlinear crystal and D1 and
D2 are photon number resolving detectors.}
\end{figure}
It is already mentioned in Chapter 1 that a state having a negative
$P$-function is referred to as a nonclassical state. Such a state
cannot be expressed as a mixture of coherent states and does not possess
a classical analogue. In contrast to these states, coherent states
are classical, but neither their finite dimensional versions \cite{miranowicz2004dissipation,alam2017higher}
nor their generalized versions are classical \cite{satyanarayana1985generalized,thapliyal2016tomograms,thapliyal2015quasiprobability,banerjee2007phase}.
Here, we would like to focus on photon added and subtracted versions
of a particular type of generalized coherent state, which is also
referred to as the displaced Fock state (DFS). To be precise, state
of the form $|\psi\rangle=D(\alpha)|n\rangle$, where $D(\alpha)$
is the displacement operator with Fock state $|n\rangle$, is referred
to as generalized coherent state (see Section \ref{sec:Quantum-states-of-4}
of Chapter \ref{cha:Introduction1}), as this state is reduced to
a coherent state in the limit $n=0$. However, from the structure
of the state it seems more appropriate to call this state as the DFS,
and {this seems to be the nomenclature usually adopted in the literature}
\cite{keil2011classical,wunsche1991displaced,blavzek1994high,moya1995generation}.
In some other works, it is referred to as displaced number state \cite{ziesel2013experimental,de2006nonlinear,dodonov2005decoherence},
but all these names are equivalent; and in what follows, we will refer
to it as DFS. This is an extremely interesting quantum state for various
reasons. Specifically, its relevance in various areas of quantum optics
is known. For example, in the context of cavity QED, it constitutes
the eigenstates of the Jaynes-Cummings systems with coherently driven
atoms \cite{alsing1992dynamic}. Naturally, various lower-order nonclassical
properties and a set of other quantum features of DFS have already
been studied. Specifically, quasiprobability distributions of DFS
were studied in \cite{wunsche1991displaced}, phase fluctuations of
DFS was investigated in \cite{zheng1992fluctuation}, decoherence
of superposition of DFS was discussed in \cite{dodonov2005decoherence},
$Q$ function, Wigner function and probability distribution of DFS
were studied in \cite{de1990properties}, Pancharatnam phase of DFS
has been studied in \cite{mendas1993pancharatnam}. Further, in the
context of non-optical DFS, various possibilities of generating DFS
from Fock states by using a general driven time-dependent oscillator
has been discussed in \cite{lo1991generating}; and in the trapped-ion
system, quantum interference effects have been studied for the superposition
of DFS \cite{marchiolli2004engineering}. Thus, DFS seems to be a
well studied quantum state, but it may be noted that {a little effort}
has yet been made to study higher-order nonclassical properties of
DFS. It is a relevant observation as in the recent past, it has been
observed that higher-order nonclassicality has various applications
\cite{hillery1999quantum,banerjee2017quantum,sharma2017quantumauction,thapliyal2017protocols},
and it can be used to detect the existence of weak nonclassical characters
\cite{thapliyal2014higher,thapliyal2014nonclassical,verma2010generalized,thapliyal2017comparison,alam2016nonclassical,alam2015approximate,thapliyal2017nonclassicality,prakash2006higher,prakash2010detection,das2018lower}.
Further, various higher-order nonclassical features have been experimentally
detected \cite{allevi2012measuring,allevi2012high,avenhaus2010accessing,perina2017higher}.
However, we do not want to restrict to DFS, rather we wish to focus
on lower- and higher-order nonclassical properties of a set of even
more general states, namely photon added DFS (PADFS) and photon subtracted
DFS (PSDFS). The general nature of these states can be visualized
easily, as in the special case that no photon is added (subtracted)
PADFS (PSDFS) would reduce to DFS. Further, for $n=0$, PADFS would
reduce to photon added coherent state which has been widely studied
(\cite{agarwal1991nonclassical,verma2008higher,thapliyal2017comparison}
and references therein) and experimentally realized \cite{zavatta2004quantum,zavatta2005single}.
Here it is worth noting that DFS has also been generated experimentally
by superposing a Fock state with a coherent state on a beam splitter
\cite{lvovsky2002synthesis}. Further, an alternative method for the
generation of DFS has been proposed by Oliveira et al., \cite{de2005alternative}.
From the fact that photon added coherent state and DFS have already
been generated experimentally, and the fact that the photon added
states can be prepared via conditional measurement on a beam splitter,
it appears that PADFS and PSDFS can also be built in the lab. In fact,
inspired by these experiments, we have proposed a schematic diagram
for generating the PADFS and PSDFS in Figure \ref{DFS} using single-mode
and two-mode squeezed vacuum states. Specifically, using three (two)
highly transmitting beam splitters, a conditional measurement of single
photons at both detectors D1 and D2 in Figure \ref{DFS} (a) (Figure
\ref{DFS} (b)) would result in single photon subtracted DFS as output
from a single-mode (two-mode) squeezed vacuum state. Similarly, to
generate PADFS conditional subtraction of photon is replaced by photon
addition, using a nonlinear crystal and heralding one output mode
for a successful measurement of a single photon to ensure generation
of single photon added DFS. This fact, their general nature, and the
fact that nonclassical properties of PADFS and PSDFS have not yet
been {accorded sufficient attention, has} motivated us to perform
this study.

Motivated by the above facts, in what follows, we investigate the
possibilities of observing lower- and higher-order sub-Poissonian
photon statistics, antibunching and squeezing in PADFS and PSDFS.
We have studied nonclassical properties of these states through a
set of other witnesses of nonclassicality, e.g., zeros of $Q$ function,
Mandel $Q_{M}$ parameter, Klyshko's criterion, and Agarwal-Tara's
criterion. These witnesses of nonclassicality successfully establish
that both PADFS and PSDFS (along with most of the states to which
these two states reduce at different limits) are highly nonclassical.
Thus, {making use of the analytical expressions of moments of creation
and annihilation operators, discussed below, facilitates an analytical
understanding for most of the nonclassical witnesses.}

\section{Higher-order moment for PADFS and PSDFS \label{sec:Quantum-states-of}}

We have already mentioned that we are interested in PADFS and PSDFS.
In what follows, we will see that various experimentally measurable
nonclassicality witnesses can be expressed as the moments of annihilation
and creation operators \cite{allevi2012measuring,allevi2012high,avenhaus2010accessing,perina2017higher,miranowicz2010testing}.
To utilize those witnesses to identify the signatures of nonclassicality,
we will compute an analytic expression for the most general moment,
$\langle\hat{a}^{\dagger q}\hat{a}^{r}\rangle$, with $q$ and $r$
being non-negative integers. This is the most general moment in the
sense that any other moment can be obtained as a special case of it.
For example, if we need $\langle\hat{a}^{2}\rangle,$ we would just
require to consider $q=2$ and $r=0$. Thus, an analytic expression
for $\langle\hat{a}^{\dagger q}\hat{a}^{r}\rangle$ would essentially
help us to obtain analytic expression for any moment-based witness
of nonclassicality. Further, the analytic expressions of moment obtained
for PADFS and PASDFS would also help us to obtain nonclassical features
in the set of states obtained in the limiting cases, like Fock state,
DFS, photon added coherent state. Keeping this in mind, we have computed
$\langle\psi_{+}(u,n,\alpha)|\hat{a}^{\dagger q}\hat{a}^{r}|\psi_{+}(u,n,\alpha)\rangle$
and $\langle\psi_{-}(v,n,\alpha)|\hat{a}^{\dagger q}\hat{a}^{r}|\psi_{-}(v,n,\alpha)\rangle$
 using Eq. (\ref{eq:PADFS}) and (\ref{eq:PSDFS}),
and provide the final analytic expressions of these moments without
going into the mathematical details to maintain the flow of the chapter.
The obtained expressions for the above mentioned moments for PADFS
and PSDFS are {\small{}
\begin{eqnarray}\label{eq:PA-expectation}
\langle\hat{a}^{\dagger q}\hat{a}^{r}\rangle_{{\rm PADFS}} & = & \langle\psi_{+}(u,n,\alpha)|\hat{a}^{\dagger q}\hat{a}^{r}|\psi_{+}(u,n,\alpha)\rangle\nonumber \\
 & = & \frac{N_{+}^{2}}{n!}\sum\limits _{p,p'=0}^{n}{n \choose p}{n \choose p'}(-\alpha^{\star})^{(n-p)}(-\alpha)^{(n-p')}\\
 & \times & \exp\left[-\mid\alpha\mid^{2}\right]\sum\limits _{m=0}^{\infty}\frac{\alpha^{m}(\alpha^{\star})^{m+p-p'-r+q}(m+p+u)!(m+p+u-r+q)!}{m!(m+p-p'-r+q)!(m+p+u-r)!},\nonumber 
\end{eqnarray}
}and {\small{}
\begin{eqnarray}\label{eq:PS-expectation}
\langle\hat{a}^{\dagger q}\hat{a}^{r}\rangle_{{\rm PSDFS}} & = & \langle\psi_{-}(v,n,\alpha)|\hat{a}^{\dagger q}\hat{a}^{r}|\psi_{-}(v,n,\alpha)\rangle\nonumber \\
 & = & \frac{N_{-}^{2}}{n!}\sum\limits _{p,p'=0}^{n}{n \choose p}{n \choose p'}(-\alpha^{\star})^{(n-p)}(-\alpha)^{(n-p')}\nonumber \\
 & \times & \exp\left[-\mid\alpha\mid^{2}\right]\sum\limits _{m=0}^{\infty}\frac{\alpha^{m}(\alpha^{*})^{m+p-p'-r+q}(m+p)!(m+p-r+q)!}{m!(m+p-p'-r+q)!(m+p-v-r)!},
\end{eqnarray}
}respectively. The values of normalization constants for PADFS and
PSDFS are already given in Eqs. (\ref{eq:norP}) and (\ref{eq:norS}),
respectively. In the following section, we shall investigate the possibilities
of observing various types lower- and higher-order nonclassical features
in PADFS and PSDFS by using Eqs. (\ref{eq:PA-expectation}) and
(\ref{eq:PS-expectation}).

\section{Nonclassical features of PADFS and PSDFS \label{sec:Nonclassicality-witnesses}}

The moments of number operators for PADFS and PSDFS states obtained
in the previous section enable us to study nonclassical properties
of these states using a set of moments-based criteria of nonclassicality
\cite{miranowicz2010testing}\cite{naikoo2018probing}. In the recent
past, an infinite set of these moments-based criteria is shown to
be equivalent to the $P$-function-based criterion, i.e., it becomes
both necessary and sufficient \cite{richter2002nonclassicality,shchukin2005nonclassical}.
However, in this section, we will use a subset of this infinite set
as witnesses of nonclassicality to investigate various nonclassical
properties of the PADFS and PSDFS. Specifically, nonclassicality will
be witnessed through Mandel $Q_{M}$ parameter, criteria of lower-
and higher-order antibunching, Agarwal-Tara's criterion, Klyshko's
criterion, criteria of higher-order sub-Poissonian photon statistics,
zeros of $Q$ function, etc. As all these criteria are already introduced
in Section \ref{sec:Nonclassicality-witnesses},\textcolor{blue}{{}
}here we may discuss the plots and results.

\subsection{Mandel $Q_{M}$ Parameter}

\begin{figure}
\centering %
\begin{tabular}{cc}
\includegraphics[width=60mm]{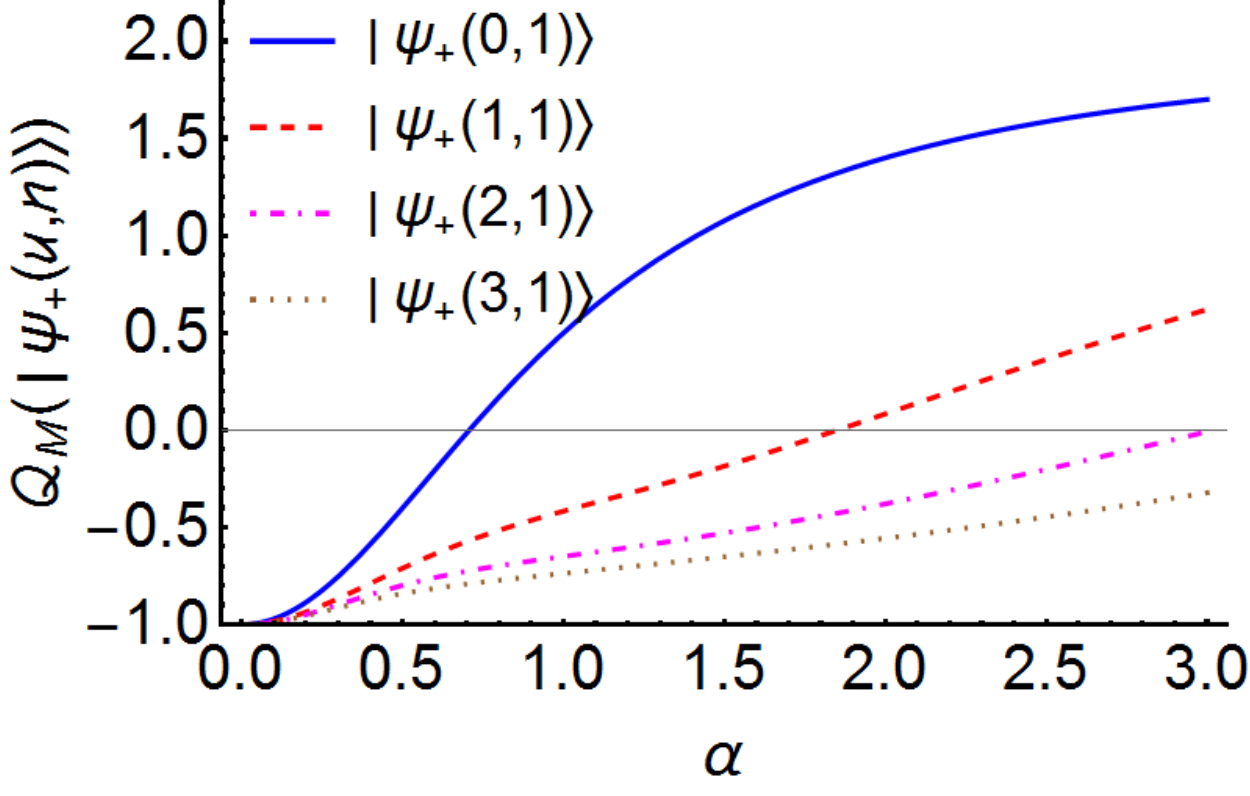}  & \includegraphics[width=60mm]{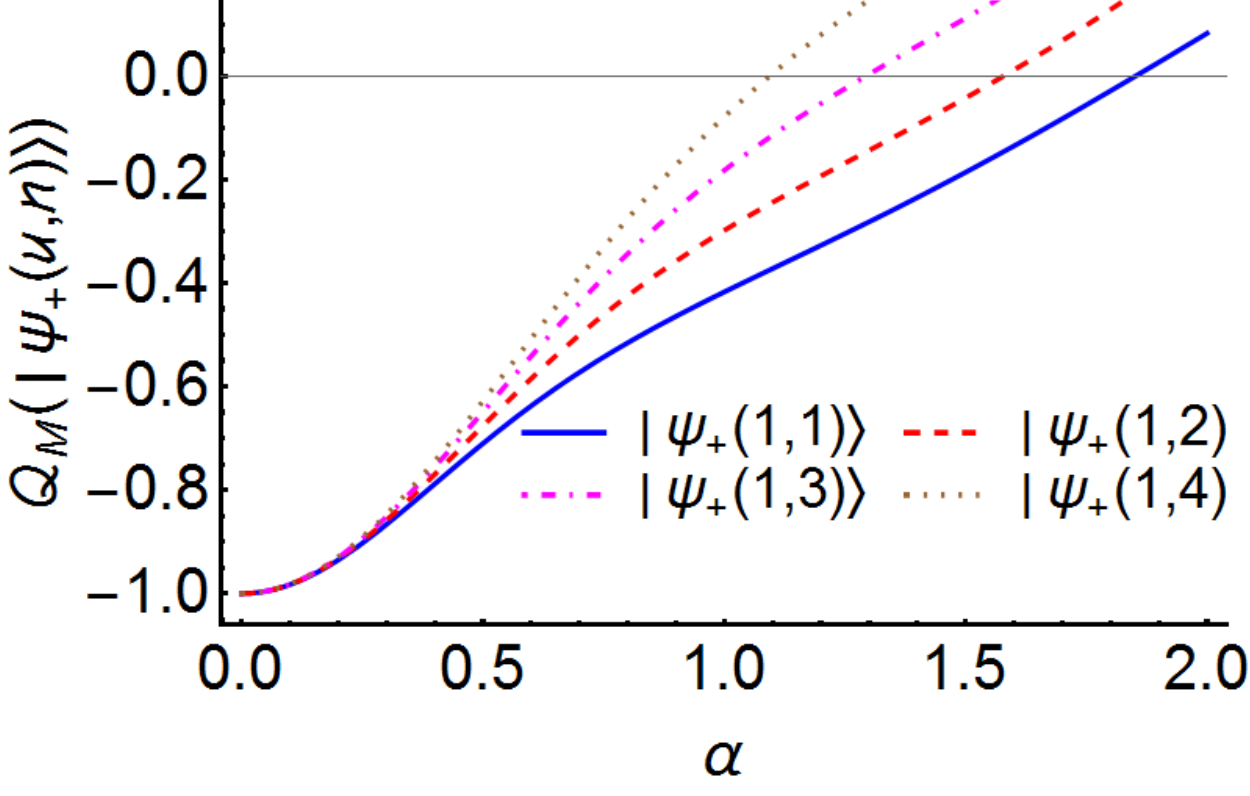} \tabularnewline
(a)  & (b) \tabularnewline
\includegraphics[width=60mm]{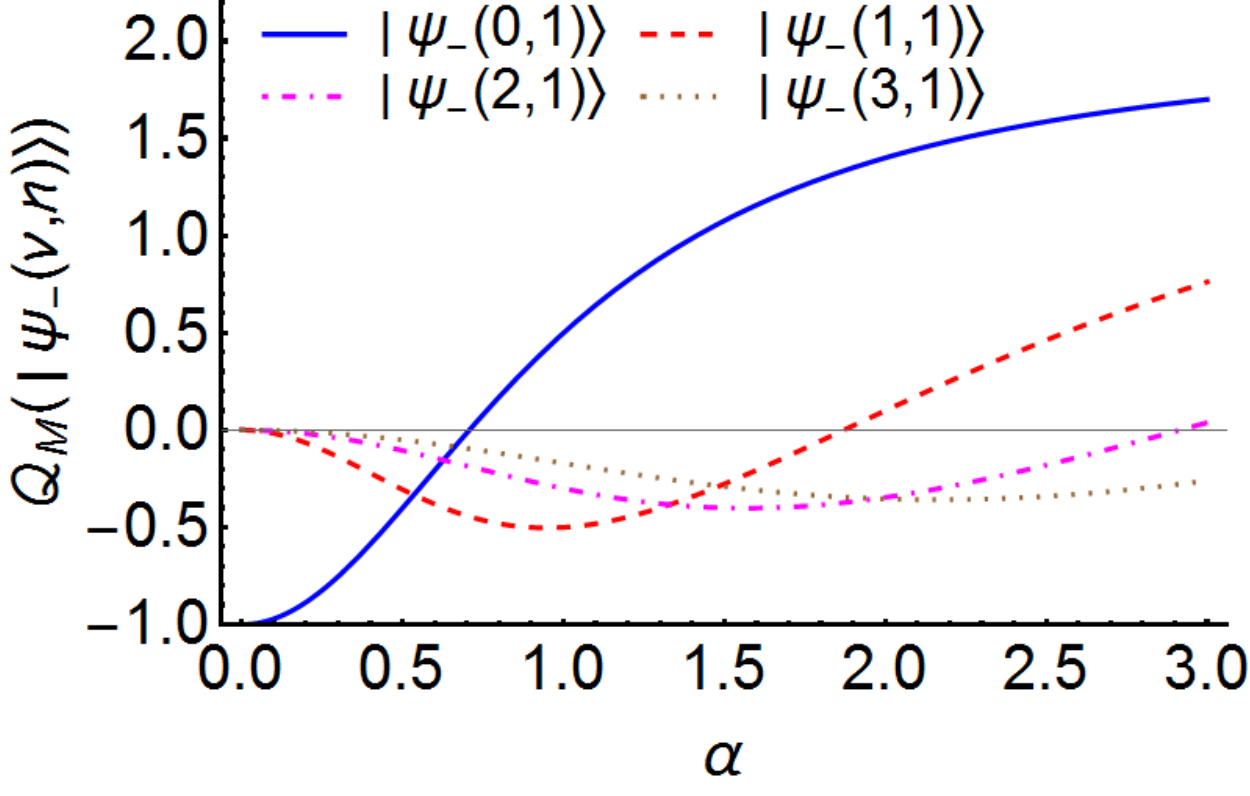}  & \includegraphics[width=60mm]{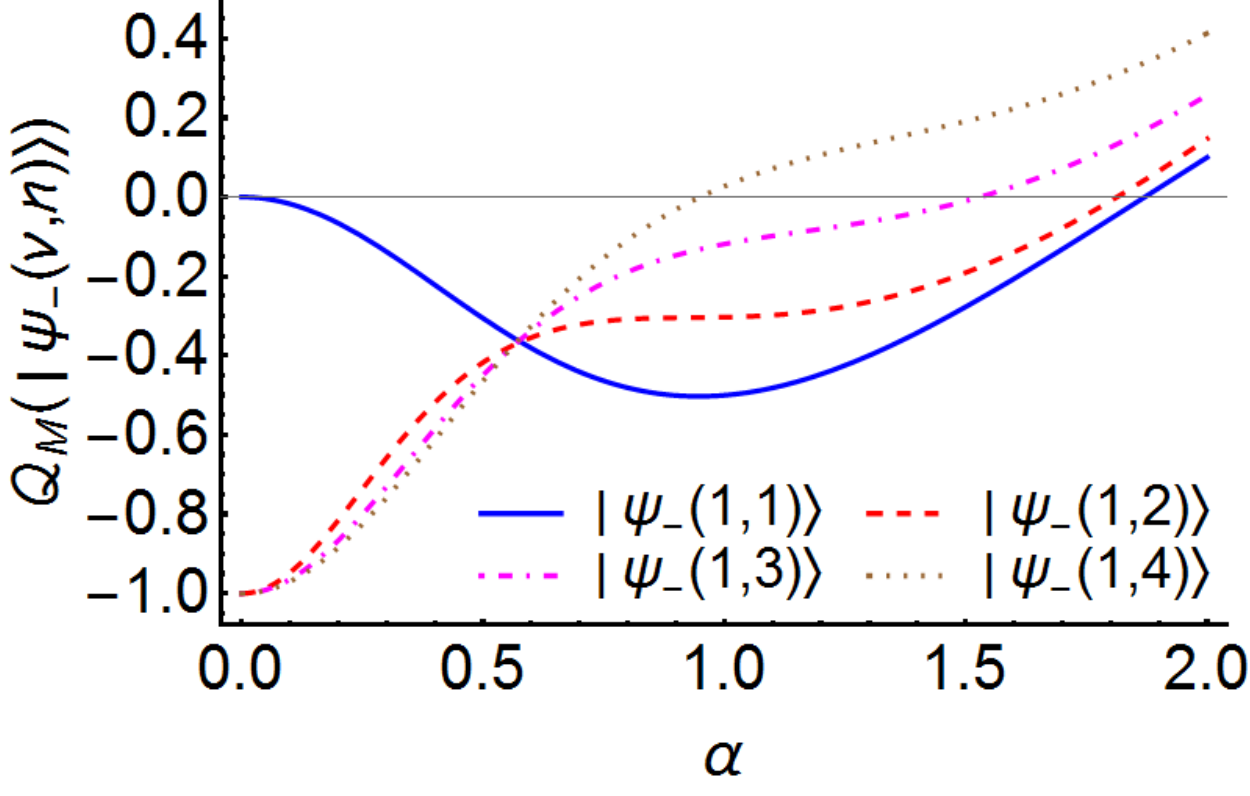} \tabularnewline
(c)  & (d) \tabularnewline
\end{tabular}\caption{\label{Mandel-Q-parameter} Variation of Mandel $Q_{M}$ parameter
for PADFS (in (a) and (b)) and PSDFS (in (c) and (d)) is shown with
displacement parameter $\alpha$. In (a) and (c), the value of number
of photons added/subtracted (i.e., $u$ or $v$) is changed for the
same initial Fock state $|1\rangle$. Different initial Fock states
$|n\rangle$ are chosen to be displaced in (b) and (d) for the single
photon addition/subtraction.  The blue curve corresponds
to vacuum for $\alpha=0$ and thus starts from $0$ unlike rest of
the states which are Fock state ($n\protect\neq0)$ in the limiting
case. Therefore, nonclassicality increases first with increasing $\alpha$
before decreasing as in the rest of the cases.}
\end{figure}

Negativity of this parameter indicates nonclassicality which can be
calculated using Eqs. (\ref{eq:PA-expectation}) and (\ref{eq:PS-expectation}).
In Figure \ref{Mandel-Q-parameter}, the dependence of $Q_{M}$ on
the state parameter $\alpha$ and non-Gaussianity inducing parameters
 (i.e., photon addition, subtraction, and Fock parameters
as they can induce non-Gaussianity in a quantum state) is shown.
Specifically, variation of $Q_{M}$ parameter for PADFS and PSDFS
is shown with state parameter $\alpha$, where the effect of the number
of photons added/subtracted and the initial Fock state is also established.
For $\alpha=0$, the PADFS with an arbitrary number of photon addition
has $Q_{M}$ parameter -1, which can be attributed to the fact that
final state, which reduces to the Fock state ($|1\rangle$ chosen
to be displaced in this case) is the most nonclassical state (cf.
Figure \ref{Mandel-Q-parameter} (a)). With increase in the number
of photons added to the DFS, the depth of nonclassicality witness
$Q_{M}$ increases. However, the witness of nonclassicality becomes less negative for higher values
of the displacement parameter. In contrast to the photon addition,
with the subtraction of photons from the DFS the $Q_{M}$ parameter
becomes almost zero for the smaller values of displacement parameter
$\alpha$ in DFS as shown in Figure \ref{Mandel-Q-parameter} (c).
This behavior can be attributed to the fact that photon subtraction
from $D\left(\alpha\right)|1\rangle$ for small values of $\alpha$
will most likely yield vacuum state. Also, with the increase in the
displacement parameter the witness of nonclassicality becomes more
negative as with a higher average number of photons in DFS photon
subtraction becomes more effective. However, for the larger values
of displacement parameter the nonclassicality disappears analogous
to the PADFS.   For large values of $\alpha$, this
parameter dominates in the behavior of the state and thus it behaves
analogous to coherent state.

As Fock states are known to be nonclassical, and photon addition and
subtraction are established as nonclassicality inducing operations,
it would be worth comparing the effect of these two independent factors
responsible for the observed nonclassical features in the present
case. To perform this study, we have shown the variation of the single
photon added (subtracted) DFS with different initial Fock states in
Figure \ref{Mandel-Q-parameter} (b) (Figure \ref{Mandel-Q-parameter}
(d)). Specifically, the nonclassicality present in PADFS decays faster
for the higher values of the Fock states with increasing displacement
parameter (cf. Figure \ref{Mandel-Q-parameter} (b)). However, such
nature was not present in PSDFS shown in Figure \ref{Mandel-Q-parameter}
(d). Note that variation of $Q_{M}$ parameter with $\alpha$ starts
from 0 (-1) iff $u\leq n$ $\left(u>n\right)$. For instance, if $u=n=1$,
i.e., corresponding to state $\hat{a}D\left(\alpha\right)|1\rangle$,
nonclassicality witness is zero for $\alpha=0$ as it corresponds
to vacuum state. Therefore, the present study reveals that photon
addition is a stronger factor for the nonclassicality present in the
state when compared to the initial Fock state chosen to be displaced.
Whereas photon subtraction is a preferred choice for large values
of displacement parameter in contrast to the higher values of Fock
states to displace with small $\alpha$. Among photon addition and
subtraction, addition is a preferred choice for the smaller values
of displacement parameter, while the choice between addition and subtraction
becomes immaterial for large $\alpha$.

\subsection{Lower- and higher-order antibunching}

\begin{figure}
\centering{} %
\begin{tabular}{ccc}
\includegraphics[width=50mm]{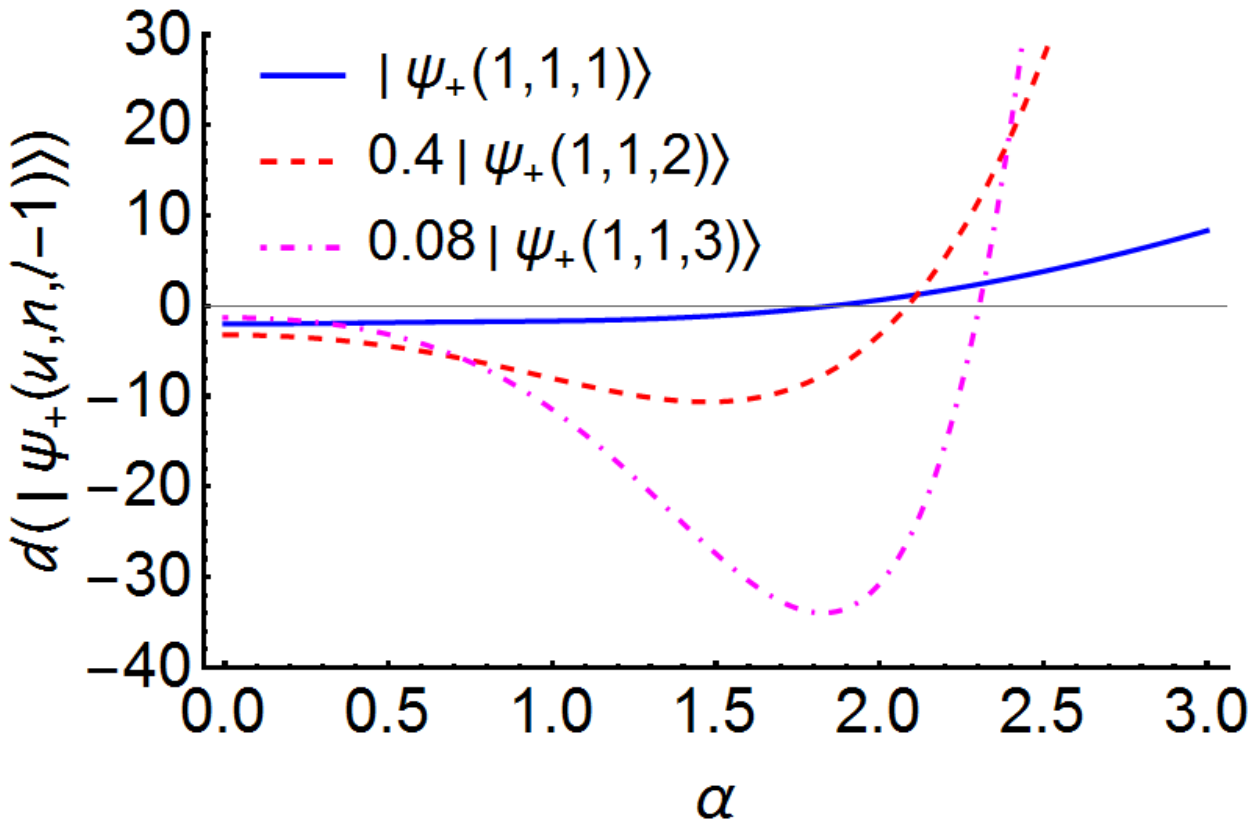}  & \includegraphics[width=50mm]{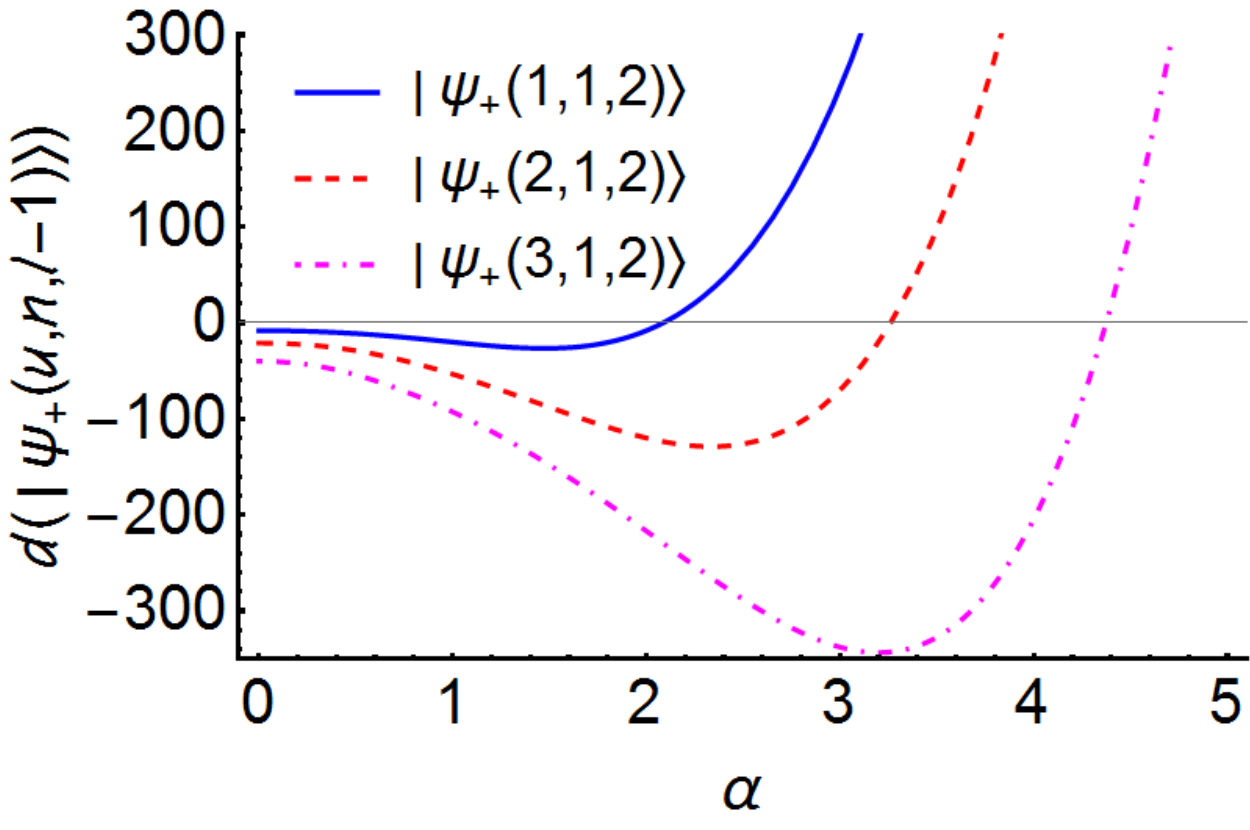}  & \tabularnewline
(a)  & (b)  & \tabularnewline
\includegraphics[width=50mm]{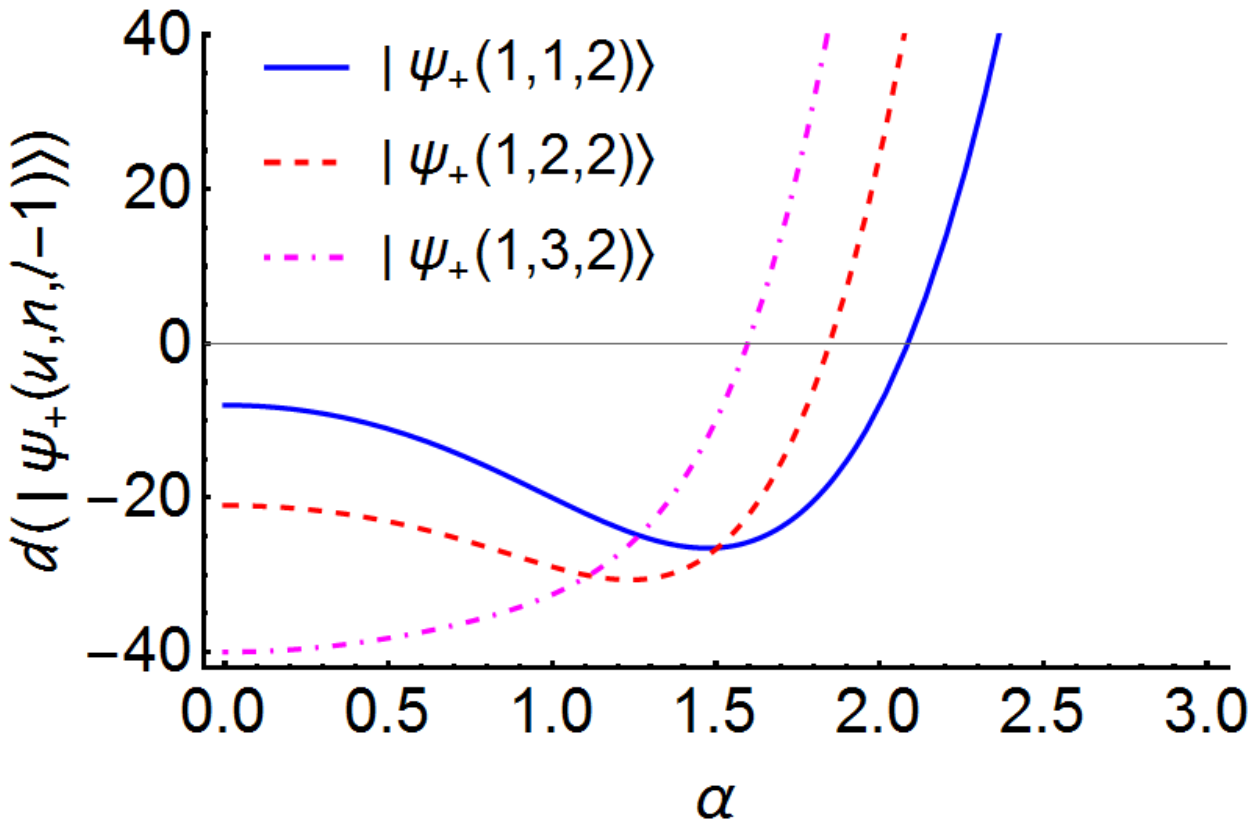}  & \includegraphics[width=50mm]{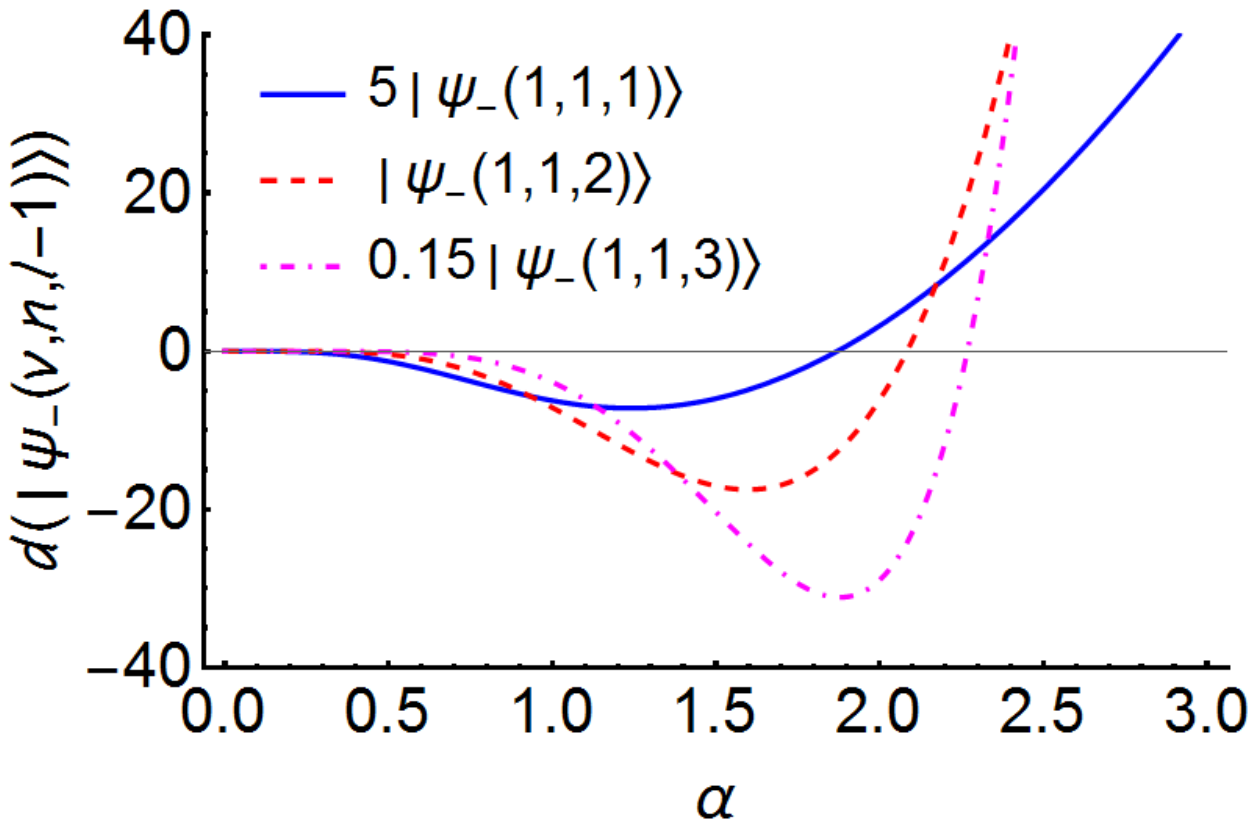} & \tabularnewline
 &  & \tabularnewline
(c)  & (d)  & \tabularnewline
\includegraphics[width=50mm]{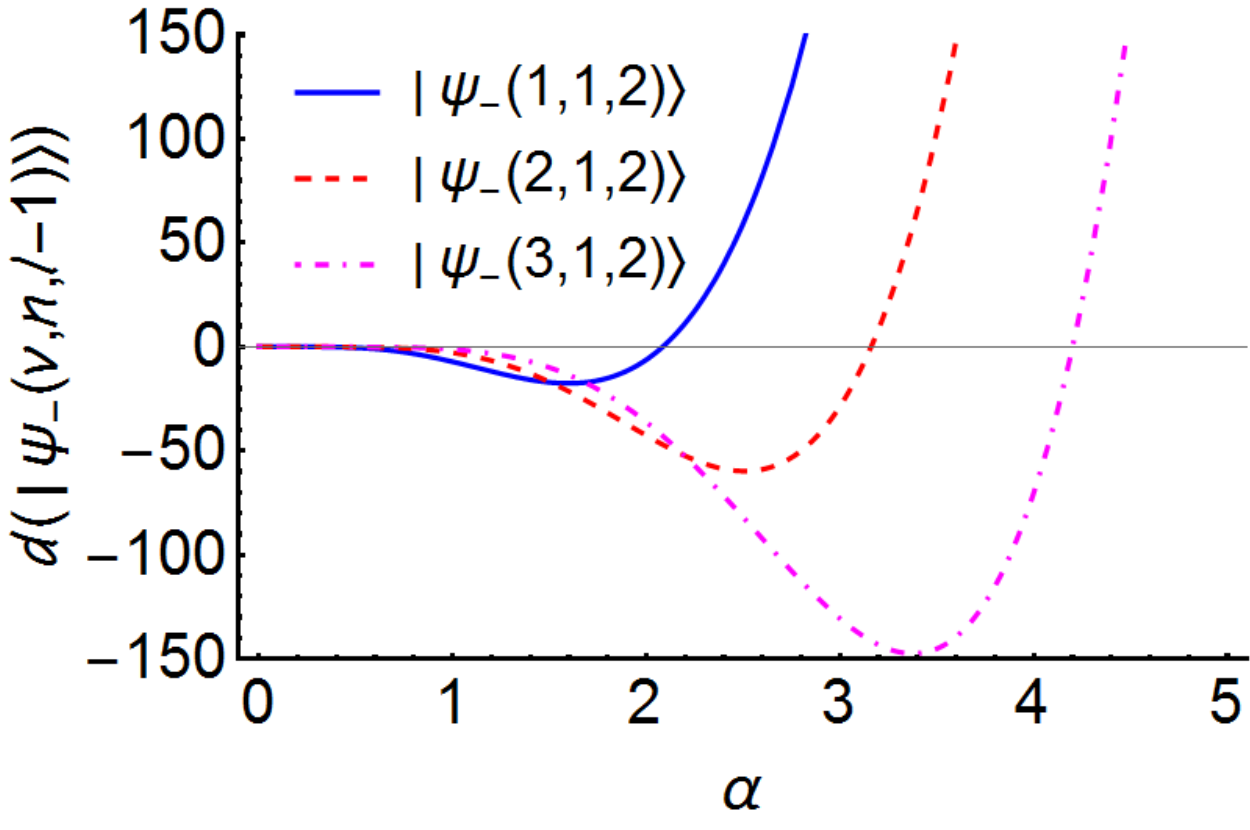}  & \includegraphics[width=50mm]{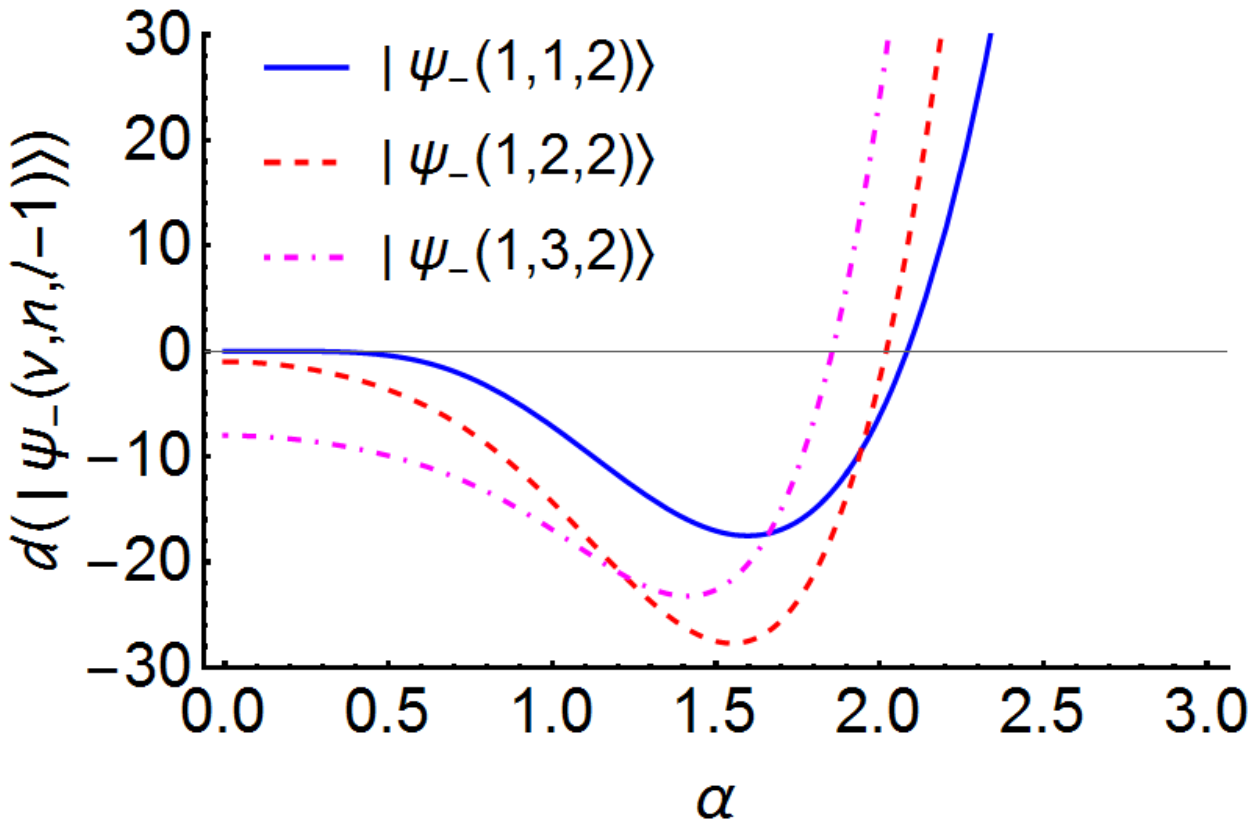}  & \tabularnewline
(e)  & (f)  & \tabularnewline
\end{tabular}\caption{\label{HOA} The presence of higher-order antibunching is shown as
a function of $\alpha$ for PADFS (in (a)-(c)) and PADFS ((d)-(f)).
Specifically, (a) and (d) illustrate comparison between lower- and
higher-order antibunching. {It should be noted that some of the curves
are multiplied by a scaling factor in order to present them in one
figure}. Figures (b) and (e) show the effect of photon addition/subtraction,
and (c) and (f) establish the effect of Fock state chosen to displace
in PADFS and PSDFS, respectively. Here, without the loss of generality, we have used the notation $\psi_{+}(u, n, l-1)$ ($\psi_{-}(u, v, l-1)$) for nonclassicality in photon added (subtracted) scenarios, and will follow this notation in subsequent figures of this chapter.}
\end{figure}

The nonclassicality reflected by the lower-order antibunching criterion
obtained here is the same as Mandel $Q_{M}$ parameter  $\left(Q_{M}=\frac{d\left(1\right)}{\left\langle \hat{a}^{\dagger}\hat{a}\right\rangle }\right)$
illustrated in Figure \ref{Mandel-Q-parameter}. Therefore we will
rather discuss here the possibility of observing higher-order antibunching
in the quantum state of our interest using Eqs. (\ref{eq:PA-expectation})
and (\ref{eq:PS-expectation}) in Eq. (\ref{eq:HOA-1}). Specifically,
the depth of nonclassicality witness can be observed to increase with
order for both PADFS and PSDFS as depicted in Figure \ref{HOA} (a)
and (d). This fact is consistent with the earlier observations (\cite{thapliyal2014higher,thapliyal2014nonclassical,thapliyal2017comparison,thapliyal2017nonclassicality,alam2017lower}
and references therein) that higher-order nonclassicality criteria
are useful in detecting weaker nonclassicality. On top of that the
higher-order antibunching can be observed for larger values of displacement
parameter $\alpha$, when lower-order antibunching is not present.
The presence of higher-order nonclassicality in the absence of its
lower-order counterpart establishes the relevance of the present study.

The depth of nonclassicality parameter  ($d\left(l-1\right)$)
was observed to decrease with an increase in the number of photons
subtracted from DFS for small values of $\alpha$ in Figure \ref{Mandel-Q-parameter}
(c). A similar nature is observed in Figure \ref{HOA} (e), which
shows that for the higher values of displacement parameter, the depth
of higher-order nonclassicality witness increases with the number
of photon subtraction. Therefore, not only the depth of nonclassicality
but the range of displacement parameter for the presence of higher-order
antibunching also increases with photon addition/subtraction (cf.
Figure \ref{HOA} (b) and (e)). With the increase in the value of
Fock state parameter $n$, the depth of higher-order nonclassicality
witness increases (decreases) for smaller (larger) values of displacement
parameter in both PADFS and PSDFS as shown in Figure \ref{HOA} (c)
and (f), respectively. Thus, we have observed that the range of $\alpha$
with the presence of nonclassicality increases (decreases) with photon
addition/subtraction (Fock state) in DFS.

\subsection{Higher-order sub-Poissonian photon statistics}

\begin{figure}
\centering{} %
\begin{tabular}{ccc}
\includegraphics[width=60mm]{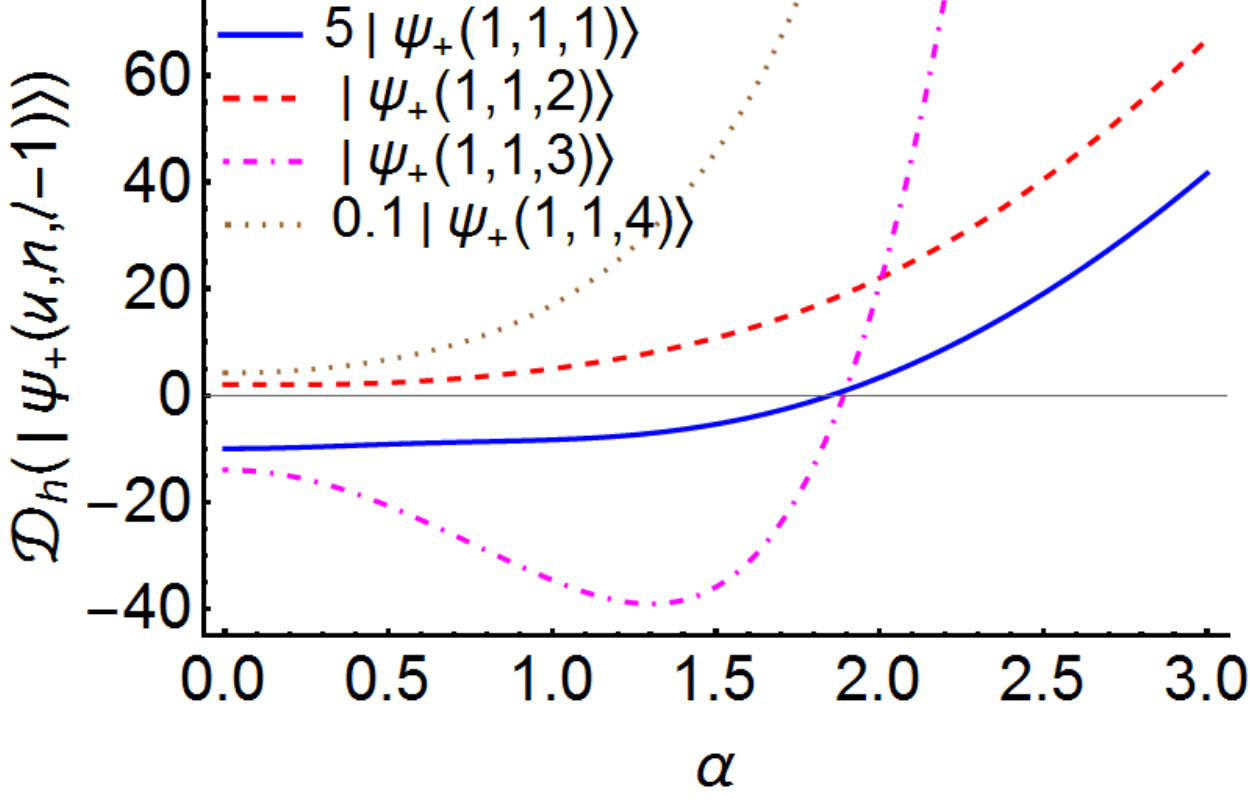}  & \includegraphics[width=60mm]{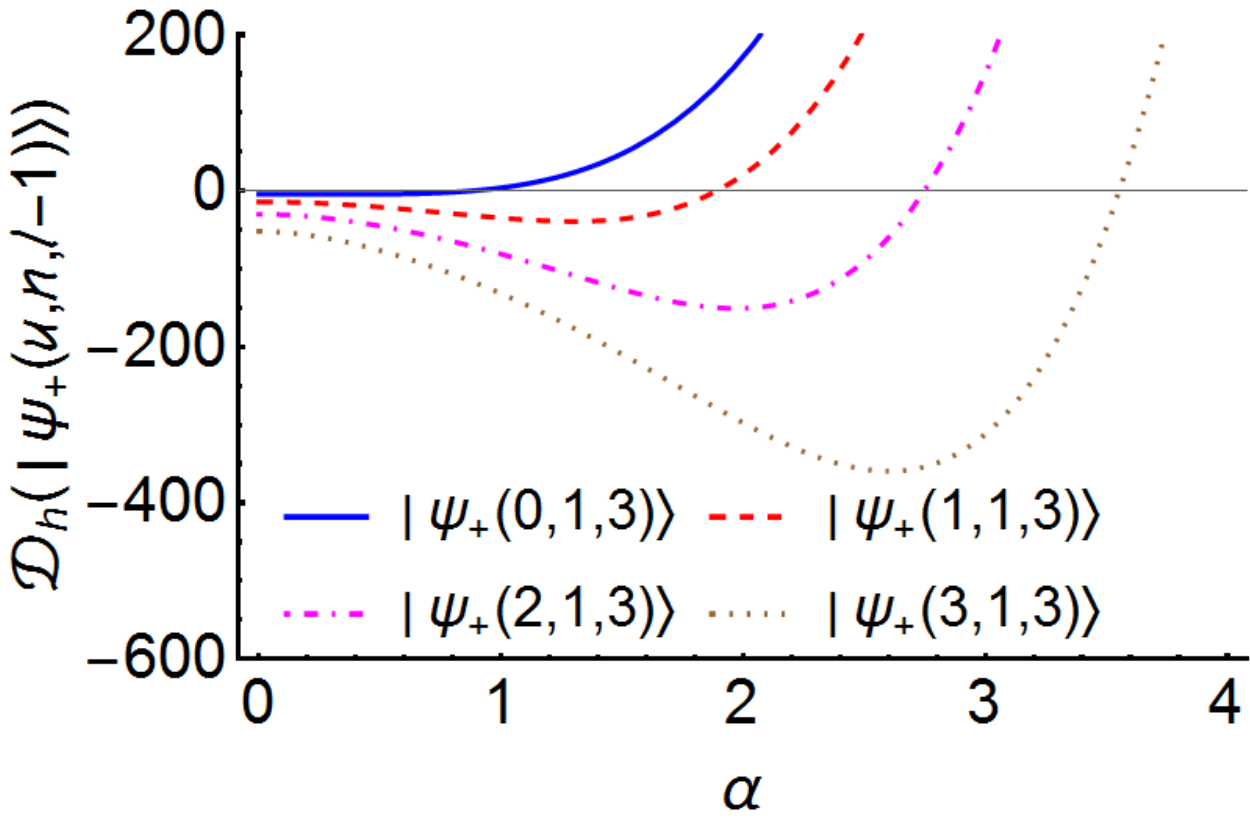}  & \tabularnewline
(a)  & (b)  & \tabularnewline
\includegraphics[width=60mm]{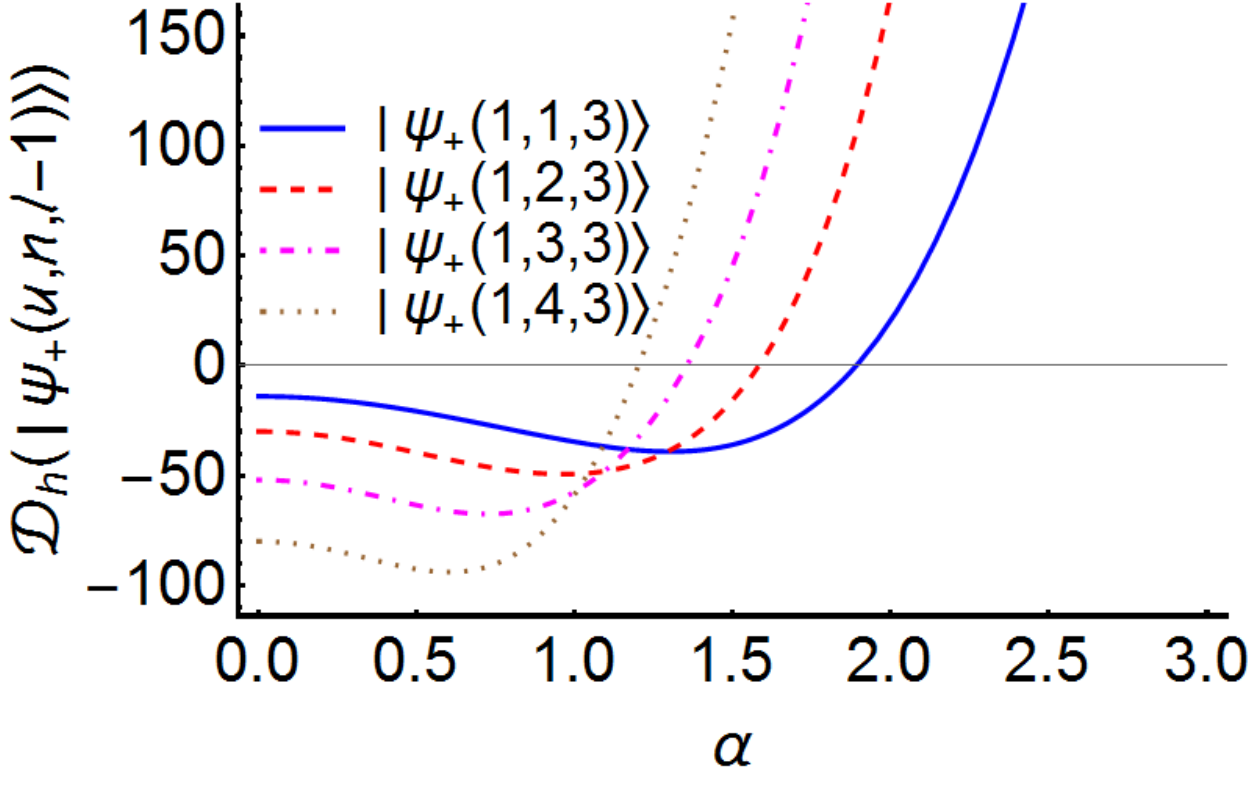}  & \includegraphics[width=60mm]{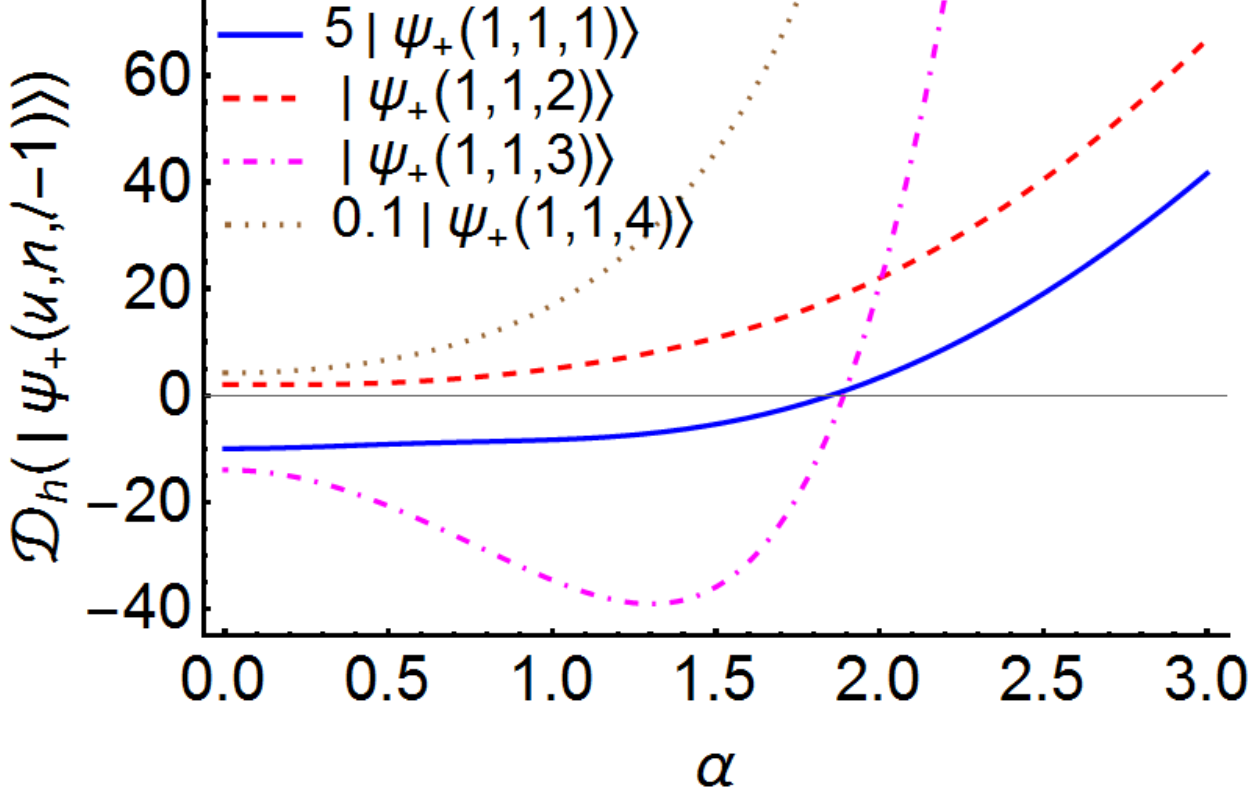}  & \tabularnewline
(c)  & (d)  & \tabularnewline
\includegraphics[width=60mm]{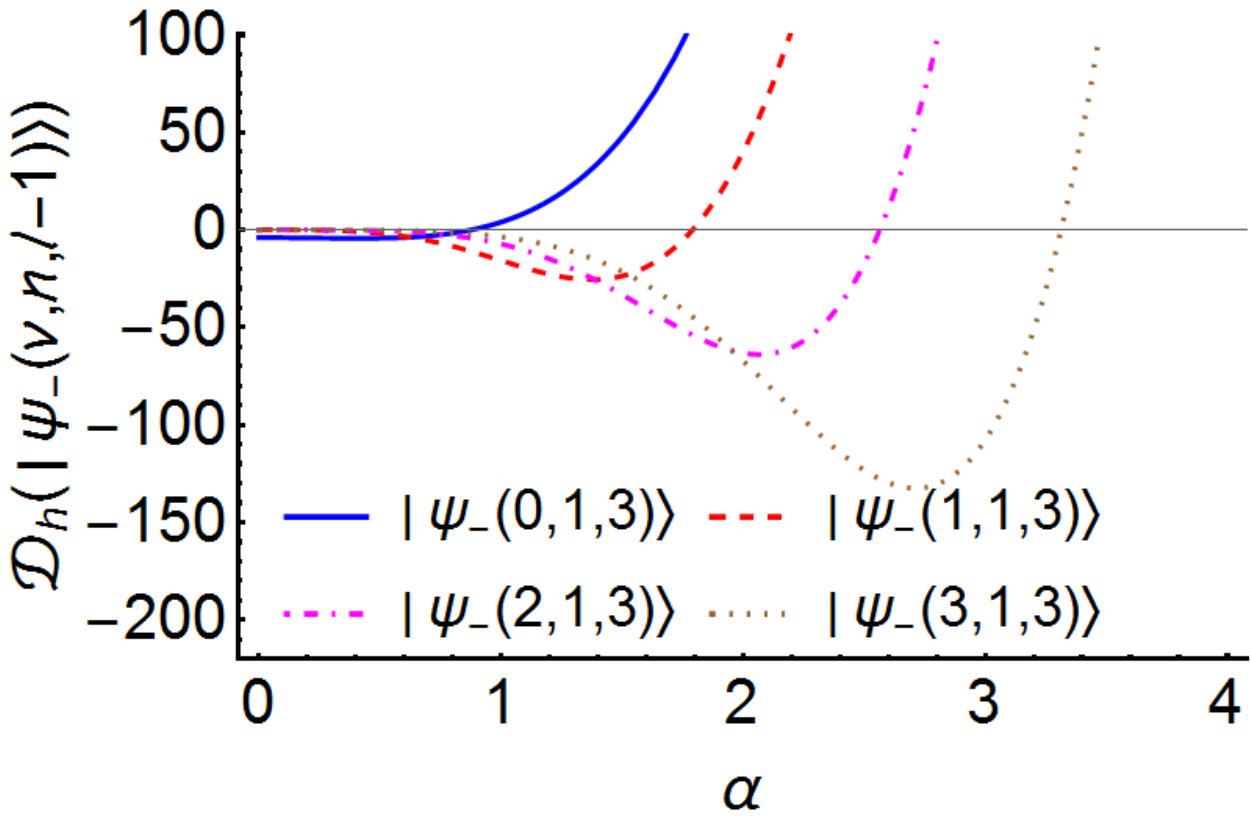}  & \includegraphics[width=60mm]{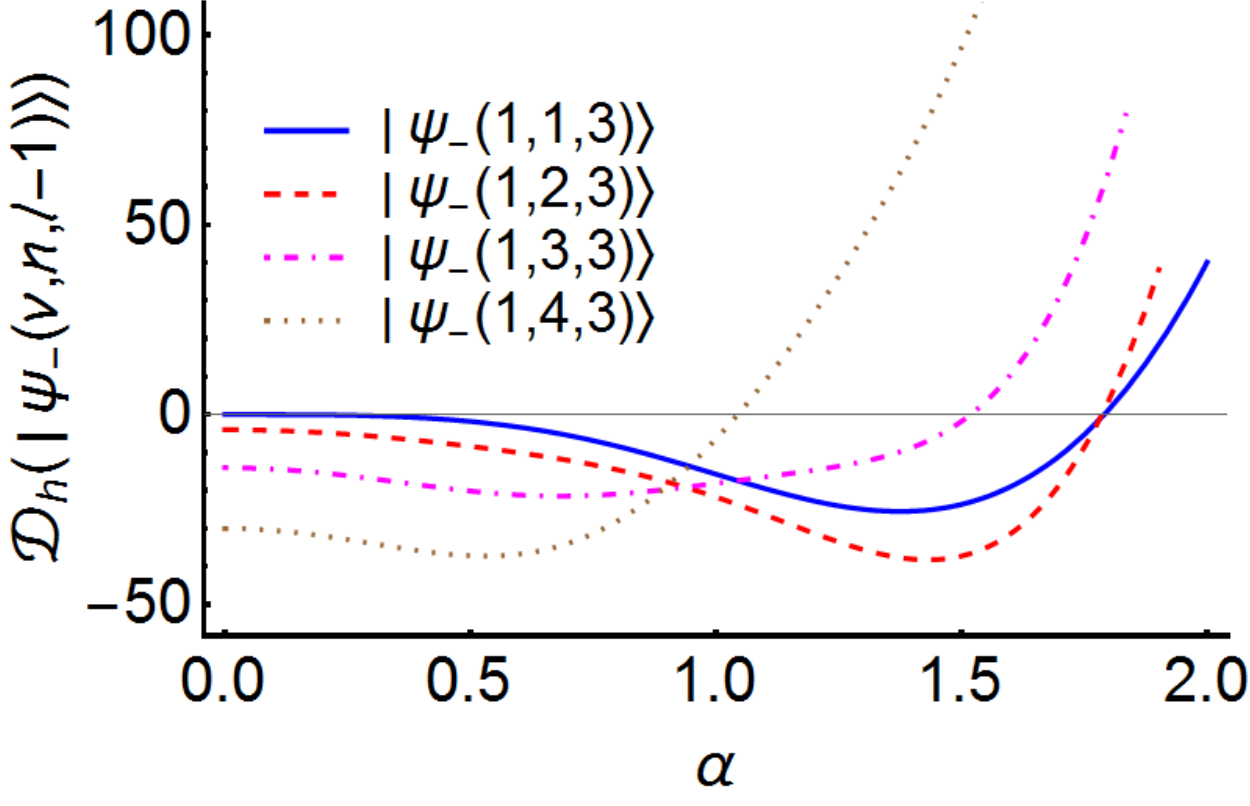}  & \tabularnewline
(e)  & (f)  & \tabularnewline
\end{tabular}\caption{\label{HOSPS} Dependence of higher-order sub-Poissonian photon statistics
on $\alpha$ for PADFS (in (a)-(c)) and PSDFS ((d)-(f)) is illustrated
here. Specifically, (a) and (d) show the increase in depth of nonclassicality
witness with order, (b) and (e) depict the effect of photon addition
and subtraction, respectively, and (c) and (f) establish the effect
of choice of Fock state to be displaced in PADFS and PSDFS, respectively. }
\end{figure}

This allows us to study higher-order sub-Poissonian photon statistics
using Eqs. (\ref{eq:PA-expectation}) and (\ref{eq:PS-expectation})
in Eq. (\ref{eq:hosps22-1}). The presence of higher-order sub-Poissonian
photon statistics (as can be seen in Figure \ref{HOSPS} (a) and (d)
for PADFS and PSDFS, respectively) is dependent on the order of nonclassicality
unlike higher-order antibunching, which is observed for all orders.
Specifically, the nonclassical feature was observed only for odd orders,
which is consistent with some of the earlier observations \cite{thapliyal2017comparison},
where nonclassicality in those cases could be induced due to squeezing.
   Along the same line, we
expect to observe the nonclassicality in such cases with appropriate
use of squeezing as a useful quantum resource, which will be discussed
elsewhere. In case of photon addition/subtraction in DFS, a behavior
analogous to that observed for higher-order antibunching is observed,
i.e., the depth of nonclassciality increases with the photon addition
while it decreases (increases) for small (large) values of $\alpha$
(cf. Figure \ref{HOSPS} (b) and (e)). Similar to the previous case,
nonclassicality can be observed to be present for larger values of
displacement parameter with photon addition/subtraction, while increase
in the value of Fock parameter has an opposite effect.

\subsection{Higher-order squeezing}

\begin{figure}
\begin{centering}
\begin{tabular}{ccc}
\includegraphics[width=45mm]{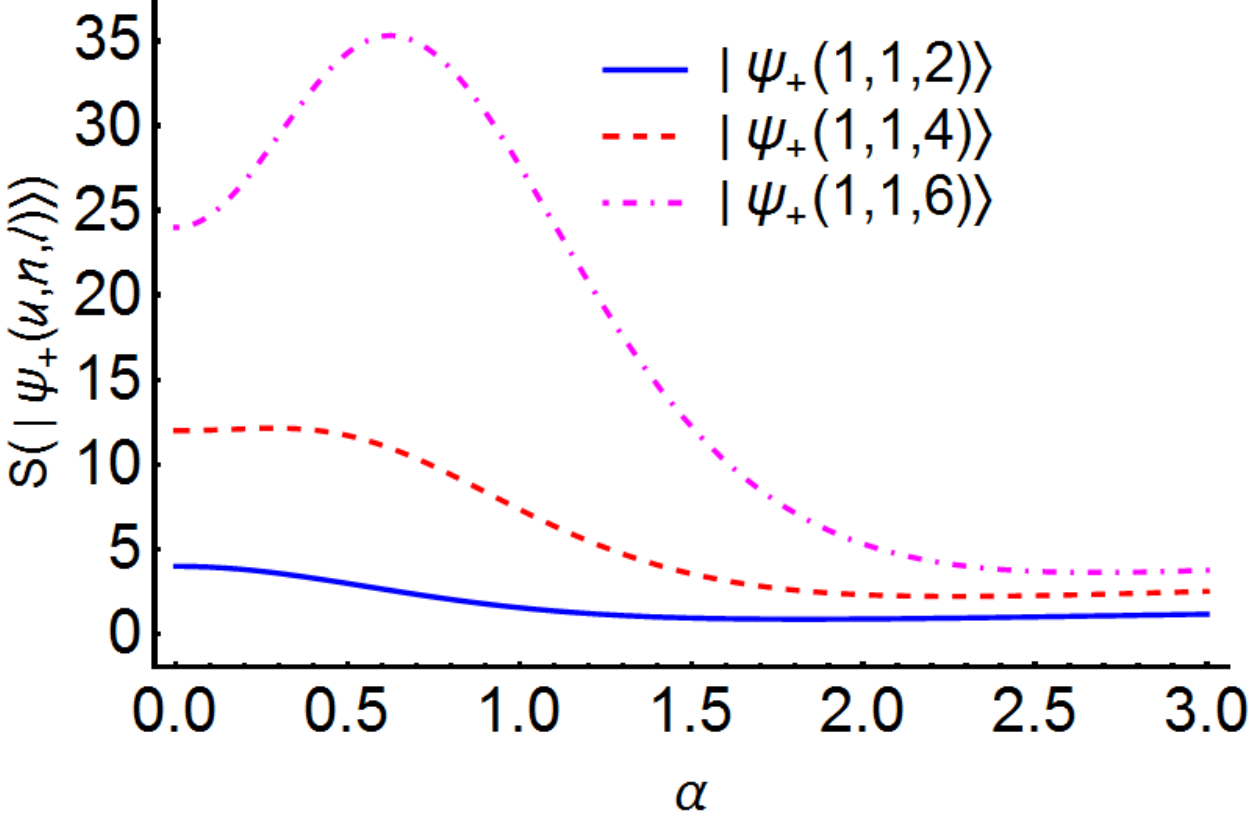}  & \includegraphics[width=45mm]{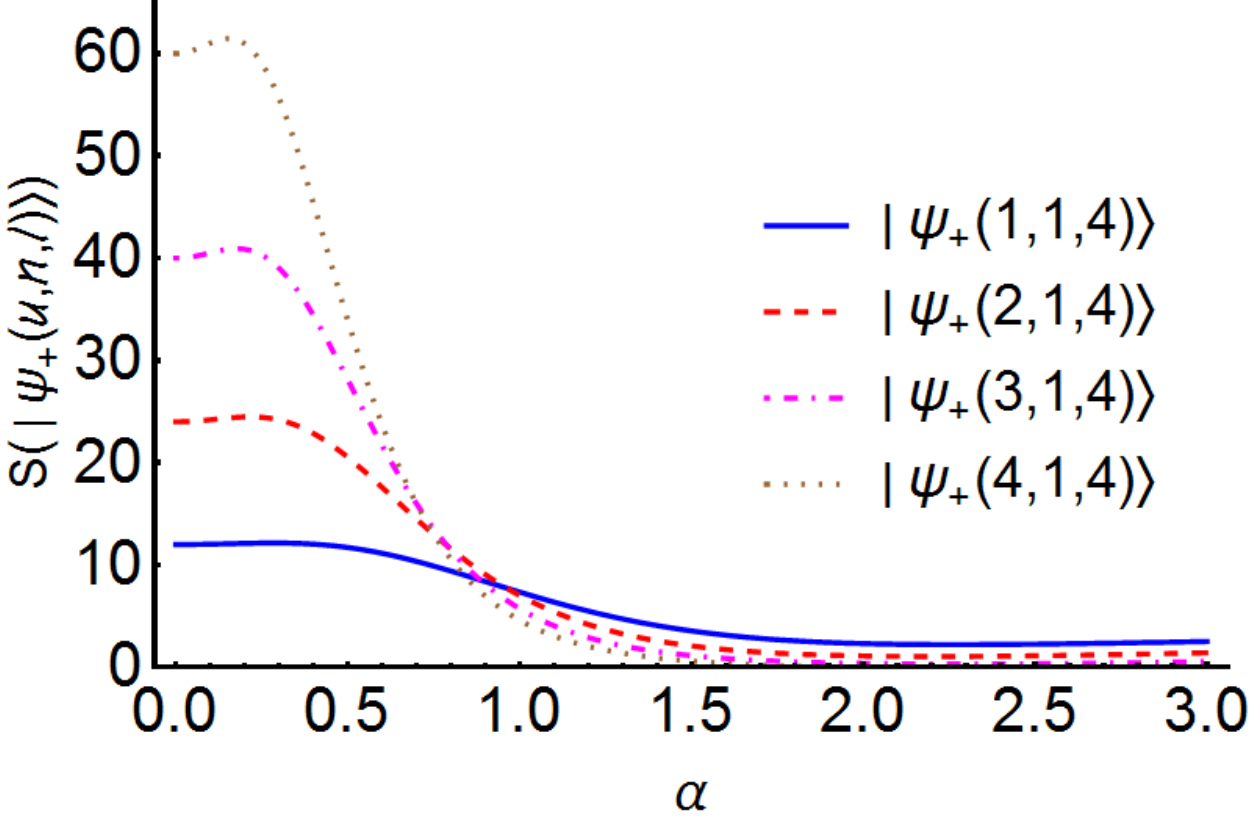}  & \includegraphics[width=45mm]{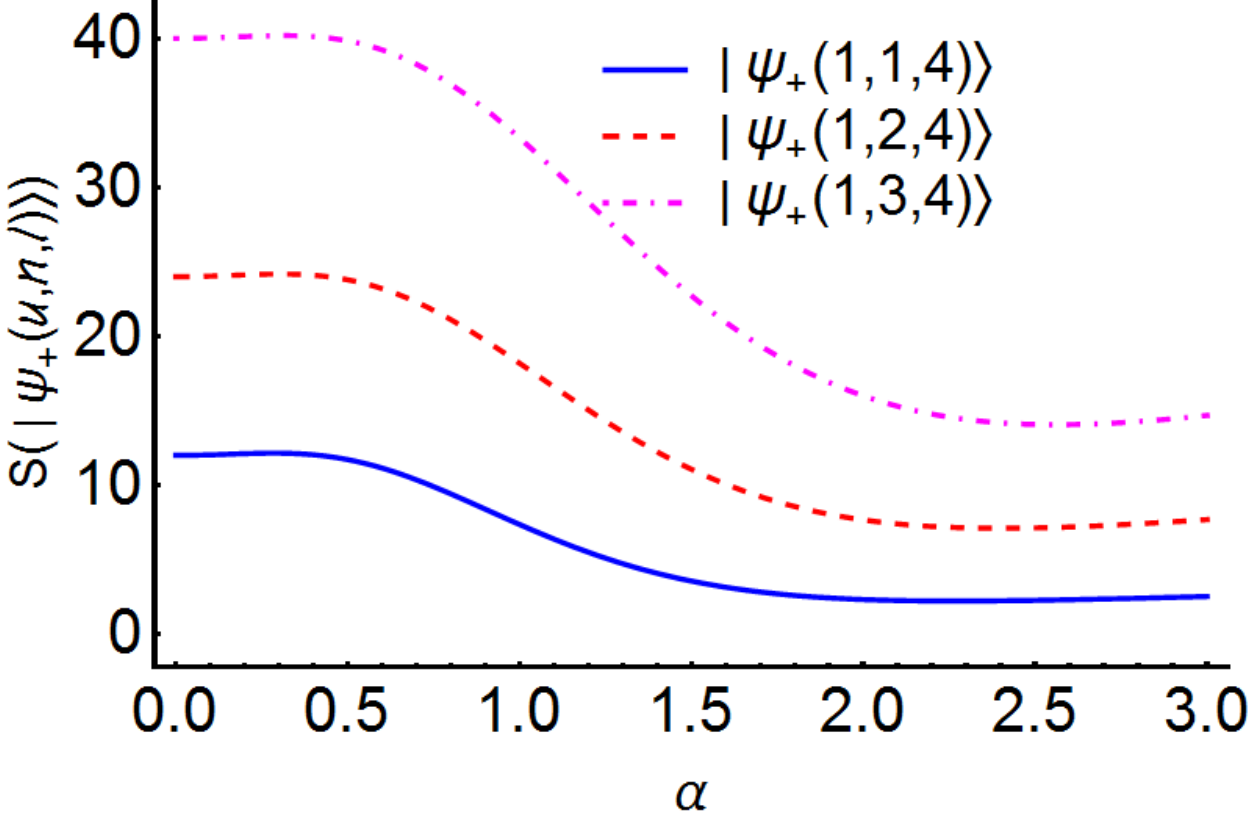} \tabularnewline
(a)  & (b)  & (c) \tabularnewline
\includegraphics[width=45mm]{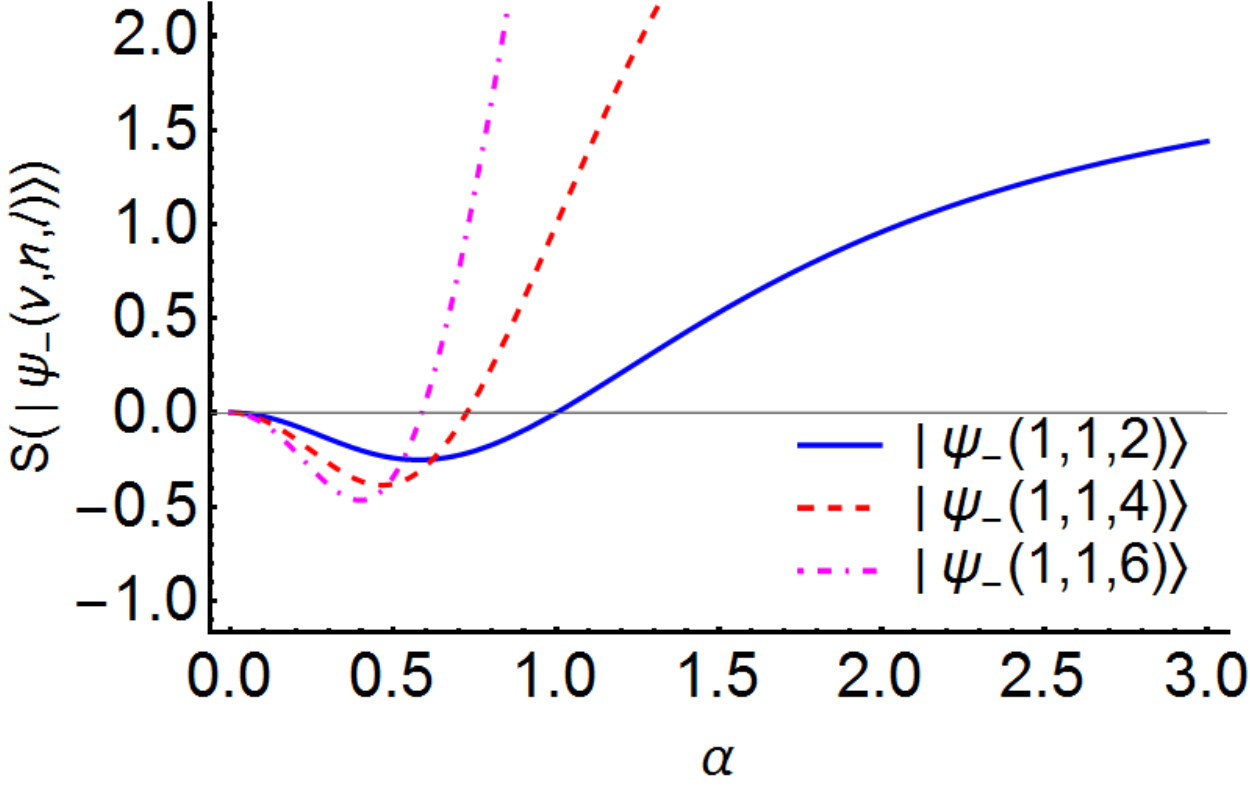}  & \includegraphics[width=45mm]{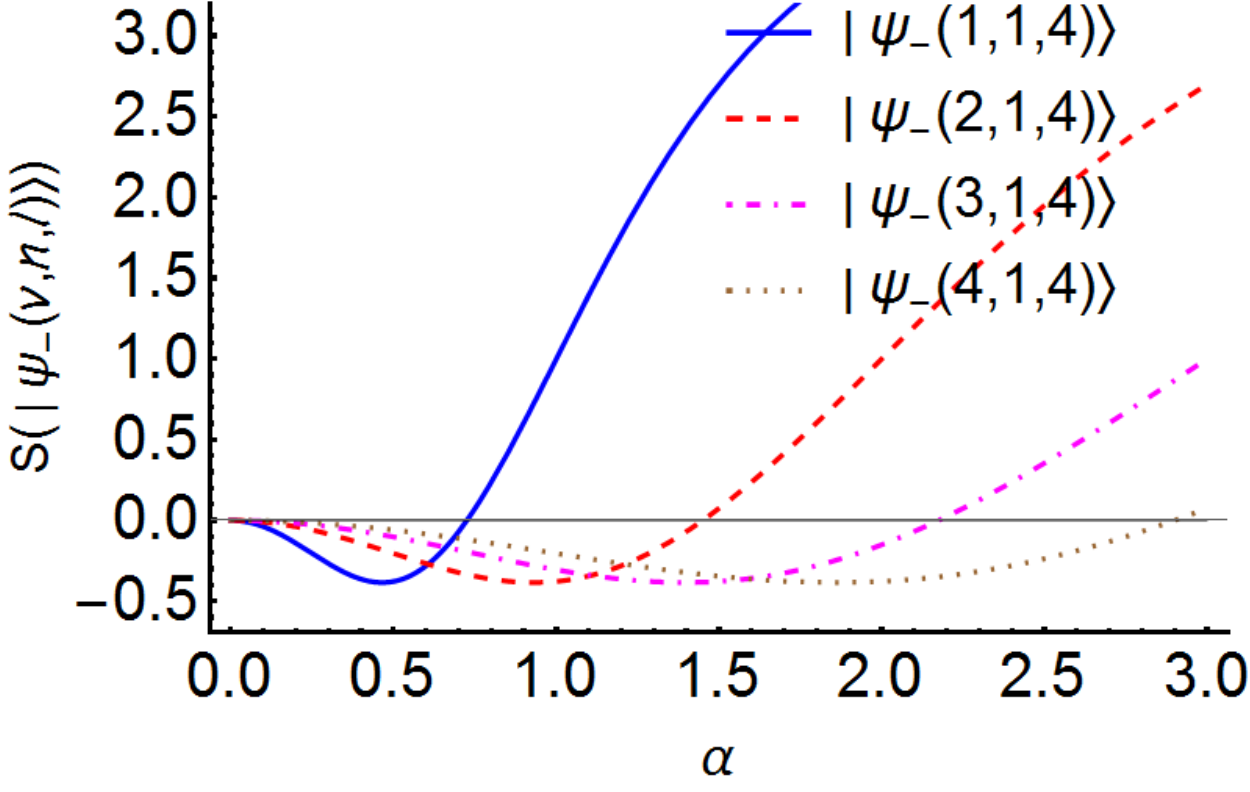}  & \includegraphics[width=45mm]{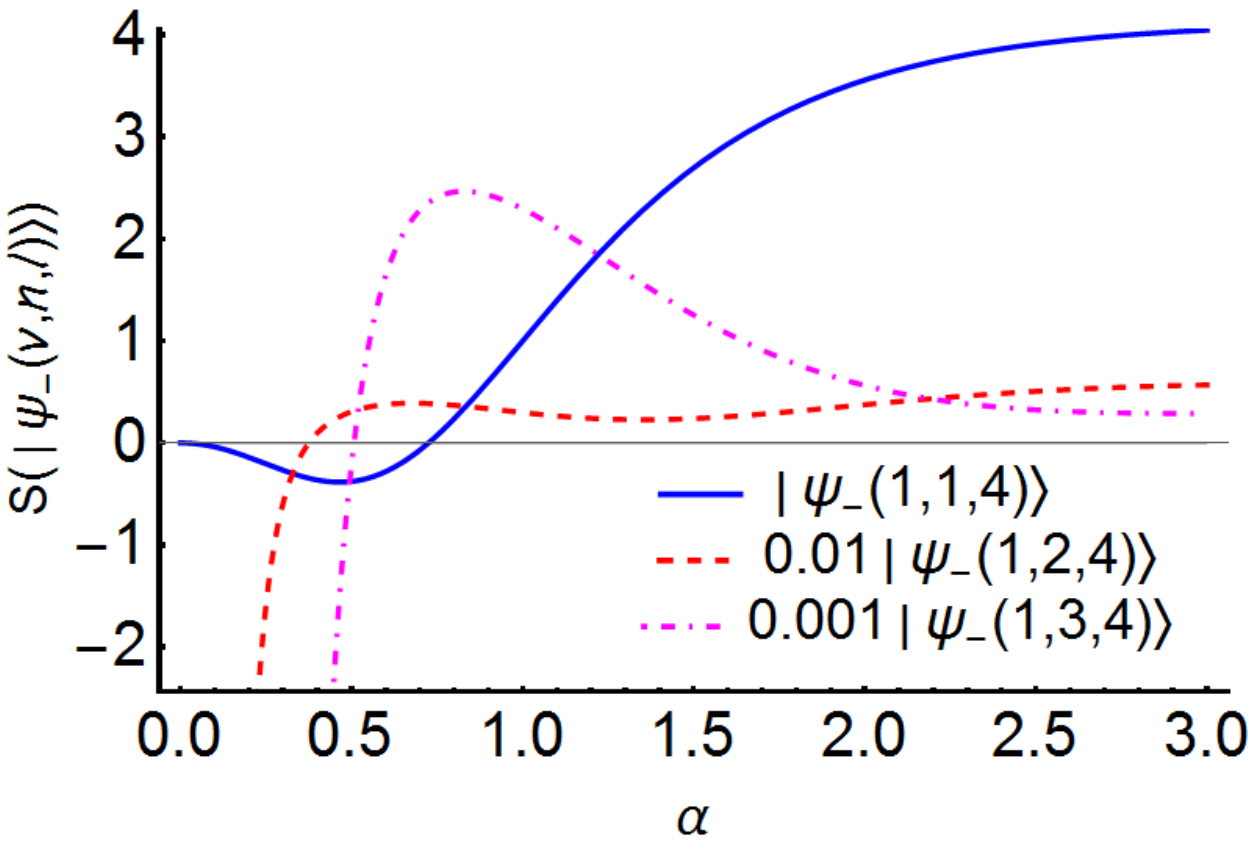} \tabularnewline
(d)  & (e)  & (f) \tabularnewline
\end{tabular}
\par\end{centering}
\caption{\label{fig:HOS} Illustration of the higher-order squeezing using
Hong-Mandel criterion as the function of displacement parameter. In
(a) and (d), dependence of the observed nonclassicality on different
orders ($l$) is shown for PADFS and PSDFS, respectively; while in
(b) and (e), the effect of variation in the number of photon added/subtracted
is shown in case of PADFS and PSDFS, respectively. In (c) and (f),
the variation due to change in the initial Fock state chosen to be
displaced is shown for PADFS and PSDFS, respectively. }
\end{figure}

\begin{figure}
\begin{centering}
\begin{tabular}{cc}
\includegraphics[width=60mm]{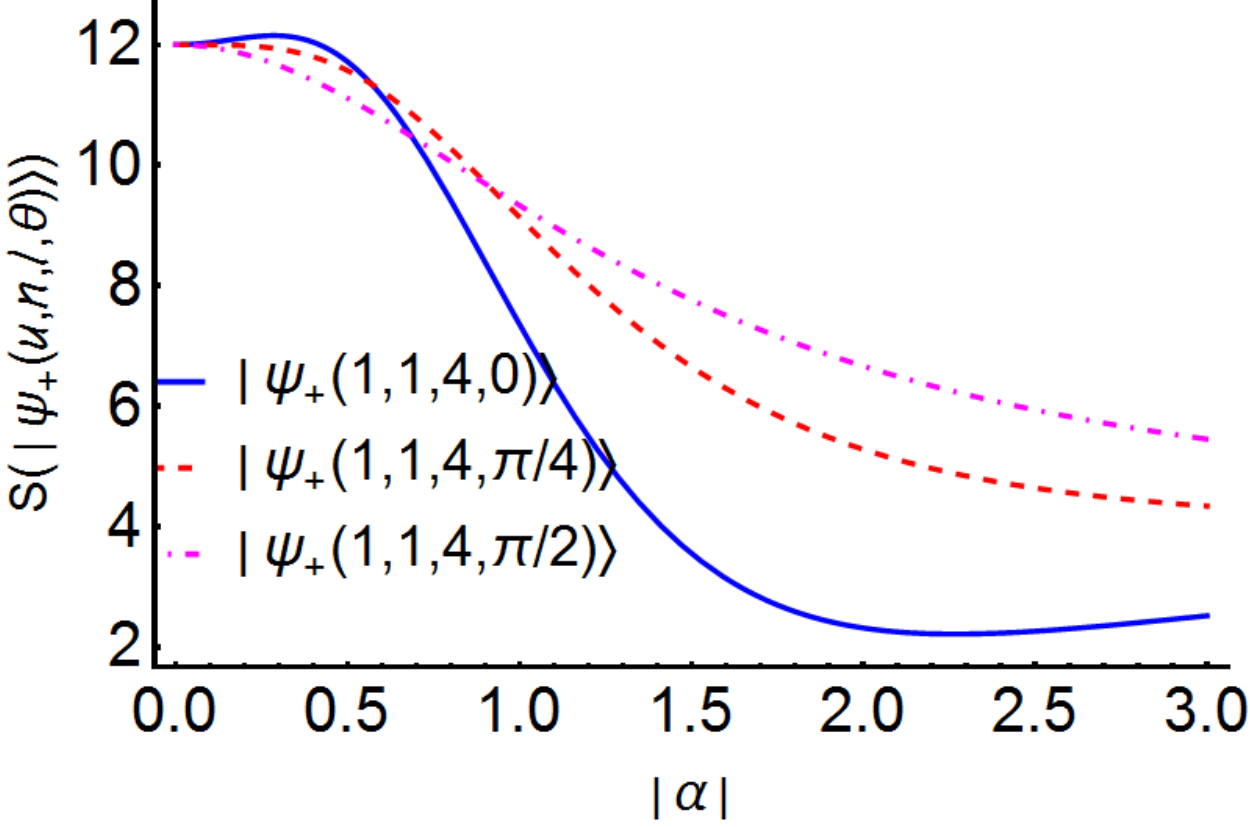}  & \includegraphics[width=60mm]{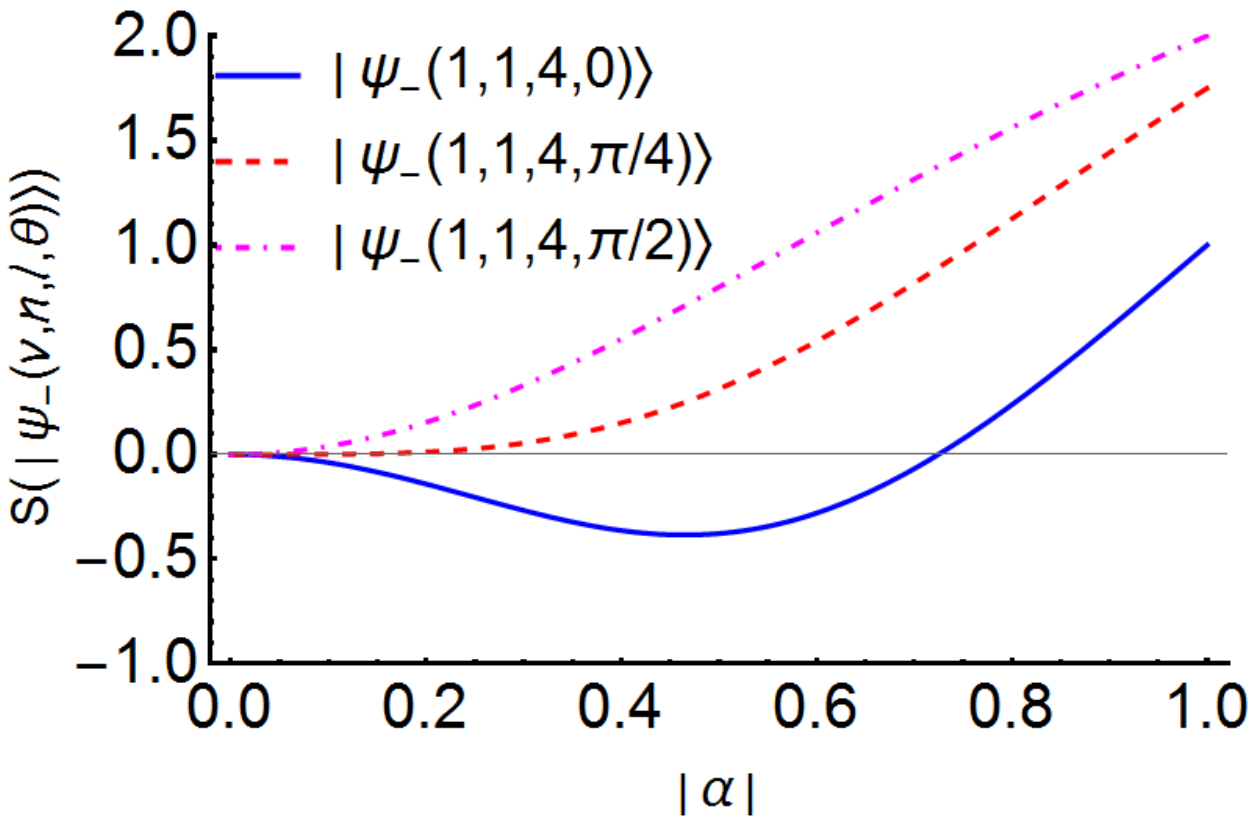} \tabularnewline
(a)  & (b) \tabularnewline
\end{tabular}
\par\end{centering}
\caption{\label{fig:HOS-diff-phase} Hong-Mandel type higher-order squeezing
for PADFS and PSDFS is shown dependent on the phase of the displacement
parameter $\alpha=|\alpha|\exp\left(i\theta\right)$ in (a) and (b),
respectively.}
\end{figure}

The analytical expressions of the nonclassicality witness of the Hong-Mandel
type higher-order squeezing criterion for PADFS and PSDFS can be obtained
with the help of Eqs. (\ref{eq:PA-expectation}), (\ref{eq:PS-expectation}),
and (\ref{eq:Hong-Def})-(\ref{eq:cond2.1-1}). We have investigated
the higher-order squeezing and depict the result in Figure \ref{fig:HOS}
assuming $\alpha$ to be real. Incidentally, we could not establish
the presence of higher-order squeezing phenomena in PADFS (Figure
\ref{fig:HOS} (a)-(c)). However, the depth of the higher-order squeezing
witness increases with order for small values of $\alpha$ as shown
in Figure \ref{fig:HOS} (a), while for the higher values of the displacement
parameter, higher-order squeezing disappear much quicker (cf. Figure
\ref{fig:HOS} (d)). With increase in the number of photons subtracted,
the presence of this nonclassicality feature can be maintained for
the higher values of displacement parameter as well (cf. Figure \ref{fig:HOS}
(e)). In general, photon subtraction is a preferred mode for nonclassicality
enhancement as far as this nonclassciality feature is concerned. The
choice of the initial Fock state is also observed to be relevant as
the depth of squeezing parameter can be seen increasing with value
of the Fock parameter for PSDFS in the small displacement parameter
region (shown in Figure \ref{fig:HOS} (f)), where this nonclassical
behavior is also shown to succumb to the higher values of Fock and
displacement parameters. Unlike the other nonclassicalities discussed
so far, the observed squeezing also depends on phase $\theta$ of
the displacement parameter $\alpha=|\alpha|\exp\left(i\theta\right)$
due to the second last term in Eq. (\ref{eq:cond2.1-1}). We failed
to observe this nonclassicality behavior in PADFS even by controlling
the value of the phase parameter (also shown in Figure \ref{fig:HOS-diff-phase}
(a)). For PSDFS, the squeezing disappears for some particular values
of the phase parameter, while the observed squeezing is maximum for
$\theta=n\pi$ (see Figure \ref{fig:HOS-diff-phase} (b)). It thus
establishes the phase parameter of the displacement operator as one
more controlling factor for nonclassicality in these engineered quantum
states.


\subsection{$Q$ function}

\begin{figure}
\begin{centering}
\begin{tabular}{cc}
\includegraphics[width=100mm]{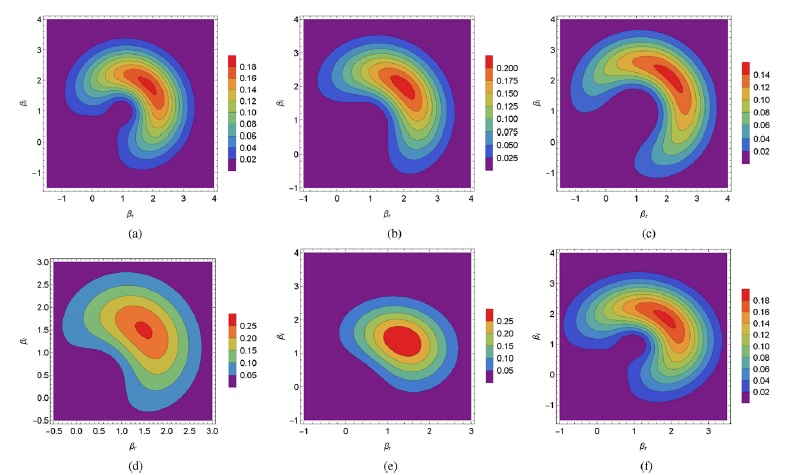} 
\end{tabular}
\par\end{centering}
\caption{\label{fig:Q-function} Contour plots of the $Q$ function for (a)
single photon added displaced Fock $|1\rangle$ state, (b) two photon
added displaced Fock $|1\rangle$ state, (c) single photon added displaced
Fock $|2\rangle$ state, (d) single photon subtracted displaced Fock
$|1\rangle$ state, (e) two photon subtracted subtracted displaced
Fock $|1\rangle$ state, (f) single photon subtracted displaced Fock
$|2\rangle$ state. In all cases, $\alpha=\sqrt{2}\exp\left(\frac{i\pi}{4}\right)$
is chosen. }
\end{figure}

Using Eqs. (\ref{eq:PADFS}) and (\ref{eq:PSDFS}) in (\ref{eq:Q-function-1}),
we obtain the analytic expressions for the Husimi $Q$ function for
PADFS and PSDFS as

\begin{equation}
\begin{array}{lcl}
Q_{+} & = & \frac{N_{+}^{2}}{\pi}\frac{\exp\left[-\mid\beta\mid^{2}\right]}{n!}\sum\limits _{p,p'=0}^{n}{n \choose p}{n \choose p'}(-\alpha^{\star})^{(n-p)}(-\alpha)^{(n-p')}\exp\left[-\mid\alpha\mid^{2}\right]\\
 & \times & \sum\limits _{m,m^{\prime}=0}^{\infty}\frac{\alpha^{m}(\alpha^{\star})^{m^{\prime}}\beta^{(m^{\prime}+p'+u)}(\beta^{\star})^{(m+p+u)}}{m!m^{\prime}!}
\end{array}\label{eq:Q-PADFS}
\end{equation}
and 
\begin{eqnarray}
\begin{array}{lcl}
Q_{-} & = & \frac{N_{-}^{2}}{\pi}\frac{\exp\left[-\mid\beta\mid^{2}\right]}{n!}\sum\limits _{p,p'=0}^{n}{n \choose p}{n \choose p'}(-\alpha^{\star})^{(n-p)}(-\alpha)^{(n-p')}\exp\left[-\mid\alpha\mid^{2}\right]\\
 & \times & \sum\limits _{m,m^{\prime}=0}^{\infty}\frac{\alpha^{m}(\alpha^{\star})^{m^{\prime}}\beta^{(m^{\prime}+p'-v)}(\beta^{\star})^{(m+p-v)}(m+p)!(m^{\prime}+p')!}{m!m^{\prime}!(m+p-v)!(m^{\prime}+p'-v)!},
\end{array}\label{eq:Q-PSDFS}
\end{eqnarray}
respectively. We failed to observe the nonclassical features reflected
beyond moments based nonclassicality criteria through a quasiprobability
distribution, i.e., zeros of the $Q$ function. We have shown the
$Q$ function in Figure \ref{fig:Q-function}, where the effect of
photon addition/subtraction and the value of Fock parameter on the
phase space distribution is shown. Specifically, it is observed that
the value of Fock parameter affects the quasidistribution function
more compared to the photon addition/subtraction.

\subsection{Agarwal-Tara's criterion}

\begin{figure}
\centering{} %
\begin{tabular}{cc}
\includegraphics[width=60mm]{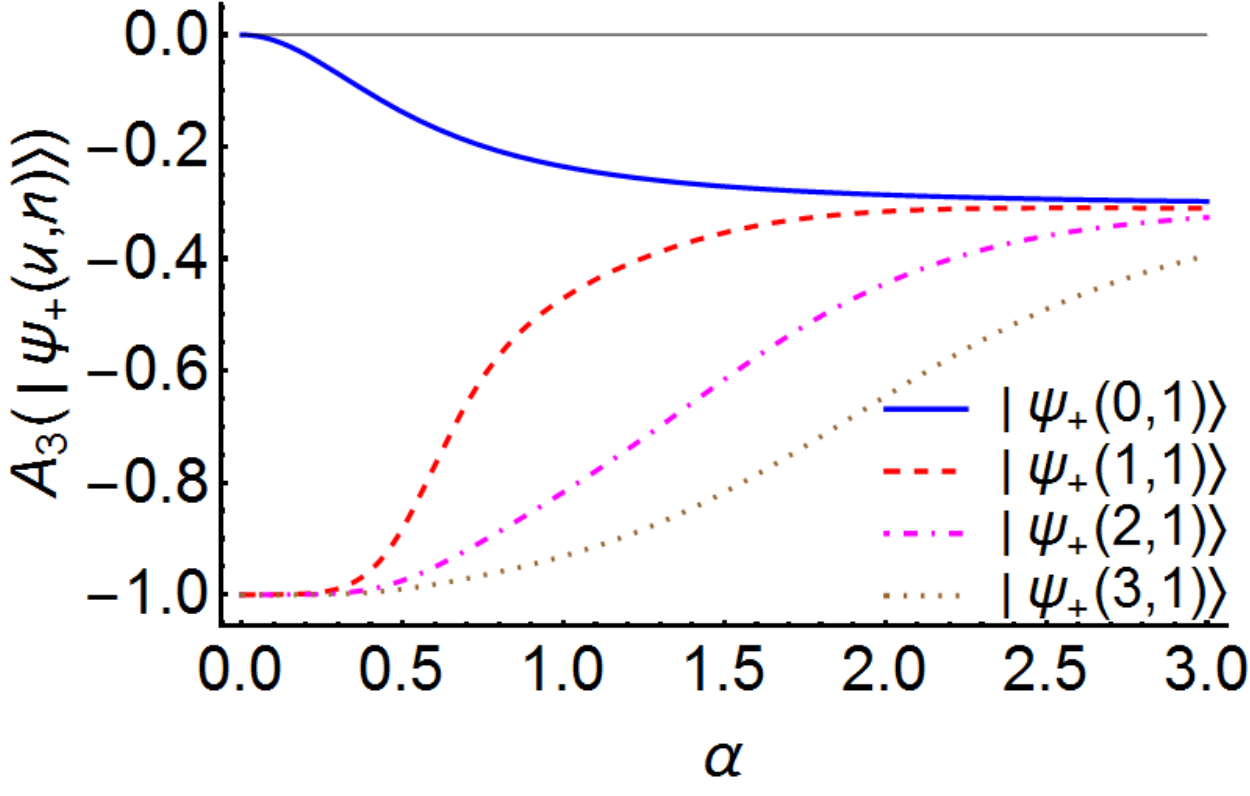}  & \includegraphics[width=60mm]{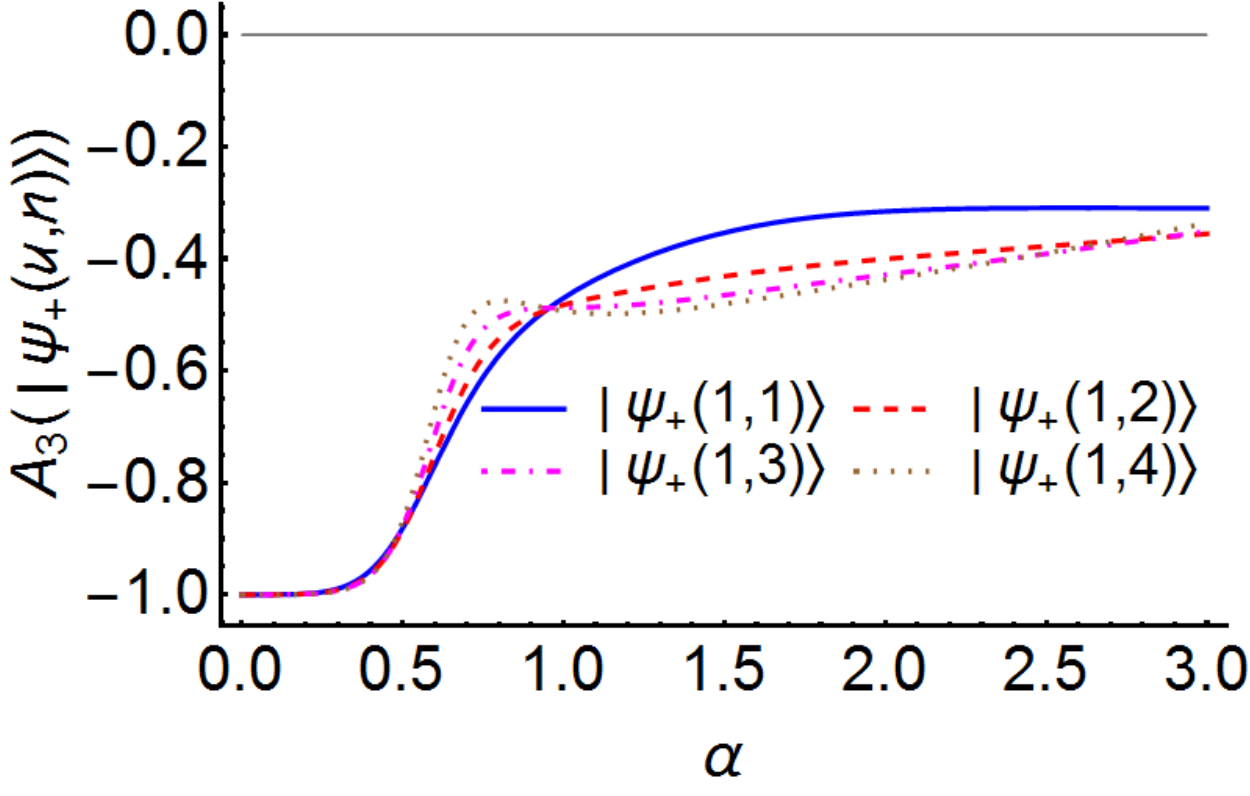} \tabularnewline
(a)  & (b) \tabularnewline
\includegraphics[width=60mm]{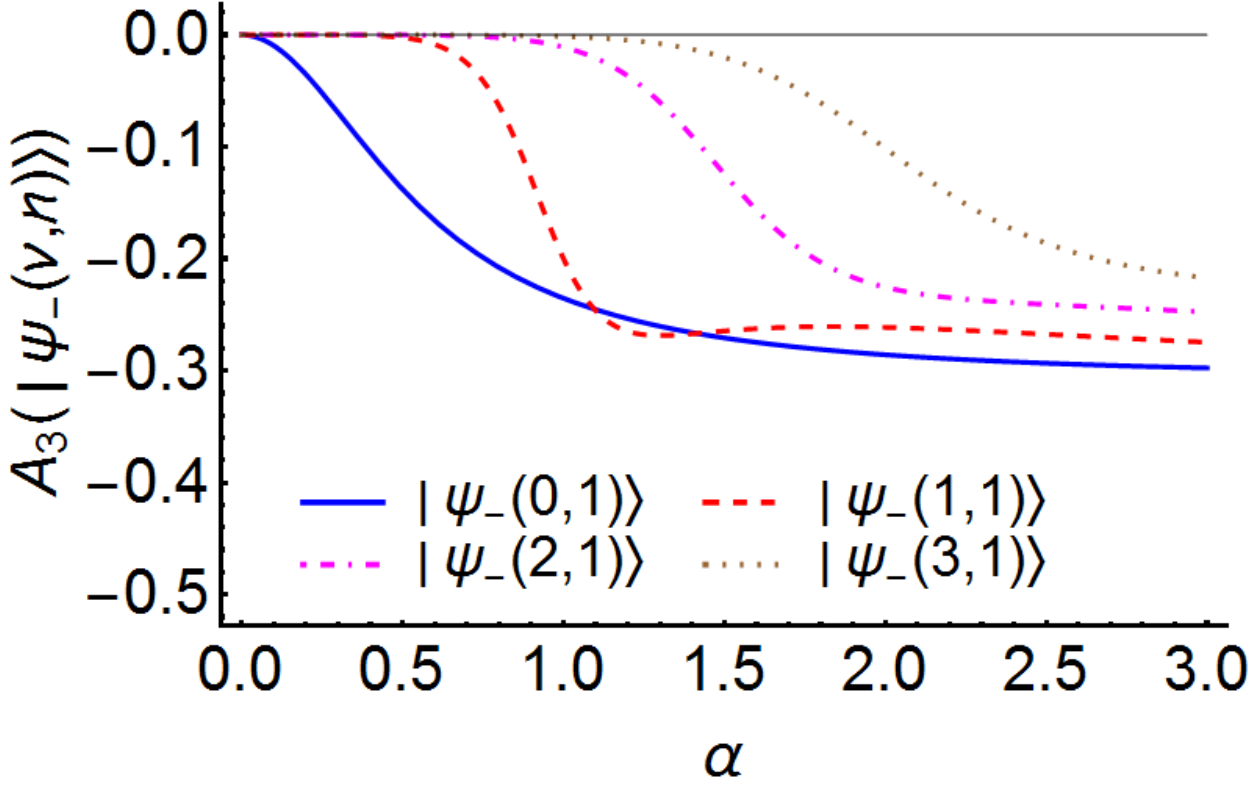}  & \includegraphics[width=60mm]{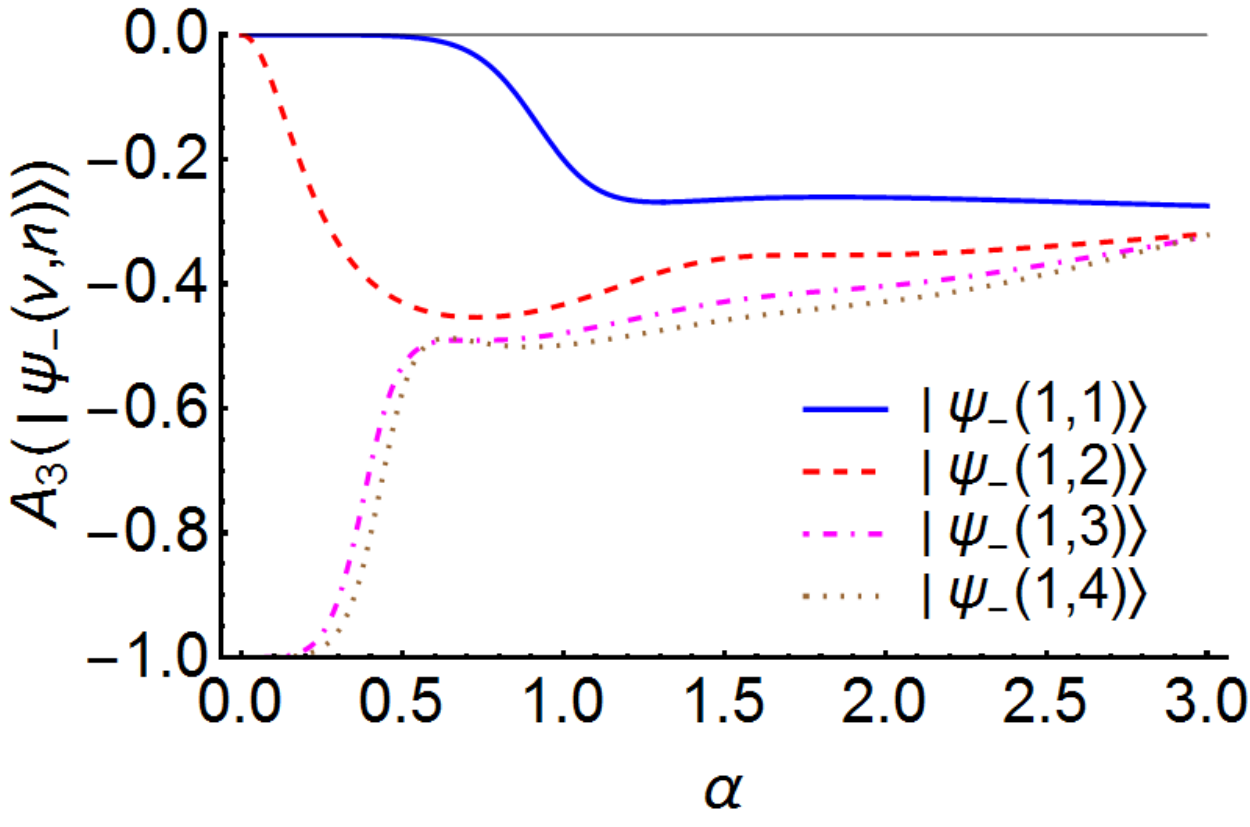} \tabularnewline
(c)  & (d) \tabularnewline
\end{tabular}\caption{\label{fig:A3} Variation of Agarwal-Tara's parameter with $\alpha$
for PADFS and PSDFS is shown in (a)-(b) and (c)-(d), respectively.
Specifically, the effect of photon addition/subtraction (in (a) and
(c)) and the choice of Fock state ((b) and (d)) on the presence of
nonclassicality in PADFS and PSDFS is illustrated.}
\end{figure}

The analytic expression of $A_{3}$ parameter defined in Eq. (\ref{eq:Agarwal-1})
{can} be obtained for PADFS and PSDFS using Eqs. (\ref{eq:PA-expectation})
and (\ref{eq:PS-expectation}). The nonclassical properties of the
PADFS and PSDFS using Agarwal-Tara's criterion are investigated, and
the corresponding results are depicted in Figure \ref{fig:A3}, which
shows highly nonclassical behavior of the {states generated by engineering}.
Specifically, the negative part of the curves, which is bounded by
-1, ensures the existence of the nonclassicality. From Figure \ref{fig:A3},
it is clear that $A_{3}$ is 0 (-1) for the displacement parameter
$\alpha=0$ because then DFS, PADFS, and PSDFS reduce to Fock state,
and $A_{3}=0$ (-1) for the Fock state parameter $n=0,\,1$ $\left(n>1\right)$.
Nonclassicality reflected through $A_{3}$ parameter increases (decreases)
with photon addition (subtraction) (shown in Figure \ref{fig:A3}
(a) and (c)). In contrast, the Fock parameter has a completely opposite
effect that it leads to decrease (increase) in the observed nonclassicality
for PADFS (PSDFS), which can be seen in Figure \ref{fig:A3} (b) and
(d). However, for larger values of displacement parameter, the depth
of nonclassicality illustrated through this parameter can again be
seen to increase (cf. Figure \ref{fig:A3} (b)).

\subsection{Klyshko's Criterion}

The analytic expression for the $m$th photon-number distribution
for PADFS and PSDFS can be calculated (using $q=r=1$) from Eqs. (\ref{eq:PA-expectation})
and (\ref{eq:PS-expectation}), respectively.

\begin{figure}
\centering %
\begin{tabular}{cc}
\includegraphics[width=60mm]{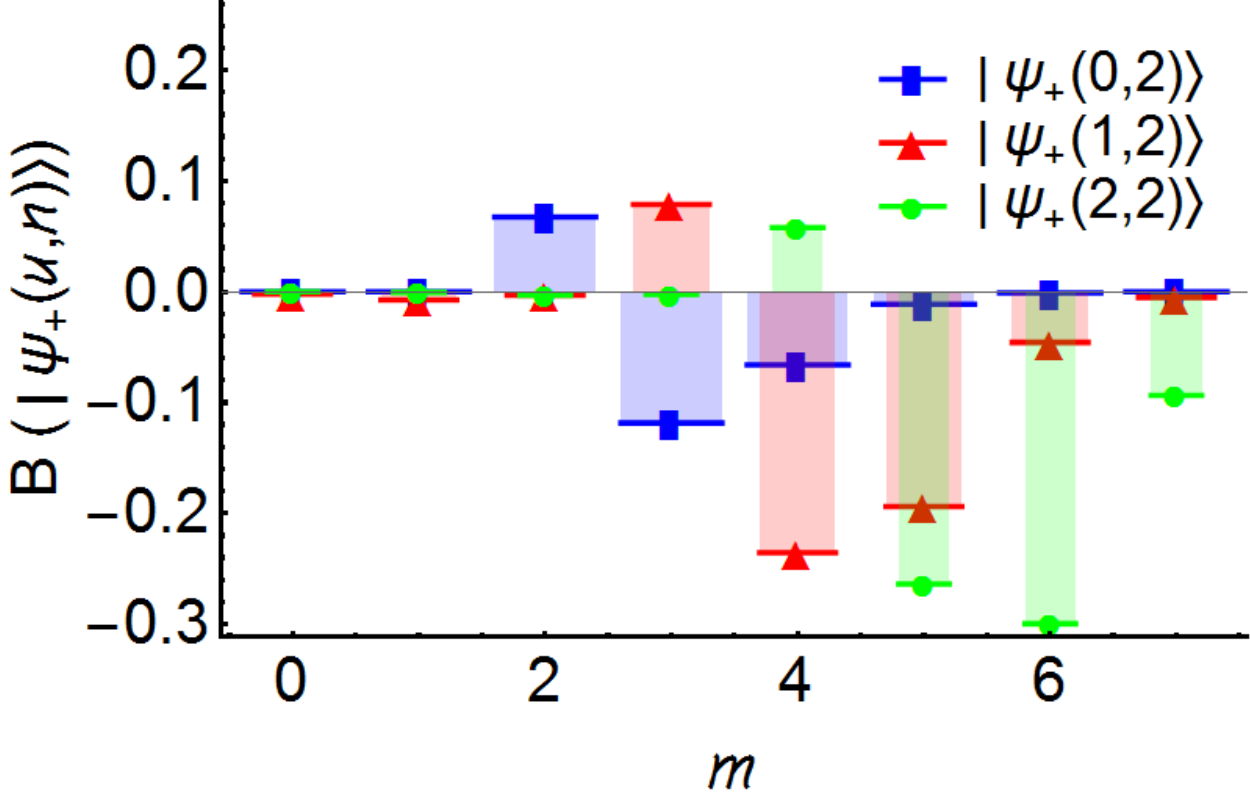}  & \includegraphics[width=60mm]{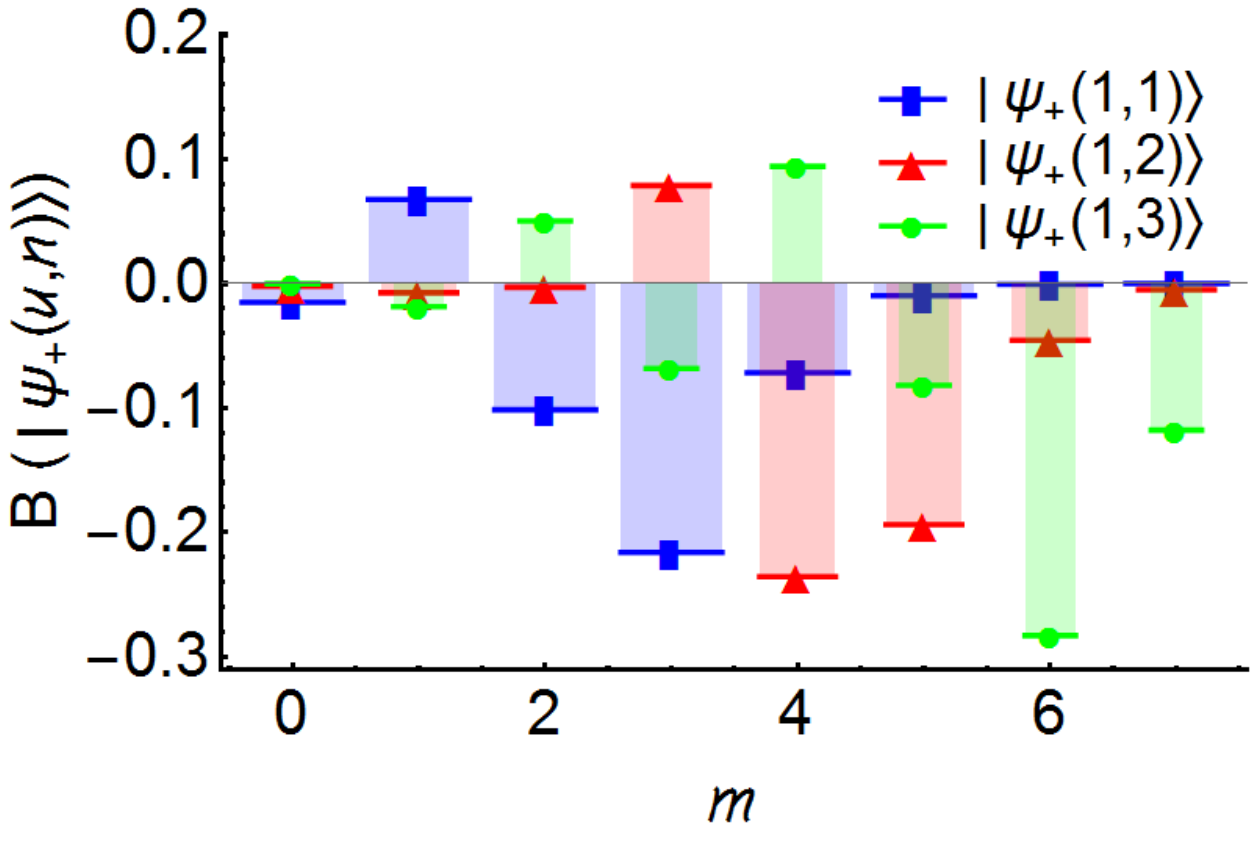} \tabularnewline
(a)  & (b) \tabularnewline
\includegraphics[width=60mm]{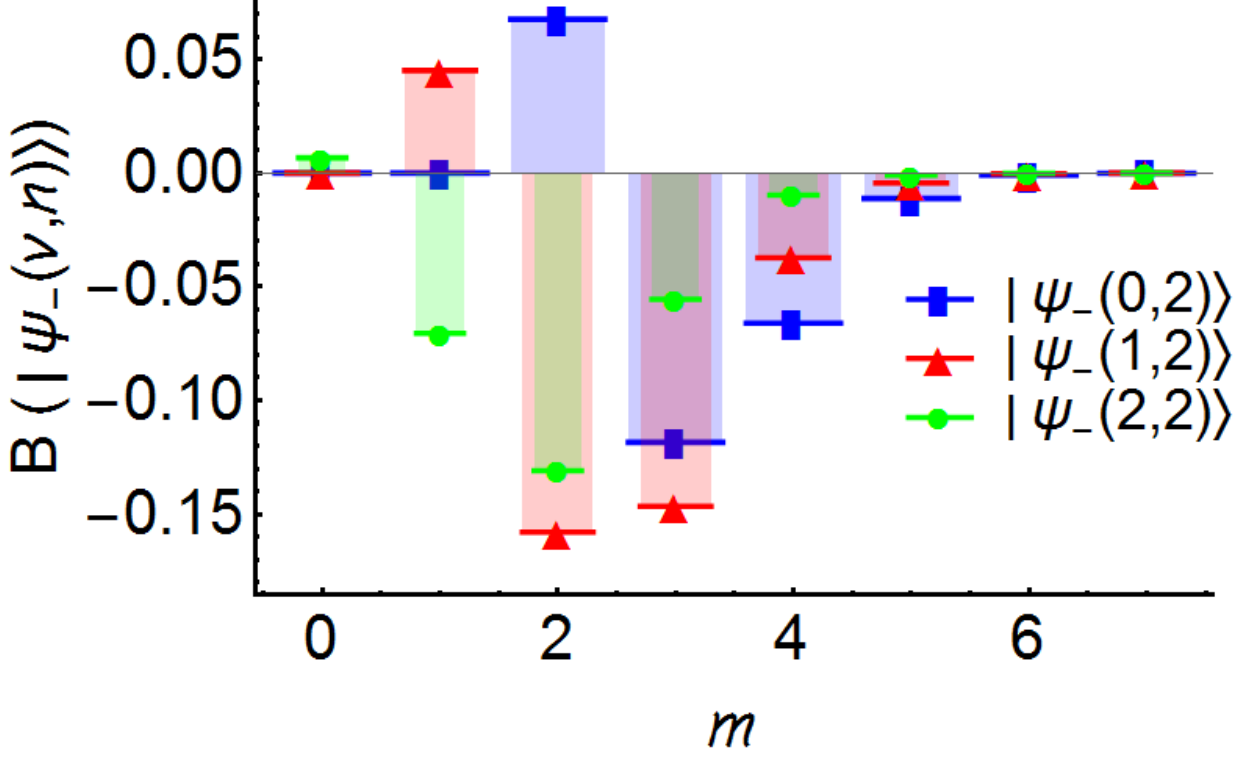}  & \includegraphics[width=60mm]{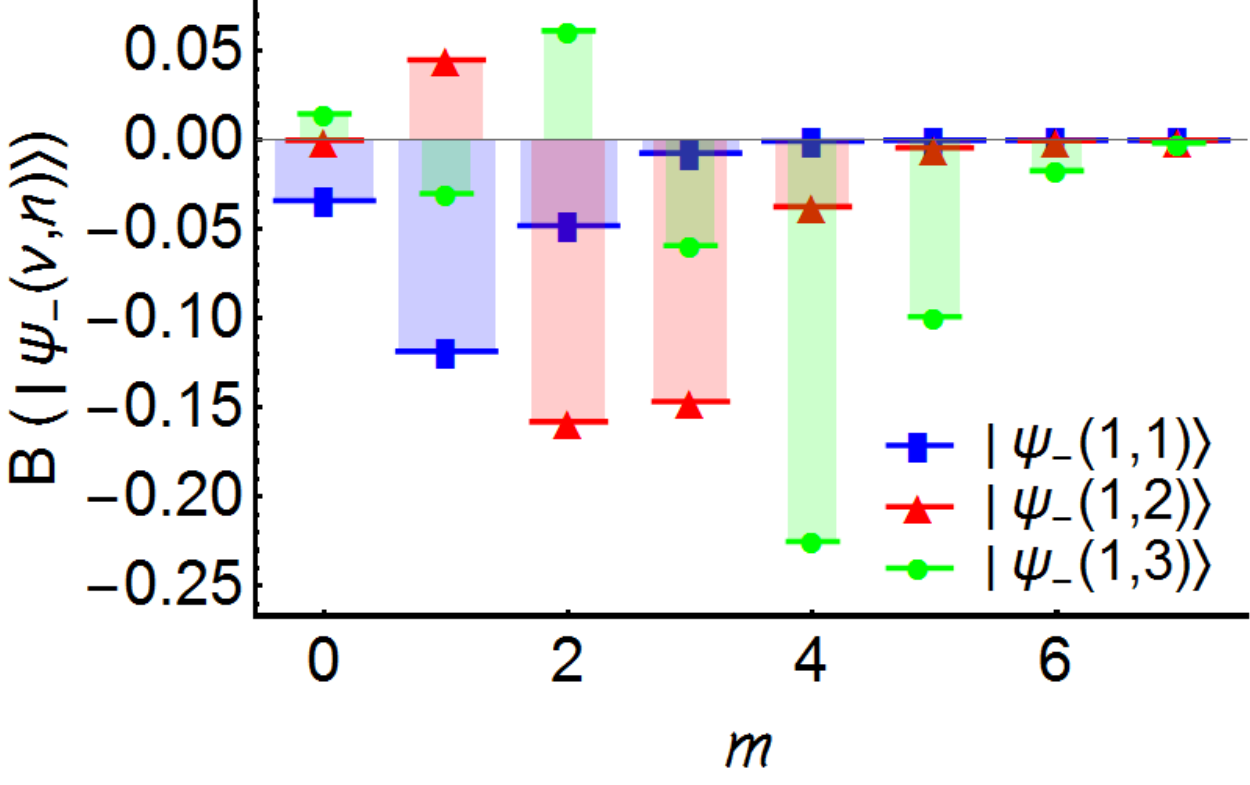}\tabularnewline
(c)  & (d) \tabularnewline
\end{tabular}\caption{\label{fig:Klyshko} Illustration of the Klyshko's criterion. Variation
of $B(m)$ with respect to $m$ (a) and (c) for different values of
the number of the photon additon/subtraction for PADFS and PSDFS,
respectively; (b) and (d) for different values of the number of the
Fock state parameter for PADFS and PSDFS, respectively. Here, we have
chosen $\alpha=1$ in all cases  because almost all
of the other criteria detected nonclassicality for this choice of
$\alpha$ so does the Klyshko's criterion.}
\end{figure}
The advantage of the Klyshko's criterion over any other existing moments
based criteria is that a very small amount of information is required.
Specifically, probability of only three successive photon numbers
is sufficient to investigate the nonclassical property. The negative
values of $B(m)$ serve as the witness of nonclassicality. Klyshko's
criterion in Eq. (\ref{eq:Klyshko-1}) is derived analytically and
the corresponding nonclassical properties for both PADFS and PSDFS
are investigated (cf. Figure \ref{fig:Klyshko}). Specifically, the
negative values of $B(m)$ are observed for different values of $m$
in case of photon addition and subtraction (cf. Figure \ref{fig:Klyshko}
(a) and (c)) being the signature of nonclassicality induced via independent
operations. Additionally, one can also visualize that due to the photon
addition (subtraction) the negative peaks in the values of $B(m)$
shift to higher (lower) photon number regime. A similar observation
is obtained for different values of the Fock state parameter for the
PADFS and PSDFS, where the negative values of the witness of nonclassicality
get amplified towards higher photon number regime and becomes more
negative, and the corresponding results are shown in Figure \ref{fig:Klyshko}
(b) and (d), respectively. This further establishes the relevance
of operations, like photon addition, subtraction, and starting with
Fock states in inducing nonclassicality in the engineered quantum
states.

\section{Conclusions \label{sec:Conclusions}}

The only Fock state that does not show any nonclassical feature is
the vacuum state \cite{miranowicz2015statistical}, and displacement
operator preserves its classicality. All the rest of the Fock states
are maximally nonclassical and they are shown to remain nonclassical
even after application of a displacement operator. Here, we set ourselves
a task: What happens when the displacement operator applied on a Fock
state is followed by addition or subtraction of photon(s)? Independently,
photon addition and subtraction are already established as nonclassicality
inducing operations in case of displaced vacuum state (i.e., coherent
state). In this chapter, we have established that photon addition/subtraction
is not only nonclassicality inducing operation, it can also enhance
the nonclassicality present in the DFS. It's expected that these operations
would increase the amount of nonclassicality (more precisely, the
depth of the nonclassicality witnessing parameter) present in other
nonclassical states, too. There is one more advantage of studying
the nonclassical features of PADFS and PSDFS. These states can be
reduced to a class of quantum states, most of which are already experimentally
realized and found useful in several applications. Inspired by the
available experimental results, we have also proposed optical designs
for generation of PADFS and PSDFS from the squeezed vacuum state.

To analyze the nonclassical features of the engineered final states,
i.e., PADFS and PSDFS, we have used a set of moments based criteria,
namely Mandel $Q_{M}$ parameter, Agarwal-Tara's $A_{3}$ parameter,
criteria for higher-order antibunching, sub-Poissonian photon statistics,
and squeezing. In addition, the nonclassical features have been investigated
through Klyshko's criterion and a quasiprobability distribution--$Q$
function. The states of interest are found to show a large variety
of nonclassical features as all the nonclassicality witnesses used
here (except $Q$ function) are found to detect the nonclassicality.
 They show that state is useful as squeezed, antibunched,
as well as generation of entangled state.

This study has revealed that the amount of nonclassicality in PADFS
and PSDFS can be controlled by the Fock state parameter, displacement
parameter, the number of photons added or subtracted. In general,
the amount of nonclassicality with respect to the witness used here
is found to increase with the number of photons added/subtracted,
while smaller values of Fock state and displacement parameters are
observed to be preferable for the presence of various nonclassical
features. On some occasions, nonclassicality has also been observed
to increase with the Fock state parameter, while larger values of
displacement parameter always affect the nonclassicality adversely.
Most of the nonclassicality criteria used here, being moments-based
criteria, could not demonstrate the effect of phase parameter of the
displacement parameter. Here, higher-order squeezing witness and $Q$
function are found to be dependent on the phase of the displacement
parameter. However, only higher-order squeezing criterion was able
to detect nonclassicality, and thus established that this phase parameter
can also be used to control the amount of nonclassicality.

Further, in the past, it has been established that higher-order nonclassicality
criteria have an advantage in detecting weaker nonclassicality. We
have also shown that the depth of nonclassicality witness increases
with order of nonclassicality thus providing an advantage in the experimental
characterization of the observed nonclassical behavior. 
%
%
%
%
%
%
\chapter{Quantum phase properties of photon added and subtracted displaced
Fock states\textsc{{\label{cha:phase}}}}

In this chapter, the motive is to observe the phase properties of
PADFS and PSDFS.  The main findings of this chapter
are published in \cite{malpani2019quantum}

\section{{Introduction\label{sec:Introduction-chap3}}}

In the previous chapter, the nonclassical properties of PADFS and
PSDFS were studied. Here, our specific interest is to study the phase
properties of PADFS and PSDFS and their limiting cases. In the recent
past, the nonclassical properties of this set of engineered quantum
states, many of which have been experimentally generated \cite{lvovsky2001quantum,lvovsky2002synthesis,zavatta2004quantum,zavatta2005single,zavatta2008subtracting},
were focus of various studies (see \cite{malpani2019lower} and references
therein). In Section \ref{subsec:Photon-added-and sub}, we have already
expressed PADFS and PSDFS as superposition of Fock states. Further,
in Section \ref{sec:Analytic-tools-forphase} we have described the
parameters used for the study of phase properties of a quantum state.
In that context we have already mentioned several applications of
quantum phase distribution and quantum phase fluctuation.

To stress on the recently reported applications of quantum phase distribution
and quantum phase fluctuation, we note that these have applications
in quantum random number generation \cite{xu2012ultrafast,raffaelli2018soi},
cryptanalysis of squeezed state based continuous variable quantum
cryptography \cite{horak2004role}, generation of solitons in a Bose-Einstein
condensate \cite{denschlag2000generating}, in phase encoding quantum
cryptography \cite{gisin2002quantum}, phase imaging of cells and
tissues for biomedical application \cite{park2018quantitative}; as
well as have importance in determining the value of transition temperature
for superconductors \cite{emery1995importance}. Keeping these applications
and the general nature of engineered quantum states PADFS and PSDFS
in mind, in what follows, we aim to study phase distribution, $Q$
phase, phase fluctuation measures, phase dispersion, and quantum phase
estimation using the concerned states and the states obtained in the
limiting cases. As PADFS and PSDFS are already described, we may begin
this study by describing limiting cases of these states as our states
of interest.

We have already mentioned that our focus would be on PADFS and PSDFS.
Due to the general form of PADFS and PSDFS, a large number of states
can be obtained in the limiting cases. Some of the important limiting
cases of PADFS and PSDFS in the present notation are summarized in
Table \ref{tab:state}. This table clearly establishes that the applicability
of the results obtained in the present study is not restricted to
PADFS and PSDFS; rather an investigation of the phase properties of
PADFS and PSDFS would also reveal phase properties of many other quantum
states of particular interest.

\begin{table}
\begin{centering}
\begin{tabular}{c>{\centering}p{3cm}c>{\centering}p{3cm}}
\hline 
Reduction of state  & Name of the state  & Reduction of state  & Name of the state\tabularnewline
\hline 
\hline 
$|\psi_{+}(u,n,\alpha)\rangle$  & $u$-PADFS  & $|\psi_{-}(v,n,\alpha)\rangle$  & $v$-PSDFS\tabularnewline
\textbar$\psi_{+}(0,n,\alpha)\rangle${}  & DFS  & \textbar$\psi_{-}(0,n,\alpha)\rangle${}  & DFS\tabularnewline
$|\psi_{+}(0,0,\alpha)\rangle$  & Coherent state  & $|\psi_{-}(0,0,\alpha)\rangle$  & Coherent state\tabularnewline
\textbar$\psi_{+}(0,n,0)\rangle${}  & Fock state  & \textbar$\psi_{-}(0,n,0)\rangle${}  & Fock state\tabularnewline
\textbar$\psi_{+}(u,0,\alpha)\rangle${}  & $u$-Photon added coherent state  & \textbar$\psi_{-}(v,0,\alpha)\rangle${}  & $v$-Photon subtracted coherent state \tabularnewline
\hline 
\end{tabular}
\par\end{centering}
\caption{\label{tab:state}Various states that can be obtained as the limiting
cases of the PADFS and PSDFS.}
\end{table}

\section{Quantum phase distribution and other phase properties \label{sec:Quantum-phase-parameters}}

Quantum phase operator $\hat{\phi}$ was introduced by Dirac based
on his assumption that the annihilation operator $\hat{a}$ can be
factored out into a Hermitian function $f(\hat{N})$ of the number
operator $\hat{N}=\hat{a}^{\dagger}\hat{a}$ and a unitary operator
$\hat{U}$ \cite{dirac1927quantum} as 
\begin{equation}
\hat{a}=\hat{U}\,f\left(\hat{N}\right),\label{eq:Dirac_pahse}
\end{equation}
where 
\begin{equation}
\hat{U}=e^{\iota\hat{\phi}}.\label{eq:phase-operator}
\end{equation}
However, there was a problem with the Dirac formalism of phase operator
as it failed to provide a meaning to the corresponding uncertainty
relation. Specifically, in the Dirac formalism, the creation ($\hat{a}^{\dagger}$)
and annihilation ($\hat{a}$) operators satisfy the bosonic commutation
relation, $\left[\hat{a},\,\hat{a}^{\dagger}\right]=1$, iff $\left[\hat{N},\,\hat{\phi}\right]=\iota$,
which leads to the number phase uncertainty relation $\Delta N\,\Delta\phi\geq1$.
Therefore, in order to satisfy the bosonic commutation relation under
Dirac formalism, the phase uncertainty should be greater than 2$\pi$
for $\Delta N$ \textless{} $\frac{1}{2\pi}$ which lacks a physical
description. Subsequently, Louisell \cite{louisell1963amplitude}
proposed some periodic phase based method, which was followed by Susskind
and Glogower formalism based on Sine and Cosine operators \cite{susskind1964quantum}.
An important contribution to this problem is the Barnett-Pegg formalism
\cite{barnett1986phase} which is used in this thesis. In what follows,
we will also briefly introduce notions, such as quantum phase distribution,
angular $Q$ phase function, phase fluctuation parameters, phase dispersion,
quantum phase estimation to study the phase properties of the quantum
states of our interest.

\section{Phase properties of PADFS and PSDFS \label{sec:phase-witnesses}}

The description of the states of our interest given in the previous
section can be used to study different phase properties and quantify
phase fluctuation in the set of quantum states listed in Table \ref{tab:state}.
Specifically, with the help of the quantum states defined in Eqs.
(\ref{eq:PADFS})-(\ref{eq:PSDFS}), we have obtained the analytic
expressions of phase distribution and other phase parameters defined
in Section \ref{sec:Quantum-phase-parameters}.

\subsection{Phase distribution function}

\begin{figure}
\centering{} \centering{} %
\begin{tabular}{cc}
\includegraphics[width=60mm]{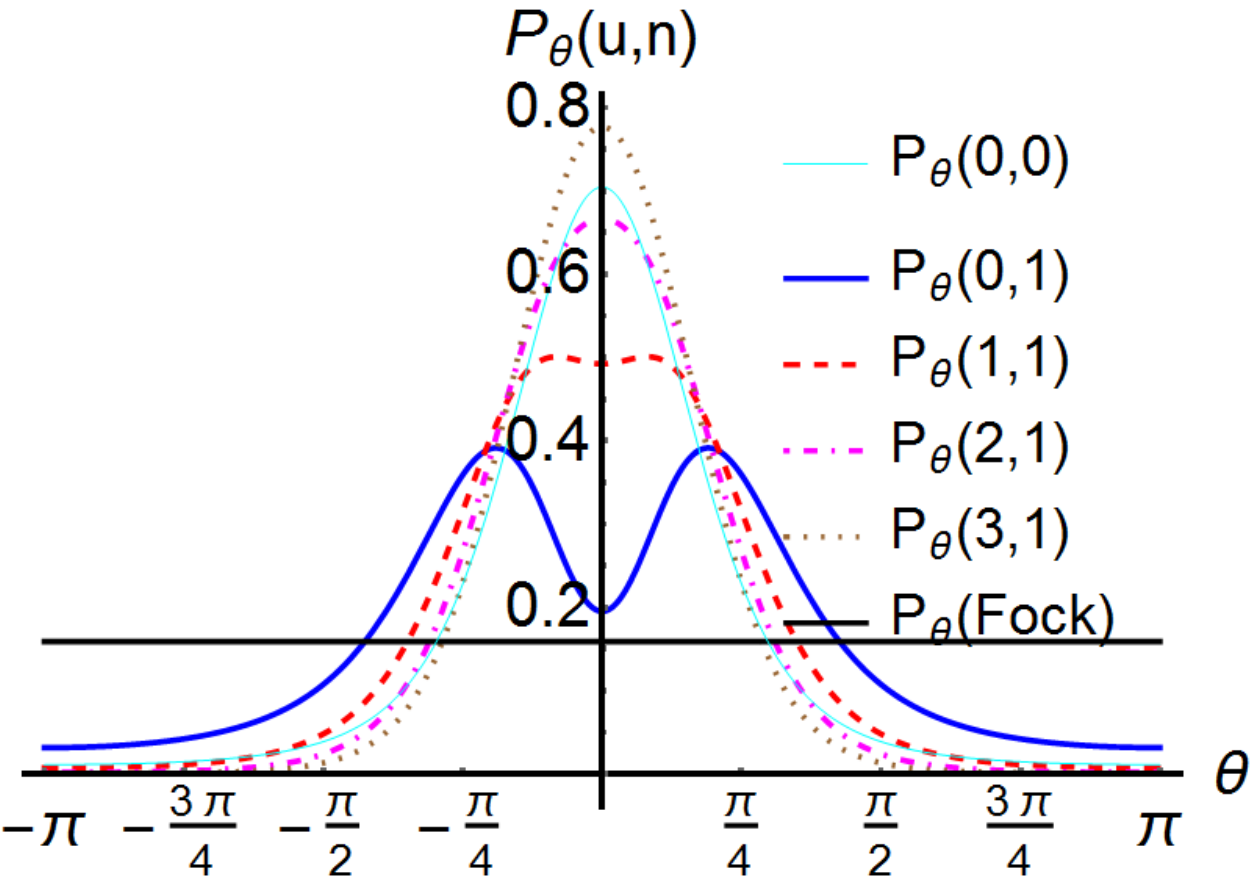}  & \includegraphics[width=60mm]{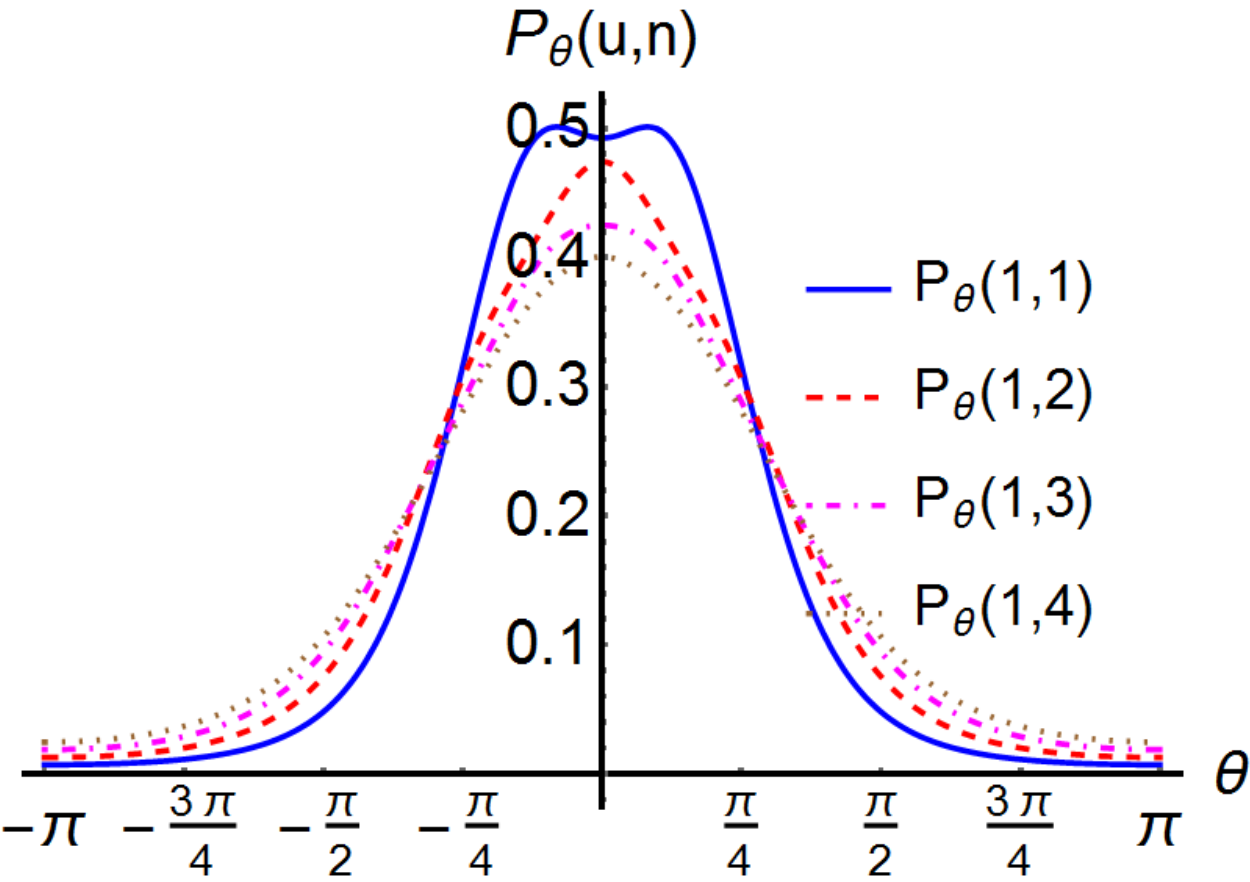} \tabularnewline
(a)  & (b) \tabularnewline
\includegraphics[width=60mm]{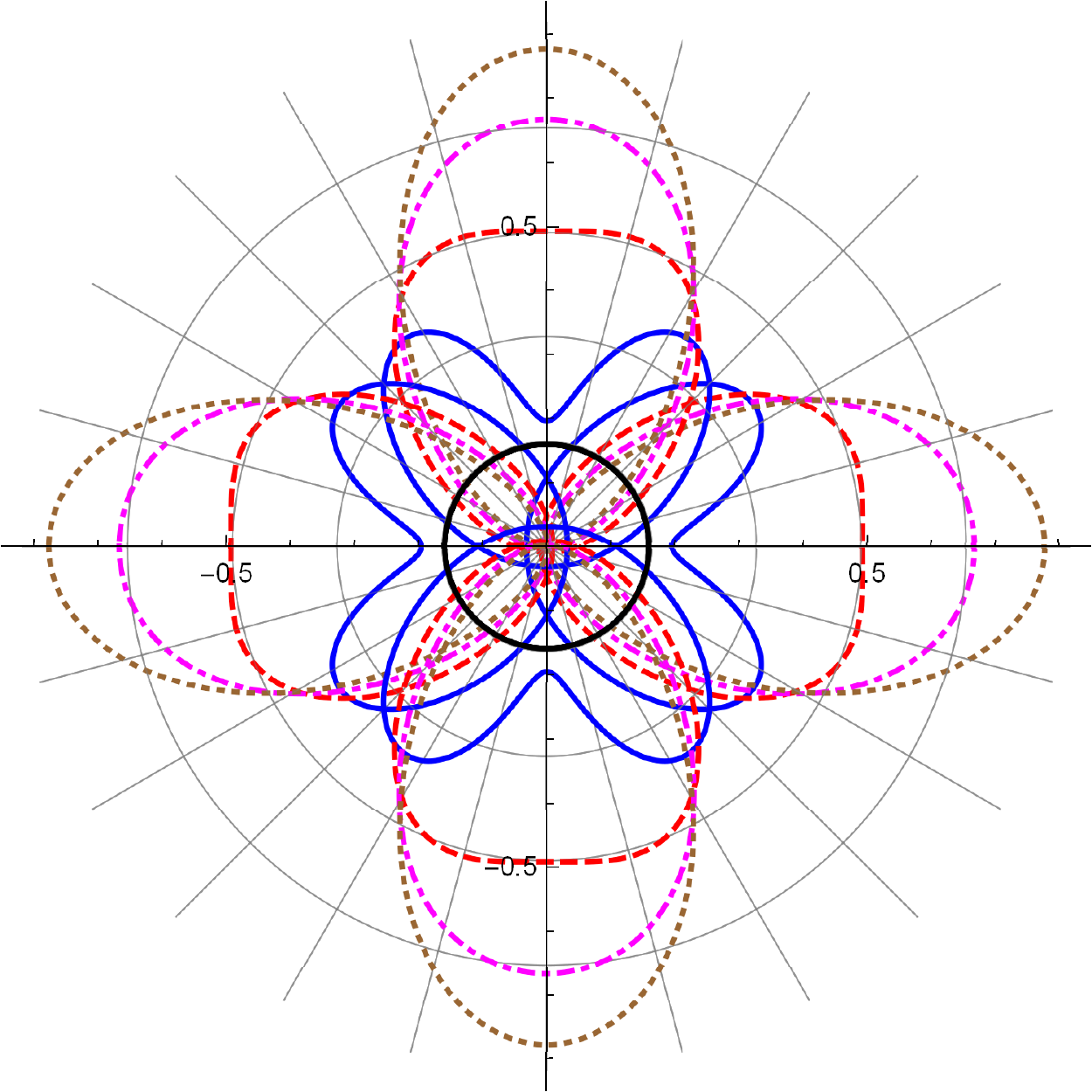}  & \includegraphics[width=60mm]{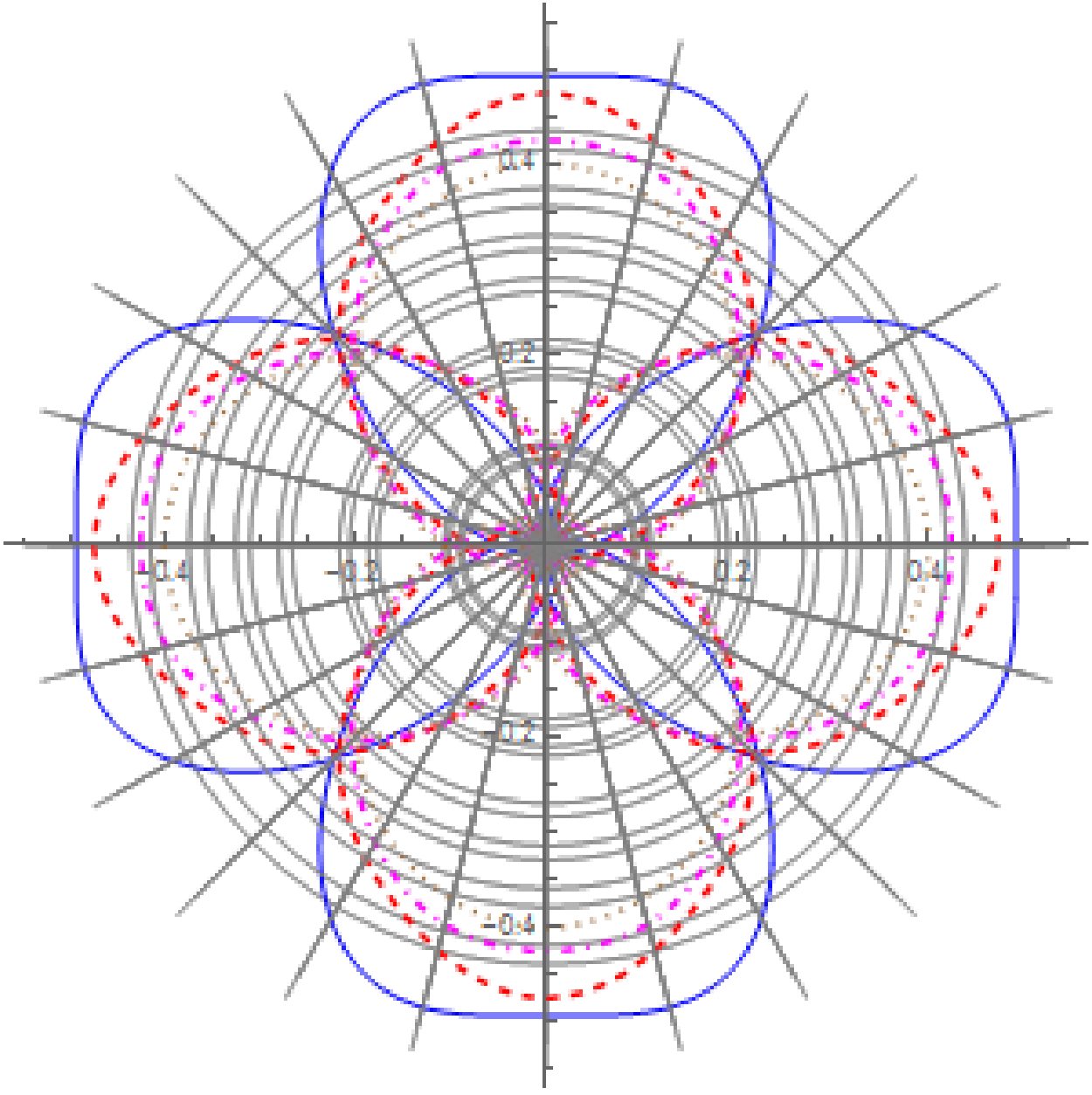}\tabularnewline
(c)  & (d) \tabularnewline
\end{tabular}\caption{\label{fig:Phase-Distribution-Function} Variation of phase distribution
function with phase parameter for PADFS with displacement parameter
$\left|\alpha\right|=1$ for different values of photon addition ((a)
and (c)) and Fock parameters ((b) and (d)). The phase distribution
is shown using both two-dimensional ((a) and (b) with $\theta_{2}=0$)
and polar ((c) and (d)) plots. In (c) and (d), $\theta_{2}=\frac{n\pi}{2}$
with integer $n\in\left[0,3\right]$, and the legends are same as
in (a) and (b), respectively. }
\end{figure}

\begin{figure}
\centering{} %
\begin{tabular}{cc}
\includegraphics[width=60mm]{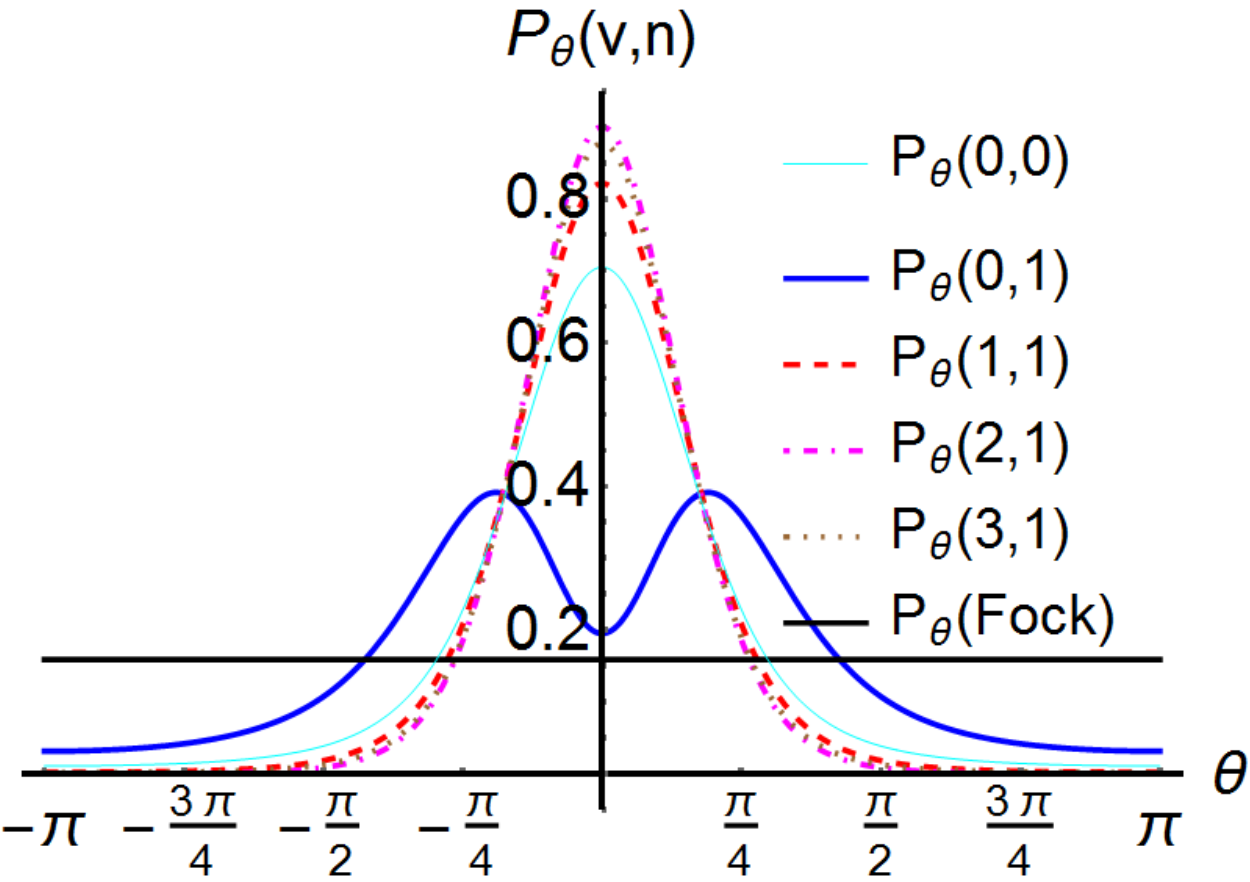}  & \includegraphics[width=60mm]{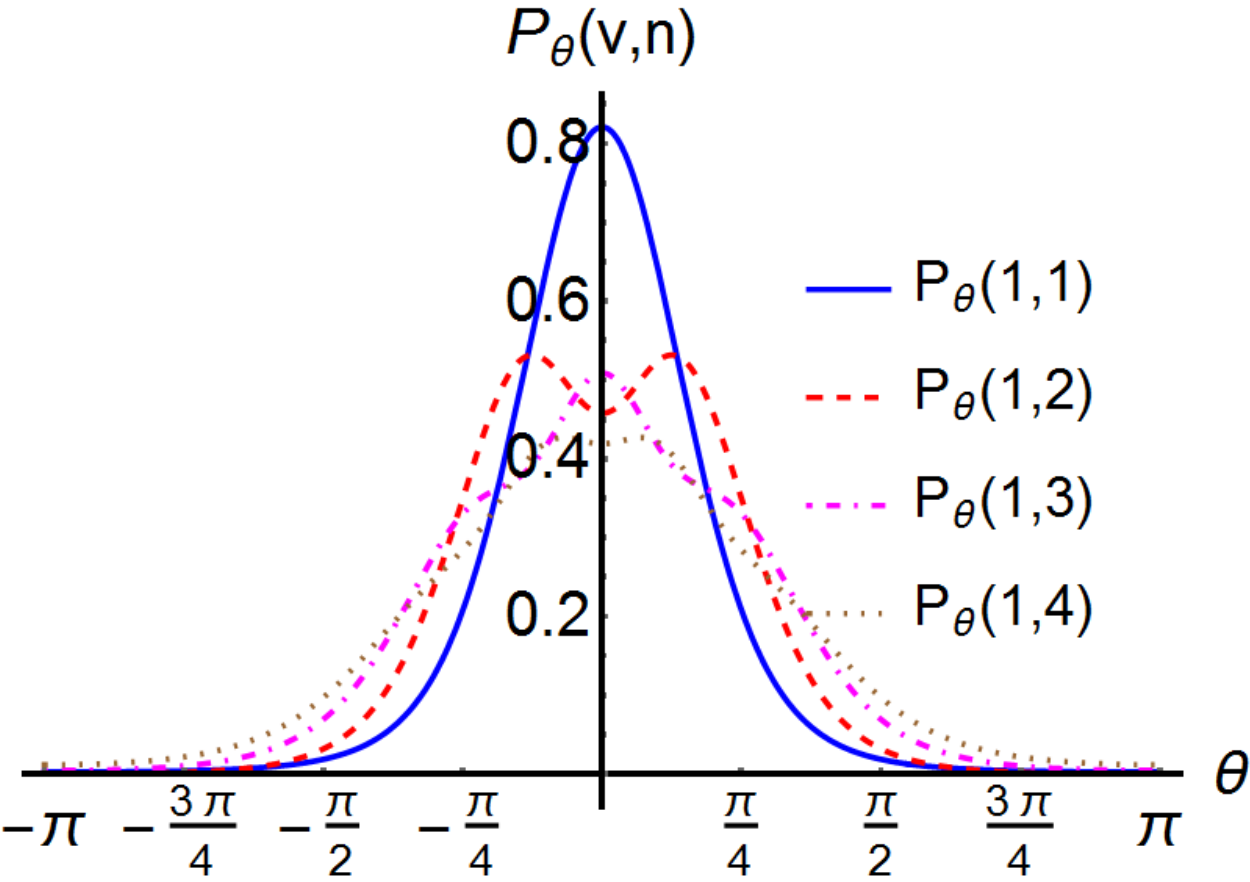} \tabularnewline
(a)  & (b) \tabularnewline
\includegraphics[width=60mm]{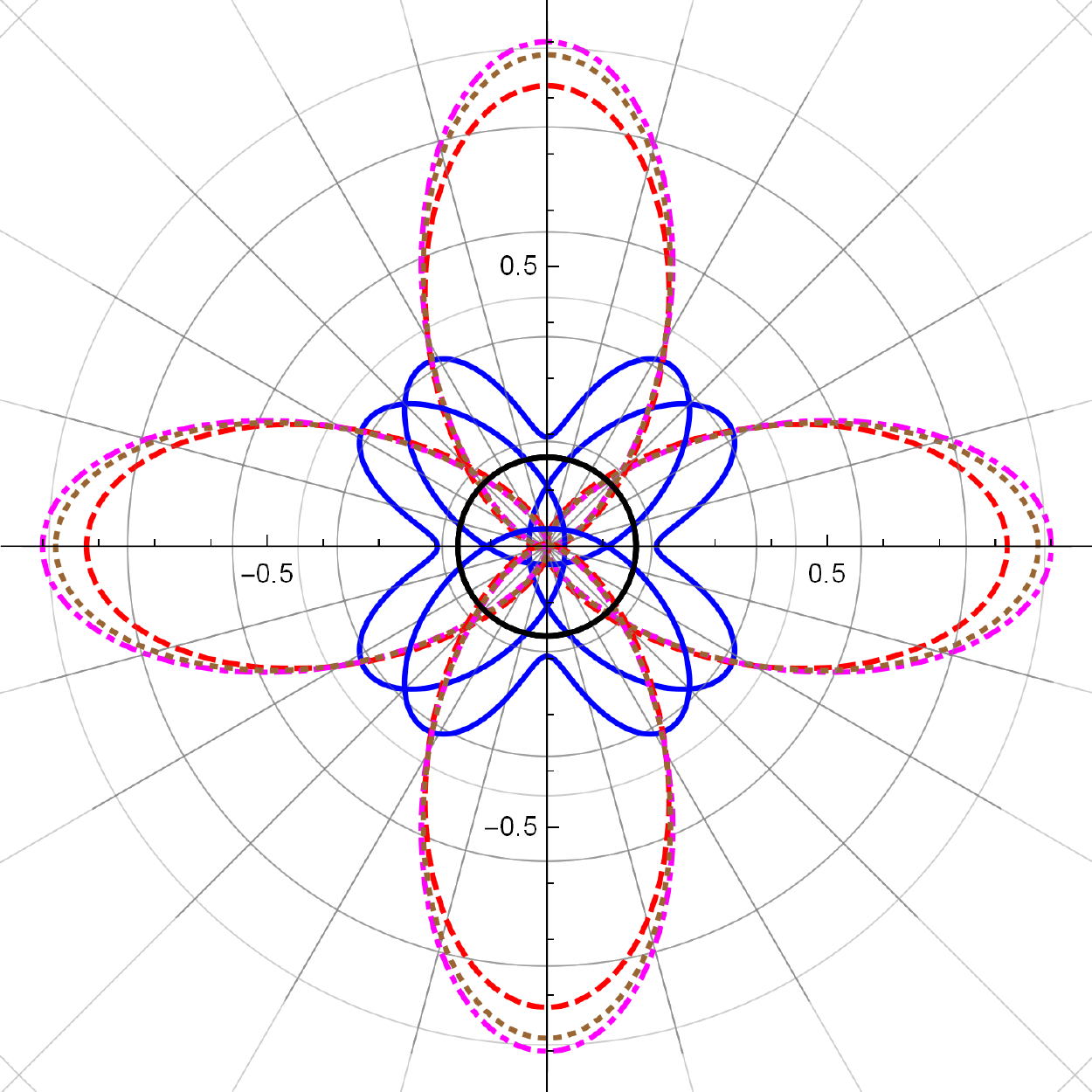}  & \includegraphics[width=60mm]{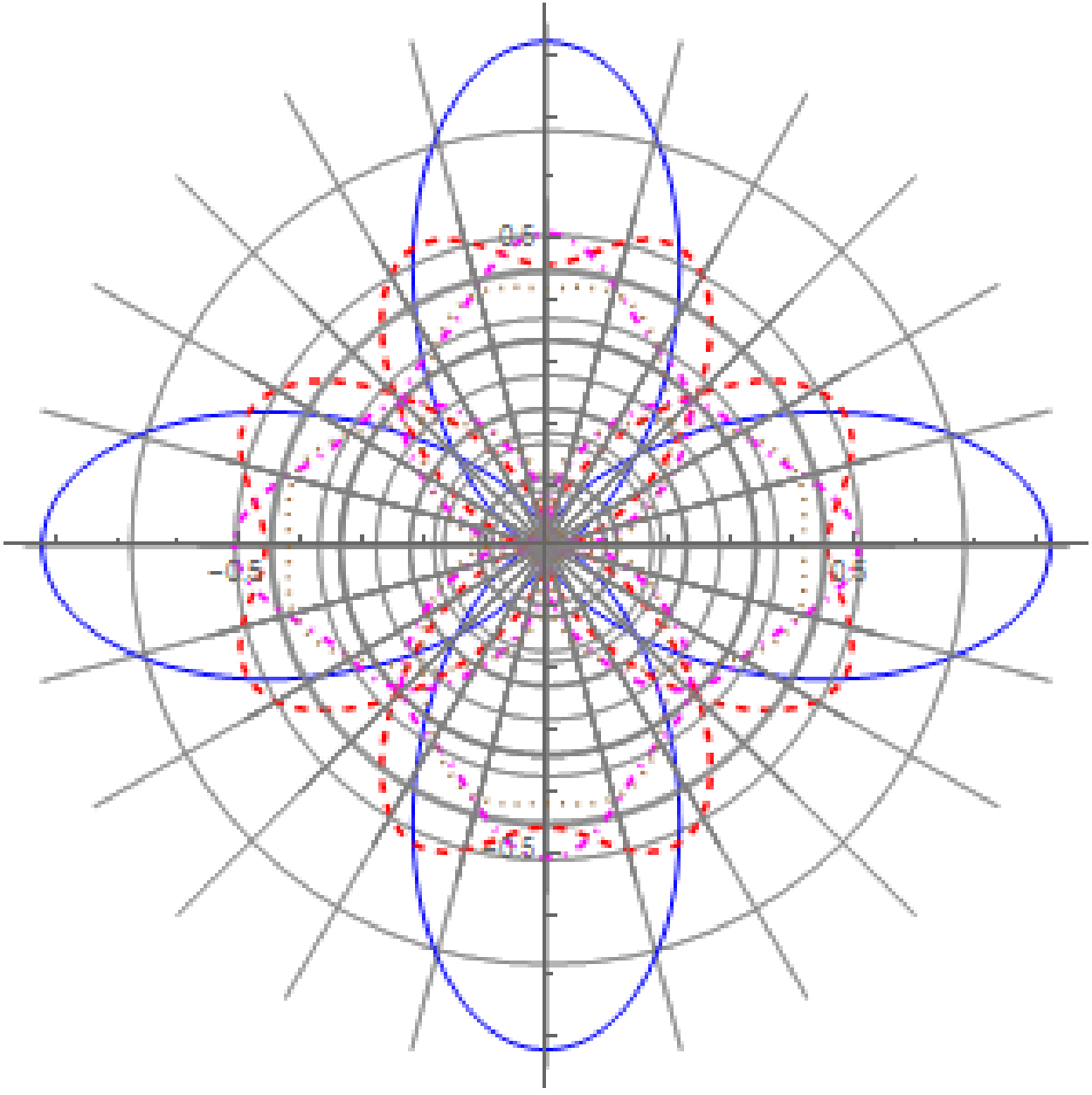}\tabularnewline
(c)  & (d) \tabularnewline
\end{tabular}\caption{\label{fig:Phase-Distribution-Function-1} Variation of phase distribution
function with phase parameter for PSDFS with displacement parameter
$\left|\alpha\right|=1$ for different values of photon subtraction
((a) and (c)) and Fock parameters ((b) and (d)). The phase distribution
is shown using both two-dimensional ((a) and (b) with $\theta_{2}=0$)
and polar ((c) and (d)) plots. In (c) and (d), $\theta_{2}=\frac{n\pi}{2}$
with integer $n\in\left[0,3\right]$, and the legends are same as
in (a) and (b), respectively.}
\end{figure}

From the definition of the phase distribution (\ref{eq:Phase-Distridution-1}),
it can be observed that for a Fock state, $P_{\theta}=\frac{1}{2\pi}$,
implying it has a uniform distribution of phase. Interestingly, the
states of our interest, PADFS and PSDFS, are obtained by displacing
the Fock state followed by photon addition/subtraction. Therefore,
we will study here what is the effect of application of displacement
operator on a uniformly phase distributed (Fock) state and how subsequent
photon addition/subtraction further alters the phase distribution.
Using phase distribution function, the information regarding uncertainty
in phase and phase fluctuation can also be obtained. To begin with,
we compute the analytic expressions of $P_{\theta}$ for the PADFS
and PSDFS, using Eq. (\ref{eq:Phase-Distridution-1}) as 
\begin{eqnarray}
\begin{array}{lcl}
P_{\theta}\left(u,n\right) & = & \frac{1}{2\pi}\dfrac{\left|N_{+}\right|^{2}}{n!}\sum\limits _{p,p^{\prime}=0}^{n}{n \choose p}{n \choose p^{\prime}}\exp\left[-\mid\alpha\mid^{2}\right]\left|\alpha\right|^{2n-p-p^{\prime}}\\
 & \times & \sum\limits _{m,m^{\prime}=0}^{\infty}\frac{(-\left|\alpha\right|)^{m+m^{\prime}}\sqrt{(m+p+u)!(m^{\prime}+p^{\prime}+u)!}}{m!m^{\prime}!}\exp[\iota\left(\theta-\theta_{2}\right)(m^{\prime}+p^{\prime}-m-p)],
\end{array}\label{eq:PA-phase}
\end{eqnarray}
and 
\begin{eqnarray}
\begin{array}{ccc}
P_{\theta}\left(v,n\right) & = & \frac{1}{2\pi}\dfrac{\left|N_{-}\right|^{2}}{n!}\sum\limits _{p,p^{\prime}=0}^{n}{n \choose p}{n \choose p^{\prime}}\exp\left[-\mid\alpha\mid^{2}\right]\left|\alpha\right|^{2n-p-p^{\prime}}\\
 & \times & \sum\limits _{m,m^{\prime}=0}^{\infty}\frac{(-\left|\alpha\right|)^{m+m^{\prime}}(m+p)!(m^{\prime}+p^{\prime})!}{m!m^{\prime}!\sqrt{(m+p-v)!(m^{\prime}+p^{\prime}-v)!}}\exp[\iota\left(\theta-\theta_{2}\right)(m^{\prime}+p^{\prime}-m-p)],
\end{array}\label{eq:PS-phase}
\end{eqnarray}
respectively.  Here, $\theta_2$ is the phase associated with the displacement parameter $\alpha$ ($ \alpha = |\alpha|e^{\iota \theta_2}$). Since the obtained expressions in Eqs. (\ref{eq:PA-phase})
and (\ref{eq:PS-phase}) are complex in nature, we depict numerical
(graphical) analysis of the obtained results in Figs. \ref{fig:Phase-Distribution-Function}
and \ref{fig:Phase-Distribution-Function-1} for PADFS and PSDFS,
respectively. Specifically, in Figure \ref{fig:Phase-Distribution-Function}
(a), we have shown the variation of phase distribution with phase
parameter $\theta$ for different number of photon added in the displaced
single photon Fock state ($D\left(\alpha\right)\left|1\right\rangle $)
for $\theta_{2}=0$. A uniform phase distribution for Fock state (with
a constant value of $\frac{1}{2\pi}$) is found to transform to one
that decreases for higher values of phase and possess a dip in the
phase distribution for $\theta=0$, which can be thought of as an
approach to the Fock state. In fact, in case of classical states,
$P_{\theta}$ has a peak at zero phase difference $\theta-\theta_{2}$,
and therefore, this contrasting behavior can be viewed as signature
of quantumness of DFS. However, with the increase in the number of
photons added to the DFS, the phase distribution of the PADFS is observed
to become narrower. In fact, a similar behavior with increase in the
mean photon number of coherent state was observed previously \cite{agarwal1992classical}.
It is imperative to state that $P_{\theta}$ in case of higher number
of photon added to DFS has similar but narrower distribution than
that of coherent state. In contrast, with increase in the Fock parameter,
the phase distribution is observed to become broader (cf. Figure \ref{fig:Phase-Distribution-Function}
(b)). Thus, the increase in the number of photons added and the increase
in Fock parameter have opposite effects on the phase distribution.
The same is also illustrated through the polar plots in Figure \ref{fig:Phase-Distribution-Function}
(c)-(d), which not only reestablish the same fact, but also illustrate
the dependence of $P_{\theta}$ on the phase of the displacement parameter.
Specifically, the obtained phase distribution remains symmetric along
the value of phase $\theta_{2}$ (i.e., $P_{\theta}$ is observed
to have a mirror symmetry along $\theta=\theta_{2}$) of the displacement
parameter. The phase distribution of Fock state is shown by a black
circle in the polar plot.

Instead of photon addition, if we subtract photons from the DFS, a
similar effect on the phase distribution to that of photon addition
is observed. Further, a comparison between photon addition and subtraction
on the phase distribution establishes that a single photon subtraction
has a prominent impact on phase distribution when compared to that
of single photon addition, i.e., the distribution can be observed
to be narrower than that of coherent state in most of the cases for
$u=v$. For instance, single photon added (subtracted) DFS is broader
(narrower) than corresponding coherent state. Similarly, with the
increase in the value of Fock parameter, we can observe more changes
on PSDFS than what was observed in PADFS, i.e., the phase distribution
broadens more with Fock parameter for PSDFS. Note that $P_{\theta}$
has a peak at $\theta=\theta_{2}$ only for photon addition $u>n$,
while in case of photon subtraction it can be observed for $v\geq n$.
With the increase in the amplitude of displacement parameter ($\left|\alpha\right|$)
initially the phase distribution becomes narrower, which is further
supported by both addition and subtraction of photons, but it becomes
broader again for very high $\left|\alpha\right|$ (figure is not
shown here). 

\subsection{Angular $Q$ function of PADFS and PSDFS}

\begin{figure}
\centering{} %
\begin{tabular}{cc}
\centering{}\includegraphics[width=60mm]{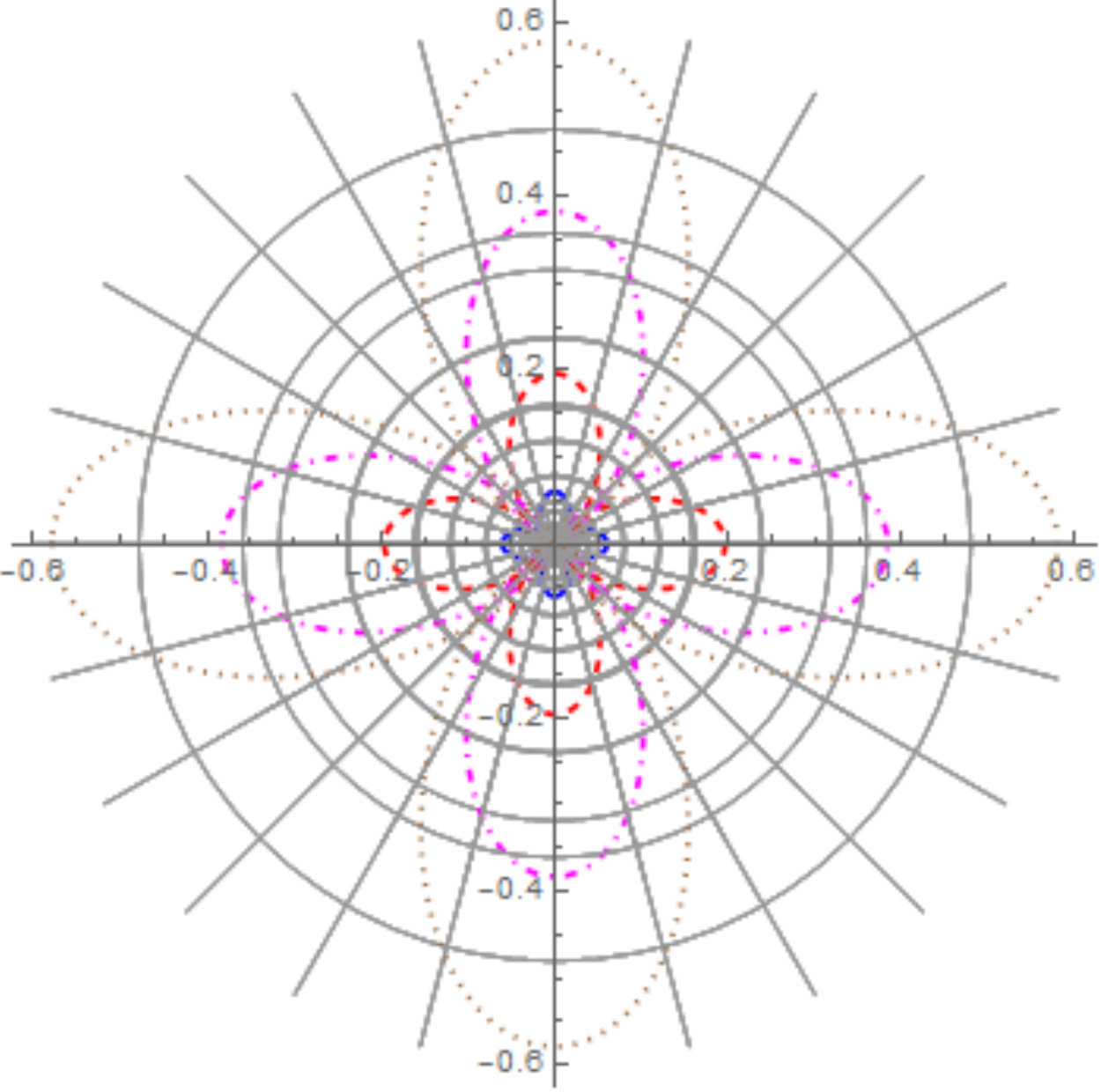}  & \includegraphics[width=60mm]{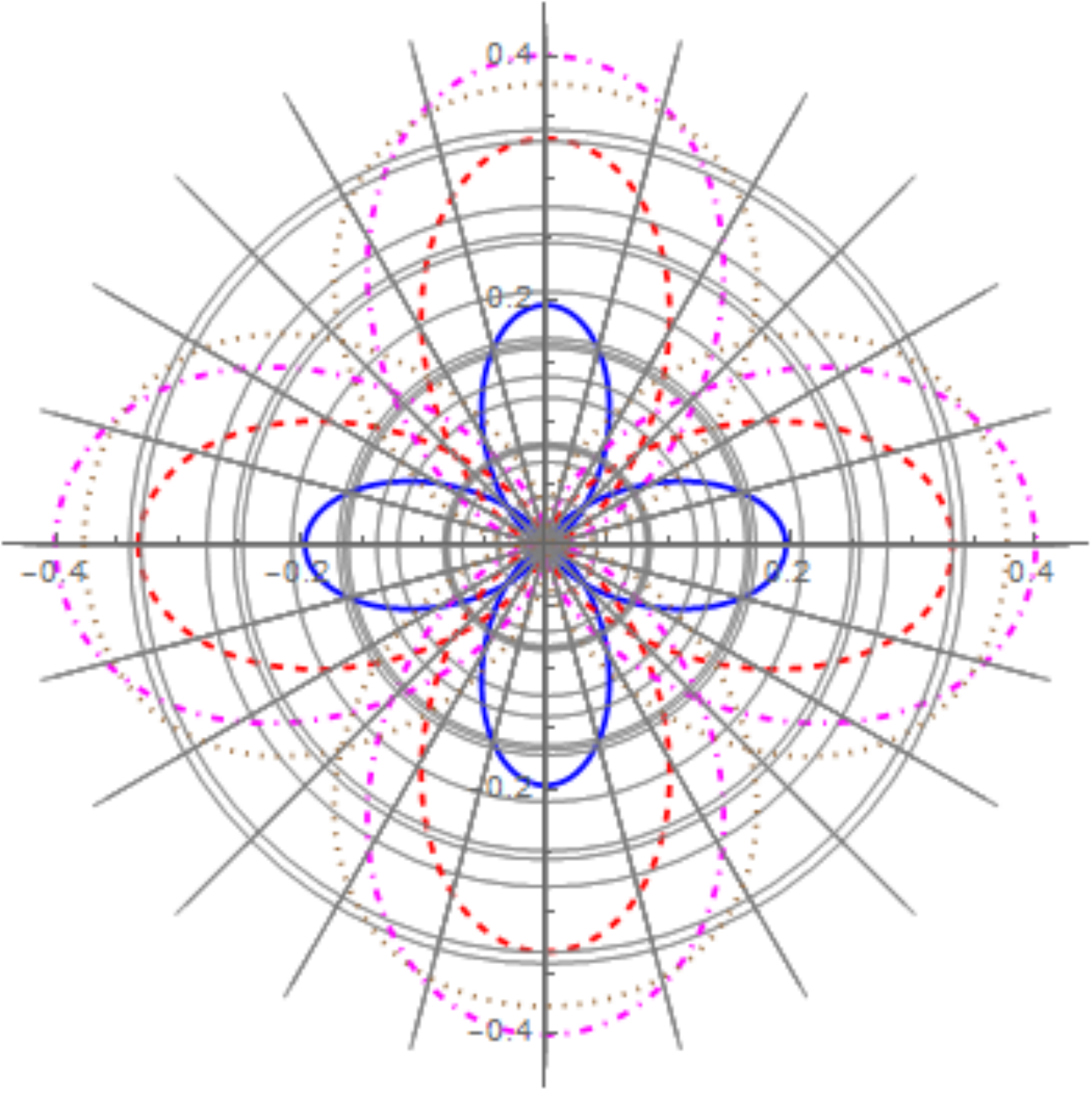}\tabularnewline
(a)  & (b) \tabularnewline
\includegraphics[width=60mm]{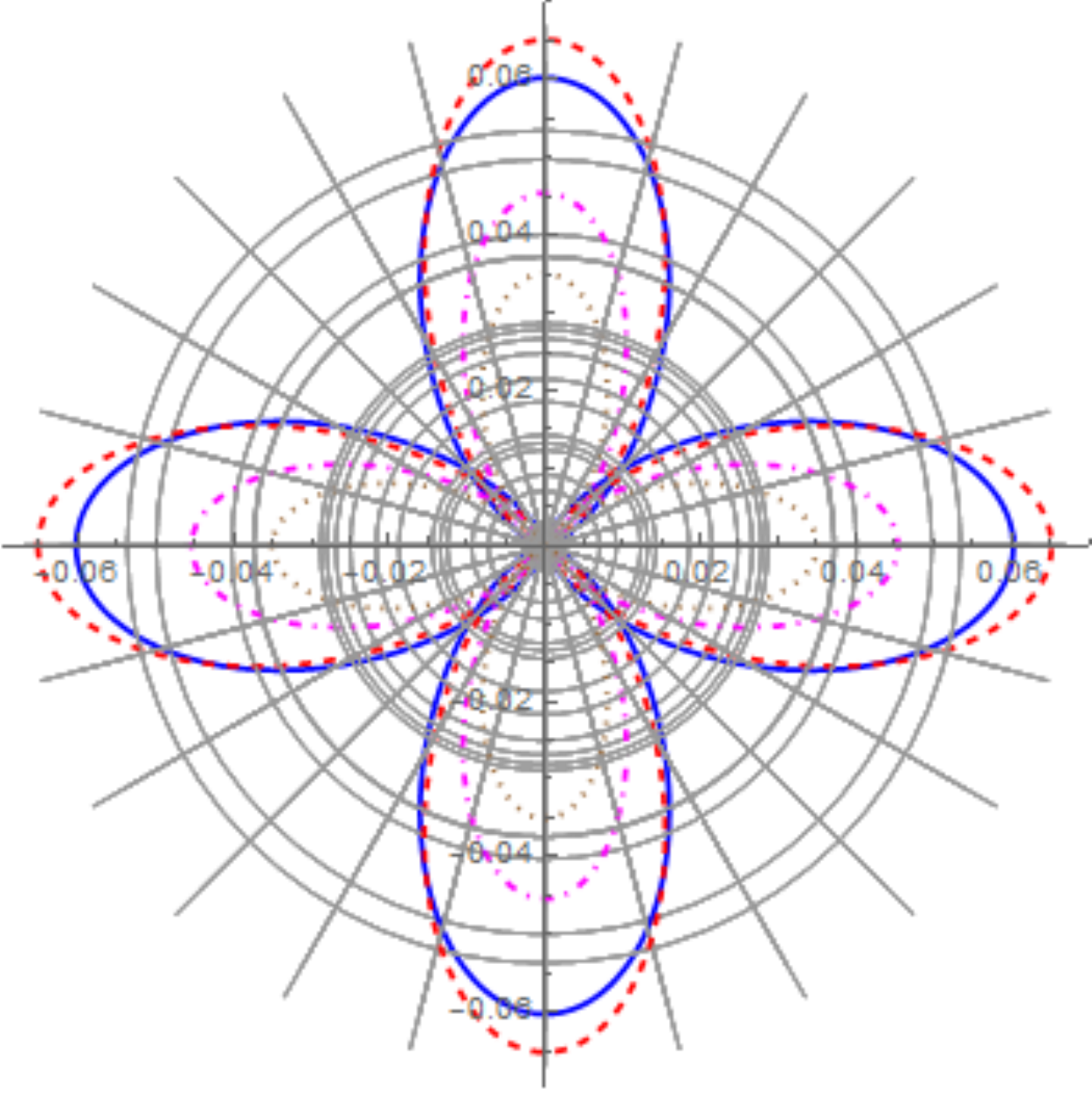}  & \includegraphics[width=60mm]{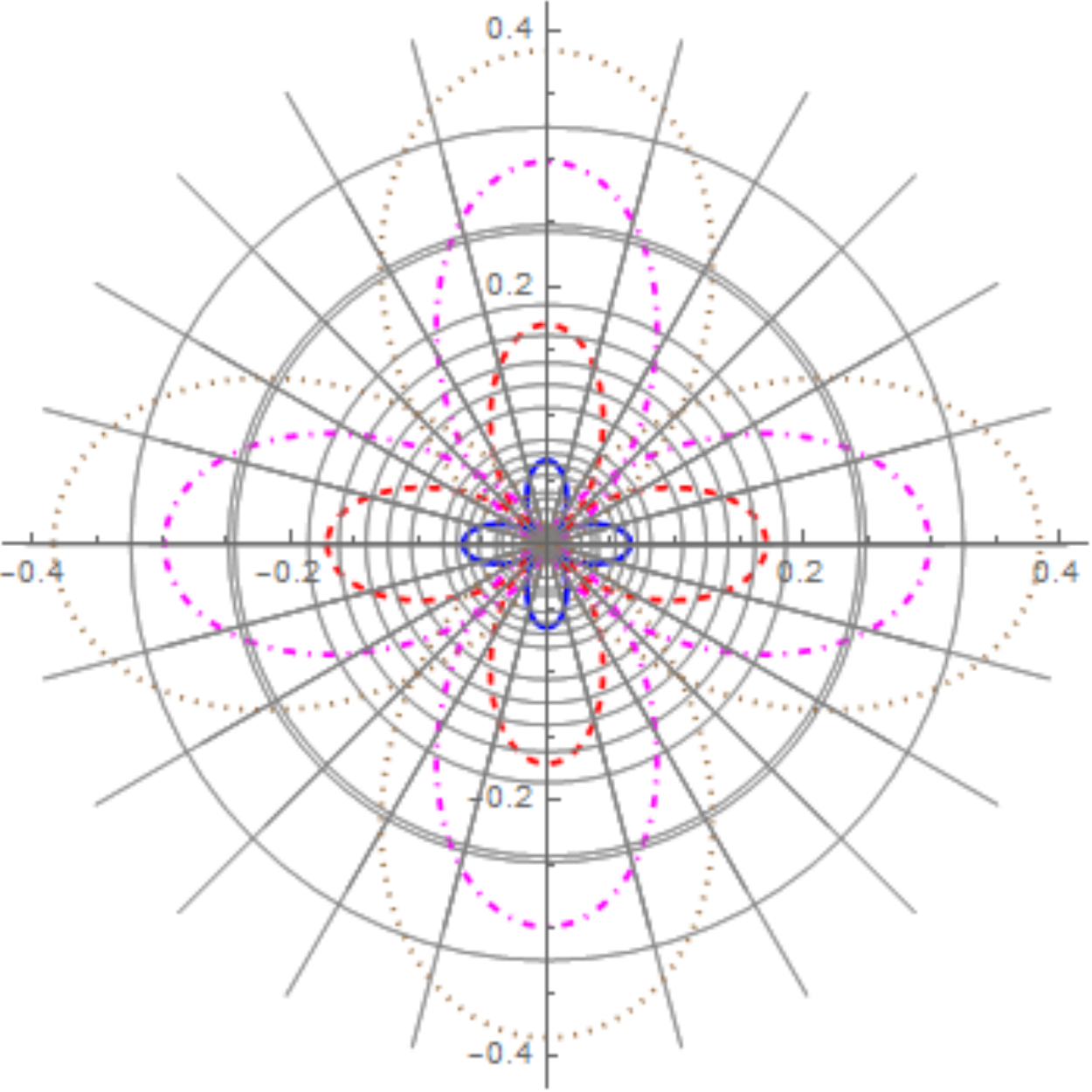}\tabularnewline
(c)  & (d) \tabularnewline
\end{tabular}\caption{\label{fig:Angular Q function} The polar plots for angular $Q$ function
for PADFS (in (a) and (b)) and PSDFS (in (c) and (d)) for displacement
parameter $\left|\alpha\right|{\rm =}1$ and $\theta_{2}=\frac{n\pi}{2}$
with integer $n\in\left[0,3\right]$ for different values of photon
addition/subtraction and Fock parameters. In (a) and (c), for $n=1$,
the smooth (blue), dashed (red), dot-dashed (magenta), and dotted
(brown) lines correspond to photon addition/subtraction 0, 1, 2, and
3, respectively. In (b) and (d), for the single photon added/subtracted
displaced Fock state, the smooth (blue), dashed (red), dot-dashed
(magenta), and dotted (brown) lines correspond to Fock parameter 1,
2, 3, and 4, respectively. }
\end{figure}

The relevance of the $Q$ function as witness of nonclassicality \cite{thapliyal2015quasiprobability}
and in state tomography \cite{thapliyal2016tomograms} is well studied.
On top of that, non-Gaussianity of the PADFS and PSDFS using $Q$
function was recently reported by us \cite{malpani2019lower}. We
further discuss a phase distribution based on $Q$ function using
Eq. (\ref{eq:ang-Qf-1}). In this particular case, we have obtained
the angular $Q$ function from the $Q$ functions of the PADFS and
PSDFS reported as Eqs. (15)-(16) in \cite{malpani2019lower}. Specifically,
we have shown the effect of photon addition on the DFS ($D\left(\alpha\right)\left|1\right\rangle $)
for a specific value of the displacement parameter in Figure \ref{fig:Angular Q function}
(a) for angular $Q$ function. One can clearly see that the polar
plots show an increase in the peak (located at $\theta_{1}=\theta_{2}$)
of the distribution with photon addition. Further, one can compare
the behavior of $Q_{\theta_{1}}$ with $P_{\theta}$ in Figure \ref{fig:Phase-Distribution-Function}
and observe that they behave quite differently (as reported in \cite{agarwal1992classical}
for the coherent states), other than increase in the peak of the distribution.
Specifically, $P_{\theta}$ has a peak at $\theta=\theta_{2}$ only
for $u>n$, while $Q_{\theta_{1}}$ is always peaked at the phase
of the displacement parameter which also becomes a line of symmetry.
Interestingly, the effect of increase in the Fock parameter of PADFS
on $Q_{\theta_{1}}$is similar but less prominent in comparison to
photon addition. This is in quite contrast of that observed for $P_{\theta}$
(in Figs. \ref{fig:Phase-Distribution-Function} and \ref{fig:Angular Q function}
(b)). In case of PSDFS, both photon subtraction and Fock parameter
have completely different effects on $Q_{\theta_{1}}$ (cf. Figure
\ref{fig:Angular Q function} (c)-(d)) which is also in contrast to
that on corresponding $P_{\theta}$ (shown in Figure \ref{fig:Phase-Distribution-Function-1}).
Specifically, with increase in photon subtraction the angular $Q$
function becomes narrower peaked at $\theta=\theta_{2}$, but for
larger number of photon subtraction the peak value decreases quickly.
However, with increasing Fock parameter (cf. Figure \ref{fig:Angular Q function}
(d)), $Q_{\theta_{1}}$ behaves much like photon addition on DFS (shown
in Figure \ref{fig:Angular Q function} (a)). The observed behavior
shows the relevance of studying both these phase distributions due
to their independent characteristics. 

\subsection{Quantum phase fluctuation of PADFS and PSDFS}

\begin{figure}
\begin{centering}
\begin{tabular}{ccc}
\includegraphics[width=60mm]{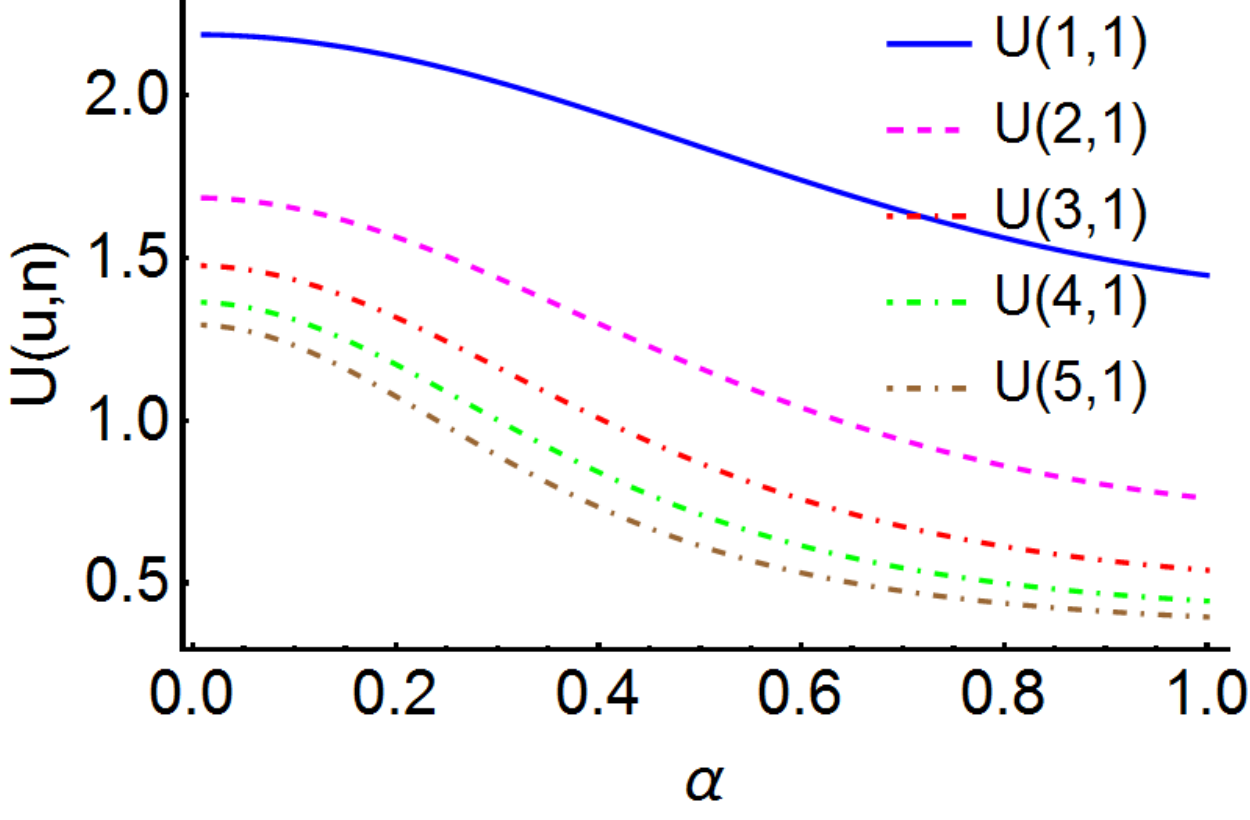}  & \includegraphics[width=60mm]{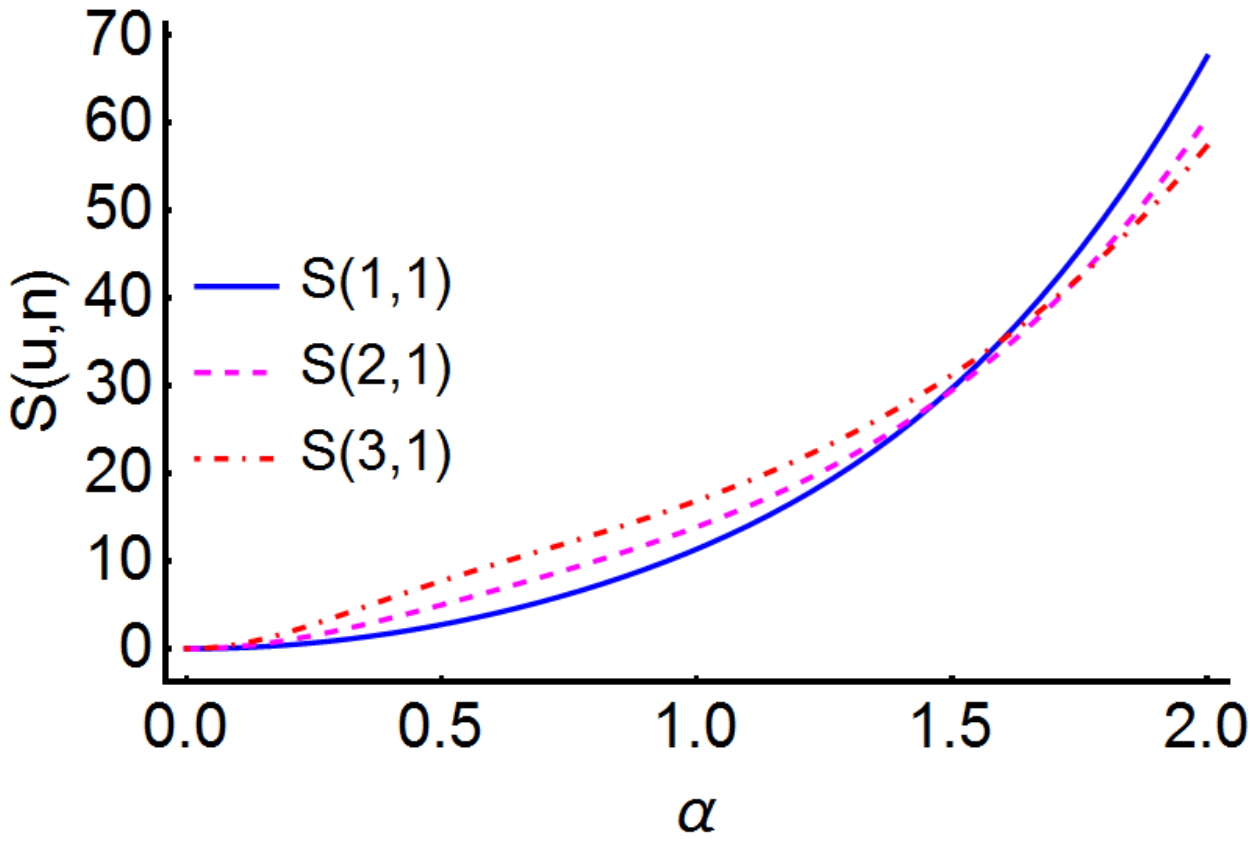} & \tabularnewline
(a)  & (b)  & \tabularnewline
\includegraphics[width=60mm]{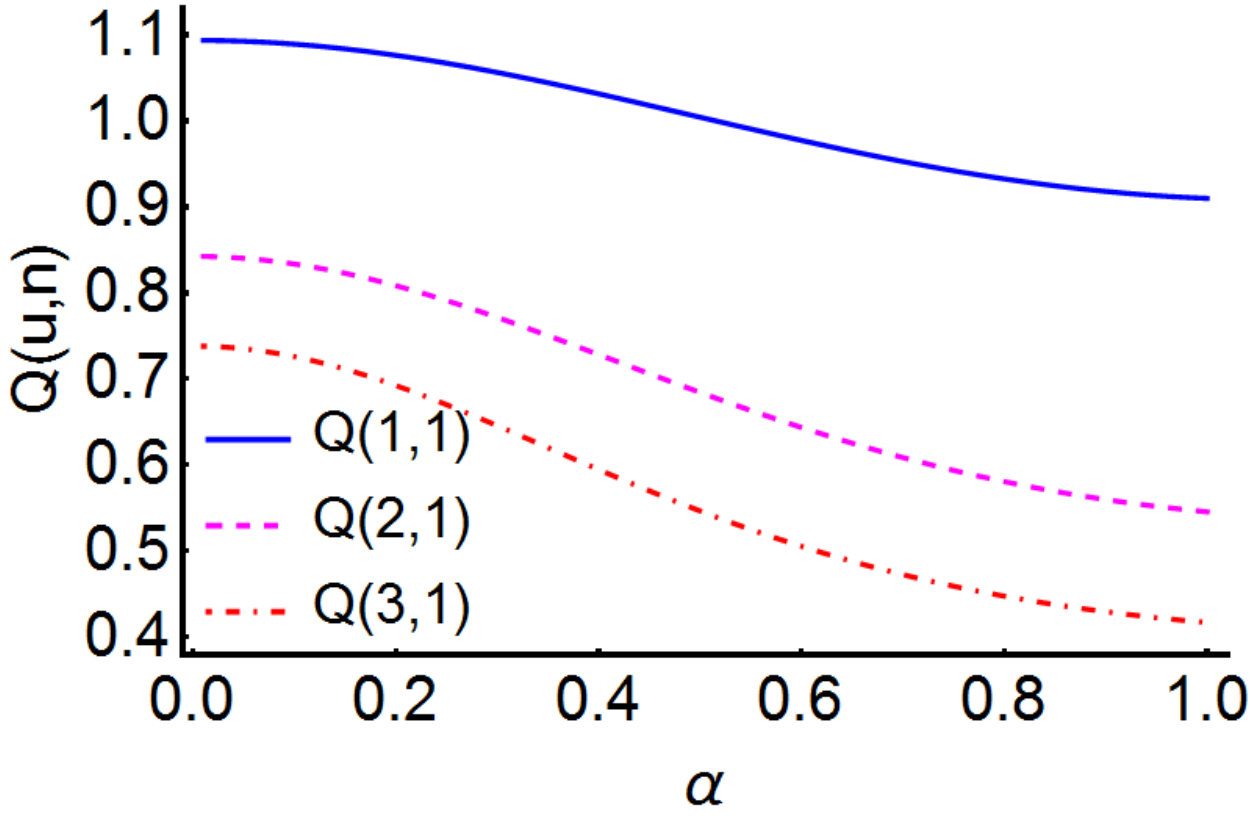}  & \includegraphics[width=60mm]{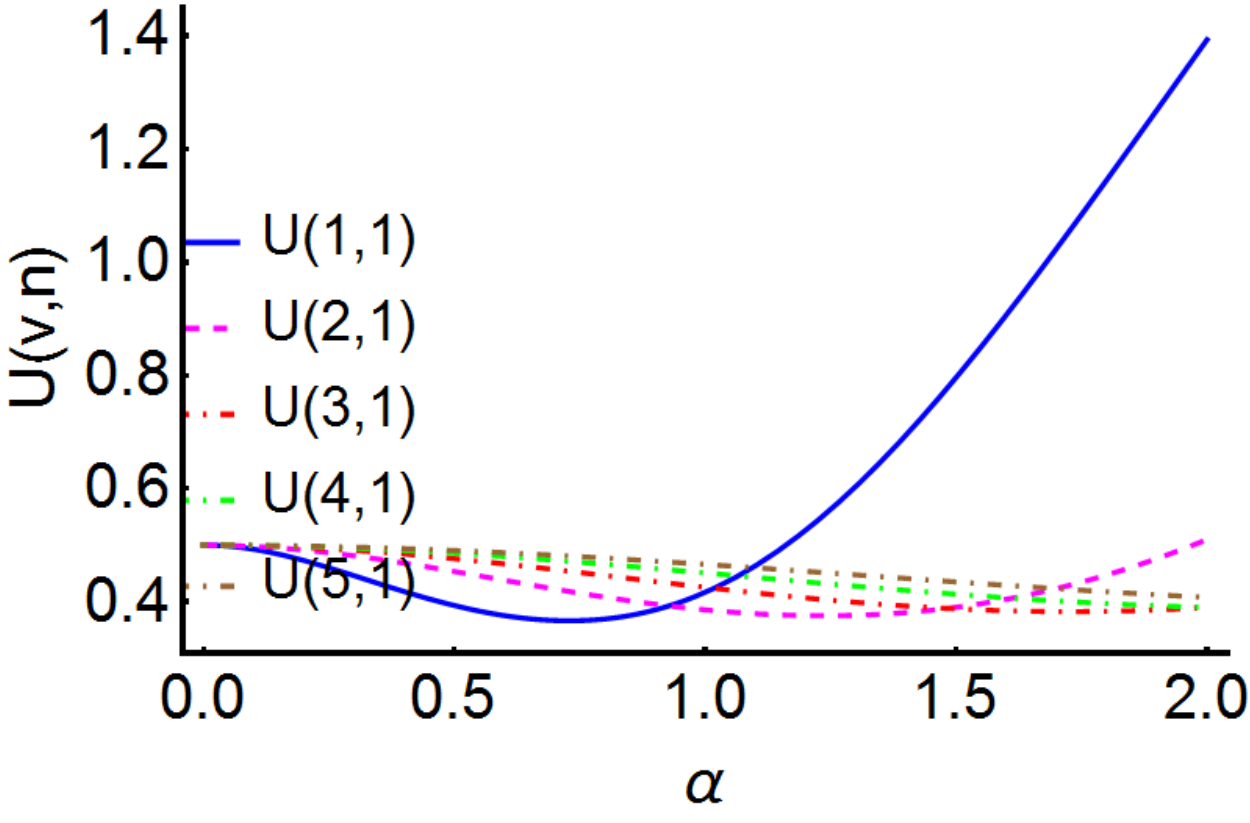} & \tabularnewline
(c)  & (d)  & \tabularnewline
\includegraphics[width=60mm]{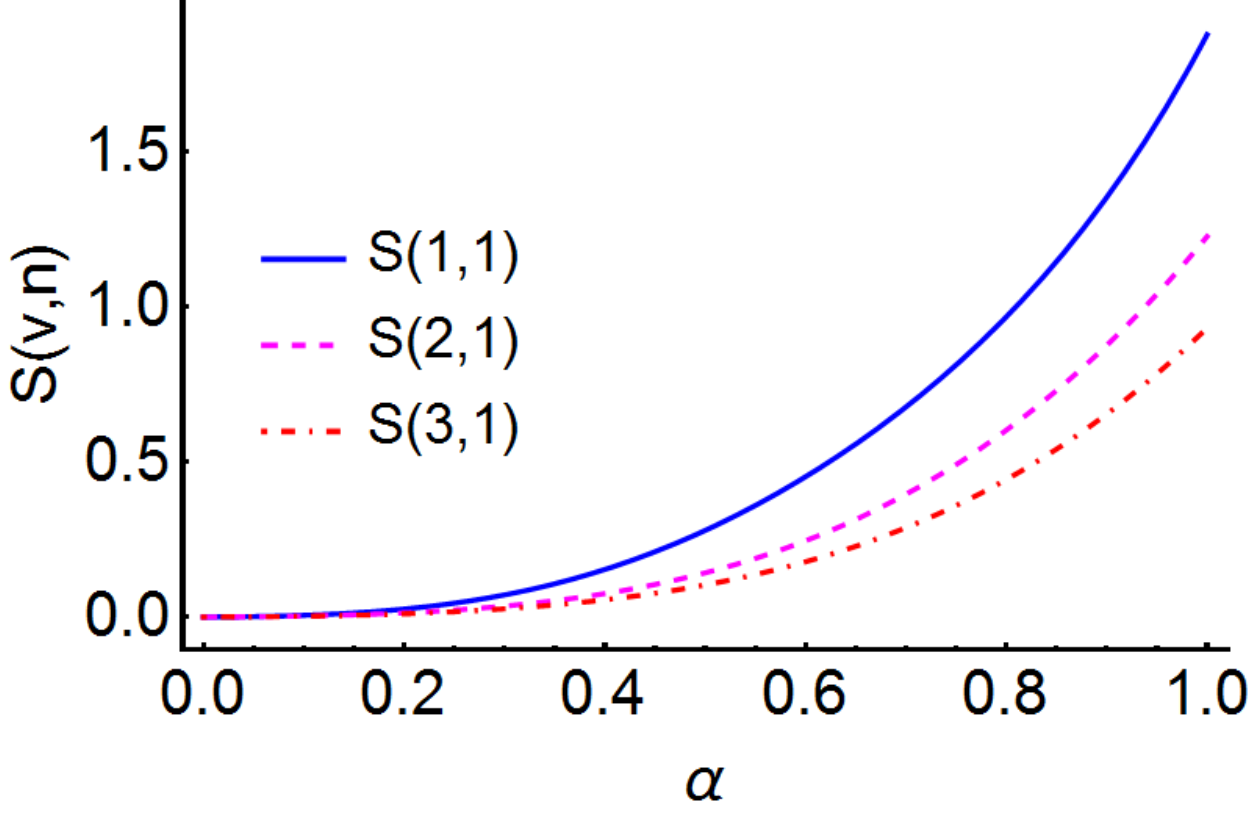}  & \includegraphics[width=60mm]{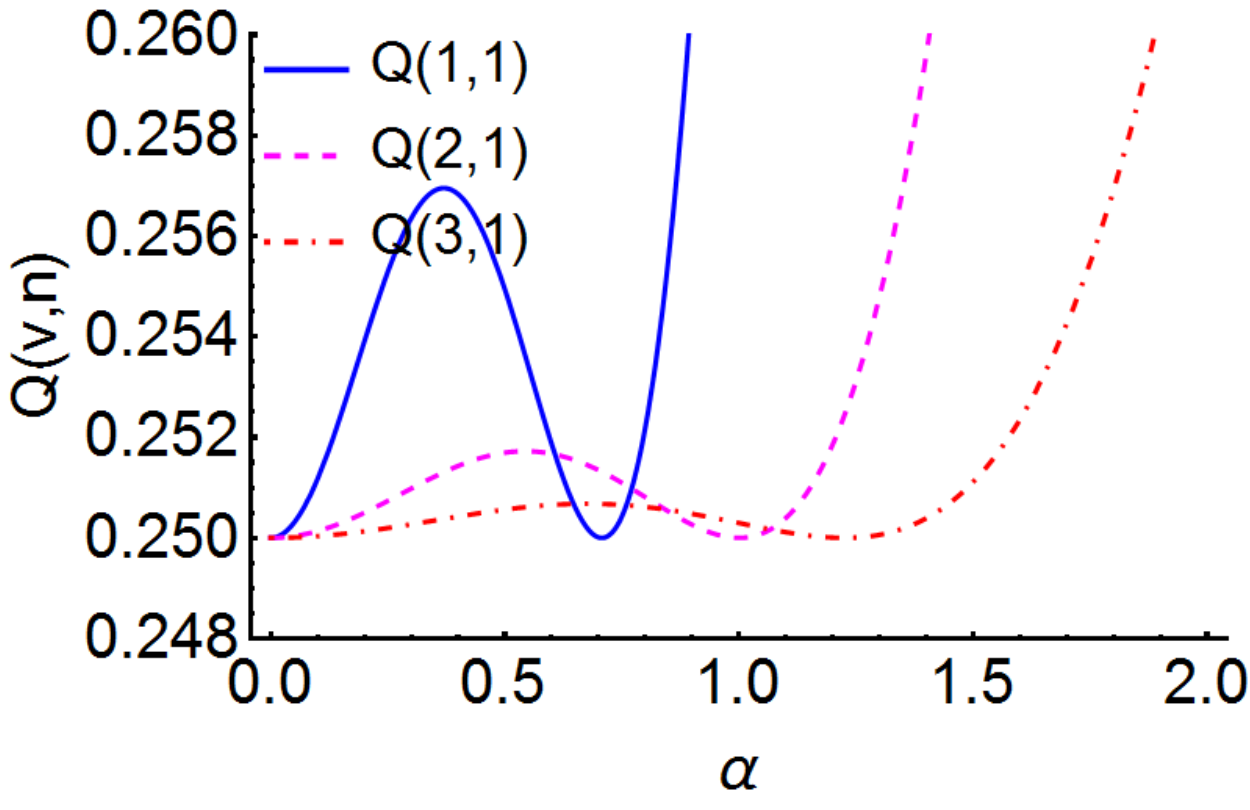}  & \tabularnewline
(e)  & (f)  & \tabularnewline
\end{tabular}
\par\end{centering}
\caption{\label{fig:phase fluctuation} Variation of three phase fluctuation
parameters introduced by Carruthers and Nieto with the displacement
parameter with $\theta_{2}=0$. The values of photon addition ($u$),
subtraction ($v$), and Fock parameter $n=1$ are given in the legends.
Parameter $U\left(i,n\right)\,\forall i\in\left\{ u,v\right\} $ also
illustrates antibunching in the states for values less than $\frac{1}{2}$. }
\end{figure}

Note that Carruthers and Nieto \cite{carruthers1968phase} had introduced
these parameters in terms of Susskind and Glogower operators \cite{susskind1964quantum};
here we use them in Barnett-Pegg formalism to remain consistent with
\cite{gupta2007reduction}, where $U$ parameter is shown relevant
as a witness of nonclassicality \cite{gupta2007reduction}. Specifically,
$U$ is 0.5 for coherent state, and reduction of $U$ parameter below
the value for coherent state can be interpreted as the presence of
nonclassical behavior \cite{gupta2007reduction}. In what follows,
we will study quantum phase fluctuations for PADFS and PSDFS by computing
analytic expressions of $U,\,S$ and $Q$ parameters in Barnett-Pegg
formalism, with a specific focus on the possibility of witnessing
nonclassical properties of these states via the reduction of $U$
parameter below the coherent state limit. Carruthers and Nieto \cite{carruthers1968phase}
introduced three parameters to study quantum phase fluctuation (\ref{eq:fluctuation3-1})-(\ref{eq:fluctuation5-1}).
It was only in the recent past that Gupta and Pathak provided a physical
meaning to one of these parameters by establishing its relation with
antibunching and sub-Poissonian photon statistics \cite{gupta2007reduction}.
Thus, the quantum phase fluctuation studied here using three parameters
will also be used to witness the nonclassical nature of the quantum
states under consideration. Here, the effect of photon addition/subtraction
and displacement parameters on these fluctuation parameters is also
studied (shown in Fig. \ref{fig:phase fluctuation}). Specifically,
Figure \ref{fig:phase fluctuation} (a)-(c) show variation of the
three parameters of quantum phase fluctuation for different values
of the number of photons added in the displaced Fock state ($D\left(\alpha\right)\left|1\right\rangle $)
with displacement parameter $\left|\alpha\right|$. It may be clearly
observed that two of the quantum phase fluctuation parameters, namely
$U\left(u,n\right)$ and $Q\left(u,n\right)$ decrease with the value
of displacement parameter, while $S\left(u,n\right)$ increases with
$\left|\alpha\right|$. Interestingly, the photon addition and increase
in the displacement parameter exhibit the same effect on all three
quantum phase fluctuation parameters for PADFS, while for higher values
of displacement parameter $S\left(u,n\right)$ show completely opposite
effect of photon addition. In contrast, $U\left(v,n\right)$ for $v$
subtracted photons from $D\left(\alpha\right)\left|1\right\rangle $
is found to increase (decrease) with photon subtraction while decrease
(increase) with the displacement parameter for small (large) value
of $\left|\alpha\right|$ (cf. Figure \ref{fig:phase fluctuation}
(d)). On the other hand, parameter $S\left(v,n\right)$ is also observed
to increase (decrease) with $\left|\alpha\right|$ ($v$) as shown
in Figure \ref{fig:phase fluctuation} (e). The third parameter $Q\left(v,n\right)$
shows slightly complex behavior for PSDFS with both $\left|\alpha\right|$
and $v$ (cf. Figure \ref{fig:phase fluctuation} (f)) as it behaves
analogous to PADFS for each subtracted photon for both small and large
values of the displacement parameter (when it increases with $\left|\alpha\right|$),
but for intermediate values the behavior is found to be completely
opposite.

As mentioned previously, $U\left(i,n\right)\,\forall i\in\left\{ u,v\right\} $
has a physical significance as a witness of antibunching for values
of this parameter less than $\frac{1}{2}$, Figure \ref{fig:phase fluctuation}
(a) and (d) can be used to perform similar studies for PADFS and PSDFS,
respectively. In case of PADFS, we can observe this relevant parameter
to become less than $\frac{1}{2}$, and thus to illustrate the presence
of antibunching, only at higher values of the displacement parameter
and photon added to the displaced Fock state. In contrast, PSDFS shows
the presence of this nonclassical feature in all cases.Thus, occurrence
of antibunching in PADFS and PSDFS is established here through this
phase fluctuation parameter. Interestingly, a similar dependence of
antibunching in PADFS and PSDFS    Eq. (\ref{eq:HOA-1}) has
been recently reported by us \cite{malpani2019lower} using a different
criterion. Further, one can observe from the expression of $U$ in
Eq. (\ref{eq:fluctuation3-1}) that it is expected to be independent
of the phase of the displacement parameter, which can also be understood
from the use of this parameter as a witness for an intensity moments
based nonclassical feature. In contrast, $S$ and $Q$ in Eqs. (\ref{eq:fluctuation4-1})-(\ref{eq:fluctuation5-1})
show dependence on the phase of displacement parameter. Here, we have
not discussed the effect of Fock parameter in detail, but in case
of photon addition, $u$ and $n$ have same (opposite) effects on
$S$ ($U$ and $Q$) parameter(s). Fock parameter has always shown
opposite effect of photon subtraction on all three phase fluctuation
parameters, and thus nonclassicality revealed by $U$ can be enhanced
with Fock parameter. The relevance of Fock parameter can also be visualized
by observing the fact that the single photon subtracted coherent state
has $U=0.5$ (which is consistent with the value zero of the antibunching
witness reported in \cite{thapliyal2017comparison}). Thus, in this
case, the origin of the induced antibunching can be attributed to
the non-zero value of Fock parameter. 

\subsection{Phase Dispersion}

\begin{figure}
\centering{} %
\begin{tabular}{cc}
\includegraphics[width=60mm]{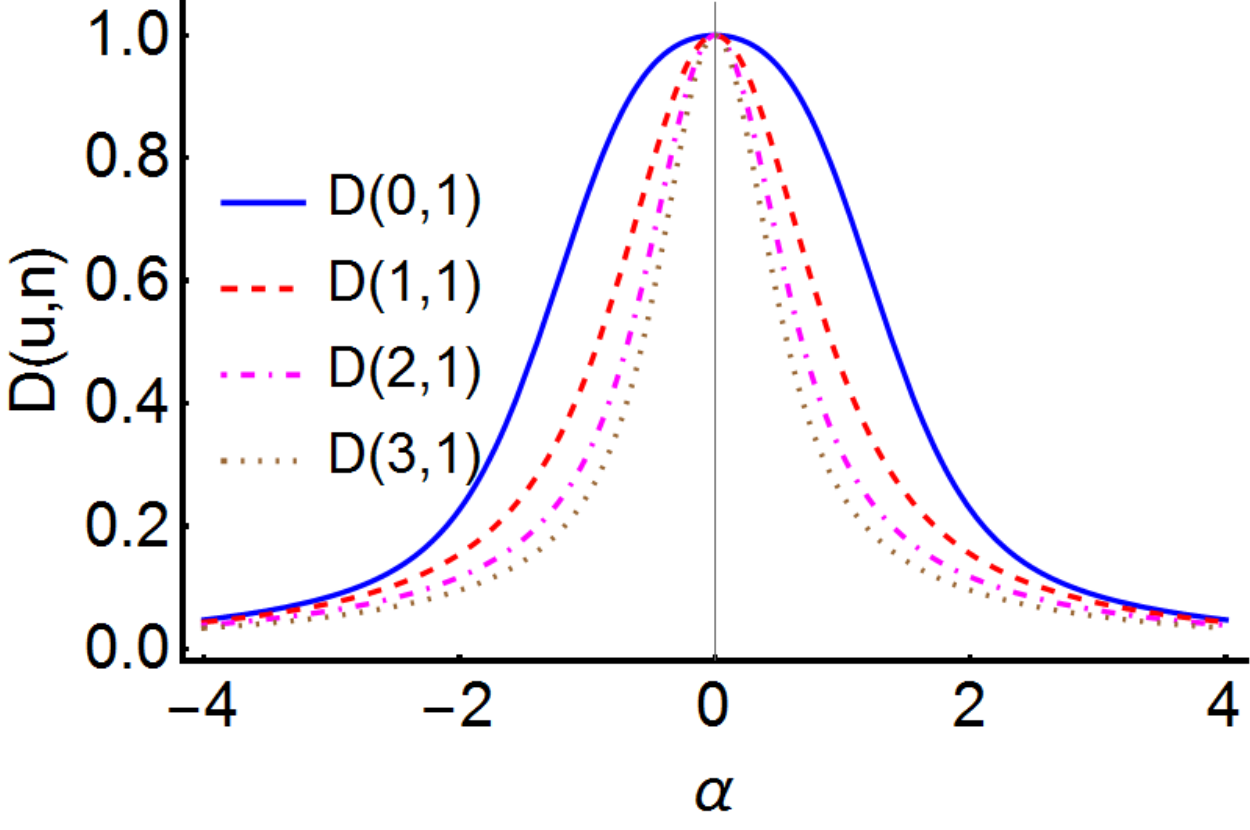}  & \includegraphics[width=60mm]{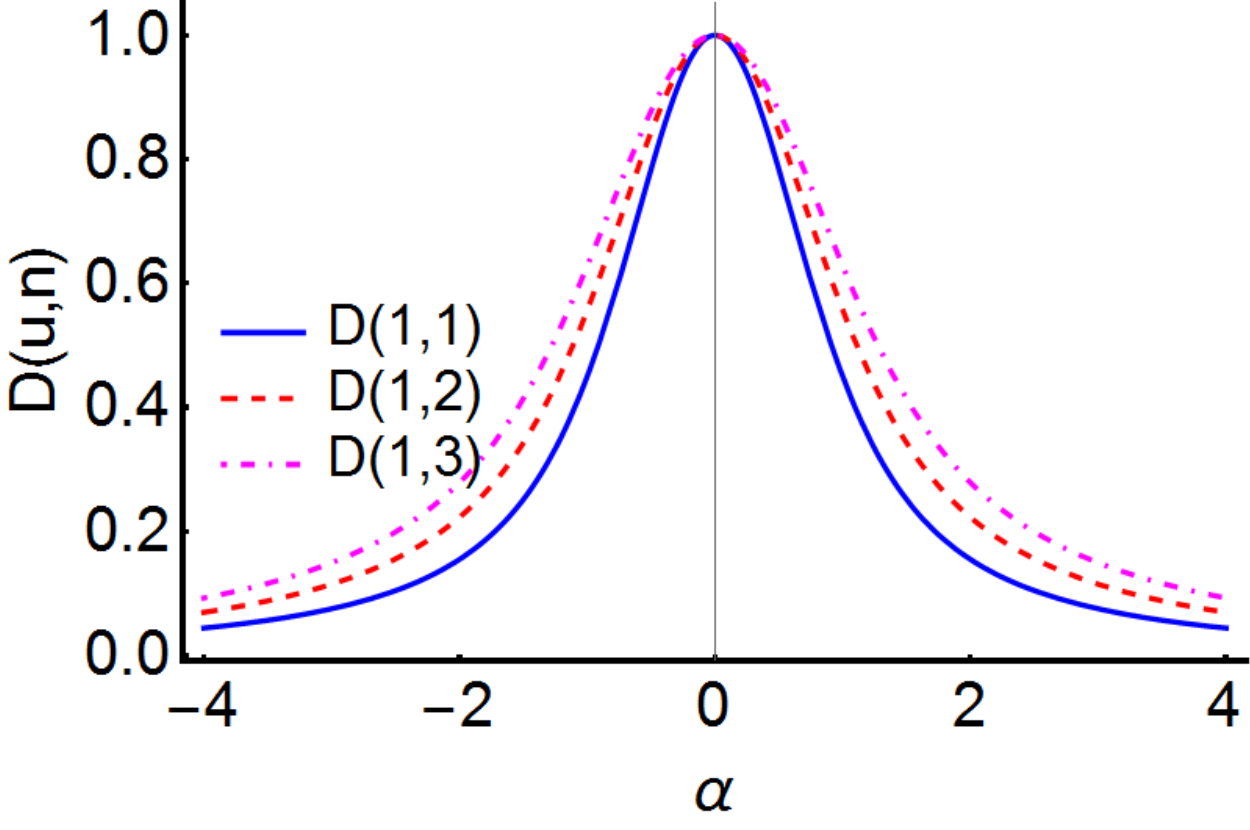}\tabularnewline
(a)  & (b) \tabularnewline
\includegraphics[width=60mm]{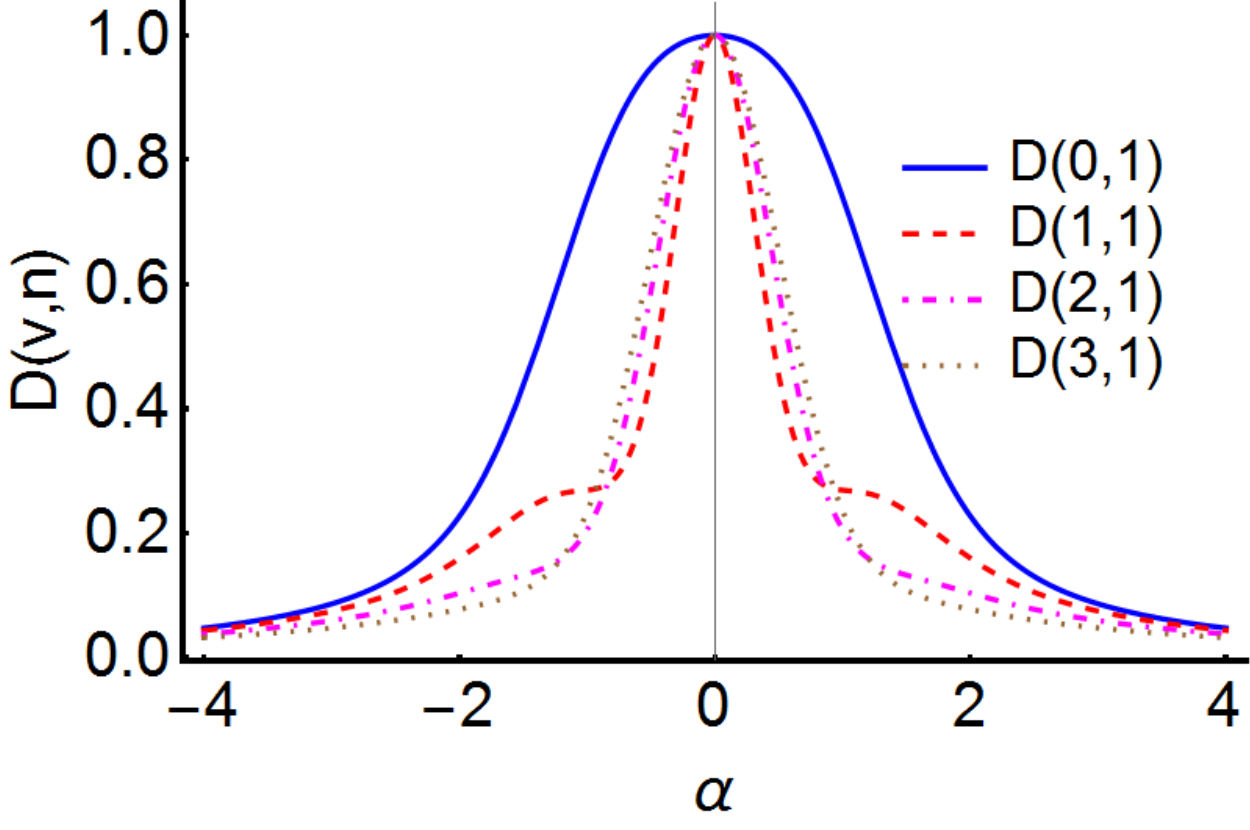}  & \includegraphics[width=60mm]{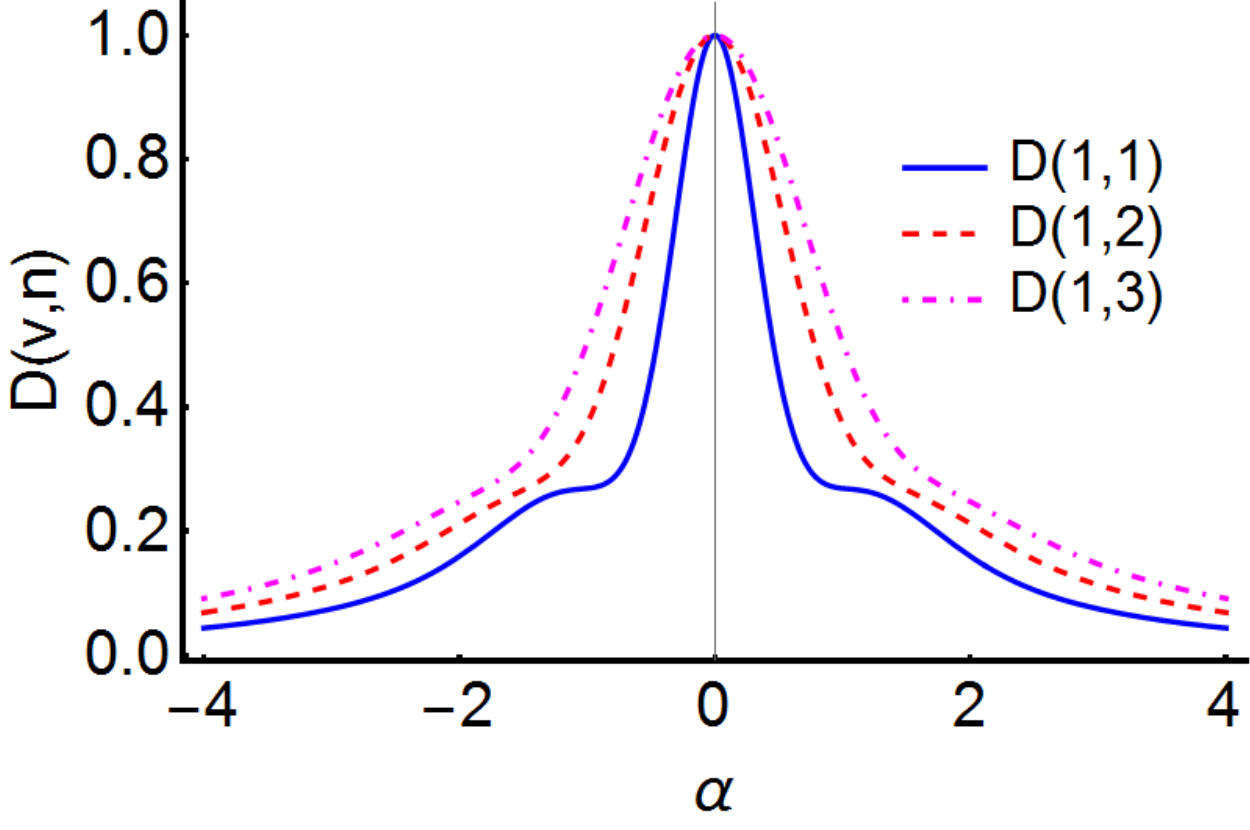}\tabularnewline
(c)  & (d) \tabularnewline
\end{tabular}\caption{\label{fig:Phase-Dispersion} Variation of phase dispersion for PADFS
(in (a) and (b)) and PSDFS (in (c) and (d)) with displacement parameter
for an arbitrary $\theta_{2}$. Dependence on different values of
photon added/subtracted and the initial Fock state $\left|1\right\rangle $
(in (a) and (c)), while on different values of Fock parameter for
single photon added/subtracted state (in (b) and (d)).}
\end{figure}

Here, it is worth stressing that both Carruthers-Nieto parameters
and phase dispersion $D$ correspond to phase fluctuation, Our primary
focus is to study phase fluctuation and further to check the correlation
between these measures of phase fluctuation. Thus, it would be interesting
to study phase fluctuation from the two perspectives. We compute a
measure of quantum phase fluctuation based on quantum phase distribution,
the phase dispersion (\ref{eq:Dispersion-1}), for both PADFS and
PSDFS to perform a comparative study between them. Specifically, the
maximum value of dispersion is 1 which corresponds to the uniform
phase distribution, i.e., $P_{\theta}=\frac{1}{2\pi}$. Both PADFS
and PSDFS show a uniform distribution for the displacement parameter
$\alpha=0$ (cf. Figure \ref{fig:Phase-Dispersion}). It is a justified
result as both the states reduce to the Fock state in this case. However,
with the increase in the value of displacement parameter quantum phase
dispersion is found to decrease. This may be attributed to the number-phase
complimentarity \cite{banerjee2010complementarity,srikanth2009complementarity,srikanth2010complementarity},
which leads to smaller phase fluctuation with increasing variance
in the number operator at higher values of displacement parameter.
Thus, with an increase in the average photon number by increasing
the displacement parameter, phase dispersion decreases for both PADFS
and PSDFS. Addition of photons in DFS leads to decrease in the value
of phase dispersion, while subtraction of photons has more complex
effect on phase dispersion (cf. Figure \ref{fig:Phase-Dispersion}
(a) and (c)). Specifically, for the smaller values of the displacement
parameter ($\left|\alpha\right|<1$), the phase dispersion parameter
behaves differently for $v\leq n$ and $v>n$. This can be attributed
to the sub-Poissonian photon statistics for $v\leq n$ with $\left|\alpha\right|<1$
as well as the small value of average photon number (Figure \ref{fig:phase fluctuation}
(d)). However, at the higher values of the displacement parameter
$D$ for the PSDFS behaves in a manner analogous to the PADFS. Interestingly,
increase in the Fock parameter shows similar effect on PADFS and PSDFS
in Figure \ref{fig:Phase-Dispersion} (b) and (d), respectively. 

\subsection{Phase sensing uncertainity for PADFS and PSDFS}

We finally discuss quantum phase estimation using Eq. (\ref{eq:PE-1}),
assuming the two mode input state in the Mach-Zehnder interferometer
as $|\psi_{i}(j,n,\alpha)\rangle\otimes|0\rangle$. The expressions
for the variance of the difference in the photon numbers in the two
output modes of the Mach-Zehnder interferometer for input PADFS and
PSDFS and the rest of the parameters required to study phase sensing
are reported in Appendix.

\begin{figure}
\centering{} %
\begin{tabular}{cc}
\includegraphics[width=60mm]{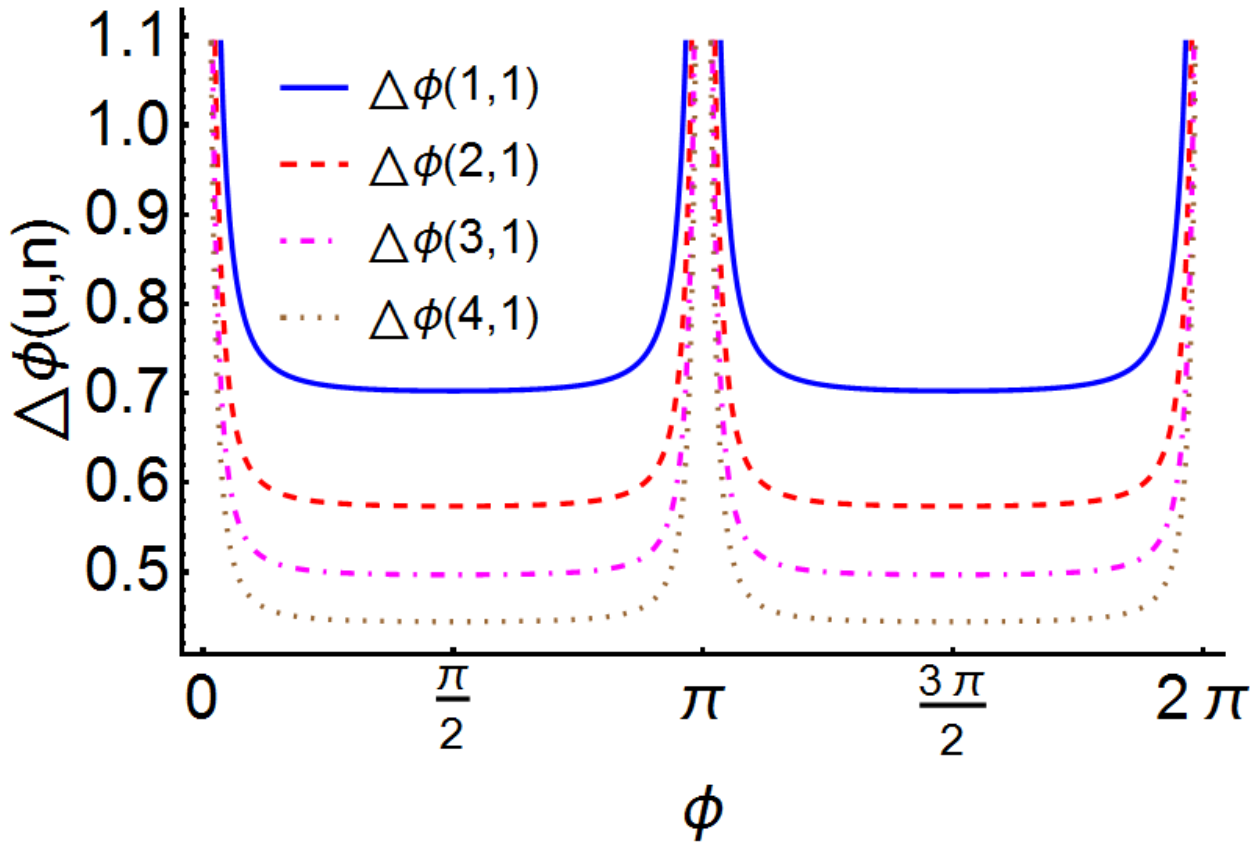}  & \includegraphics[width=60mm]{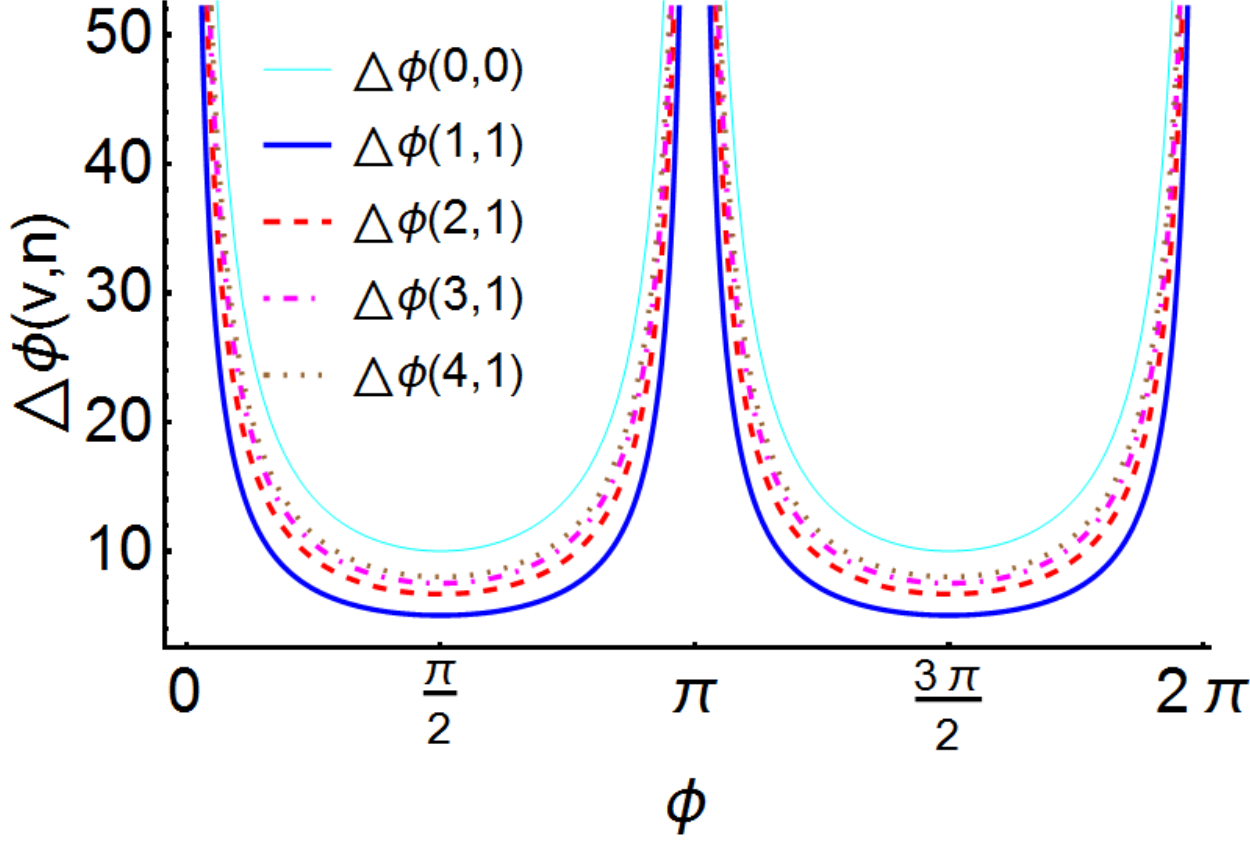}\tabularnewline
(a)  & (b) \tabularnewline
\includegraphics[width=60mm]{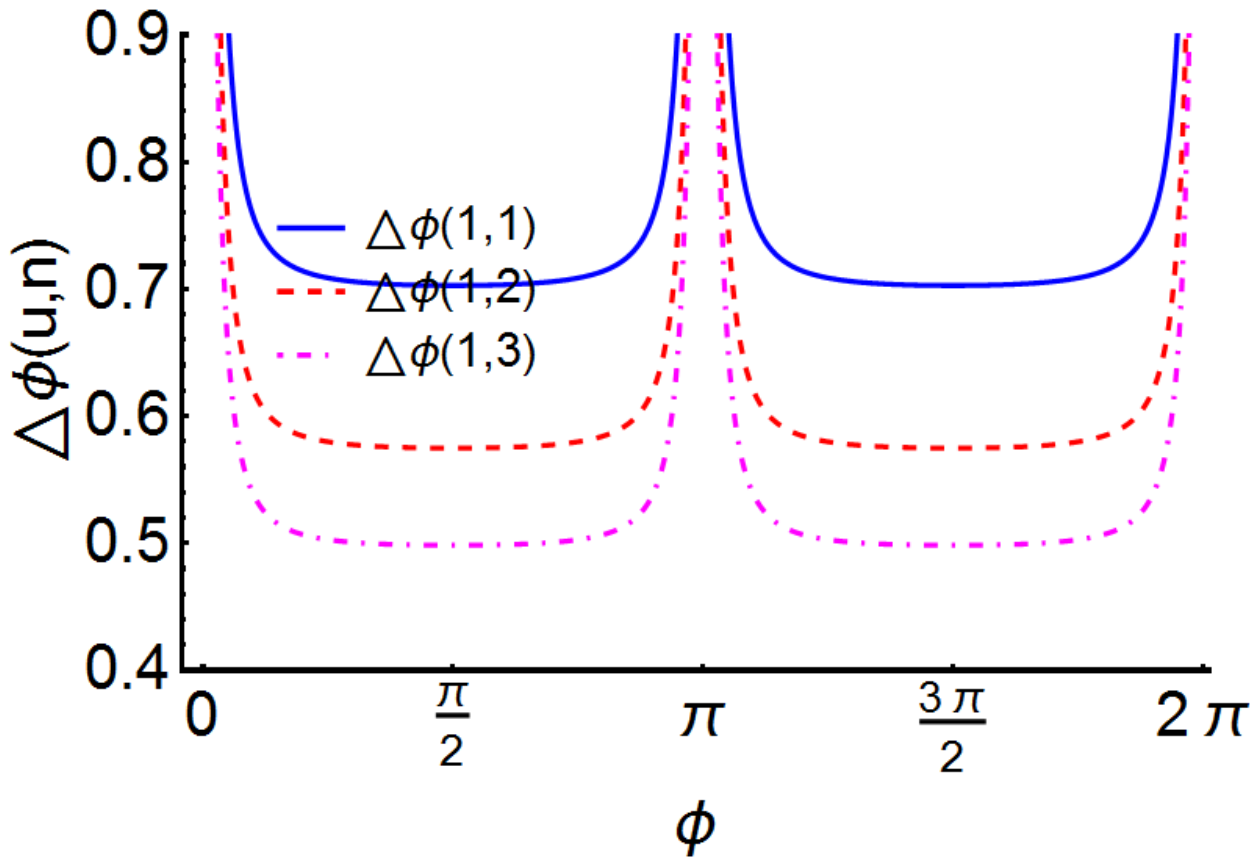}  & \includegraphics[width=60mm]{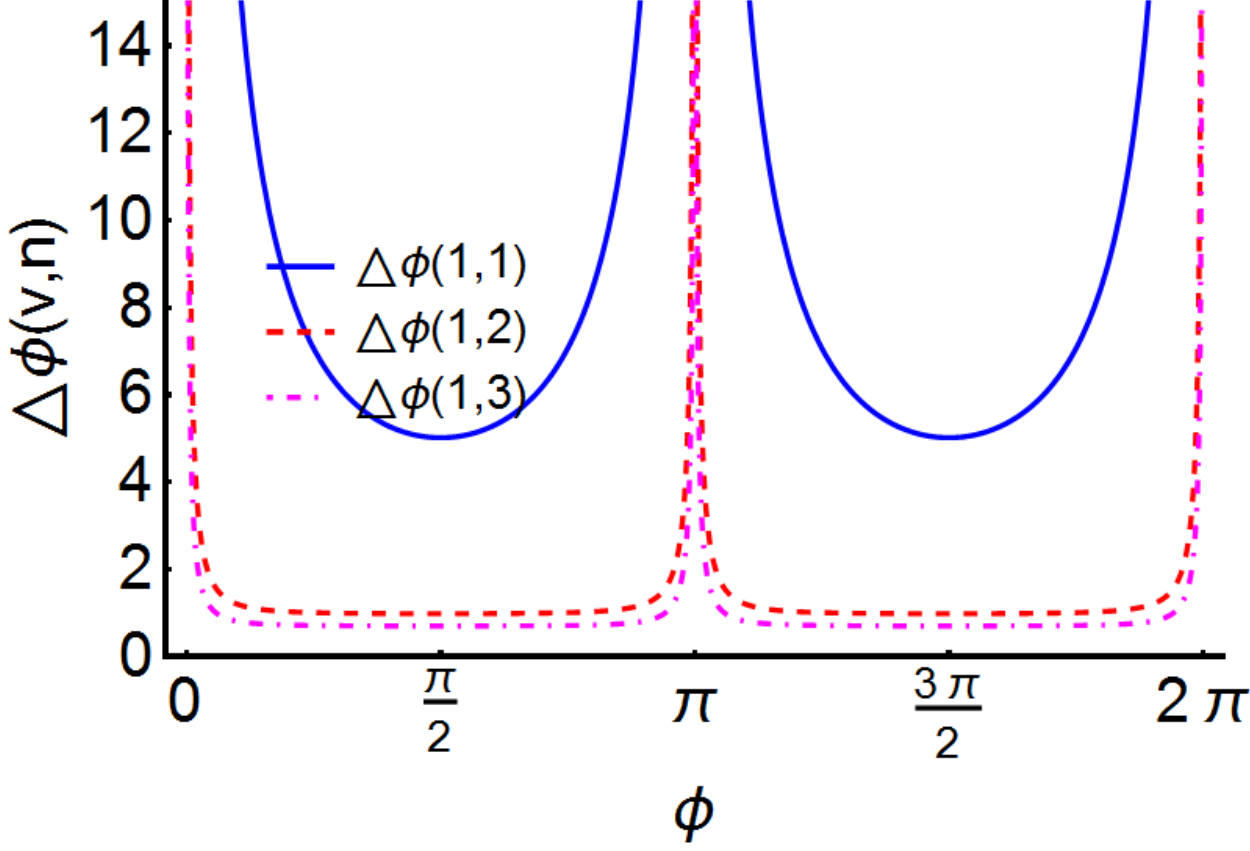} \tabularnewline
(c)  & (d) \tabularnewline
\end{tabular}\caption{\label{fig:Phase sensing uncertainity}Phase sensing uncertainty for
(a) PADFS and (b) PSDFS as a function of phase to be estimated $\phi$
for different number of photon addition/subtraction with $n=1$. The
dependence for (c) PADFS and (d) PSDFS is also shown for different
values of Fock parameters with $u=1$ and $v=1$, respectively. In
all cases, we have chosen $\alpha=0.1$.}
\end{figure}

The obtained expressions allow us to study the optimum choice of state
parameters for quantum phase estimation using PADFS and PSDFS. The
variation of these parameters is shown in Figure \ref{fig:Phase sensing uncertainity}.
Specifically, we have shown that PSDFS is preferable over coherent
state for phase estimation (cf. Figure \ref{fig:Phase sensing uncertainity}
(b)). However, with the increase in the photon subtraction this phase
uncertainty parameter is found to increase although remaining less
than corresponding coherent state value. In contrast, with photon
addition, advantage in phase estimation can be attained as the reduction
of the phase uncertainty parameter allows one to perform more precise
measurement. This advantage can be enhanced further by choosing large
values of photon addition and Fock parameter (cf. Figure \ref{fig:Phase sensing uncertainity}
(a) and (c)). In a similar sense, appropriate choice of Fock parameter
would also be advantageous in phase estimation with PSDFS as it decreases
the phase uncertainty parameter, but still PADFS remains preferable
over PSDFS. This can further be controlled by an increase in $\left|\alpha\right|$
which decreases (increases) phase uncertainty parameter for PADFS
(PSDFS). 

\section{Conclusions \label{sec:Conclusions-1}}

A set of engineered quantum states can be obtained as the limiting
cases from the PADFS and PSDFS, e.g., DFS, coherent state, photon
added/subtracted coherent state, and Fock state. Specifically, PADFS/PSDFS
are obtained by application of unitary (displacement) and non-unitary
(addition and subtraction of photons) operations on Fock state. In
view of the fact that the Fock states have uniform phase distribution,
the set of unitary and non-unitary quantum state engineering operations
are expected to affect the phase properties of the generated state.
Therefore, here we have calculated quantum phase distribution, which
further helped in quantifying phase fluctuation as phase dispersion.
We have also computed the phase distribution as the angular $Q$ function.
We have further studied phase fluctuation using three Carruthers and
Nieto parameters, and have used one of them to reveal the existence
of antibunching in the quantum states of our interest.

Both the phase distribution and angular $Q$ functions are found to
be symmetric along the value of the phase of the displacement parameter.
The phase distribution is observed to become narrow and peak(s) to
increase with the amplitude of the displacement parameter ($\left|\alpha\right|$),
which further becomes broader for higher values of $\left|\alpha\right|$.
Further, photon addition/subtraction and Fock parameters are observed
to have opposite effects on phase distribution, i.e., distribution
function becomes narrower (broader) with photon addition/subtraction
(Fock parameter). Among photon addition and subtraction operations,
subtracting a photon alters the phase properties more than that of
photon addition. Specifically, at the small values of the displacement
parameter ($\left|\alpha\right|<1$), the phase properties of PSDFS
for $v\leq n$ and $v>n$ behave differently. This can be attributed
to the fact that for $v\leq n$ with $\left|\alpha\right|<1$, the
average photon number becomes very small. Further, the peak of the
phase distribution remains at the phase of displacement parameter
only when the number of photons added/subtracted is more than that
of the Fock parameter. However, in case the number of photons subtracted
(added) is same as the Fock parameter, the peak of the phase distribution
is observed (not observed) at the phase of displacement parameter.
The angular $Q$ function can be observed to show similar dependence
on various parameters, but the peak of the distribution remains located
at the value of phase of the displacement parameter. The three phase
fluctuation parameters introduced by Carruthers and Nieto \cite{carruthers1968phase}
show phase properties of PADFS and PSDFS, while one of them, $U$
parameter also reveals antibunching in both PADFS and PSDFS. In this
case, the role of Fock parameter as antibunching inducing operation
in PSDFS is also discussed. Phase dispersion quantifying phase fluctuation
remains unity for Fock state reflecting uniform distribution, which
can be observed to decrease with increasing displacement parameter.
This may be attributed to the number-phase complimentarity as the
higher values of variance with increasing displacement parameter lead
to smaller phase fluctuation. Fock parameter and photon addition/subtraction
show opposite effects on the phase dispersion as it increases (decreases)
with $n$ ($u/v$).

Finally, we have also discussed the advantage of the PADFS and PSDFS
in quantum phase estimation and obtained the set of optimized parameters
in the PADFS/PSDFS. Both photon addition and Fock parameter decrease
the uncertainty in phase estimation, while photon subtraction, though
performs better than coherent state is not as advantageous as $u$
or $n$. In \cite{ou1997fundamental}, it was established that signal-to-noise
ratio is significant only when the phase shift to measure is of the
same order as multiplicative inverse of the average photon number.
Therefore, in case of PADFS this limitation of quantum measurement
is expected to play an important role. Thus, we have shown here that
state engineering tools can be used efficiently to control the phase
properties of the designed quantum states for suitable applications.
The study performed in this chapter can be extended for other such
operations, like squeezing, photon addition followed by subtraction
or vice versa. 
%
%
%
%
%
%
\chapter{Impact of photon addition and subtraction on nonclassical and phase
properties of a displaced Fock state \textsc{\label{cha:PASDFS}}}

In this chapter, we aim to observe the nonclassical and phase properties
of a PASDFS.  The work done in this chapter is published
in ~\cite{malpani2020impact}.

\section{Introduction \label{sec:Intro-chap-4}}

In Chapter 2, we have already studied nonclassical properties of PADFS
and PSDFS. In Chapter 3, we have investigated phase properties of
the same set state. In both chapters, we have obtained various exciting
observations  such as photon addition and subtraction
enhance nonclassical properties of non-Gaussian DFS. Motivated by
these, here we aim to study both nonclassical and phase properties
for a more general quantum state. To be specific, in this chapter,
we aim to study the nonclassical (both lower- and higher-order) and
phase properties of a PASDFS. The reason behind selecting this particular
state lies in the fact that this is a general state in the sense that
in the limiting cases, this state reduces to different quantum states
having known applications in continuous variable quantum cryptography
(this point will be further elaborated in the next section).

As it appears from the above discussion, this investigation has two
facets. Firstly, we wish to study nonclassical features of PASDFS,
namely Klyshko's \cite{klyshko1996observable}, Agarwal-Tara's \cite{agarwal1992nonclassical},
Vogel's \cite{shchukin2005nonclassical} criteria, lower- and higher-order
antibunching \cite{pathak2006control}, squeezing \cite{hillery1987amplitude,hong1985generation,hong1985higher},
and sub-Poissonian photon statistics (HOSPS) \cite{zou1990photon}.
We subsequently study the phase properties of PASDFS by computing
phase distribution function \cite{agarwal1996complementarity,beck1993experimental},
phase fluctuation parameters \cite{carruthers1968phase,barnett1986phase},
and phase dispersion \cite{perinova1998phase}. A detailed analysis
of the obtained results will also be performed to reveal the usefulness
of the obtained results. 

\section{Moments of the field operators for the quantum states of our interest }

\label{sec:Quantum-states-of-3}

As mentioned in the previous section, this work is focused on PASDFS.
A PASDFS as a Fock superposition state has already been expressed
in Eq.(\ref{eq:PADFS-1}). To study nonclassical and phase properties
of this state, we have used nonclassicality witnesses introduced in
Section \ref{sec:Nonclassicality-witnesses} and phase parameters
in Section \ref{sec:Analytic-tools-forphase}, we would require analytic
expression for moments of the field operators. A bit of computation
yields the expression for higher-order moment of annihilation and
creation operator as 
\begin{eqnarray}\label{eq:PA-expepectation-1}
\langle\hat{a}^{\dagger t}\hat{a}^{j}\rangle & = & \langle\psi(k,q,n,\alpha)|\hat{a}^{\dagger t}\hat{a}^{j}|\psi(k,q,n,\alpha)\rangle\nonumber \\
 & = & \frac{N^{2}}{n!}\sum\limits _{p,p'=0}^{n}{n \choose p}{n \choose p'}(-\alpha^{\star})^{(n-p)}(-\alpha)^{(n-p')}\\
 & \times & \exp\left[-\mid\alpha\mid^{2}\right]\sum\limits _{m=0}^{\infty}\frac{\alpha^{m}(\alpha^{\star})^{m+p-p'-j+t}(m+p+k)!(m+p+k-j+t)!}{m!(m+p-p'-j+t)!(m+p+k-q-j)!}.\nonumber 
\end{eqnarray}
For different values of $t$ and $j$, moments of any order can be
obtained, and the same may be used to investigate the nonclassical
properties of PASDFS and its limiting cases by using various moments-based
criteria of nonclassicality. The same will be performed in the following
section, but before proceeding, it would be apt to briefly state our
motivation behind the selection of this particular state for the present
study (or why do we find this state as interesting?).

Due to the difficulty in realizing single photon on demand sources,
the unconditional security promised by various QKD schemes, like BB84
\cite{bennett1984quantum} and B92 \cite{bennett1992quantumBBM},
does not remain unconditional in the practical situations This is
where continuous variable QKD (CVQKD) becomes relevant as they do
not require single photon sources. Special cases of PASDFS has already
been found useful in the realization of CVQKD. For example, protocols
for CVQKD have been proposed using photon added coherent state ($k=1,\,q=0,\,n=0$)
\cite{pinheiro2013quantum,wang2014quantum}, photon added then subtracted
coherent states ($k=1,\,q=1,\,n=0$) \cite{borelli2016quantum,srikara2019continuous},
and coherent state ($k=0,\,q=0,\,n=0$) \cite{grosshans2002continuous,hirano2017implementation,huang2016long,ma2018continuous}.
Further, boson sampling with displaced single photon Fock states and
single photon added coherent state \cite{seshadreesan2015boson} has
been reported, and an $m$ photon added coherent state ($k=m,\,q=0,\,n=0$)
has been used for quantum teleportation \cite{pinheiro2013quantum}.
Apart from these schemes of CVQKD, which can be realized by using
PASDFS or its limiting cases, the fact that the photon addition and/or
subtraction operation from a classical or nonclassical state can be
performed experimentally using the existing technology \cite{parigi2007probing,zavatta2004quantum}
has enhanced the importance of PASDFS.

\section{Nonclassicality witnesses and the nonclassical features of PASDFS
witnessed through those criteria \label{sec:Nonclassicality-witnesses-2}}

The negative value of the Glauber-Sudarshan $P$-function characterizes
nonclassicality of an arbitrary state \cite{glauber1963coherent,sudarshan1963equivalence}.
As $P$-function is not directly measurable in experiments, many witnesses
of nonclassicality have been proposed, such as, negative values of
Wigner function \cite{wigner1932quantum,kenfack2004negativity}, zeroes
of $Q$ function \cite{husimi1940some,lutkenhaus1995nonclassical},
several moments-based criteria \cite{miranowicz2010testing,naikoo2018probing}.
An infinite set of such moments-based criteria of nonclassicality
is equivalent to $P$-function in terms of necessary and sufficient
conditions to detect nonclassicality \cite{richter2002nonclassicality}.
Here, we would discuss some of these moments-based criteria of nonclassicality
and $Q$ function (in Section \ref{sec:Qfn}) to study nonclassical
properties of the state of our interest.

\subsection{Lower- and higher-order antibunching}

\begin{figure}
\begin{centering}
\begin{tabular}{cc}
\includegraphics[width=60mm]{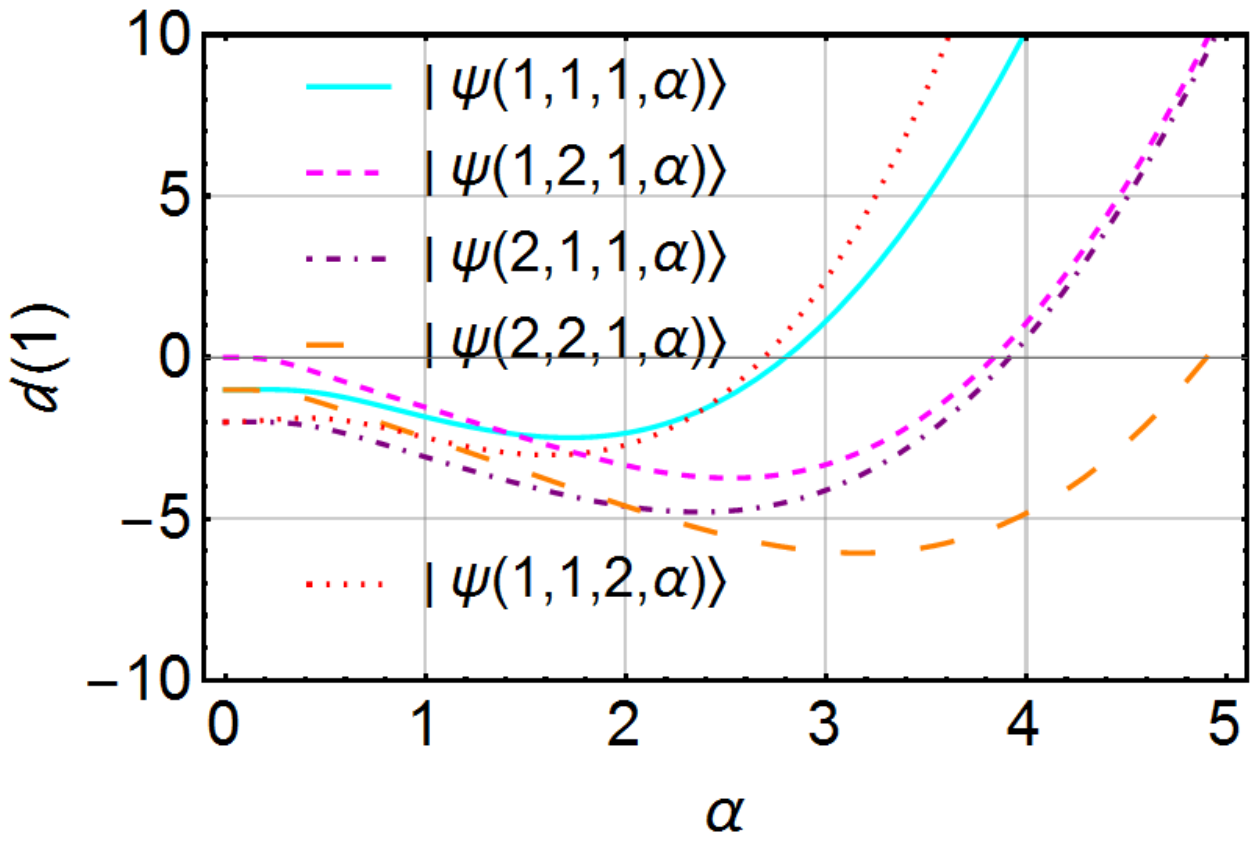}  & \includegraphics[width=60mm]{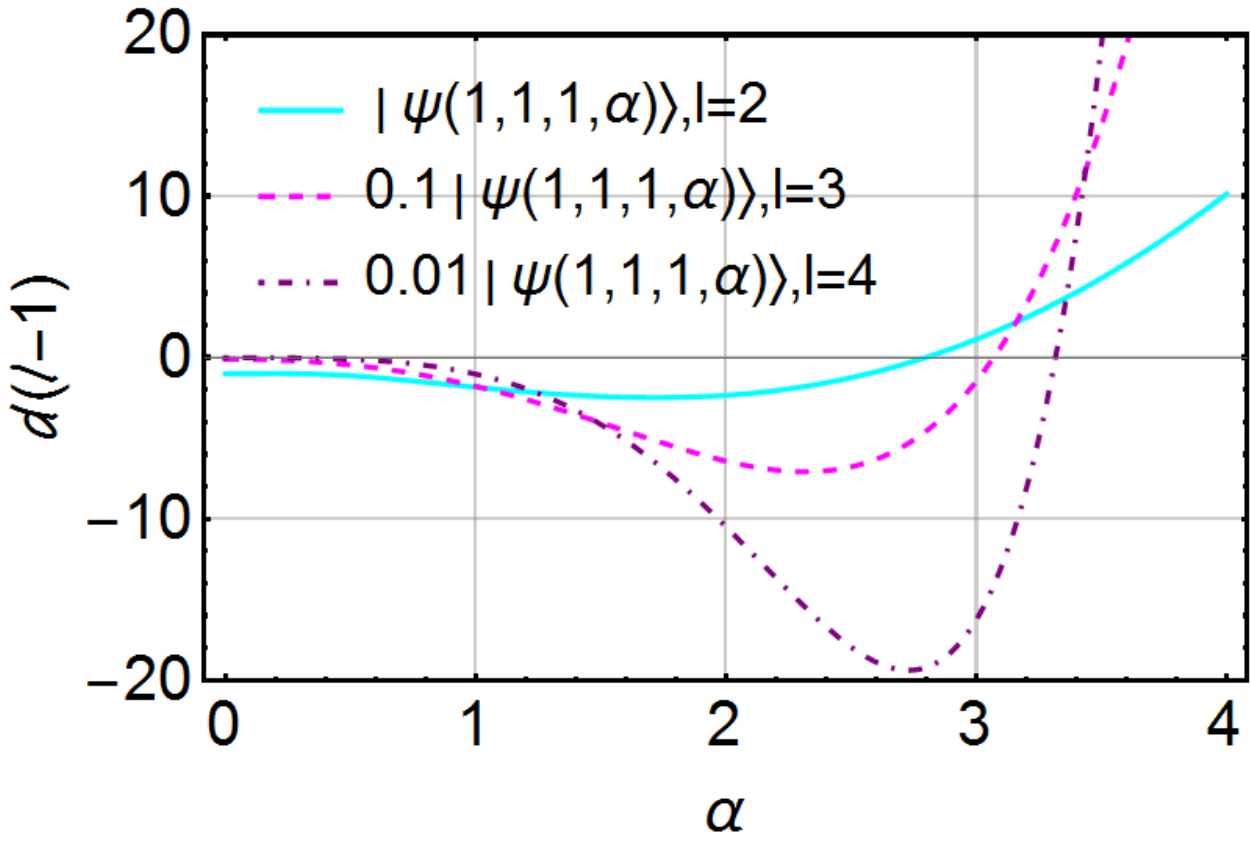}\tabularnewline
(a)  & (b) \tabularnewline
\includegraphics[width=60mm]{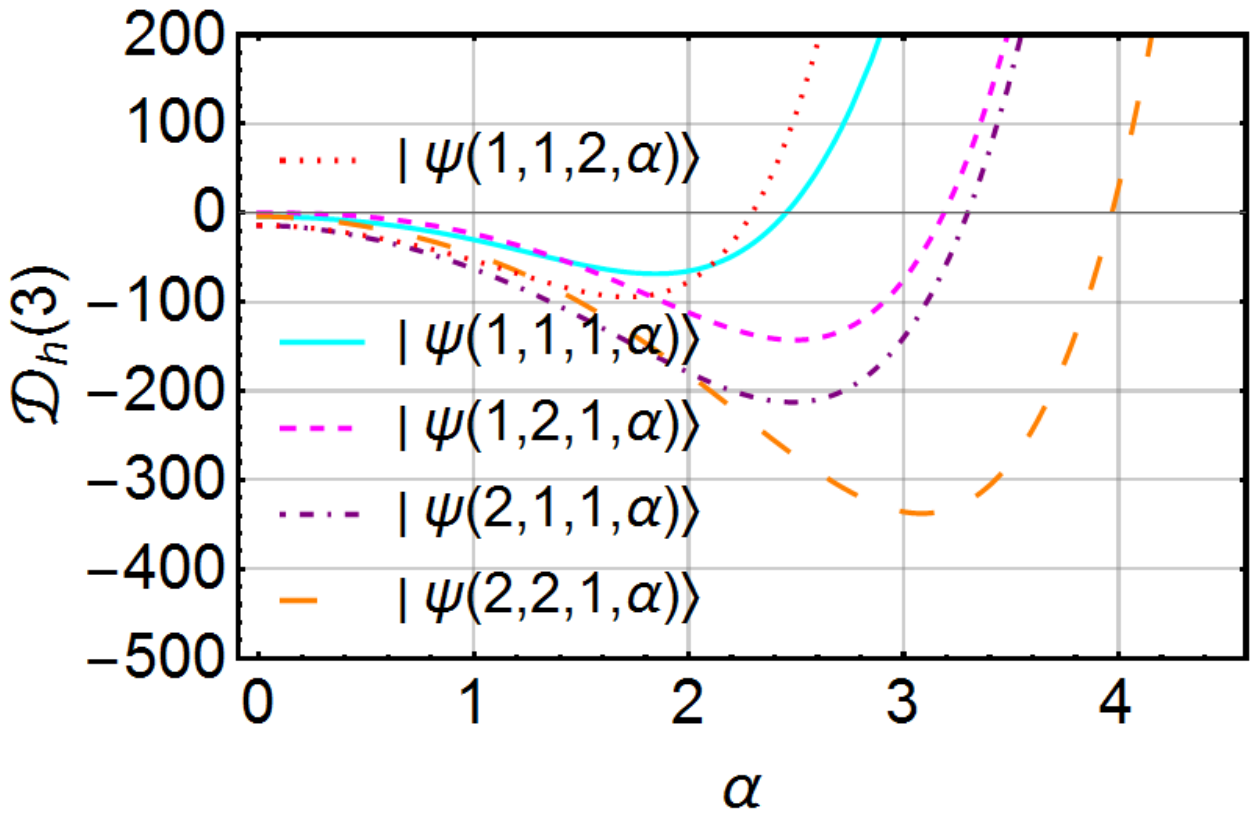}  & \includegraphics[width=60mm]{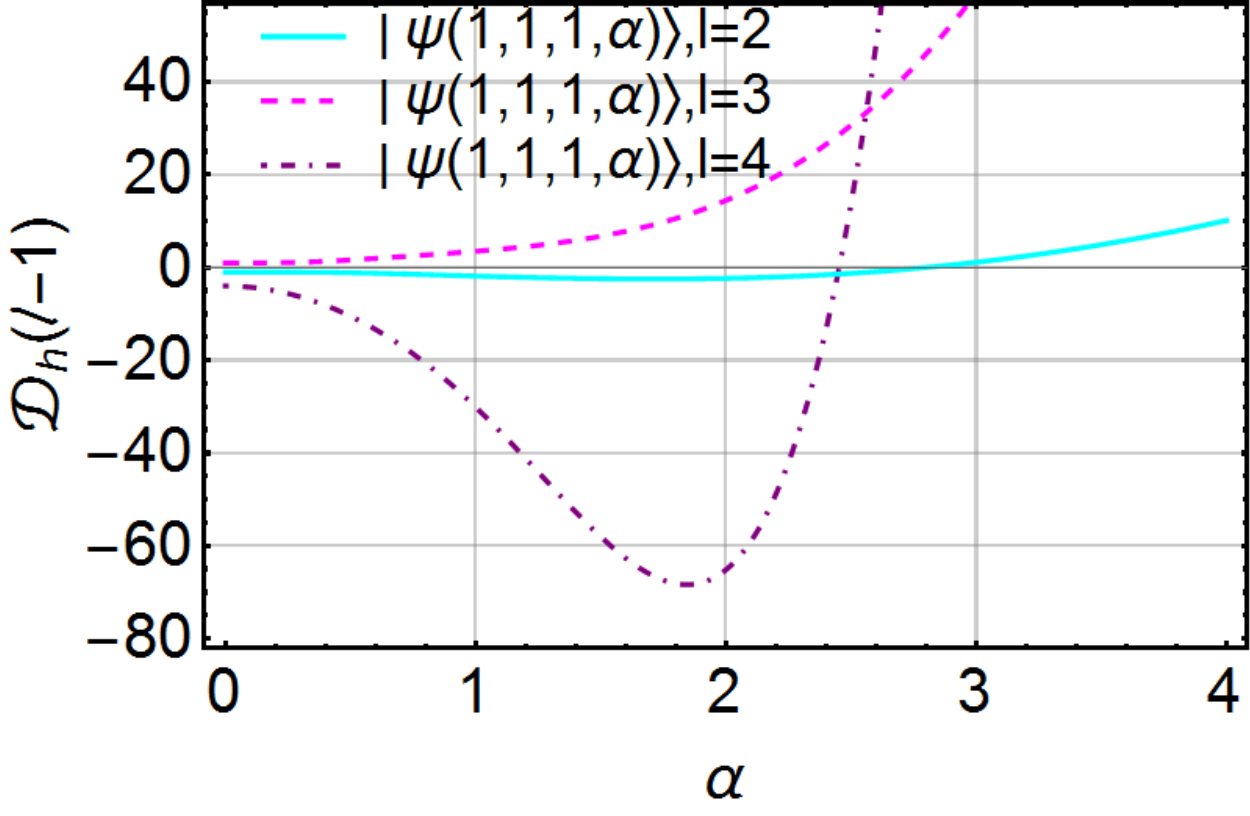}\tabularnewline
(c)  & (d) \tabularnewline
\end{tabular}
\par\end{centering}
\caption{\label{fig:HOSPS-1} For PASDFS the lower-and higher-order antibunching
is given as a function of displacement parameter $\alpha$. (a) {Lower-order antibunching for }different values
of parameters of the state. (b) {Higher-order
antibunching for particular state.} HOSPS for PASDFS for {different
}values of (c) state parameters and (d) {order of nonclassicality.}}
\end{figure}

The relevance of photon addition, photon subtraction, Fock, and displacement
parameters in the nonclassical properties of the class of PASDFSs
is studied here rigorously. Specifically, using Eq. (\ref{eq:PA-expepectation-1})
with the criterion of antibunching (\ref{eq:HOA-1}) we can study
the possibilities of observing lower- and higher-order antibunching
in the quantum states of PASDFS class, where the class of PASDFSs
refers to all the states that can be reduced from state (\ref{eq:PADFS-1})
in the limiting cases. The outcome of such a study is illustrated
in Figure \ref{fig:HOSPS-1}. It is observed that the depth of lower-
and higher-order nonclassicality witnesses can be increased by increasing
the value of the displacement parameter, but large values of $\alpha$
deteriorate the observed nonclassicality (cf. Figure \ref{fig:HOSPS-1}
(a)-(b)). The nonclassicality for higher-values of displacement parameter
$\alpha$ can be induced by subtracting photons at the cost of reduction
in the depth of nonclassicality witnessed for smaller $\alpha$, as
shown in Figure \ref{fig:HOSPS-1} (a). However, photon addition is
always more advantageous than subtraction. Therefore, both addition
and subtraction of photons illustrate these collective effects by
showing nonclassicality for even higher values of $\alpha$ at the
cost of that observed for the small values of displacement parameter.
Fock parameter has completely opposite effect of photon subtraction
as it shows the advantage (disadvantage) for small (large) values
of displacement parameter. Figure \ref{fig:HOSPS-1} (b) shows benefit
of studying higher-order nonclassicality as depth of corresponding
witness of nonclassicality can be observed to increase with the order.
The higher-order nonclassicality criterion is also able to detect
nonclassicality for certain values of displacement parameter for which
the corresponding lower-order criterion failed to do so.

\subsection{Higher-order sub-Poissonian photon statistics}

Variation of HOSPS nonclassicality witness for class of PASDFSs obtained
by different nonclassicality inducing operations show the same effect
as that of antibunching witness for all the odd orders of HOSPS, and
as depicted in Figure \ref{fig:HOSPS-1} (c). However, this nonclassical
feature disappears for even orders of HOSPS (cf. Figure \ref{fig:HOSPS-1}
(d)). In case of the odd orders of HOSPS, though the depth of nonclassicality
witness increases with the order, higher-order criterion is found
to fail to detect nonclassicality for certain values of $\alpha$
when corresponding HOSPS criterion for smaller values of orders shown
the nonclassicality.

\subsection{Lower- and higher-order squeezing}

\begin{figure}
\begin{centering}
\begin{tabular}{cc}
\includegraphics[width=60mm]{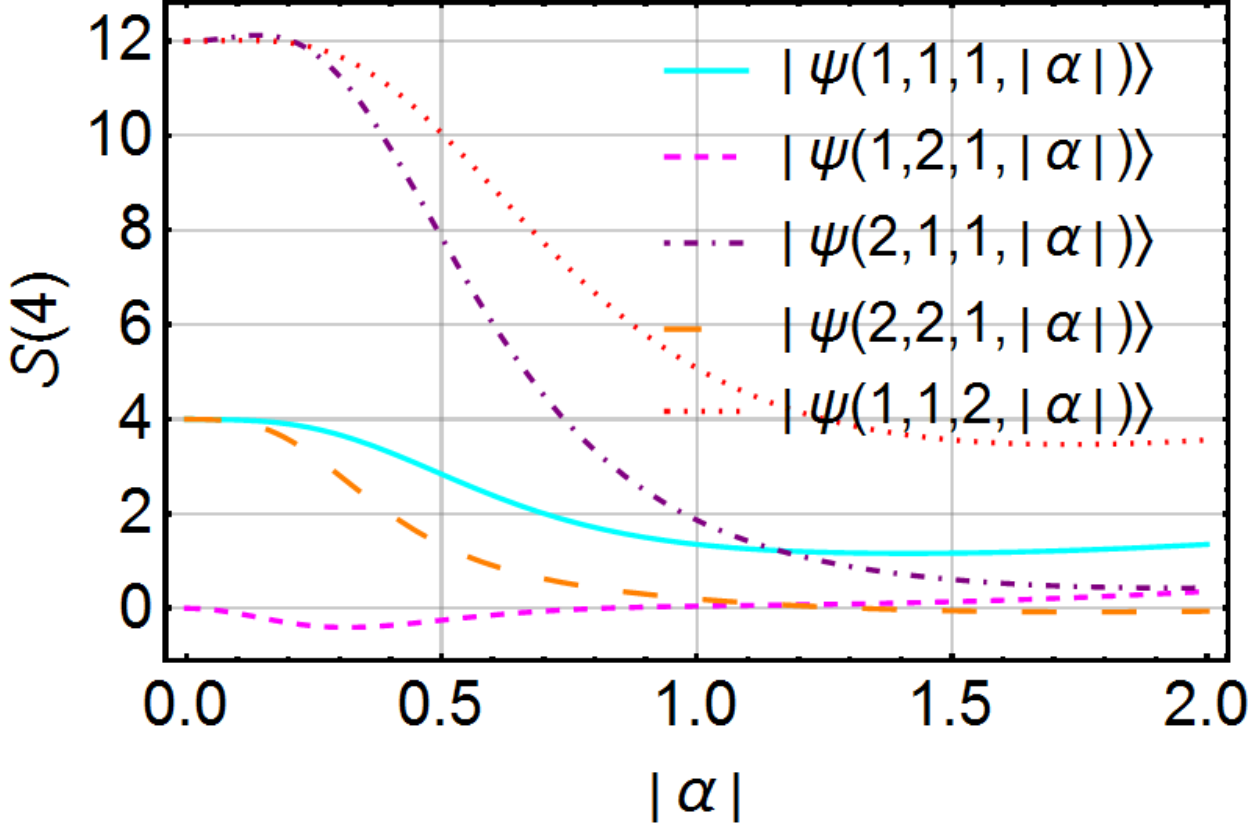}  & \includegraphics[width=60mm]{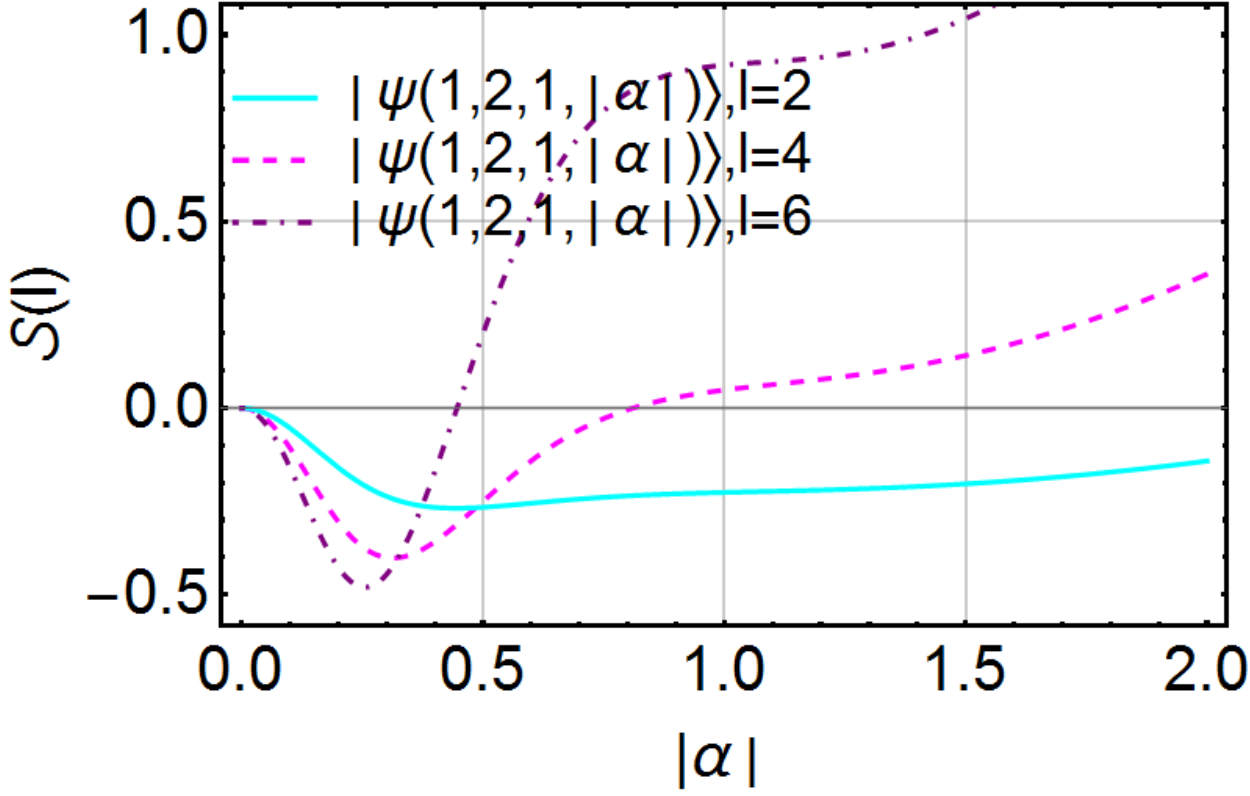}\tabularnewline
(a)  & (b) \tabularnewline
\includegraphics[width=60mm]{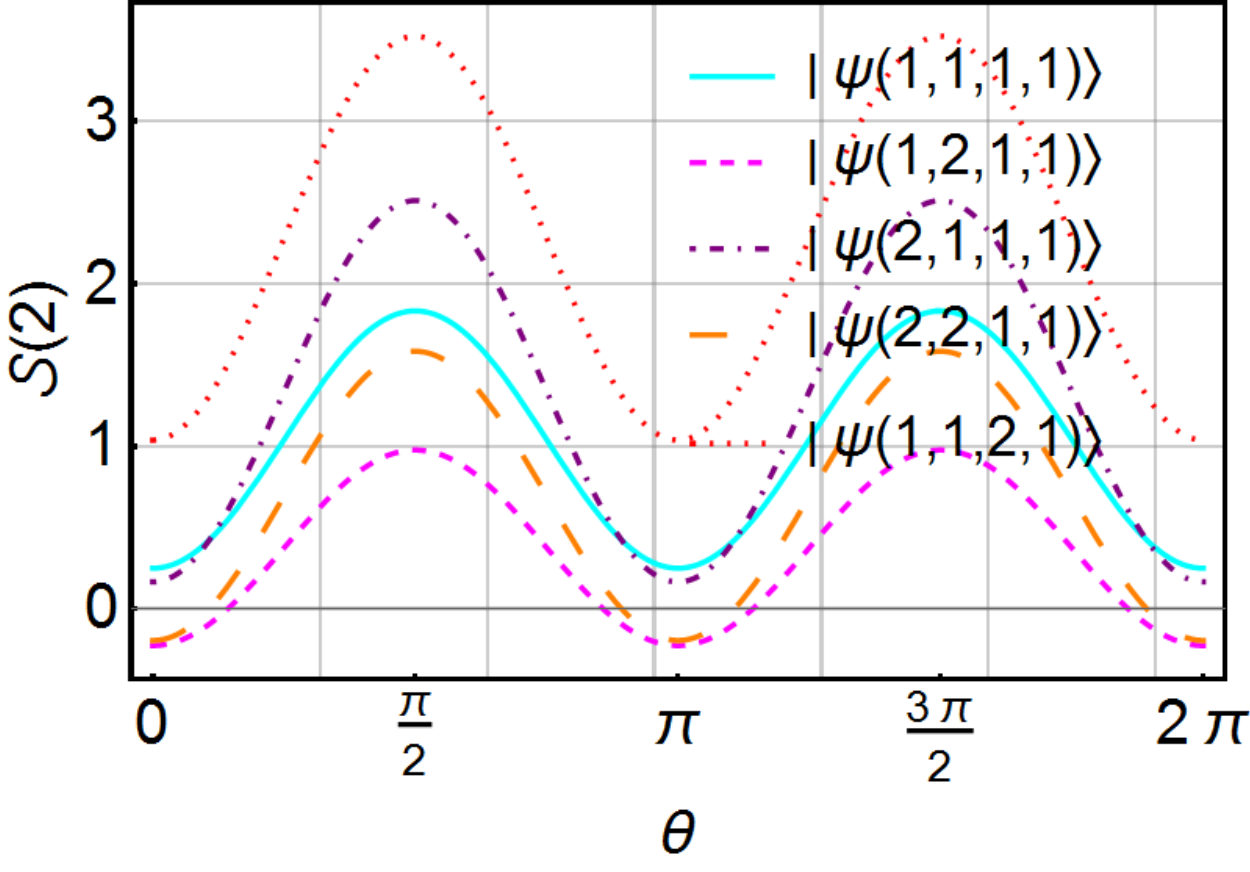} & \tabularnewline
(c)  & \tabularnewline
\end{tabular}
\par\end{centering}
\caption{\label{fig:HOS-2} Dependence of the Hong-Mandel-type higher-order
squeezing witness on displacement parameter for (a) different state
parameters and (b) order of squeezing. (c) Lower-order squeezing as
a function of phase parameter of the state, i.e., phase $\theta$
of displacement parameter $\alpha=1.\exp[\iota\theta].$}
\end{figure}

Out of all the nonclassicality inducing operations used in PASDFS
only photon subtraction is squeezing inducing operation as shown in
Figure \ref{fig:HOS-2}, which is consistent with some of our recent
observations \cite{malpani2019lower}. With photon addition higher-order
squeezing can be induced for large values of modulus of displacement
parameter at the cost of squeezing observed for small $\left|\alpha\right|$
as long as the number of photon subtracted is more than the value
of Fock parameter. As far as higher-order squeezing is concerned,
the observed nonclassicality disappears for large values of real displacement
parameter with increase in the depth of the nonclassicality witness.
Squeezing being a phase dependent nonclassical feature depends on
the phase $\theta$ of the displacement parameter $\alpha=\left|\alpha\right|\exp[\iota\theta]$
(shown in Figure \ref{fig:HOS-2} (c) for lower-order squeezing).

\begin{figure}
\begin{centering}
\begin{tabular}{c}
\includegraphics[width=80mm]{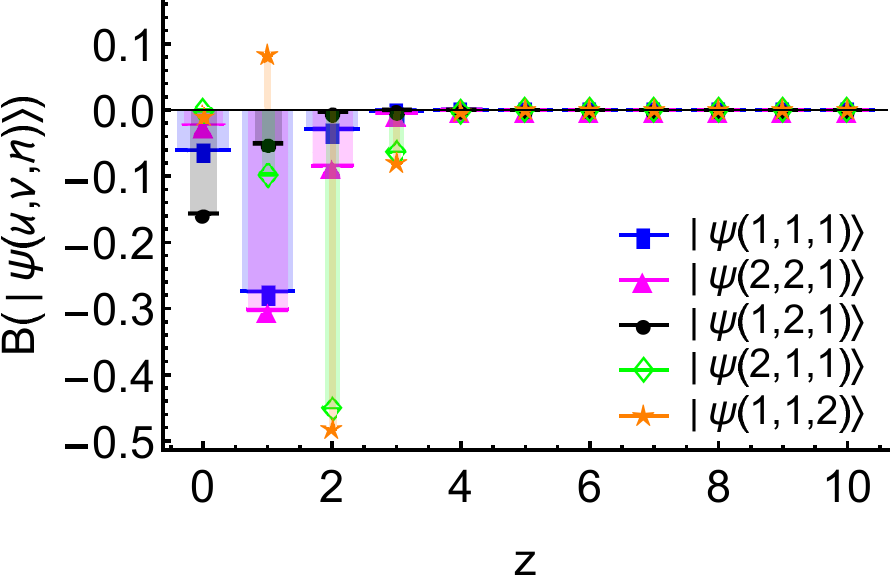}\tabularnewline
(a) \tabularnewline
\includegraphics[width=80mm]{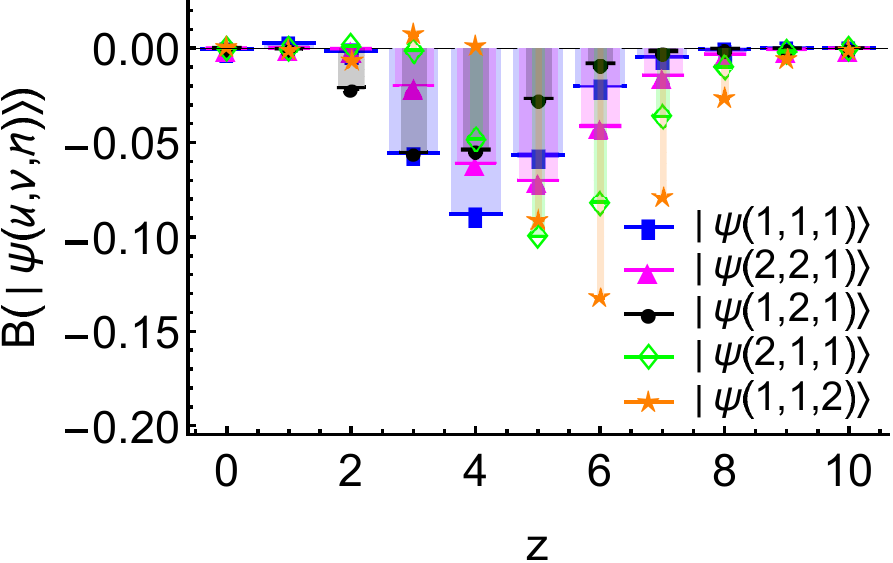} \tabularnewline
(b) \tabularnewline
\end{tabular}
\par\end{centering}
\caption{\label{fig:Klyshko-1}  Illustration of Klyshko's
parameter $B\left(z\right)$ with respect to the photon number $z$  for different values of state parameters  with
(a) $\alpha=0.5$ and (b) $\alpha=1$.}
\end{figure}

\subsection{Klyshko's Criterion}

For PASDFS $p_{z}$ can be obtained from Eq. (\ref{eq:PA-expepectation-1}).
Nonclassicality reflected through Klyshko's criterion can be controlled
by all the state engineering operations used here as shown in Figure
\ref{fig:Klyshko-1}. The depth of this nonclassicality witness increases
at higher values of photon numbers $z$ due to increase in photon
addition and/or Fock parameter. In contrast, depth of witness increases
at smaller photon numbers $z$ due to photon subtraction. The Klyshko's
nonclassicality witness is positive for some photon numbers only if
$k+n>q$. Additionally, with increase in displacement parameter the
depth of nonclassicality witness decreases, and the weight of the
distribution of witness shift to higher values of $z$.

\subsection{Agarwal-Tara's criterion}

\begin{figure}
\begin{tabular}{c}
\includegraphics[width=80mm]{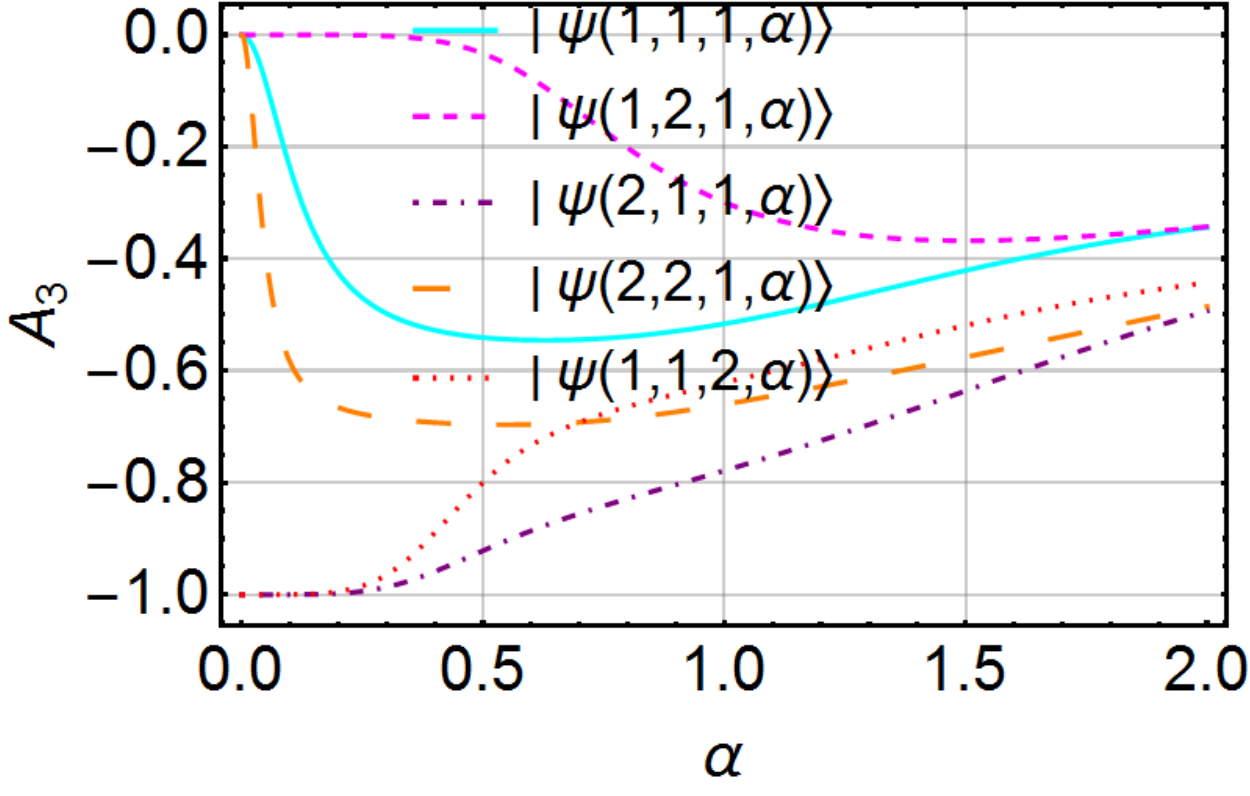}\tabularnewline
(a) \tabularnewline
\includegraphics[width=80mm]{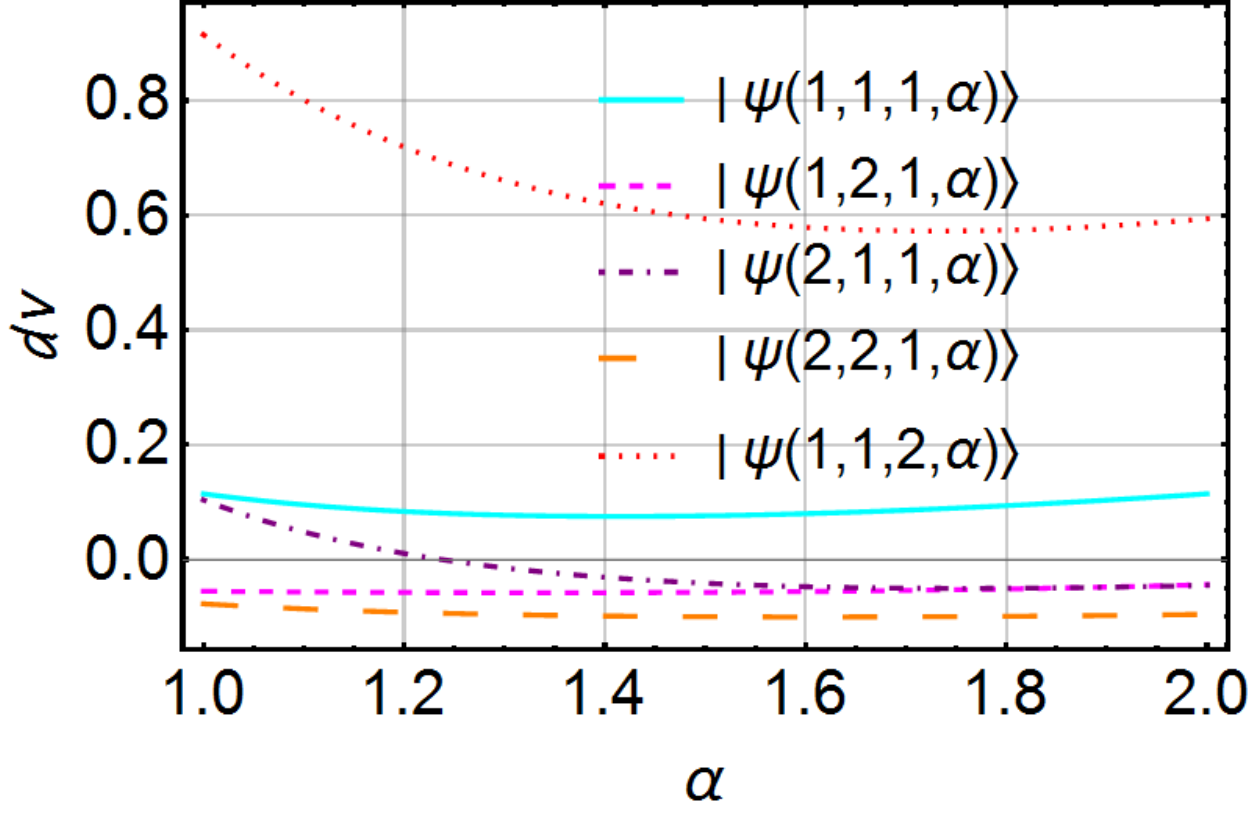}\tabularnewline
(b) \tabularnewline
\end{tabular}\caption{Nonclassicality reflected through the negative values of (a) Agarwal-Tara's
and (b) Vogel's criteria as a function of $\alpha$ or different state
parameters.}
\label{Vogel's criteria} 
\end{figure}

This nonclassicality witness is able to detect nonclassicality in
all the quantum states in the class of PASDFSs (cf. Figure \ref{Vogel's criteria}
(a)). Note that for $|\psi\left(1,2,1,\alpha\right)\rangle$ with
small $\alpha$, $A_{3}$ parameter is close to zero, which is due
to very high probability for zero photon states.

\subsection{Vogel's criterion}

The negative value of the determinant $dv$ of matrix $v$ in Eq.
(\ref{eq:vogel}) is signature of nonclassicality. Fock parameter
has adverse effect on the nonclassicality in PASDFS detected by this
criterion. This averse effect can be compensated by photon subtraction
and can be further controlled by photon addition (as shown in Figure
\ref{Vogel's criteria} (b)). Notice that the nonclassical behavior
illustrated by Agarwal-Tara's (Vogel's) criterion is related to higher-order
antibunching (squeezing) criterion. However, nonclassicality witness
of Vogel's criterion is a phase independent property unlike squeezing.

\section{Phase properties of PASDFS\label{sec:Phase-properties-of}}

The nonclassicality inducing operations are also expected to impact
the phase properties of a quantum state \cite{banerjee2007phase}.
Recently, we have reported an extensive study on the role that such
quantum state engineering tools can play in application oriented studies
on quantum phase \cite{malpani2019quantum}. Specifically, relevance
in quantum phase estimation, phase fluctuation, and phase distribution
were discussed which can play an important role in quantum metrology
\cite{giovannetti2011advances}. Here, we briefly discuss some of
the phase properties of the class of PASDFSs.

\subsection{Phase distribution function}

The analytical expression for phase distribution function for PASDFS
can be computed as

\begin{eqnarray}
\begin{array}{lcl}
P(\theta) & = & \frac{1}{2\pi}\dfrac{N^{2}}{n!}\sum\limits _{p,p'=0}^{n}{n \choose p}{n \choose p'}(-\alpha^{\star})^{(n-p)}(-\alpha)^{(n-p')}\exp\left[-\mid\alpha\mid^{2}\right]\\
 & \times & \sum\limits _{m,m^{\prime}=0}^{\infty}\frac{\alpha^{m}(\alpha^{\star})^{m^{\prime}}(m+p+k)!(m^{\prime}+p'+k)!}{m!m^{\prime}!\sqrt{\left(m+p+k-q\right)!\left(m^{\prime}+p'+k-q\right)!}}\exp[\iota\theta(m^{\prime}+p'-m-p)].
\end{array}\label{eq:PA-phase-1}
\end{eqnarray}

\begin{figure}
\begin{centering}
\begin{tabular}{ccc}
\includegraphics[width=40mm]{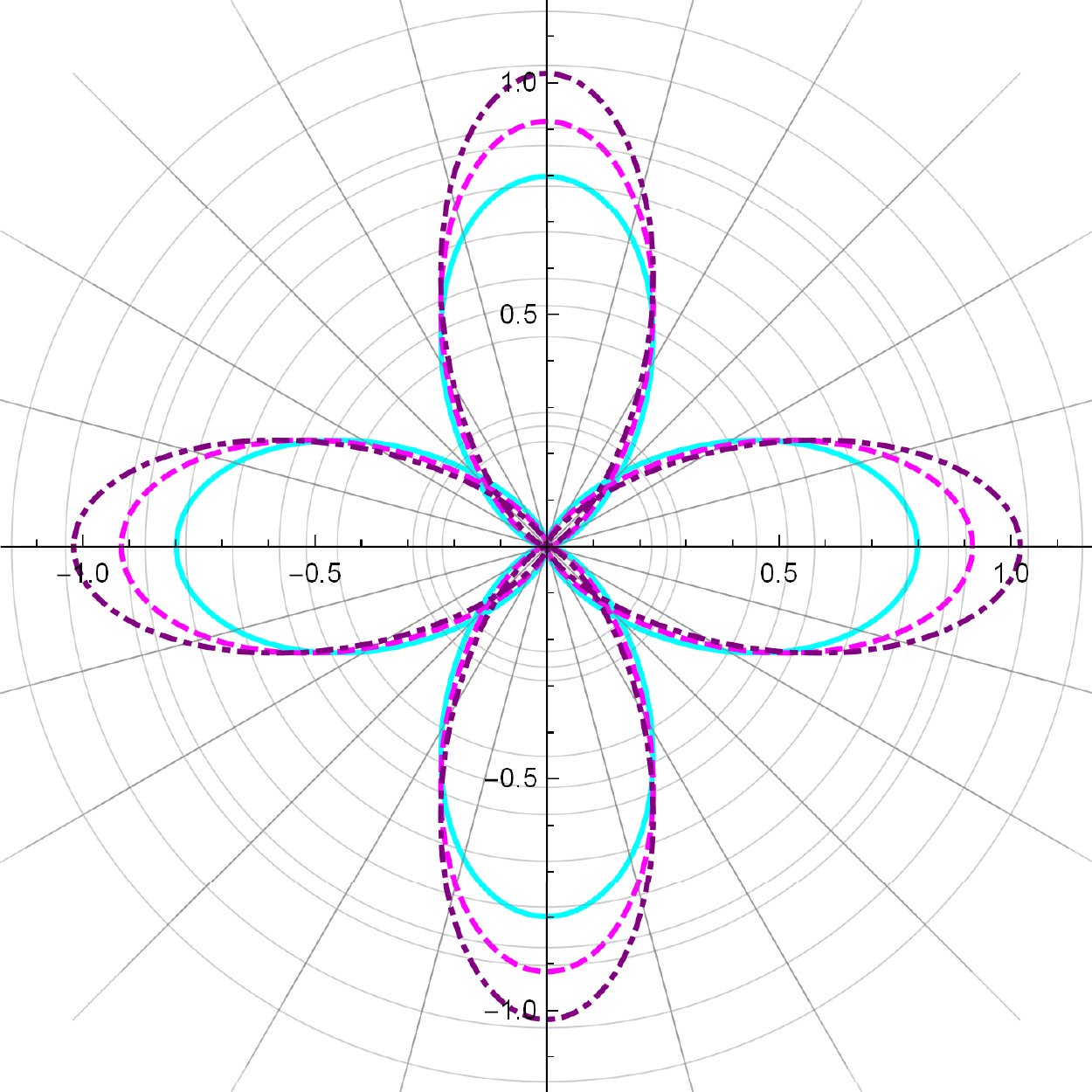}  & \includegraphics[width=40mm]{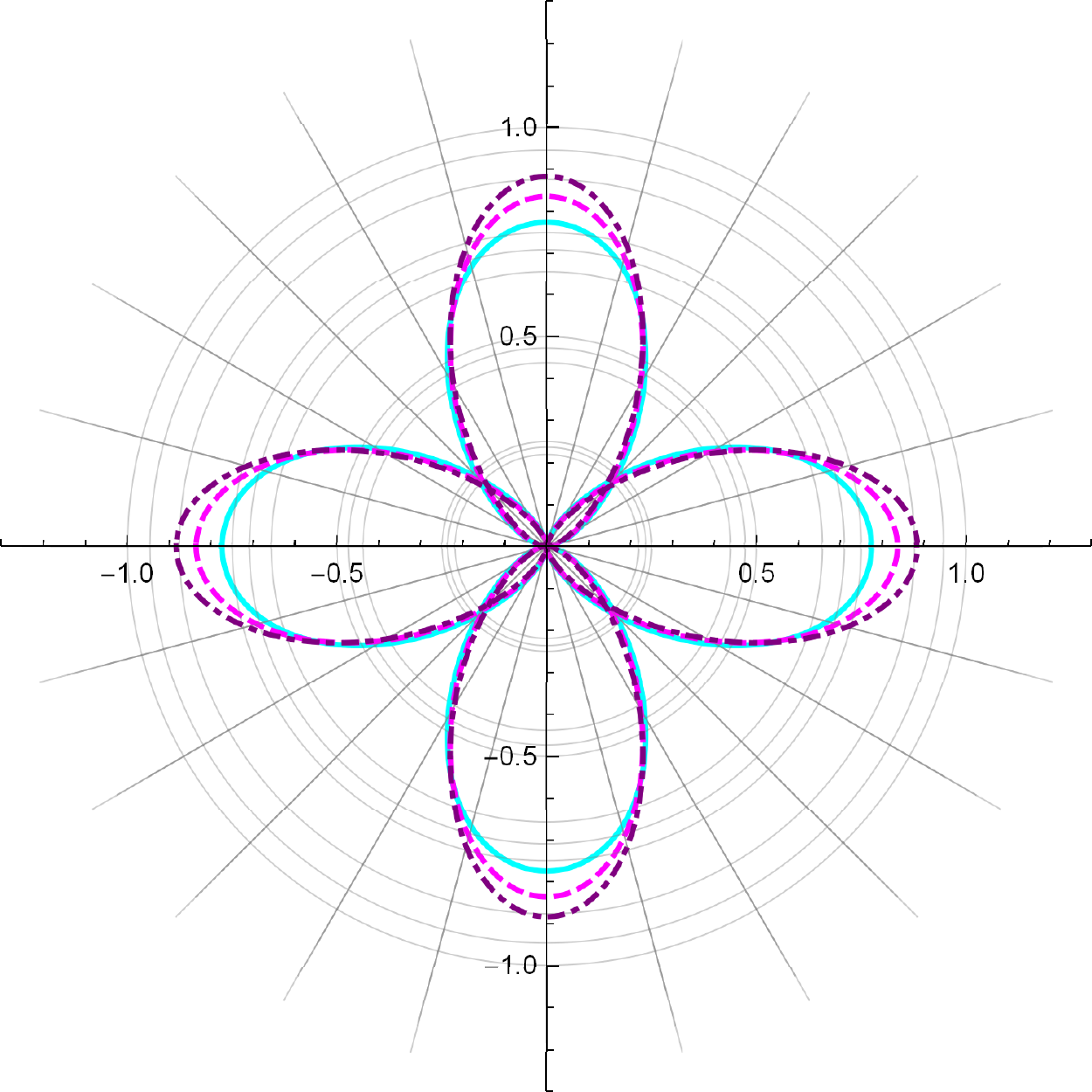}  & \includegraphics[width=40mm]{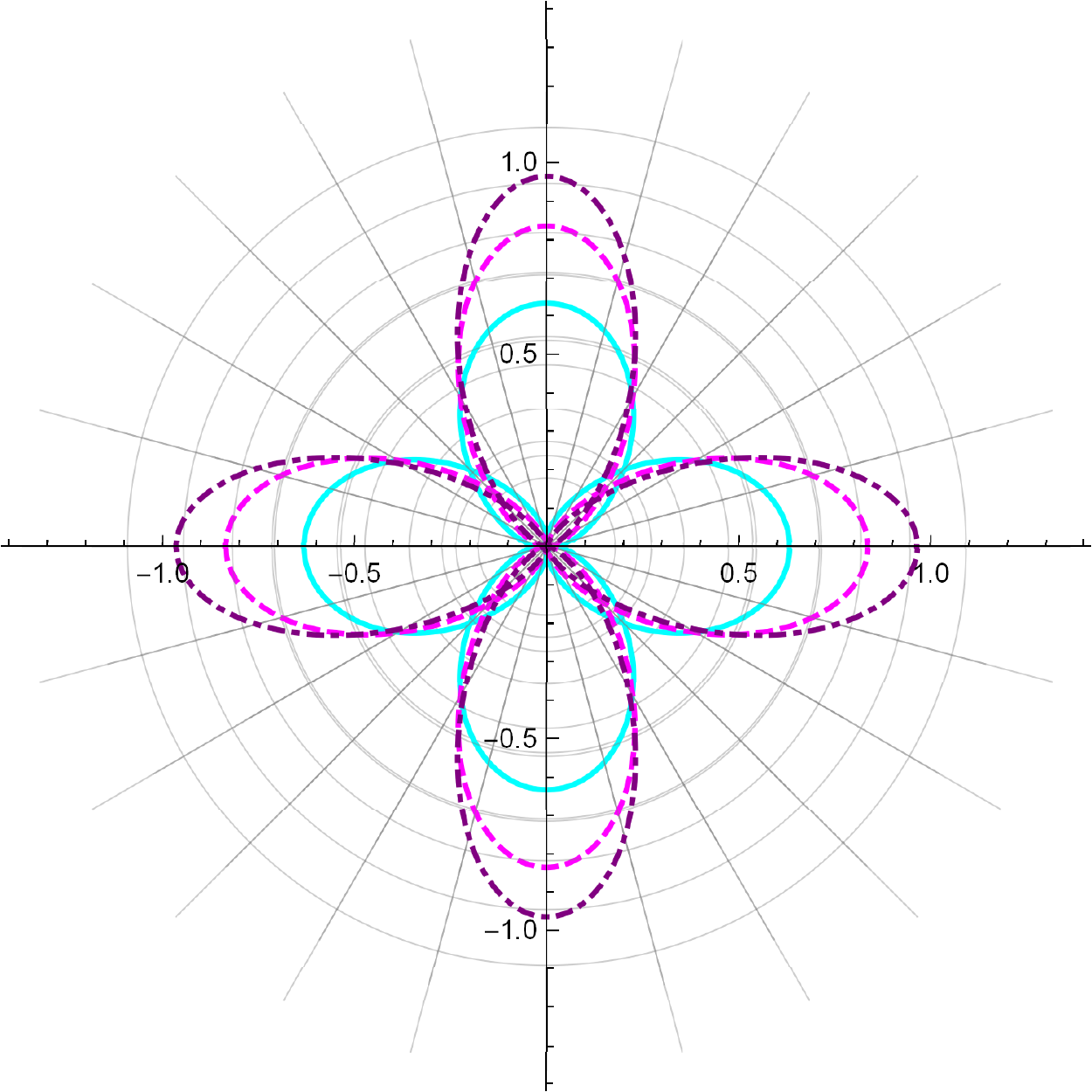}\tabularnewline
(a)  & (b)  & (c) \tabularnewline
\end{tabular}
\par\end{centering}
\caption{\label{fig:Phase distribution function}Polar plot of phase distribution
function for PASDFS $|\psi\left(k,q,n,\alpha\right)\rangle$ with
respect to variation in displacement parameter for (a) $n=1,\,k=2$
and $q=1,$ 2, and 3 represented by the smooth (cyan), dashed (magenta),
and dot-dashed (purple) lines, respectively; (b) $n=2,\,q=2$ and
$k=1,$ 2, and 3 illustrated by the smooth (cyan), dashed (magenta),
and dot-dashed (purple) lines, respectively; and (c) $n=1$ with $k=q=1,$
2, and 3 shown by the smooth (cyan), dashed (magenta), and dot-dashed
(purple) lines, respectively.}
\end{figure}

Photon subtraction can be observed to be a more effective tool to
alter phase properties of PASDFS than photon addition, as shown in
Figure \ref{fig:Phase distribution function}. Interestingly, photon
addition shows similar behavior, though less prominent, as photon
subtraction, Fock parameter has opposite effect.

\begin{figure}
\begin{centering}
\begin{tabular}{c}
\includegraphics[width=60mm]{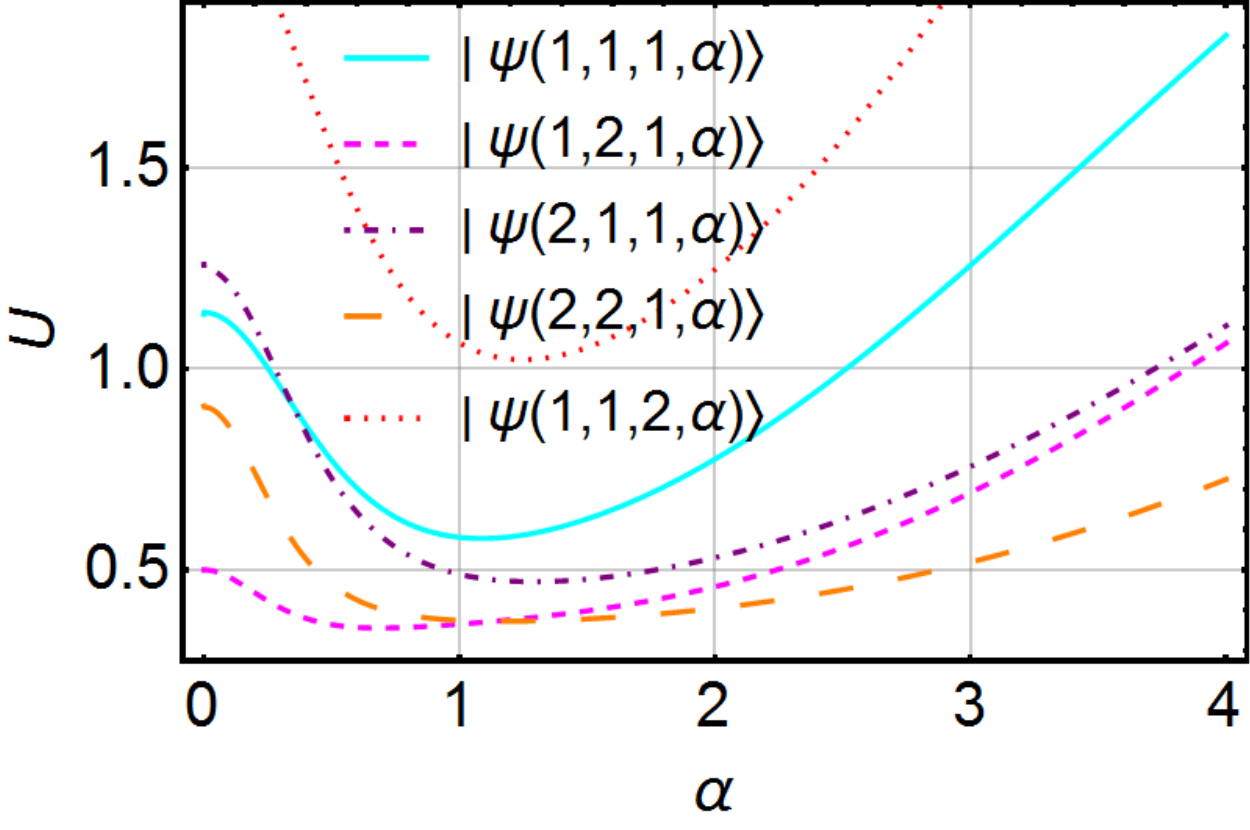} \tabularnewline
\tabularnewline
\end{tabular}
\par\end{centering}
\caption{\label{fig:Phase fluctuation}\textcolor{green}{{}} Variation of
phase fluctuation parameter with displacement parameter\textcolor{green}{{}
}for various state parameters in PASDFS.}
\end{figure}

\subsection{Phase Fluctuation}

Here, we focus only on the first phase fluctuation parameter $U$,
which is related to antibunching if $U$ is below its value for coherent
state (i.e., 0.5), remaining consistent with Barnett-Pegg formalism
\cite{gupta2007reduction,pathak2000phase}. One can observe that the
phase fluctuation parameter is able to detect nonclassicality (specifically
antibunching) only in three cases where the role of the photon subtraction
is relevant (cf. Figure \ref{fig:Phase fluctuation}). The observation
can be seen analogous to that observed for Vogel's nonclassicality
criterion.

\begin{figure}
\begin{centering}
\begin{tabular}{cc}
\includegraphics[width=70mm]{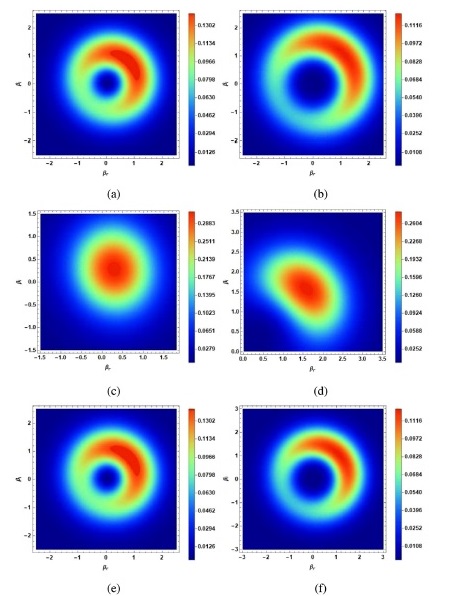}  
\end{tabular}
\par\end{centering}
\caption{\label{fig:Q function} $Q$ function for PASDFS $|\psi\left(k,q,n,\alpha\right)\rangle$
with (a) $k=q=n=1,$ (b) $k=2,\,q=n=1,$ and (c) $q=2,\,k=n=1$ with
$\alpha=\frac{1}{5\sqrt{2}}\exp\left(\iota\pi/4\right)$. (d) Similarly,
$Q$ function of PASDFS with $q=2,\,k=n=1$ and $\alpha=\sqrt{2}\exp\left(\iota\pi/4\right)$.
$Q$ function for $|\psi\left(k,q,n,\alpha\right)\rangle$ with (e)
$k=q=1,\,n=2$ and (f) $q=1,\,k=n=2$ for $\alpha=\frac{1}{5\sqrt{2}}\exp\left(\iota\pi/4\right)$.}
\end{figure}

\section{Quasidistribution function: $Q$ function \label{sec:Qfn}}

Here, We will establish non-Gaussianity inducing behavior of photon
addition and Fock parameter (cf. Figure \ref{fig:Q function}), which
are so far illustrated as nonclassicality inducing and phase altering
operations. Clearly, with photon addition tendency of quasidistribution
away from Gaussian behavior is visible, while with photon subtraction
squeezing along particular phase angle chosen by displacement parameter
can be observed. This squeezing can be noticed to be more appreciable
for higher values of displacement parameter (cf. Figure \ref{fig:Q function}
(c)-(d)). From Figure \ref{fig:Q function} (e)-(f), it can be observed
that Fock parameter and photon addition have a similar effect in the
phase space. As zeros of $Q$ function are signature of nonclassicality,
PASDFS shows nonclassicality in Figure \ref{fig:Q function} (b),
(e), and (f).  This establishes that use of more
than one state engineering tool may be helpful in generation of nonclassical
states. It would be interesting  to verify whether one more tools (say
squeezing or photon catalysis) may further enhance the nonclassical
properties.  

\section{Conclusions \label{sec:Conclusions-5}}

In this chapter, we have investigated the nonclassical behavior of
PASDFS using different witnesses of lower- and higher-order nonclassicality.
The significance of this choice of state underlies the fact that a
class of engineered quantum states can be achieved as the reduced
case of PASDFS $|\psi\left(k,q,n,\alpha\right)\rangle$, like photon
added DFS $\left(q=0\right)$, photon subtracted DFS $\left(k=0\right)$,
DFS $\left(q=k=0\right)$, photon added coherent state $\left(n=q=0\right)$,
photon subtracted coherent state $\left(n=k=0\right)$, coherent state
$\left(n=k=q=0\right)$, and Fock state $\left(n=k=q=\alpha=0\right)$.
Some of the reduced states have been experimentally realized and in
some cases optical schemes for generation have been proposed, so this
family of states is apt for various challenging tasks to establish
quantum dominance. The state under consideration requires various
non-Gaussianity inducing quantum engineering operations and thus our
focus here was to analyze the relevance of each operation independently
in the nonclasscial features (listed in Table \ref{tab:Properties of PASDFS})
observed in PASDFS. To study the nonclassical properties of PASDFS,
a set of moments-based criteria for Klyshko's, Agrwal-Tara's, and
Vogel's criteria, as well as lower- and higher-order antibunching,
HOSPS, and squeezing. Further, phase properties for the same state
are also studied using phase distribution function and phase fluctuation.
Finally, non-Gaussianity and nonclassicality of PASDFS is also studied
using $Q$ function.

\begin{table}
\begin{centering}
\begin{tabular}{ccc}
\toprule 
S. No.  & Nonclassical Properties  & Observed in PASDFS\tabularnewline
\midrule 
1  & Lower-order and higher-order Antibunching  & yes\tabularnewline
2  & Higher-order sub Poissionian photon statistics  & yes\tabularnewline
3  & Lower-order and higher-order squeezing  & yes\tabularnewline
4  & Klyshko's criterion  & yes\tabularnewline
5  & Agarwal-Tara's criterion  & yes\tabularnewline
6  & Vogel's criterion  & yes\tabularnewline
7  & Phase distribution function  & -\tabularnewline
8  & Phase fluctuation  & yes\tabularnewline
9  & $Q$ function  & yes\tabularnewline
\bottomrule
\end{tabular}
\par\end{centering}
\caption{\label{tab:Properties of PASDFS} Summary of the nonclassical properties
of PASDFS. }
\end{table}
The present study reveals that with an increase in the order of nonclassicality
the depth of nonclassicality witnesses increase. Additionally, higher-order
nonclassicality criteria were able to detect nonclassicality in the
cases when corresponding lower-order criteria failed to do so. Different
nonclassical features are observed for smaller values of displacement
parameter, which can be sustained for higher values by increasing
the number of subtracted photon. Photon addition generally improves
nonclassicality, and this advantage can be further enhanced for the
higher (smaller) values of displacement parameter using photon subtraction
(Fock parameter). The HOSPS nonclassical feature is only observed
for the odd orders. As far as squeezing is concerned, only photon
subtraction could induce this nonclassicality. Large number of photon
addition can be used to observe squeezing at higher values of displacement
parameter at the cost of that present for smaller $\alpha$. Photon
subtraction alters the phase properties more than photon addition,
while Fock parameter has an opposite effect of the photon addition/subtraction.
The nonclassicality revealed through phase fluctuation parameter shows
similar behavior as Vogel's criterion. Finally, we have shown the
nonclassicality and non-Gaussianity of PASDFS with the help of a quasidistribution
function, namely $Q$ function.
\chapter{Manipulating nonclassicality via quantum state engineering processes:
Vacuum filtration and single photon addition\textsc{\label{cha:QSE-1}}}

In this chapter, the objective is to study nonclassical properties
associated with two different quantum state engineering processes
with a specific focus on nonclassicality witnesses and measures. The
work done in this chapter is published in ~\cite{malpani2019filter}.

\section{Introduction \label{cha:QSE}\label{introduction:chapter5}}

So far, we have discussed nonclassical properties of engineered quantum
states in detail. Here, we wish to extend the discussion and investigate
the possibilities of manipulating or controlling the nonclassicality
present in the system by using two specific processes of quantum state
engineering. Precisely, by using vacuum filtration and single photon
addition processes. To introduce the idea of these quantum state engineering
operations, we can write the photon number distribution of an arbitrary
quantum state in terms of Glauber-Sudarshan $P\left(\alpha\right)$
function as 
\begin{equation}
p_{n}=\int P\left(\alpha\right)\left|\langle n|\alpha\rangle\right|^{2}d^{2}\alpha.\label{eq:pnd}
\end{equation}
If $p_{n}$ vanishes for a particular value of Fock state parameter
$n$, we refer to that as a ``hole'' or a hole in the photon number
distribution at position $n$ \cite{escher2004controlled}. Notice
that $p_{n}=0$ reveals that $P\left(\alpha\right)<0$ for some $\alpha$,
which is the signature of nonclassicality. Thus, the existence of
a hole in the photon number distribution implies that the corresponding
state is nonclassical, and corresponding technique of quantum state
engineering to create hole is called hole burning \cite{gerry2002hole}.
Interestingly, this result also implies that qudits which are $d$-dimensional
(finite dimensional) quantum states are always nonclassical as we
can see that in such a state $p_{d}=p_{d+1}=\ldots=0$. In principle,
the hole can be created for an arbitrary $n$, but here for the sake
of a comparative study, we restrict ourselves to the situation where
the hole is created at $n=0,$ i.e., the desired engineered state
has zero probability of getting vacuum state on measurement in Fock
basis (in other words, $p_{0}=0).$ In fact, Lee \cite{lee1995theorem}
had shown that a state with $p_{n}=0$ is a maximally nonclassical
as long as the nonclassicality is quantitatively measured using nonclassical
depth. Such a state can be constructed in various ways. To elaborate
on this, we describe an arbitrary pure quantum state as a superposition
of the Fock states 
\begin{equation}
|\psi\rangle=\sum\limits _{n=0}^{\infty}c_{n}|n\rangle,\label{eq:fock-superposition}
\end{equation}
where $c_{n}$ is the probability amplitude of state $|n\rangle$.
A hole can be created at $n=0$ by adding a single photon to obtain
\begin{equation}
|\psi_{1}\rangle=N_{1}a^{\dagger}|\psi\rangle,\label{eq:a-dagger-for-hole}
\end{equation}
where $N_{1}=\left(\langle\psi|aa^{\dagger}|\psi\rangle\right)^{-1/2}$
is the normalization constant. If we consider the initial quantum
state $|\psi\rangle$ as a coherent state, the addition of a single
photon would lead to a photon added coherent state which has been
experimentally realized \cite{zavatta2004quantum} and extensively
studied \cite{hong1999nonclassical} because of its interesting nonclassical
properties and potential applications. Thus, in quantum state engineering,
techniques for photon addition are known \cite{thapliyal2017comparison,malpani2019lower,malpani2019quantum}
and experimentally realized.

An alternative technique to create a hole at vacuum is vacuum filtration.
The detailed procedure of this technique is recently discussed in
\cite{Meher2018}. Vacuum filtration implies removal of the coefficient
of the vacuum state, $c_{0}$, in Eq. (\ref{eq:fock-superposition})
and subsequent normalization. Clearly this procedure would yield 
\begin{equation}
|\psi_{2}\rangle=N_{2}\sum\limits _{n=1}^{\infty}c_{n}^{\prime}|n\rangle,\label{eq:c0-is-removed}
\end{equation}
where the normalization constant $N_{2}=\left(1-\left|c_{0}\right|^{2}\right)^{-1/2}$.
Both these states (i.e., $|\psi_{1}\rangle,$ and $\psi_{2}\rangle$)
are maximally nonclassical as far as Lee's result related to nonclassical
depth is concerned \cite{lee1991measure}. However, recently lower-order
nonclassical properties of $|\psi_{1}\rangle$ and $|\psi_{2}\rangle$
in (\ref{eq:a-dagger-for-hole})-(\ref{eq:c0-is-removed}) are reported
to be different for $|\psi\rangle$ chosen as coherent state \cite{Meher2018}.
This led to several interesting questions, like- What happens if the
initial state on which addition of photon or vacuum filtration process
is to be applied is already nonclassical (specifically, pure state
other than coherent state \cite{hillery1985classical})? How do these
processes affect the higher-order nonclassical properties of the quantum
states? How does the depth of nonclassicality corresponding to a particular
witness of nonclassicality changes with the parameters of the quantum
state for these processes? The present chapter aims to answer these
questions through a comparative study using a set of interesting quantum
states $|\psi\rangle$ (and the corresponding single photon added
$|\psi_{1}\rangle$ and vacuum filtered $|\psi_{2}\rangle$ states),
each of which can be reduced to many more states. Specifically, in
what follows, we would study the lower- and higher-order nonclassical
properties of single photon addition and vacuum filtration of even
coherent state (ECS), binomial state (BS) and Kerr state (KS). In
fact, the quantum state engineering processes described mathematically
in Eqs. (\ref{eq:a-dagger-for-hole})-(\ref{eq:c0-is-removed}) can
be used to prepare a set of engineered quantum states, namely vacuum
filtered ECS (VFECS), vacuum filtered BS (VFBS), vacuum filtered KS
(VFKS), photon added ECS (PAECS), photon added BS (PABS) and photon
added (PAKS). We aim to look at the nonclassical properties of these
states with a focus on higher-order nonclassical properties and subsequently
quantify the amount of nonclassicality in all these states. In what
follows, the higher-order nonclassical properties are illustrated
through the criteria of HOA, HOS and HOSPS with brief discussion of
lower-order antibunching and squeezing.
\section{Quantum states of interest\label{sec:Quantum-states-of-1}}

In this chapter, we have selected a set of three widely studied and
important quantum states- (i) ECS, (ii) BS and (iii) KS. We subsequently
noted that these states can further be engineered to generate corresponding
vacuum filtered states and single photon added states. For example,
one can generate VFBS and PABS from BS by using vacuum filtration
\cite{Meher2018} and photon addition \cite{zavatta2004quantum} processes,
respectively. In a similar manner, these processes can also generate
VFECS and PAECS from ECS, and VFKS and PAKS from KS. In this section,
we briefly describe ECS, BS, KS, VFBS, PABS, VFECS, PAECS, VFKS and
PAKS. Specifically, we describe three parent states as Fock superposition
states. Similarly, the six engineered states are also expressed as
Fock superposition states for the convenience of identifying the corresponding
photon number distributions (each of which essentially contains a
hole at the vacuum). In the rest of the study, we wish to compare
the impact of these two quantum state engineering processes (i.e.,
vacuum filtration and photon addition processes) on the nonclassical
properties of the engineered states.In the above, we have described
six (three) quantum states of our interest as Fock superposition states
having (without) holes at vacuum. In what follows, these expressions
will be used to study the nonclassical properties of these states
using a set of witnesses of nonclassicality. Specifically, we will
use a set of witnesses of nonclassicality which are based on moments
of annihilation and creation operators. Keeping this in mind, in the
following subsection, we report the general form of such moments for
all the six engineered states of our interest and the corresponding
three parent states (thus overall nine states).

\subsection{Expressions for moments of annihilation and creation operators}

In 1992, Agarwal and Tara \cite{agarwal1992nonclassical} introduced
a criterion of nonclassicality in the form of a matrix of moments
of creation and annihilation operators. This criterion was further
modified to propose a moment-based criteria of entanglement \cite{shchukin2005inseparability}
and nonclassicality \cite{miranowicz2010testing,miranowicz2009inseparability}.
Therefore, it is convenient to find out the expectation value of the
most general term describing higher-order moment $\langle\hat{a}^{\dagger j}\hat{a}^{k}\rangle$
for a given state to investigate the nonclassicality using the set
of moment-based criteria.

\subsubsection{Expectation values for even coherent states and the corresponding
engineered states}

The analytic expression of $\langle\hat{a}^{\dagger j}\hat{a}^{k}\rangle_{i}$
is obtained for the quantum states $i\in\{{\rm ECS},{\rm VFECS,PAECS\}}$
using Eqs. (\ref{eq:VFECS-EXPANDED}) and (\ref{eq:PAECS}). For ECS
and VFECS, expressions of the moments can be written in a compact
form as

\begin{equation}
\begin{array}{lcl}
\langle\hat{a}^{\dagger j}\hat{a}^{k}\rangle_{{\rm ECS}} & = & \frac{\exp\left[-\mid\alpha\mid^{2}\right]}{2\left(1+\exp\left[-2\mid\alpha\mid^{2}\right]\right)}\sum\limits _{n=0}^{\infty}\frac{\alpha^{n}(\alpha^{\star})^{n-k+j}}{\left(n-k\right)!}\mathcal{G}_{n,j,k}.\end{array}\label{eq:ECS-moment}
\end{equation}
and 
\begin{equation}
\begin{array}{lcl}
\langle\hat{a}^{\dagger j}\hat{a}^{k}\rangle_{{\rm VFECS}} & = & \left\{ \begin{array}{l}
N_{{\rm VFECS}}^{2}\sum\limits _{n=1}^{\infty}\frac{\alpha^{n}(\alpha^{\star})^{n-k+j}}{\left(n-k\right)!}\mathcal{G}_{n,j,k}\,\,\,\,\mathrm{for}\,\,k\leq j,\\
N_{{\rm VFECS}}^{2}\sum\limits _{n=1}^{\infty}\frac{\alpha^{\star n}\alpha^{n+k-j}}{\left(n-j\right)!}\mathcal{G}_{n,j,k}\,\,\,\mathrm{for}\,\,k>j,
\end{array}\right.\end{array}\label{eq:VFECS-moment}
\end{equation}
respectively. Similarly, we obtained analytic expression for $\langle\hat{a}^{\dagger j}\hat{a}^{k}\rangle_{{\rm PAECS}}$
for PAECS as 
\begin{equation}
\begin{array}{lcl}
\langle\hat{a}^{\dagger j}\hat{a}^{k}\rangle_{{\rm PAECS}} & = & N_{{\rm PAECS}}^{2}\sum\limits _{n=0}^{\infty}\frac{\alpha^{n}(\alpha^{\star})^{n-k+j}\left(n+1\right)\left(n-k+j+1\right)}{\left(n+1-k\right)!}\mathcal{G}_{n,j,k}.\end{array}\label{eq:PAECS-moment}
\end{equation}
Here, $\mathcal{G}_{n,j,k}=\left(1+\left(-1\right)^{n}\right)\left(1+\left(-1\right)^{n-k+j}\right)$.
The above mentioned quantities are also functions of displacement
parameter of ECS used to generate the engineered states, which will
be used as a control parameter while discussion of nonclassicality
induced due to engineering operations.

\subsubsection{Expectation values for binomial state and the corresponding engineered
states}

Similarly, the compact analytic form of $\langle\hat{a}^{\dagger t}\hat{a}^{r}\rangle_{{\rm BS}}$
can be written as 
\begin{equation}
\begin{array}{lcl}
\langle\hat{a}^{\dagger t}\hat{a}^{r}\rangle_{{\rm BS}} & = & \sum\limits _{n=0}^{M}\mathcal{I}_{p,M,n,r,t}\frac{M!}{(n-r)!}.\end{array}\label{eq:BS-moment}
\end{equation}
In case of VFBS and PABS, the analytic form of $\langle\hat{a}^{\dagger t}\hat{a}^{r}\rangle_{{i}}$
is obtained as 
\begin{equation}
\begin{array}{lcl}
\langle\hat{a}^{\dagger t}\hat{a}^{r}\rangle_{{\rm VFBS}} & = & \left\{ \begin{array}{l}
N_{{\rm VFBS}}^{2}\sum\limits _{n=1}^{M}\mathcal{I}_{p,M,n,r,t}\frac{M!}{(n-r)!}\,\,\,\,\,{\rm {for}\,r\leq t,}\\
N_{{\rm VFBS}}^{2}\sum\limits _{n=1}^{M}\mathcal{I}_{p,M,n,-r,-t}\frac{M!}{(n-t)!}\,\,\,\,\,{\rm {for}\,r>t,}
\end{array}\right.\end{array}\label{eq:VFBS-moment}
\end{equation}
and 
\begin{equation}
\begin{array}{lcl}
\langle\hat{a}^{\dagger t}\hat{a}^{r}\rangle_{{\rm PABS}} & = & N_{{\rm PABS}}^{2}\sum\limits _{n=0}^{M}\mathcal{I}_{p,M,n,r,t}\frac{M!(n+1)!(n+1-r+t)!}{n!(n+1-r)!(n-r+t)!},\end{array}\label{eq:PABS-moment}
\end{equation}
respectively, with $\mathcal{I}_{p,M,n,r,t}=\left[\frac{p^{2n-r+t}\left(1-p\right)^{2M-2n+r-t}}{(M-n)!(M-n+r-t)!}\right]^{1/2}$.
Here, the obtained average values of moments are also dependent on
BS parameters, which will be used to enhance/control the nonclassicality
features in the generated states.

\subsubsection{Expectation values for Kerr state and the corresponding engineered
states}

For Kerr state, Vacuum filtered kerr state and Photon added kerr state,
we use the same approach to obtain a compact generalized forms of
$\langle\hat{a}^{\dagger q}\hat{a}^{s}\rangle_{{i}}$; and our computation
yielded 
\begin{equation}
\begin{array}{lcl}
\langle\hat{a}^{\dagger q}\hat{a}^{s}\rangle_{{\rm KS}} & = & \exp\left[-\mid\alpha\mid^{2}\right]\sum\limits _{n=0}^{\infty}\frac{\alpha^{n}(\alpha^{\star})^{n-s+q}}{\left(n-s\right)!}\mathcal{F}_{n,s,q},\end{array}\label{eq:kS-moment}
\end{equation}
\begin{equation}
\begin{array}{l}
\begin{array}{l}
\langle\hat{a}^{\dagger q}\hat{a}^{s}\rangle_{{\rm VFKS}}=\left\{ \begin{array}{l}
N_{{\rm VFKS}}^{2}\sum\limits _{n=1}^{\infty}\frac{\alpha^{n}(\alpha^{\star})^{n-s+q}}{\left(n-s\right)!}\mathcal{F}_{n,s,q},\,{\rm {for}\,\,s\leq q,}\\
N_{{\rm VFKS}}^{2}\sum\limits _{n=1}^{\infty}\frac{\alpha^{\star n}\alpha^{n+s-q}}{\left(n-q\right)!}\mathcal{F}_{n,-s,-q}^{\star},\,{\rm {for}\,\,s>q,}
\end{array}\right.\end{array}\end{array}\label{eq:VFkS-moment}
\end{equation}
and 
\begin{equation}
\begin{array}{lcl}
\langle\hat{a}^{\dagger q}\hat{a}^{s}\rangle_{{\rm PAKS}} & = & N_{{\rm PAKS}}^{2}\sum\limits _{n=0}^{\infty}\frac{\alpha^{n}(\alpha^{\star})^{n-s+q}\left(n+1\right)!\left(n-s+q+1\right)!}{n!\left(n-s+q\right)!(n+1-s)!}\mathcal{F}_{n,s,q}.\end{array}\label{eq:PAKS-moment}
\end{equation}
Here, $\mathcal{F}_{n,s,q}=\exp\left[\iota\chi\left(q-s\right)\left(2n+q-s-1\right)\right]$.
From the above expressions, it is clear that when $q=s$, there is
no role of $\chi$ and the behavior of KS is similar to that of a
coherent state. So the effect of this parameter $\left(\chi\right)$
can be observed only in HOS which also depends on the higher-order
moments other than moments of number operator, i.e., $\langle\hat{a}^{\dagger q}\hat{a}^{s}\rangle_{{i}}\,:q\neq s$.
In what follows, we use the expressions of moments given in Eqs. (\ref{eq:VFECS-moment})-(\ref{eq:PAKS-moment})
to study various lower- and higher-order nonclassicality witnesses.

\section{Nonclassicality witnesses\label{sec:Nonclassicality-witnesses-1}}

There are various criteria of nonclassicality, most of them are sufficient
but not necessary in the sense that satisfaction of such a criterion
can identify a nonclassical feature, but failure does not ensure that
the state is classical. Further, most of the criteria (specially,
all the criteria studied here) do not provide any quantitative measure
of nonclassicality present in a state, and so they are referred to
as witnesses of nonclassicality.
These witnesses are based on either quasiprobability distribution
or moments of annihilation and creation operators. In the present
work, we have used a set of moment-based criteria to investigate nonclassical
properties of our desired engineered quantum states. Specifically,
we have investigated the possibilities of observing lower-order squeezing
and antibunching as well as HOA, HOSPS, and HOS for all the states
of our interest. To begin the investigation and the comparison process,
let us start with the study of antibunching.

\subsection{Lower- and higher-order antibunching}

\begin{figure}
\centering{} %
\begin{tabular}{cc}
\centering{}\includegraphics[width=60mm]{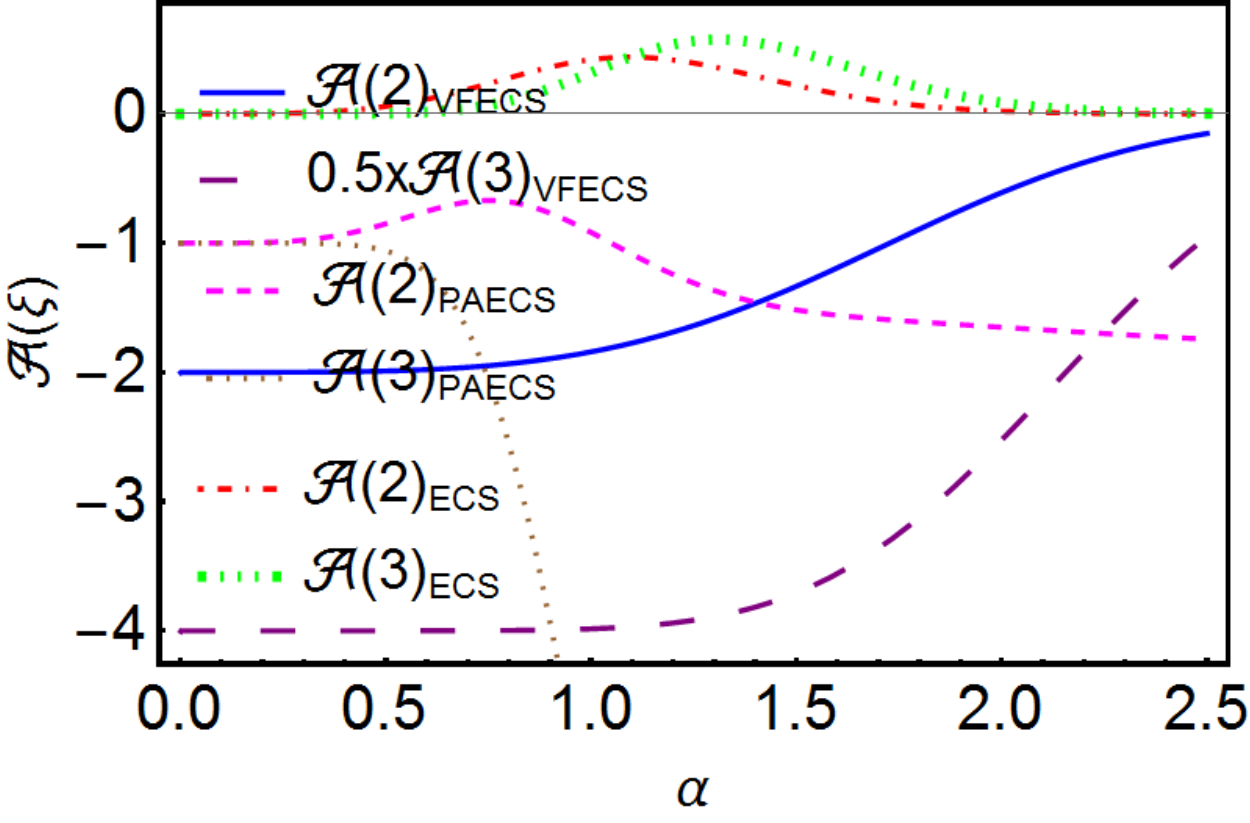}  & \includegraphics[width=60mm]{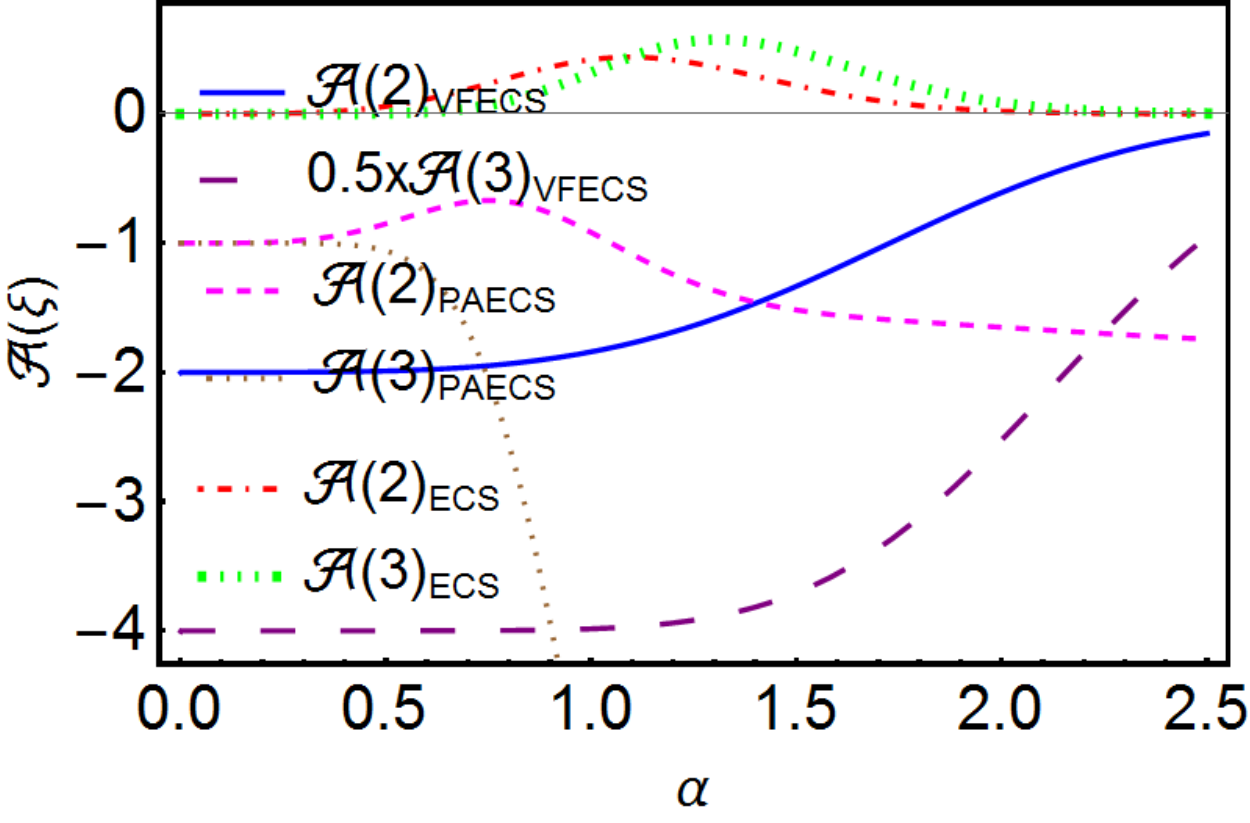}\tabularnewline
(a)  & (b) \tabularnewline
\includegraphics[width=60mm]{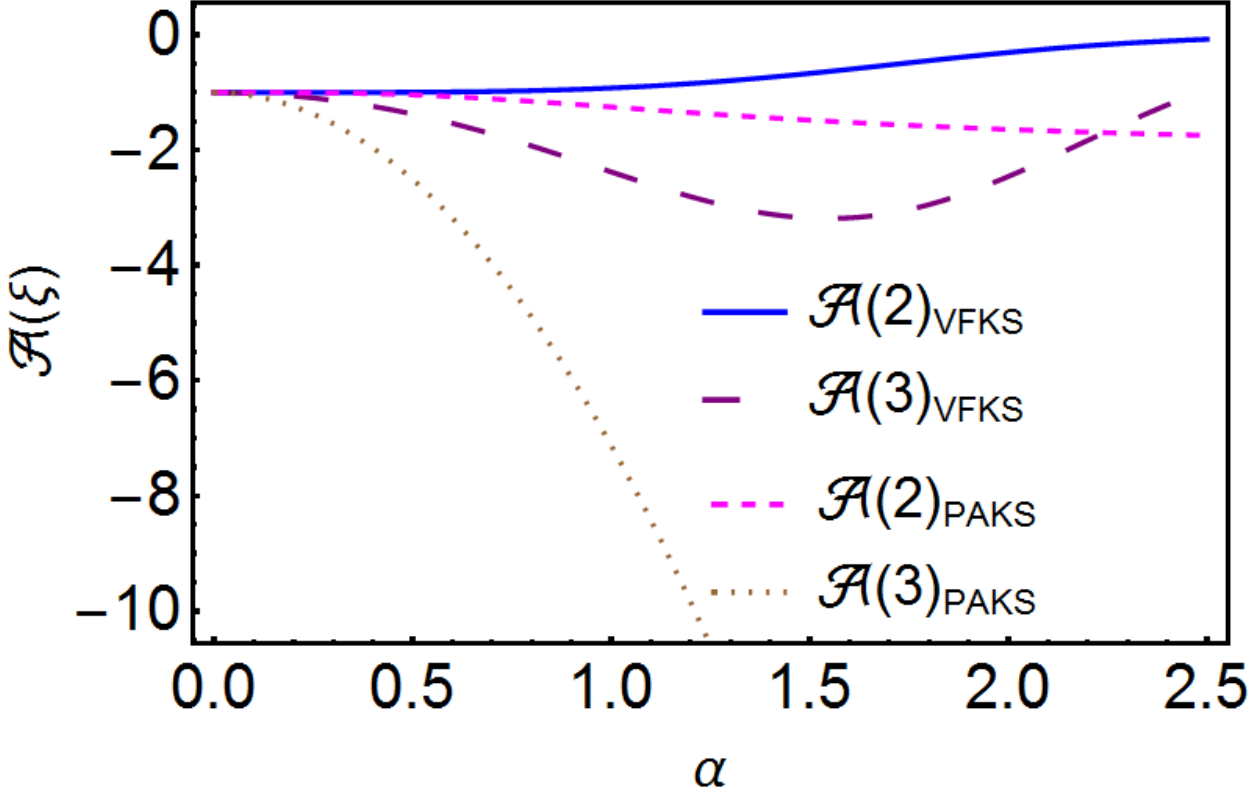} & \tabularnewline
(c)  & \tabularnewline
\end{tabular}\caption{\label{fig:HOA} Lower- and higher-order antibunching witnesses as
functions of displacement parameter $\alpha$ (for ECS and KS) and
probability $p$ (for BS with parameter $M=10$) for (a) ECS, PAECS
and VFECS, (b) BS, PABS and VFBS, and (c) KS, PAKS and VFKS. The quantities
shown in all the plots are dimensionless.}
\end{figure}

The phenomenon of lower-order antibunching is closely associated with
the lower-order sub-Poissonian photon statistics \cite{brown1956correlation}.
However, they are not equivalent \cite{zou1990photon}. The concept
of HOA also plays an important role in identifying the presence of
weaker nonclassicality \cite{allevi2012measuring,hamar2014non}. It
was first introduced in 1990 based on majorization technique \cite{lee1990higher}
followed by some of its modifications \cite{an2002multimode,pathak2006control}.
In this section, we study the generalized HOA criterion introduced
by Pathak and Garcia \cite{pathak2006control} to investigate lower-order
antibunching and HOA. The analytic expressions of moments (\ref{eq:VFECS-moment})-(\ref{eq:PAKS-moment})
can be used to investigate the nonclassicality using inequality (\ref{eq:HOA-1})
for the set of states. The obtained results are illustrated in Figure
\ref{fig:HOA} where we have compared the results between the vacuum
filtered and single photon added states. During this attempt, we also
discuss the nonclassicality present in the quantum states used for
the preparation of the engineered quantum states (cf. Figs. \ref{fig:HOA}
(a)-(c)). In Figs. \ref{fig:HOA} (b)-(c), we have shown the result
for photon added and vacuum filtered BS and KS, where it can be observed
that the depths of both lower- and higher-order witnesses in the negative
region are larger for photon added BS and KS in comparison with the
vacuum filtered BS and KS, respectively. However, an opposite nature
is observed for ECS where the depth of lower- and higher-order witnesses
is more for the vacuum filtration in comparison with the photon addition
if the values of $\alpha$ remain below certain values; whereas for
the photon addition the depth of lower- and higher-order antibunching
witnesses is found to be greater than that for vacuum filtration for
the higher values of $\alpha$ (cf. Figure \ref{fig:HOA} (a)). However,
HOA is not observed for the ECS and KS and thus both operations can
be ascribed as nonclassicality inducing operations as far as this
nonclassical feature is concerned.

\subsection{Lower- and higher-order squeezing}

The concept of squeezing originates from the uncertainty relation.
There is a minimum value of an uncertainty relation involving quadrature
operators where the variance of two non-commuting quadratures (say
position and momentum) are equal and their product satisfies minimum
uncertainty relation. Such a situation is closest to the classical
scenario, in the sense that there is no uncertainty in the classical
picture and this is the closest point that one can approach remaining
within the framework of quantum mechanics. Coherent state satisfies
this minimum uncertainty relation and is referred to as a classical
(or more precisely closest to classical state). If any of the quadrature
variances reduces below the corresponding value for a minimum uncertainty
(coherent) state (at the cost of increase in the fluctuations in the
other quadrature) then the corresponding state is called squeezed
state.

\begin{figure}
\centering{} %
\begin{tabular}{cc}
\centering{}\includegraphics[width=60mm]{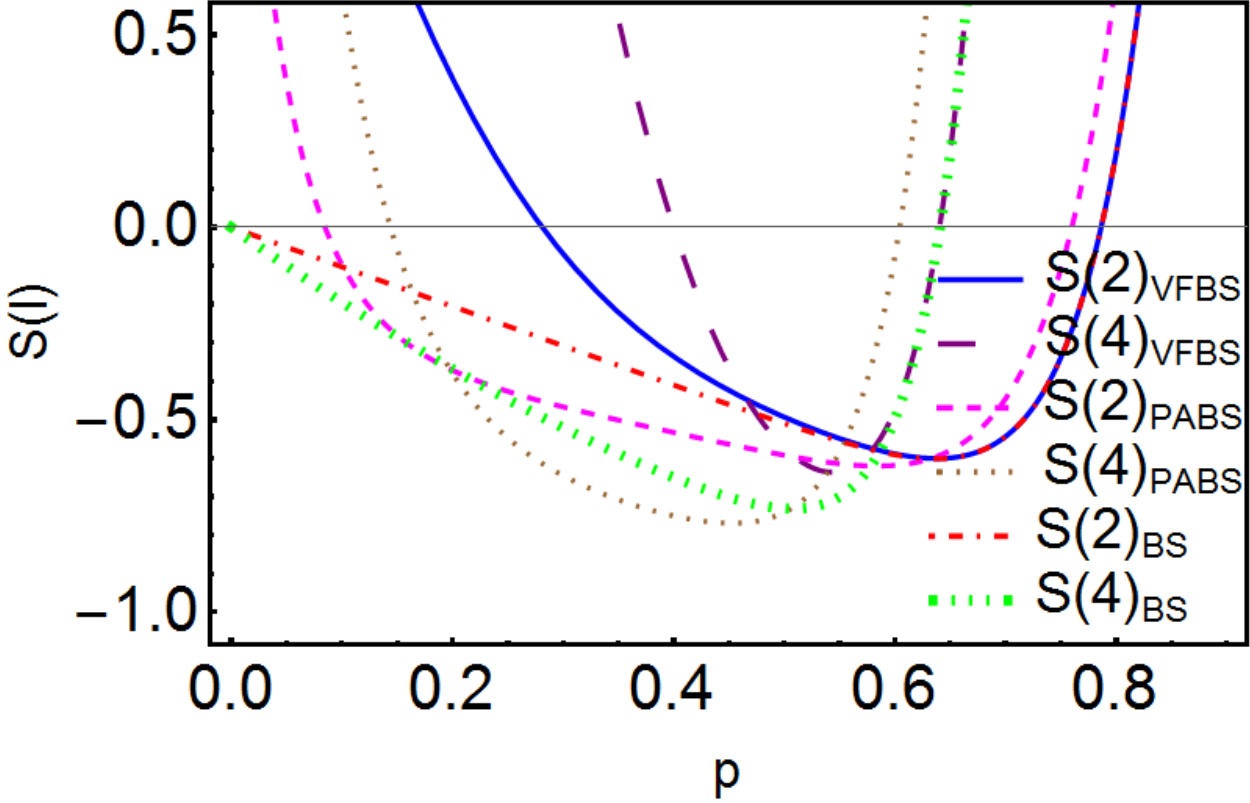}  & \includegraphics[width=60mm]{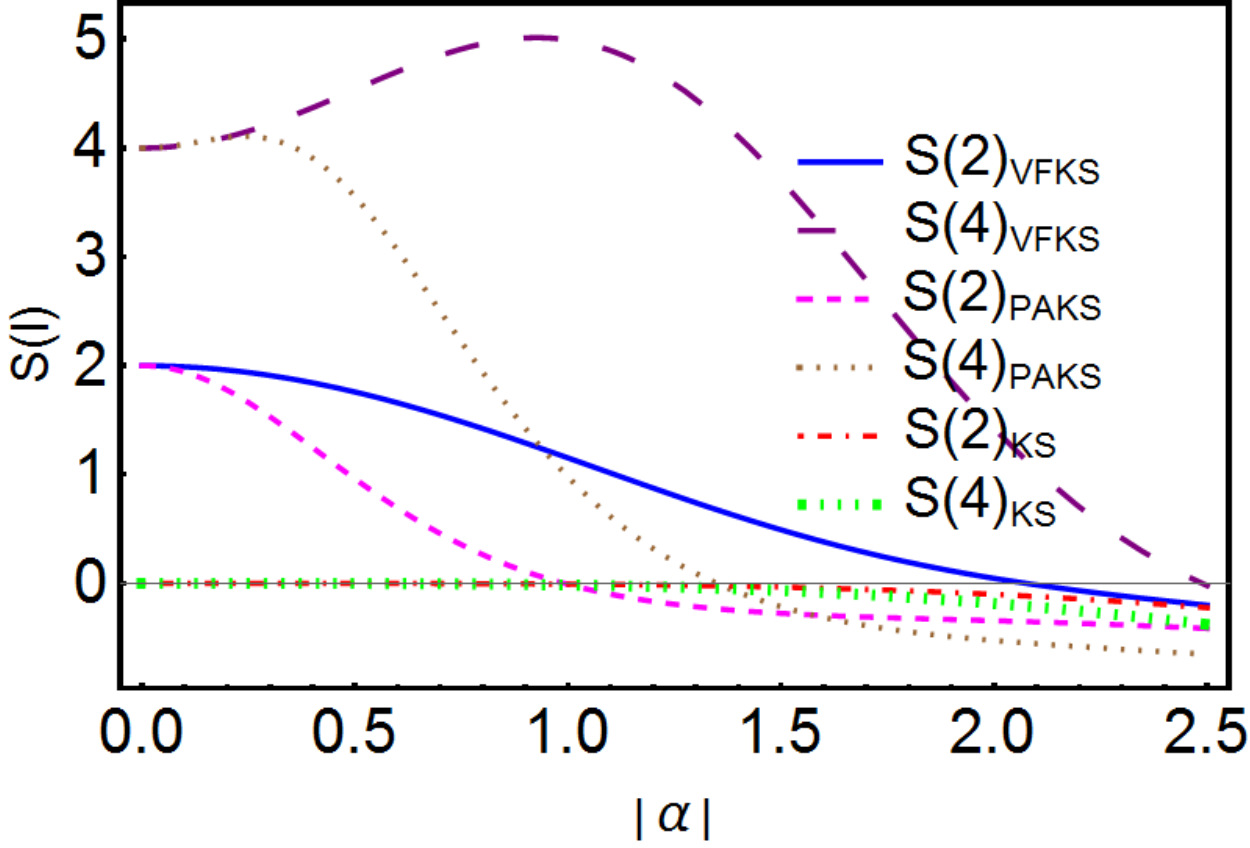}\tabularnewline
(a)  & (b) \tabularnewline
\includegraphics[width=60mm]{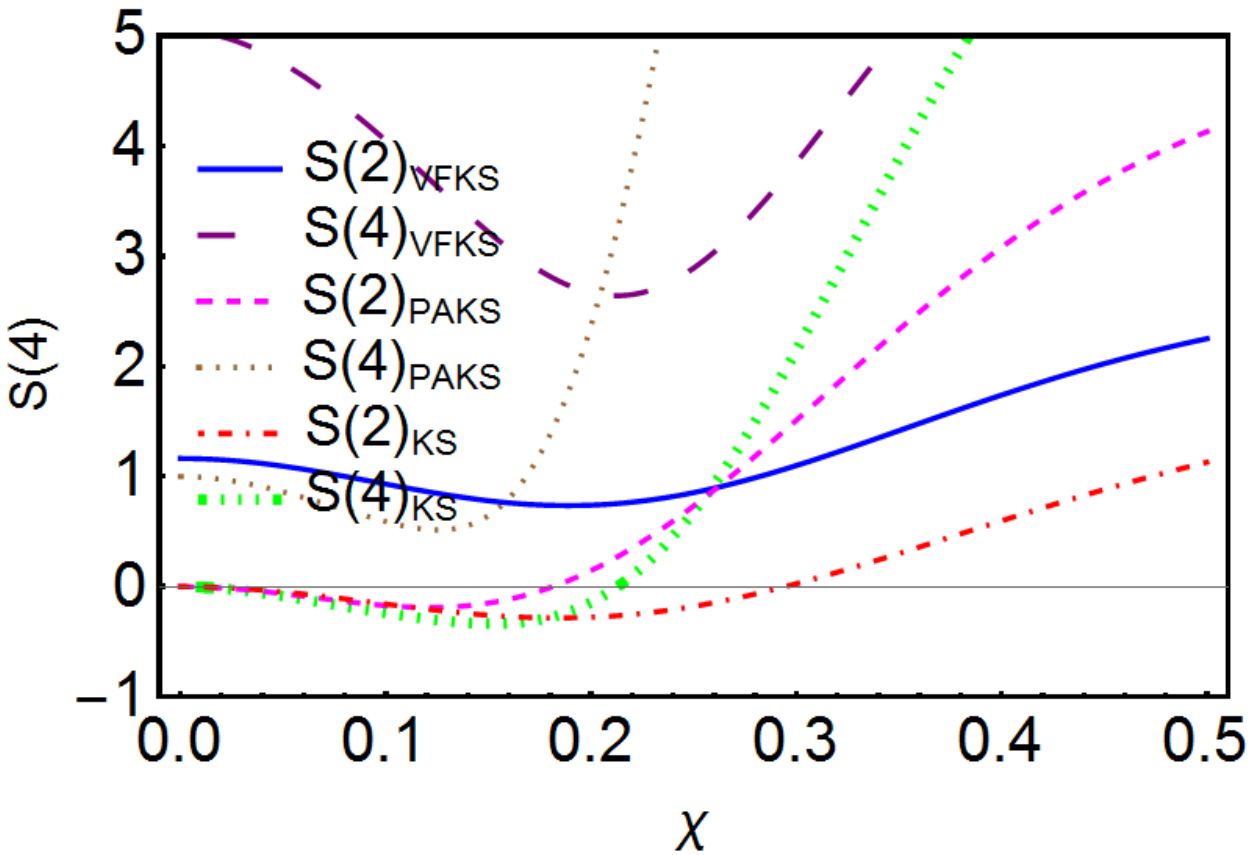} & \tabularnewline
(c)  & \tabularnewline
\end{tabular}\caption{\label{fig:HOS-1} Illustration of lower- and higher-order squeezing
for (a) BS, PABS and VFBS; (b) KS, PAKS and VFKS at the fixed value
of $\chi=0.02$; (c) KS, VFKS and PAKS as a function of $\chi$ with
$\alpha=1$. The negative regions of the curves illustrate the presence
of squeezing.}
\end{figure}

The higher-order nonclassical properties can be investigated by studying
HOS. There are two different criteria for HOS \cite{hong1985generation,hong1985higher,hillery1987amplitude}:
Hong-Mandel criterion \cite{hong1985higher} and Hillery criterion
\cite{hillery1987amplitude}. The concept of the HOS was first introduced
by Hong and Mandel using higher-order moments of the quadrature operators
\cite{hong1985higher}. According to this criterion, it is observed
if the higher-order moment for a quadrature operator for a quantum
state is observed to be less than the corresponding coherent state
value. Another type of HOS was introduced by Hillery who introduced
amplitude powered quadrature and used variance of this quadrature
to define HOS \cite{hillery1987amplitude}. Here, we aim to analyze
the possibility of HOS using Hong-Mandel criterion for $l$th order
squeezing. We have investigated the possibility of observing HOS analytically
using Eqs. (\ref{eq:VFECS-moment})-(\ref{eq:PAKS-moment}) and inequality
(\ref{eq:Hong-def2-2}) for all engineered quantum states of our interest
and have shown the corresponding results in Figs. \ref{fig:HOS-1}
(a)-(c) where we have compared the HOS in the set of quantum states
and the states obtained by photon addition and vacuum filtration.
These operations fail to induce this nonclassical feature in the engineered
states prepared from ECS, which also did not show signatures of squeezing.
In Figure \ref{fig:HOS-1} (a), we illustrate Hong-Mandel type HOS
with respect to parameter $p$ where we have shown the existence HOS
for BS, VFBS and PABS. It can be observed that the state engineering
operations fail to increase this particular feature of nonclassicality
in BS. Additionally, higher-order nonclassicality is absent for higher
values of $p$ when corresponding lower-order squeezing is present.
In case of KS, PAKS and VFKS, we have observed that HOS is observed
when the values of $\alpha$ are greater than certain values for the
individual curves of the corresponding states (cf. Figure \ref{fig:HOS-1}
(b)). Note that photon addition may provide some advantage in this
case, but vacuum filtration would not as for the same value of displacement
parameter KS and PAKS (VFKS) have (has not) shown squeezing. Interestingly,
the presence of squeezing also depends upon the Kerr nonlinearity
parameter $\chi$, which is shown in Figure \ref{fig:HOS-1} (c).
Similar to Figure \ref{fig:HOS-1} (b) photon addition shows advantage
over KS which disappears for larger values of $\chi$, while vacuum
filtering is not beneficial.

\begin{figure}
\centering \includegraphics[width=120mm]{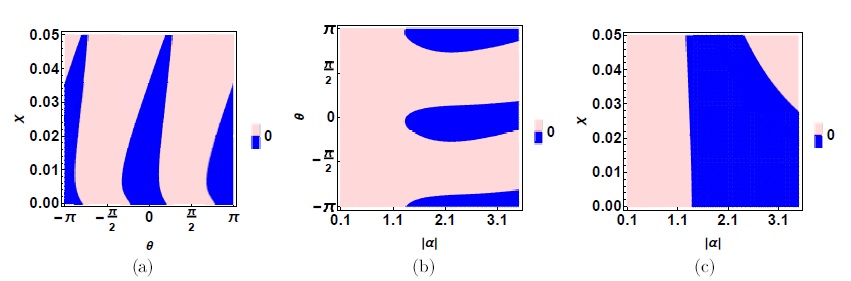}
\caption{\label{fig:HOS-contour}The dependence of HOS witness ($l=4$) on
Kerr parameter $\chi$ and displacement parameters $\left|\alpha\right|$
and $\theta$ for PAKS with (a) $\left|\alpha\right|=3$, (b) $\chi=0.02$,
(c) $\theta=0$.}
\end{figure}

In Figure \ref{fig:HOS-contour}, we have shown using the dark (blue)
color in the contour plots of the HOS witness for PAKS that squeezing
can be observed for higher values of $|\alpha|$ and smaller values
of $\chi$. Additionally, the phase parameter $\theta$ of $\alpha$
is also relevant for observing the nonclassicality as squeezing occurs
in the vicinity of $\theta=m\pi$, while disappears for $\theta=\frac{m\pi}{2}$
with integer $m$. Similar behavior is observed in KS and VFKS (not
shown here).

\subsection{Higher-order sub-Poissonian photon statistics}

\begin{figure}
\centering{} %
\begin{tabular}{cc}
\centering{}\includegraphics[width=60mm]{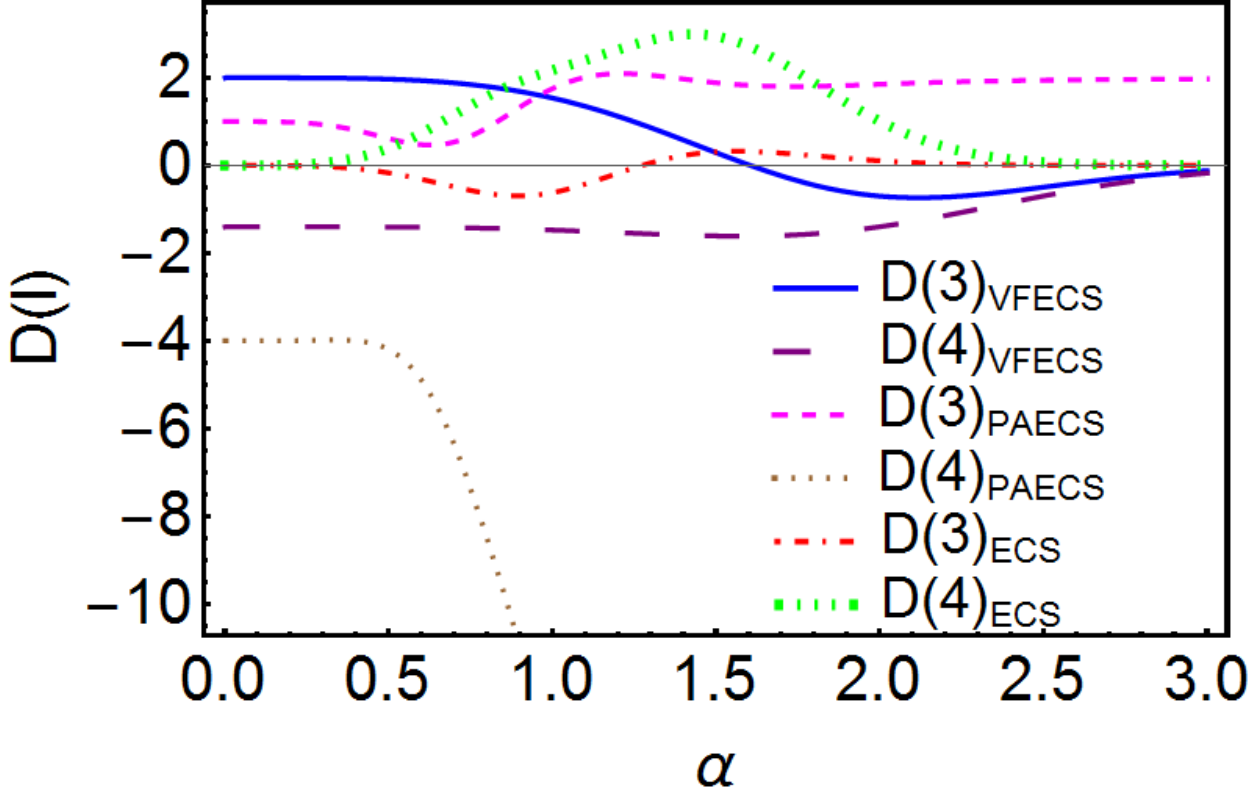}  & \includegraphics[width=60mm]{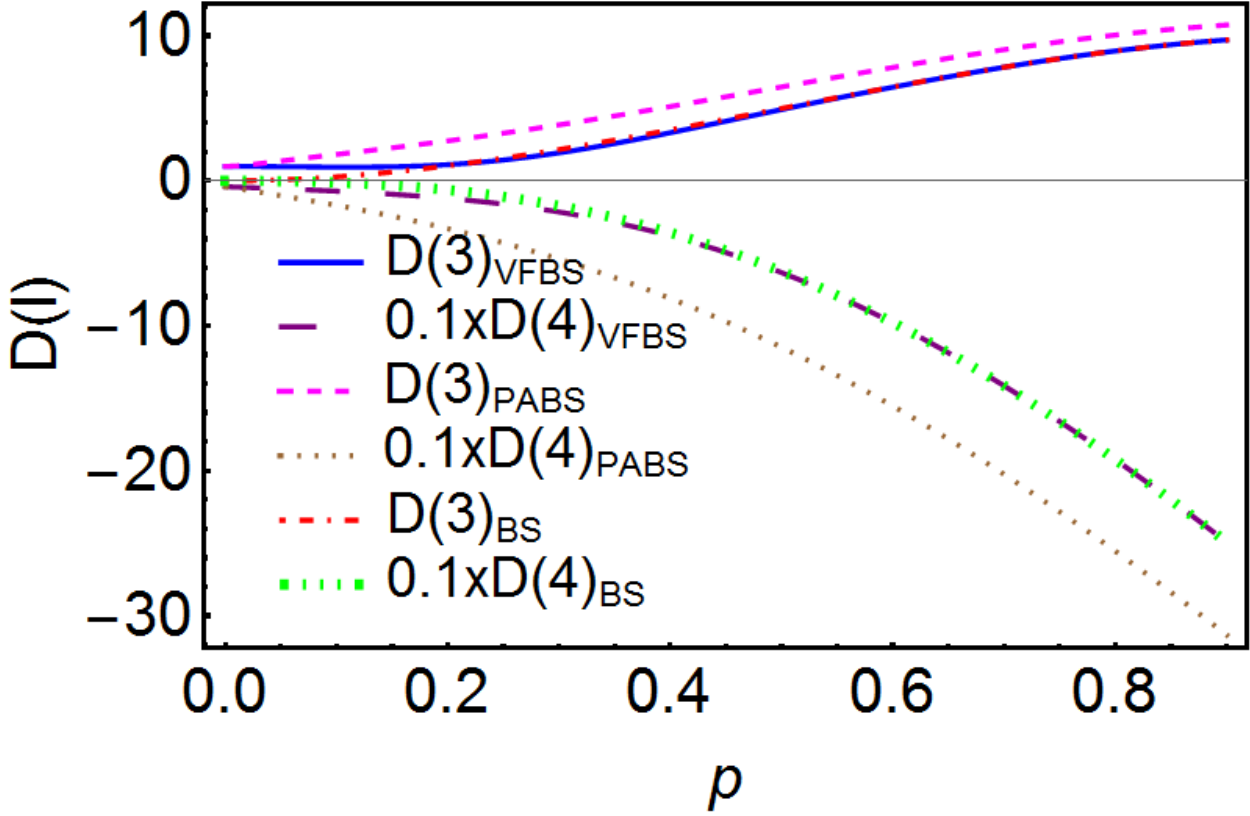}\tabularnewline
(a)  & (b) \tabularnewline
\includegraphics[width=60mm]{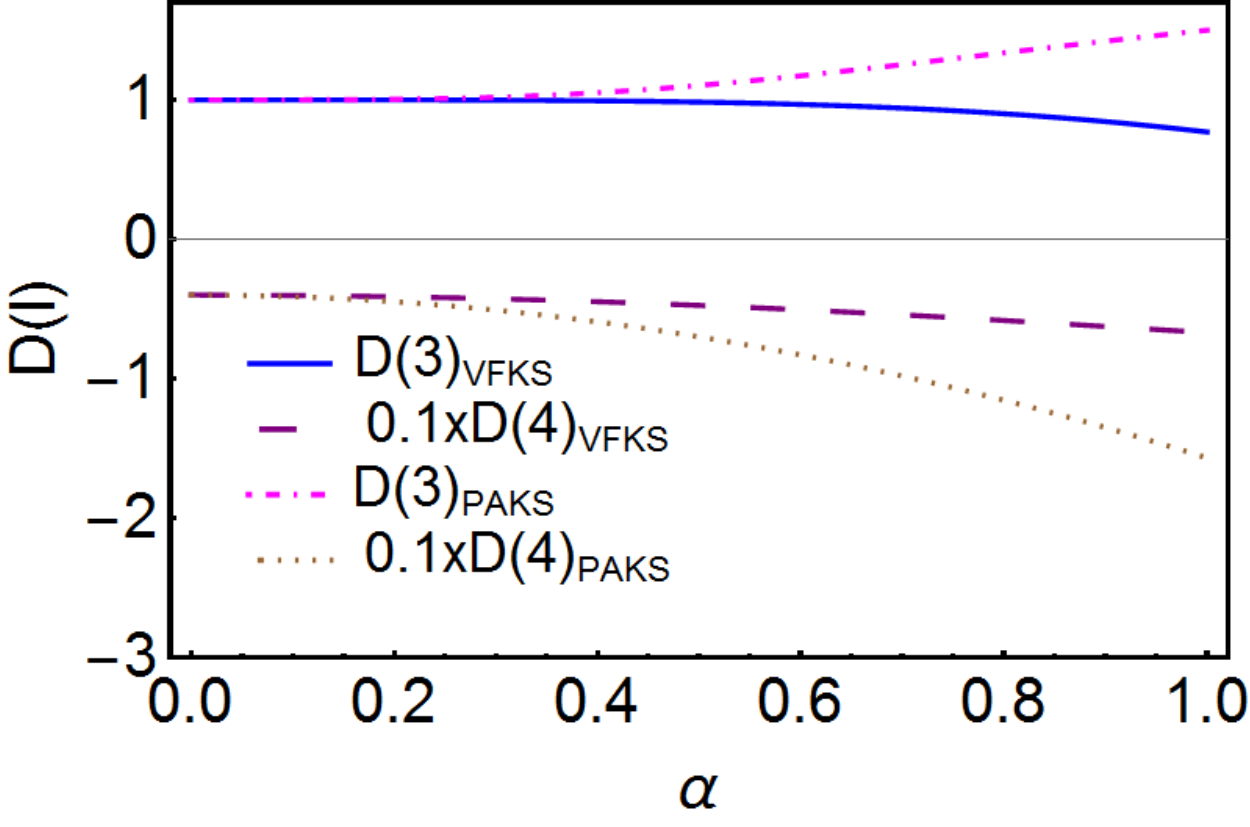} & \tabularnewline
(c)  & \tabularnewline
\end{tabular}\caption{\label{fig:HOSPS} Illustration of HOSPS as a function of displacement
parameter $\alpha$ (for ECS and KS) and probability $p$ (for BS)
for (a) ECS, (b) BS, and (c) KS and corresponding engineered states.
HOSPS is not observed in KS.}
\end{figure}

The higher-order moments in Eqs. (\ref{eq:VFECS-moment})-(\ref{eq:PAKS-moment})
are used to calculate the above inequality $(\ref{eq:hosps22-1})$
with the help of (\ref{eq:HOA-1}) for states obtained after vacuum
filtration and photon addition in ECS, BS and KS as well as the parent
states, and the corresponding results are depicted in Figure \ref{fig:HOSPS}.
Nonclassicality is not revealed by HOSPS criteria of even orders in
case of ECS, while corresponding engineered states show nonclassicality.
Additionally, nonclassicality is induced by vacuum filtration for
odd orders while it was not observed in the parent state (cf. Figure
\ref{fig:HOSPS} (a)). This clearly shows the role of hole burning
operations in inducing nonclassicality for odd orders. However, in
case of even orders, the same operations are also observed to destroy
the nonclassicality in the parent state. From Figs. \ref{fig:HOSPS}
(b) and (c), it is observed that BS and KS do not show HOSPS for the
odd values of $l$ even after application of state engineering operations.
Additionally, HOSPS is not observed for the KS for even values of
$l$, too. Consequently, the nonclassical feature witnessed through
the HOSPS criterion in PAKS can be attributed solely to the hole burning
process.

Nonclassicality in the engineered quantum states can also be studied
using quasidistribution functions \cite{thapliyal2015quasiprobability}
but here we are going to quantify the amount of nonclassicality in
these states. Further, the effect of decoherence on the observed nonclassicality
\cite{banerjee2007dynamics,banerjee2010dynamics,banerjee2010entanglement,naikoo2018probing}
and phase diffusion \cite{banerjee2007phaseQND,banerjee2007phase}
can be studied.

\begin{figure}
\centering %
\begin{tabular}{cc}
\includegraphics[width=60mm]{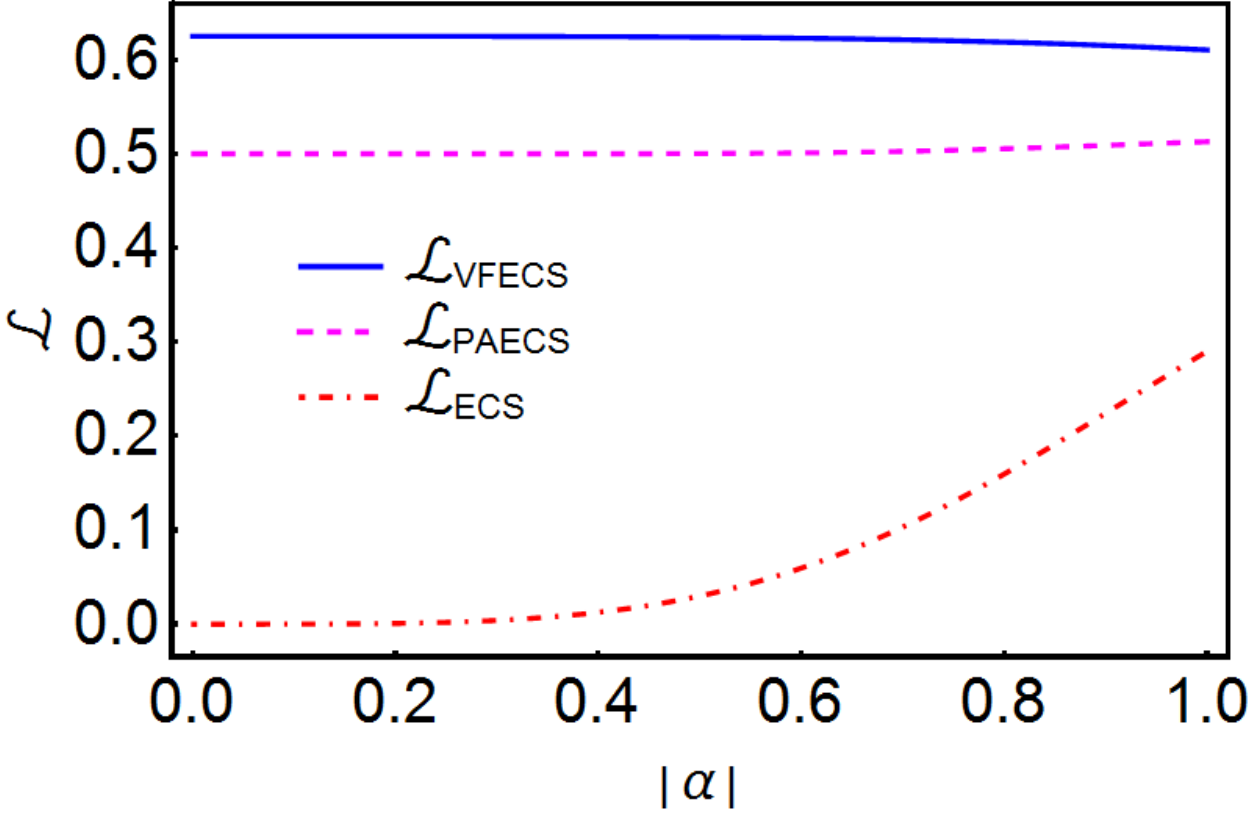}  & \includegraphics[width=60mm]{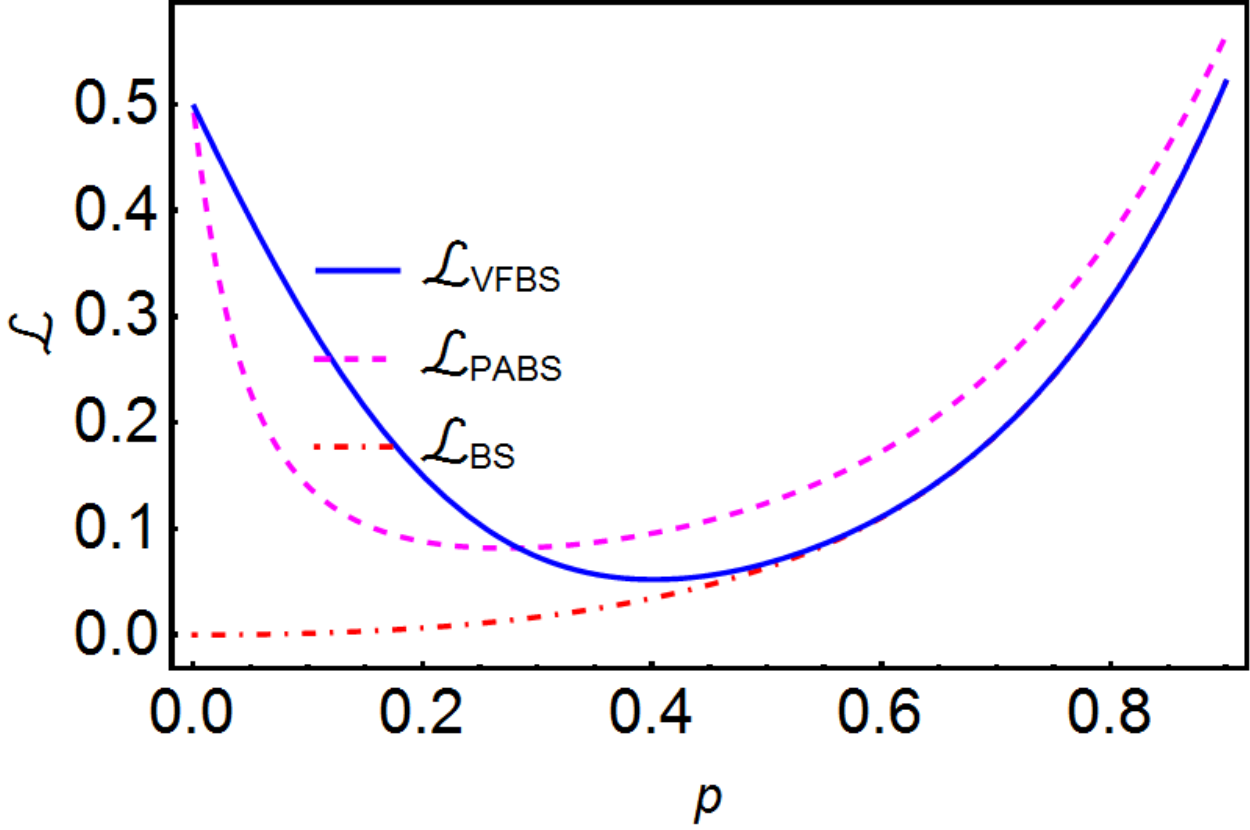}\tabularnewline
(a)  & (b) \tabularnewline
\includegraphics[width=60mm]{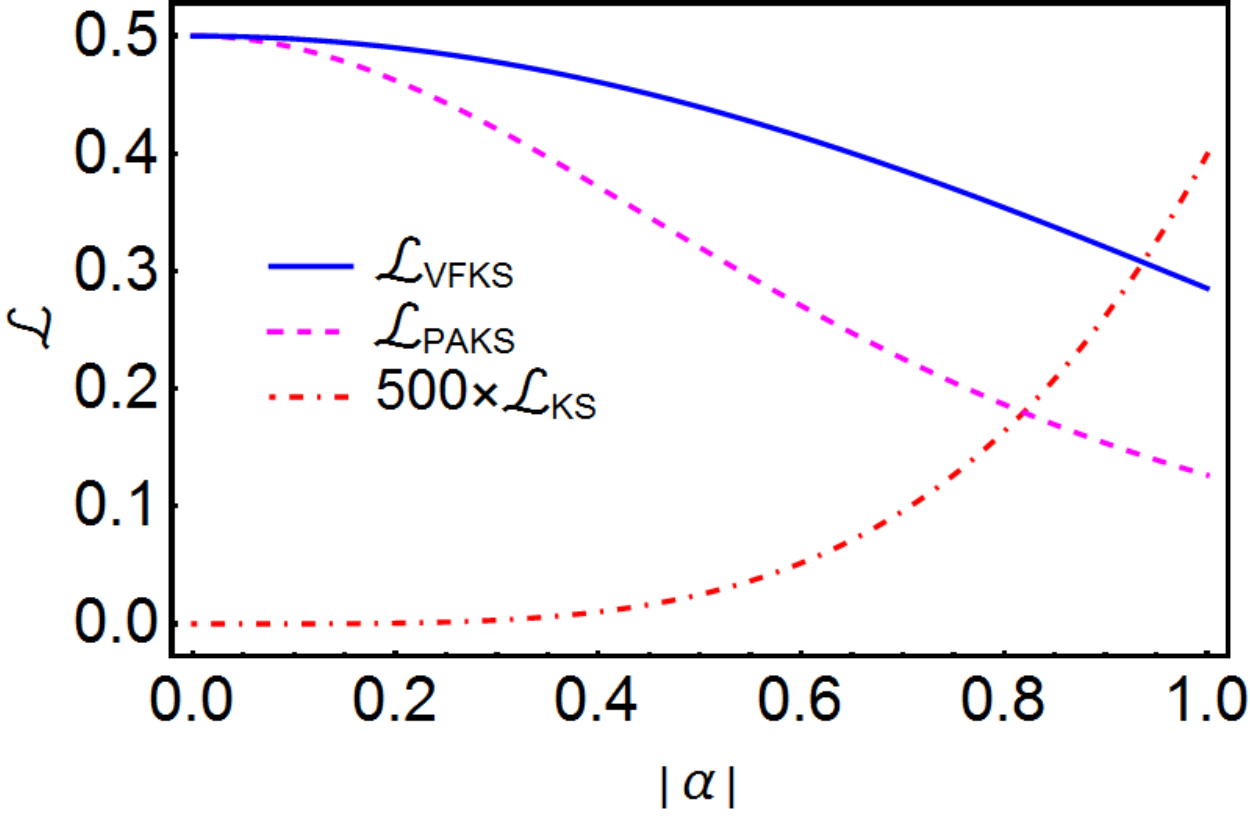}  & \includegraphics[width=60mm]{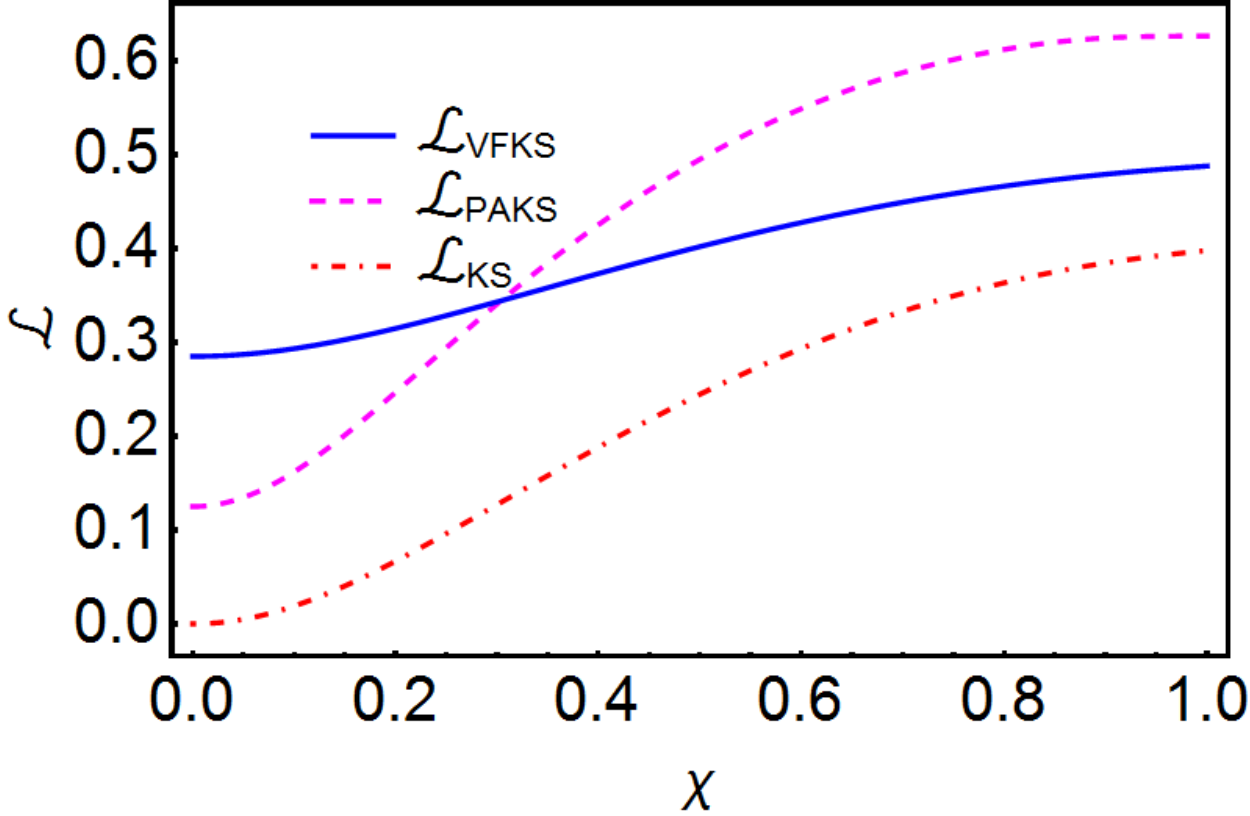}\tabularnewline
(c)  & (d) \tabularnewline
\end{tabular}\caption{\label{fig:Linear-Entropy} Illustration of linear entropy for (a)
ECS, PAECS and VFECS, (b) BS, PABS and VFBS, (c) KS, PAKS and VFKS
with $\alpha$ or $p$ for $\chi=0.02$. (d) Dependence of nonclassicality
in KS, PAKS and VFKS on $\chi$ for $\alpha=1$.}
\end{figure}

\section{Nonclassicality measure}

We have obtained the analytic expressions of linear entropy for ECS,
KS, BS and corresponding engineered states which are given as 
\begin{equation}
\begin{array}{lcl}
\mathcal{L}_{{\rm ECS}} & = & 1-\frac{\exp\left[-2\mid\alpha\mid^{2}\right]}{4\left(1+\exp\left[-2\mid\alpha\mid^{2}\right]\right)^{2}}\sum\limits _{n,m,r=0}^{\infty}f_{n,m,r}\sum\limits _{k_{1}=0}^{n}\frac{^{n}C_{k_{1}}\,{}^{r}C_{r+k_{1}-m}}{2^{n+r}},\end{array}\label{eq:LE-ECS}
\end{equation}
for VFECS 
\begin{equation}
\begin{array}{lcl}
\mathcal{L}_{{\rm VFECS}} & = & 1-\left(N_{{\rm VFECS}}\right)^{4}\sum\limits _{n,m,r=1}^{\infty}f_{n,m,r}\sum\limits _{k_{1}=0}^{n}\frac{^{n}C_{k_{1}}\,{}^{r}C_{r+k_{1}-m}}{2^{n+r}},\end{array}\label{eq:LE-VFECS}
\end{equation}
and for PAECS 
\begin{equation}
\begin{array}{lcl}
\mathcal{L}_{{\rm PAECS}} & = & 1-\left(N_{{\rm PAECS}}\right)^{4}\sum\limits _{n,m,r=0}^{\infty}f_{n,m,r}\left(m+1\right)\left(n-m+r+1\right)\\
 & \times & \sum\limits _{k_{1}=0}^{n+1}\frac{^{n+1}C_{k_{1}}\,{}^{r+1}C_{r+k_{1}-m}}{2^{n+r+2}},
\end{array}\label{eq:LE-PAECS}
\end{equation}
where $f_{n,m,r}=\frac{\mid\alpha\mid^{2n+2r}\left(1+\left(-1\right)^{n}\right)\left(1+\left(-1\right)^{m}\right)\left(1+\left(-1\right)^{r}\right)\left(1+\left(-1\right)^{n+r-m}\right)}{n!r!}$.

Similarly, analytical expression for linear entropy of BS 
\begin{equation}
\begin{array}{lcl}
\mathcal{L}_{{\rm BS}} & = & 1-\sum\limits _{n,m,r=0}^{M}g_{p,M,n,m,r}\sum\limits _{k_{1}=0}^{n}\frac{^{n}C_{k_{1}}\,{}^{r}C_{r+k_{1}-m}}{2^{n+r}},\end{array}\label{eq:LE-BS}
\end{equation}
for VFBS 
\begin{equation}
\begin{array}{lcl}
\mathcal{L}_{{\rm VFBS}} & = & 1-\left(N_{{\rm VFBS}}\right)^{4}\sum\limits _{n,m,r=1}^{M}g_{p,M,n,m,r}\sum\limits _{k_{1}=0}^{n}\frac{^{n}C_{k_{1}}\,{}^{r}C_{r+k_{1}-m}}{2^{n+r}},\end{array}\label{eq:LE-VFBS}
\end{equation}
and for PABS 
\begin{equation}
\begin{array}{lcl}
\mathcal{L}_{{\rm PABS}} & = & 1-\left(N_{{\rm PABS}}\right)^{4}\sum\limits _{n,m,r=0}^{M}g_{p,M,n,m,r}\left(m+1\right)\left(n-m+r+1\right)\\
 & \times & \sum\limits _{k_{1}=0}^{n+1}\frac{^{n+1}C_{k_{1}}\,{}^{r+1}C_{r+k_{1}-m}}{2^{n+r+2}}
\end{array}\label{eq:LE-PABS}
\end{equation}
are obtained. Here, $g_{p,M,n,m,r}=\frac{1}{n!r!}\left[\frac{(M!)^{4}p^{2(n+r)}(1-p)^{4M-2n-2r}}{(M-n)!(M-m)!(M-r)!(M-n-r+m)!}\right]^{1/2}$.

Finally, analytical expression for linear entropy of KS, VFKS, PAKS
can be given as 
\begin{equation}
\begin{array}{lcl}
\mathcal{L}_{{\rm KS}} & = & 1-\exp\left[-2\mid\alpha\mid^{2}\right]\sum\limits _{n,m,r=0}^{\infty}h_{n,m,r}\sum\limits _{k_{1}=0}^{n}\frac{^{n}C_{k_{1}}\,{}^{r}C_{r+k_{1}-m}}{2^{n+r}},\end{array}\label{eq:LS-KS}
\end{equation}
\begin{equation}
\begin{array}{lcl}
\mathcal{L}_{{\rm VFKS}} & = & 1-\left(N_{{\rm VFKS}}\right)^{4}\sum\limits _{n,m,r=1}^{\infty}h_{n,m,r}\sum\limits _{k_{1}=0}^{n}\frac{^{n}C_{k_{1}}\,{}^{r}C_{r+k_{1}-m}}{2^{n+r}},\end{array}\label{eq:LE-VFKS}
\end{equation}
and 
\begin{equation}
\begin{array}{lcl}
\mathcal{L}_{{\rm PAKS}} & = & 1-\left(N_{{\rm PAKS}}\right)^{4}\sum\limits _{n,m,r=0}^{\infty}h_{n,m,r}\left(m+1\right)\left(n-m+r+1\right)\\
 & \times & \sum\limits _{k_{1}=0}^{n+1}\frac{^{n+1}C_{k_{1}}\,{}^{r+1}C_{r+k_{1}-m}}{2^{n+r+2}},
\end{array}\label{eq:LE-PAKS}
\end{equation}
respectively, with $h_{n,m,r}=\frac{\left|\alpha\right|^{2n+2r}\exp\left[2\iota\chi(m-n)(m-r)\right]}{n!r!}$.

In general, significance of hole burning operations can be clearly
established through corresponding results shown in Figure \ref{fig:Linear-Entropy}.
Specifically, one can clearly see the amount of nonclassciality (revealed
through the amount of entanglement it can generate at a beam splitter)
increases due to these operations.

\begin{figure}
\centering{}\includegraphics[width=120mm]{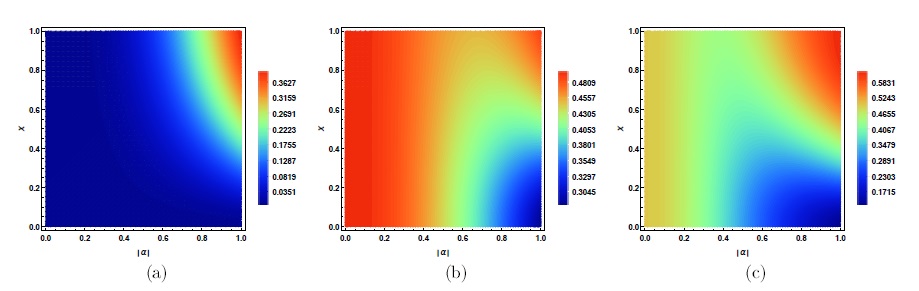}
\caption{\label{fig:LE} Illustration of linear entropy for (a) KS (b) VFKS
(c) PAKS.}
\end{figure}

From Figs. \ref{fig:Linear-Entropy} (a) and (c), it can be observed
that vacuum filtered ECS and KS are more nonclassical than corresponding
photon added counterparts. However, in case of BS and its engineered
states, it is observed that only up to a certain value of $p$ VFBS
is more nonclassical than PABS (cf. Figure \ref{fig:Linear-Entropy}
(a)). In fact, the amount of additional nonclassicality induced due
to filtration decreases with $p$ and eventually becomes zero (i.e.,
the amount of nonclassicality of VFBS becomes equal to that of BS
as far as linear entropy is considered as a measure of nonclassicality).
It is interesting to observe the effect of Kerr coupling parameter
$\chi$ on the amount of nonclassicality induced due to nonlinearity.
It is observed that for small (relatively large) values of $\chi$
nonclassicality present in VFKS (PAKS) is more than that in PAKS (VFKS)
(cf. Figure \ref{fig:Linear-Entropy} (d)). This dependence is more
clearly visible in Figure \ref{fig:LE}, where one can observe strong
nonclassicality in PAKS and VFKS (KS) favor (favors) smaller (higher)
values of $\alpha$ and large $\chi$.

\section{Conclusion \label{sec:Conclusion}}

In summary, this chapter is focused on the comparison of the effects
of two processes (vacuum state filtration and single photon addition)
used in quantum state engineering to burn hole at vacuum as far as
the higher-order nonclassical properties of the quantum states prepared
using these two processes are concerned. Specifically, various quantum
state engineering processes for burning holes at vacuum lead to different
$\sum_{m=1}c_{m}|m\rangle$ as far as the values of $c_{m}$s are
concerned (even when the parent state is the same). To study its significance
in nonclassical properties of the engineered states, we considered
a small set of finite and infinite dimensional quantum states (namely,
ECS, BS, and KS). This provided us a set of six engineered quantum
states, namely VFECS, PAECS, VFBS, PABS, VFKS, and PAKS and three
parent states for our analysis. This set of engineered quantum states
can have a great importance in quantum information processing and
quantum optics as they are found to be highly nonclassical. Especially
when some exciting applications of their parent states are already
investigated with relevance to continuous variable quantum information
processing and/or quantum optics. The present study also addresses
the significance of these hole burning processes in inducing (enhancing)
particular nonclassical features in the large set of engineered and
parent quantum states.

The general expressions for moments of the set of states are reported
in the compact analytic form, which are used here to investigate nonclassical
features of these states using a set of criteria of higher-order nonclassicality
(e.g., criteria of HOA, HOS and HOSPS). The obtained expressions can
be further used to study other moment-based criteria of nonclassicality.
The hole burning operations are found to be extremely relevant as
the states studied here are found to be highly nonclassical when quantified
through a measure of nonclassicality (entanglement potential). 
 In brief, both the vacuum filtration and photon addition operations
can be ascribed as antibunching inducing operations in KS and ECS
while antibunching enhancing operations for BS. As far as HOS is concerned
no such advantage of these operations is visible as these operations
fail to induce squeezing in ECS and often decrease the amount of squeezing
present in the parent state. Additionally, the operations are successful
in inducing HOSPS in KS and enhances this feature in the rest of the
parent states. The relevance of higher-order nonclassicality in the
context of the present study can be understood from the fact that
these hole burning operations show an increase in the depth of HOA
witness and decrease in the amount of HOS with order. While in case
of HOSPS even orders show nonclassicality whereas odd orders fail
to detect it. Finally, the measure of nonclassicality reveals vacuum
filtration as a more powerful tool than photon addition for enhancing
nonclassicality in the parent state, but photon addition is observed
to be advantageous in some specific cases.
\chapter{Conclusions And Scope For Future Work\textsc{\label{cha:Conclusions-and-Scope}}}
This concluding chapter aims to briefly summarize the results obtained
in this thesis work, and it also aims to provide some insights into
the scope of future works. To begin with, we may note that this thesis
is a theoretical work focused on nonclassical and phase properties
of  some of the  engineered
quantum states of radiation field. Here, lower- and higher-order nonclassical
properties of PADFS, PSDFS, PASDFS, ECS, VFECS, PAECS, BS, VFBS, PABS,
KS, VFKS and PAKS have been witnessed through lower- and higher-order
antibunching, higher-order sub-Poissonian photon statistics, higher-order
squeezing, Klyshko's criterion, Vogel's criterion, Agarwal-Tara's
criterion, $Q$ function, Mandel $Q_{M}$ parameter, etc. Further,
phase properties of these states have been investigated with the help
of phase distribution function, phase dispersion, phase fluctuation,
phase uncertainty parameter and angular $Q$ function. These investigations
have revealed that the state engineering processes may help us to
introduce and manipulate the nature and amount of nonclassicality
present in a quantum state. Keeping this in mind, at the end of the
thesis two quantum state engineering processes, which can be used
to generate holes at vacuum in photon number distribution, have been
compared. This systematic and rigorous study of the nonclassical and
phase properties of the above mentioned engineered quantum states
have led to many new findings, some of them are already mentioned
in the end of the individual chapters. In what follows, we list the
major findings of the present thesis.

\section{Conclusion}

The main observations of the present thesis may be summarized as follows: 
\begin{enumerate}
\item It is observed that photon addition and subtraction are not only non-gaussianity
and nonclassicality inducing operations but they can also boost the
nonclassicality present in the DFS. 
\item The results indicate that the amount of nonclassicality in PADFS and
PSDFS can be controlled by the Fock state parameter, displacement
parameter, the number of photon addition and/or subtraction. 
\item Higher-order squeezing witness and $Q$ function are observed to be
dependent on the phase of the displacement parameter. However, only
higher-order squeezing criterion was found to be able to detect nonclassicality,
and thus established that this phase parameter can also be used to
control the amount of nonclassicality. 
\item It is observed that the depth of nonclassicality witnesses increases
with order of nonclassicality. 
\item The phase distribution and angular $Q$ functions are found to be
symmetric along the value of the phase of the displacement parameter. 
\item Photon addition/subtraction and Fock parameters are found to induce
opposite effects on phase distribution. Between photon addition and
subtraction operations, subtracting a photon modifies the phase properties
more than photon addition. Interestingly, phase
properties are associated with average photon number of the state
as well. Photon subtraction increases the average photon number as
photon addition does. However, photon addition creates a hole at vacuum
unlike photon subtraction.
\item The three phase fluctuation parameters given by Carruthers and Nieto
reveal phase properties of PADFS and PSDFS, although one of them,
$U$ parameter indicates antibunching in both PADFS and PSDFS. 
\item Phase dispersion quantifying phase fluctuation remains unity for Fock
state reflecting uniform distribution, which can be observed to decrease
with increasing displacement parameter. This may be attributed to
the number-phase complimentarity as the higher values of variance
with increasing displacement parameter lead to smaller phase fluctuation. 
\item The present investigation has revealed the advantage of the PADFS
and PSDFS in quantum phase estimation and has obtained the set of
optimized parameters in the PADFS/PSDFS. 
\item The nonclassicality and non-Gaussianity of PASDFS viewed with the
help of a quasidistribution function, namely $Q$ function is shown
in the present thesis. 
\item The present study also provides a flavor of the significance of the
hole burning processes in inducing particular nonclassical features
in the family of engineered and parent quantum states. The hole burning
operations are observed to be potentially relevant as the quantum
states studied in this work are observed to be highly nonclassical
when quantification is done through a measure of nonclassicality. 
\end{enumerate}

\section{Scope for future work}

The works reported in the present thesis give us a general idea for
the investigation of phase properties and nonclassical features present
in a family of engineered quantum states. This work can be further
extended in various ways. Some of the possible extension of the present
work are listed below with a focus on the possibilities that may be
realized in the near future. 
\begin{enumerate}
\item The work may be continued to find out the non-Gaussianity of the studied
states. Subsequently, the nonclassicality and non-Gaussianity observed
in these states can be used to realize various applications in quantum
information processing tasks. Therefore, specific
applications of the non-classical properties of the aforesaid states
which (the applications) are otherwise impossible to achieve using
other types of states (classical/nonclassical).
\item A major part of the results presented here can be experimentally verified
using the available technology.  Along this line,
it would be interesting to perform resource comparison (e.g., total
number of beam splitters, photo detectors, nonlinear gadgets, etc.)
in generation of the aforesaid nonclassical states using quantum state
engineering methods.
\item The work can be extended to quantify the amount of nonclassicality
present in quantum states using different nonclassicality measures. 
\item The methods adopted here and the results obtained here can be helpful
in further theoretical studies on nonclassical and phase properties
of other engineered quantum states (both finite as well as infinite
dimensional).  There could be many such states using
several other quantum state engineering tools, for instance, squeezing,
photon catalysis, etc. 
\item Attempts can be made to observe the effect of noise in these states.
 Specifically, further study of the robustness of
observed nonclassical properties of PADFS, PSDFS, PASDFS, BS, VFBS,
PABS, KS, VFKS, PAKS under photon loss as well as inefficiency of
photo-detectors.
\end{enumerate}
We expect that theoretical work done in this thesis will be performed
experimentally and that will lead to some important applications.
We also hope this work will be very useful in quantum optics. With
these hopes this thesis is concluded. 

\backmatter
\pagenumbering{roman}\setcounter{page}{1}
	\addcontentsline{toc}{chapter}{References}
\renewcommand\bibname{\bf References}
\setlength{\bibsep}{0pt}

\end{document}